\documentclass[useAMS,usenatbib]{mn2e}
\usepackage{graphicx,natbib,times,amssymb}
\usepackage[T1]{fontenc} 
\usepackage{lmodern}
\usepackage{amsmath}
\usepackage{amssymb}
\usepackage{mathrsfs}
\usepackage{footnote}
\usepackage{listings}
\usepackage{anysize}
\usepackage{natbib}
\usepackage{aecompl}
\usepackage{enumitem}
\usepackage{hyperref}
\usepackage{tikz}
\usepackage{pdflscape}
\usetikzlibrary{arrows,shapes, positioning, trees, shapes}
\usepackage{caption}
\usepackage{booktabs}
\providecommand{\e}[1]{\ensuremath{\times 10^{#1}}}

\title[$z\gtrsim5$ AGN in the CDF-S]{The systematic search for $z\gtrsim5$ active galactic nuclei in the \textit{Chandra} Deep Field South}
\author[Anna K. Weigel et al.]{
 \parbox[t]{18cm}{
 Anna K. Weigel$^{1}$\thanks{E-mail: anna.weigel@phys.ethz.ch}, Kevin Schawinski$^{1}$, Ezequiel Treister$^{2}$,\\ C. Megan Urry$^{3}$, Michael Koss$^{1}$, Benny Trakhtenbrot$^{1}$ \\
 }\\
$^{1}$Institute for Astronomy, Department of Physics, ETH Zurich, Wolfgang-Pauli-Strasse 27, CH-8093 Zurich, Switzerland\\
$^{2}$Universidad de Concepci\'{o}n, Departamento de Astronom\'{i}a, Casilla 160-C, Concepci\'{o}n, Chile\\
$^{3}$Physics Department and Yale Center for Astronomy and Astrophysics, Yale University, New Haven, CT 06511, USA\\
}

\begin{document}

\def\Chandra{\textit{Chandra}}
\def\Spitzer{\textit{Spitzer}}
\def\Athena{\textit{Athena}}

\def\LOIII{$L[\mbox{O\,{\sc iii}}]$}
\def\Ledd{${L/L_{\rm Edd}}$}
\def\Mbh{$M_{\rm BH}$}
\def\Msigma{$M_{\rm BH} --- \sigma$}
\def\Ms{$M_{\rm *}$}
\def\Msun{$M_{\odot}$}
\def\Msunyr{$M_{\odot}yr^{-1}$}

\def\ergs{$~\rm erg~s^{-1}$}
\def\kms{$~\rm km~s^{-1}$}

\def\NH{$N_\mathrm{H}$}
\def\MBH{$M_\mathrm{BH}$}


\def\stackingfour{35}
\def\stackingfive{5}
\def\stackingsix{3}
\def\stackingseven{13}

\def\ccnotclass{49}
\def\ccfour{2}
\def\ccfive{2}
\def\ccsix{5}

\date{}

\pagerange{\pageref{firstpage}--\pageref{lastpage}} \pubyear{2013}

\maketitle

\label{firstpage}

\begin{abstract}
We investigate early black hole (BH) growth through the methodical search for $z\gtrsim5$ AGN in the \Chandra\ Deep Field South. 
We base our search on the \Chandra\ 4-Ms data with flux limits of $9.1~\times~10^{-18}$ (soft, $0.5 - 2~\mathrm{keV}$) and $5.5~\times~10^{-17}~\mathrm{erg}~\mathrm{s}^{-1}~\mathrm{cm}^{-2}$ (hard, $2 - 8~\mathrm{keV}$). At $z\sim5$ this corresponds to luminosities as low as $\sim10^{42}$ ($\sim10^{43}$) $\mathrm{erg}~\mathrm{s}^{-1}$ in the soft (hard) band and should allow us to detect Compton-thin AGN with \MBH\ $>10^7$\Msun\ and Eddington ratios > 0.1. 
Our field ($0.03~\mathrm{deg}^2$) contains over $600~z\sim5$ Lyman Break Galaxies. Based on lower redshift relations we would expect $\sim20$ of them to host AGN. 
After combining the \Chandra\ data with GOODS/ACS, CANDELS/WFC3 and \Spitzer/IRAC data, the sample consists of $58$ high-redshift candidates. We run a photometric redshift code, stack the GOODS/ACS data, apply colour criteria and the Lyman Break Technique and use the X-ray Hardness Ratio.
We combine our tests and using additional data find that all sources are most likely at low redshift. We also find five X-ray sources without a counterpart in the optical or infrared which might be spurious detections.
We conclude that our field does not contain any convincing $z\gtrsim5$ AGN. 
Explanations for this result include a low BH occupation fraction, a low AGN fraction, short, super-Eddington growth modes, BH growth through BH-BH mergers or in optically faint galaxies.
By searching for $z\gtrsim5$ AGN we are setting the foundation for constraining early BH growth and seed formation scenarios.   
\end{abstract}


\section{Introduction}
\label{sec:intro}
Massive black holes (BHs) with masses >$10^{6}$\Msun\ reside in most galaxies, including our own \citep{Genzel:1996aa, Magorrian:1998aa, Schodel:2003aa,Ghez:1998aa, Ghez:2000aa, Ghez:2008aa}. Luminous quasars with \MBH\ $\sim10^{9}$\Msun\ (\citealt{Barth:2003ab}: $z = 6.4$, \citealt{Willott:2003aa}: $z = 6.41$, \citealt{Trakhtenbrot:2011aa}: $z\sim4.8$) have been detected at $z\sim5-7$ (\citealt{Fan:2000aa, Fan:2001aa}: $z\sim6$, \citealt{Mortlock:2011aa}: $z = 7.085$). The BHs powering these quasars must therefore build up their mass in less than one billion years. Depending on the assumed seed formation model, almost constant Eddington accretion or even super-Eddington episodes are required to match these observations \citep{Volonteri:2005aa, Volonteri:2014aa, Alexander:2014aa}.
In our current understanding, BHs grew out of $\sim100-10^{5}$\Msun\ seeds by accreting infalling matter or merging with a second BH \citep{Rees:2007aa}.  
Two seed formation models are currently favored. One scenario predicts that the remnants of massive Population III (Pop III) stars constitute BH progenitors \citep{Madau:2001aa, Haiman:2001aa, Bromm:2002aa, Alvarez:2009aa, Johnson:2012aa}. The second model is based upon the direct gravitational collapse of massive gas clouds \citep{Loeb:1994aa, Bromm:2009aa, Volonteri:2010aa, Latif:2013ab}. Both models include uncertainties and predict markedly different BH growth histories. More exotic scenarios, including BH seed formation via stellar dynamical, rather than gas dynamical processes, have also been suggested ( see \citealt{Volonteri:2010aa, Bromm:2011aa} and references therein). These scenarios primarily focus on reproducing the high-redshift quasar population. We must however also be able to explain the existence of less massive and luminous, but more abundant BHs that we find in galaxies such as the Milky Way \citep{Treister:2011aa, Volonteri:2010aa}. A first step towards constraining seed formation models and determining if they are also valid for such 'normal' BHs, is measuring the BH luminosity function at high redshift.

\begin{figure}
\begin{centering}
\includegraphics[width=0.49\textwidth]{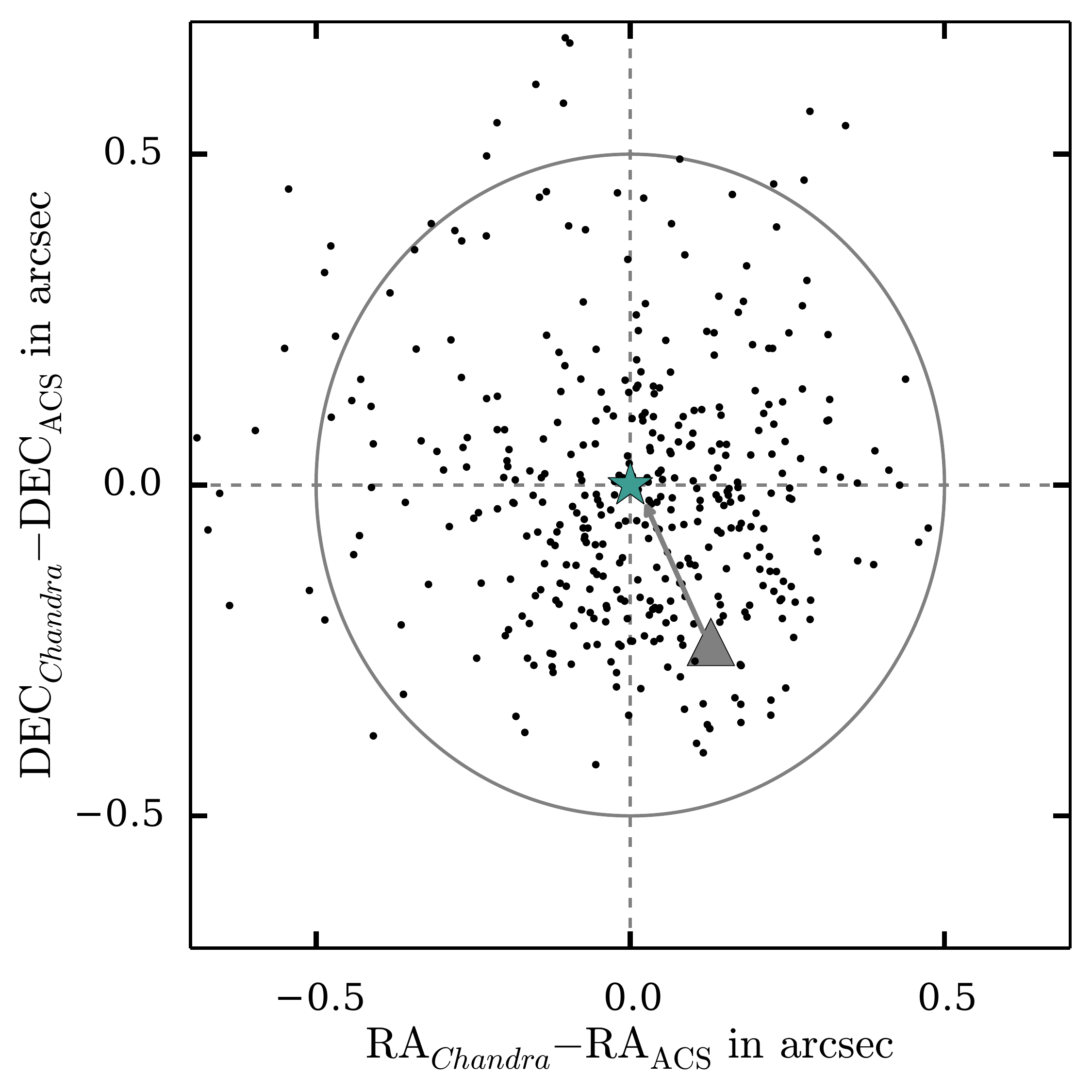}
\end{centering}
\caption{\label{fig:offset}Offset between GOODS/ACS and \Chandra\ positions after the correction has been applied. Out of the 740 \Chandra\ X-ray sources 408 possess an optical counterpart. We determine a mean offset of (0.128$^{\prime\prime}$, -0.237$^{\prime\prime}$) for these 408 objects and correct for it by shifting the \Chandra\ positions. The grey triangle indicates the mean displacement before the correction. The green star illustrates the mean offset after our correction. The black points show the corrected object positions. The grey circle illustrates an offset of 0.5$^{\prime\prime}$.}
\end{figure}

The \Chandra\ 4-Ms catalog \citep{Xue:2011aa} provides X-ray counts for the soft (0.5 keV - 2 keV), hard (2 keV - 8keV) and full (0.5 keV - 8 keV) band for the \Chandra\ Deep Field South (CDF-S). The on-axis flux limits lie at $9.1 \times 10^{-18}$, $5.5 \times 10^{-17}$ and $3.2 \times 10^{-17} \mathrm{erg}\ \mathrm{s}^{-1}\ \mathrm{cm}^{-2}$ for the soft, hard and full band, respectively. The CDF-S covers a $0.11$ $\mathrm{deg}^2$ area. For our analysis we not only use the \Chandra\ 4-Ms data, but also require coverage by the CANDELS wide and deep surveys. The effective area of our field is hence 0.03 $\mathrm{deg}^2$ \footnote{README CANDELS GOODS-S Data Release v1.0 \url{http://archive.stsci.edu/pub/hlsp/candels/goods-s/gs-tot/v1.0/hlsp_candels_hst_acs-wfc3_gs-tot_readme_v1.0.pdf}}. 

\cite{T13} used the \Chandra\ 4-Ms data in their search for high-redshift ($z > 6$) AGN. Using a sample of preselected $z = 6 - 8$ Lyman Break dropout and photometrically selected sources from the HUDF and CANDELS, \cite{T13} showed that none of these sources are detected individually in the X-rays. Stacking the X-ray observations does not produce a significant detection either. \cite{T13} suggested different processes that could account for the lack of X-ray counterparts to these high-redshift sources. 
The sample could be contaminated by a large number of low-redshift interlopers. 
A low BH occupation fraction could explain the lack of X-ray counterparts. 
It is possible that BH growth only occurs in dusty and/or small galaxies which were not included in this analysis because they lie below the detection threshold. 
The Treister et al. results can also be explained if large amounts of gas and dust obscure the X-ray emission of actively accreting BHs, 
as previously proposed by \cite{Treister:2009aa} and \cite{Fiore:2009aa}.  
Additionally, it is possible that accretion is not the dominant BH growth mode in the early universe. If BHs are primarily gaining mass by merging with other BHs, X-ray radiation might not probe BH activity. 
(See \citealt{T13} for a more detailed discussion of these possible scenarios.)
In addition to the scenarios described above, short, super-Eddington growth episodes, as proposed by \cite{Madau:2014aa} and \cite{Volonteri:2014aa}, also present a possible solution. 
In comparison to constant Eddington accretion, the amount of matter that is accreted during these super-Eddington growth phases is the same. However, \cite{Madau:2014aa} showed that, if we allow super-Eddington accretion, a duty cycle of $20\%$ is enough to grow a non-rotating $100$ \Msun\ seed BH into a $10^9$ \Msun\ object by $z\sim7$. One could imagine that the seed BH grows via five $20\ \mathrm{Myr}$ long $\dot{m}/\dot{m}_\mathrm{Edd}=4$ growth modes, each followed by a $100\ \mathrm{Myr}$ phase of quiescence. For short, super-Eddington growth episodes, we would thus expect to find fewer BHs that are actively accreting at the same time. The \cite{T13} sample could therefore not contain any BHs that are actively accreting at the time of observation. 

\label{sec:AnalysisSteps}
\begin{figure*}
\begin{centering}
\begin{center}
  \scriptsize
  \begin{tikzpicture}[auto,
  comments/.style ={rectangle, draw=none, thick, fill=white,
      text width=7.5em, text ragged, minimum height=2em, inner sep=6pt},
  block_center/.style ={rectangle, draw=black, thick, fill=white,
      text width=18em, text centered,
      minimum height=2em},
  discarded/.style ={rectangle, draw=black, thick, fill=gray!20,
      text width=7em, text ragged, minimum height=2em, inner sep=6pt, anchor = center},
  decision/.style ={rectangle, thick, text centered, draw=black,
  minimum height=.6em, text width=4.5em, anchor = center},
  block_tree/.style ={rectangle, draw=black, thick, fill=white,
      text width=8em, text centered, anchor = north,
      minimum height=2em},
  block_tree_noborder/.style ={rectangle, draw=none, thick, fill=white,
      text width=8em, text ragged, anchor = center,
      minimum height=2em},
  every child/.append style={edge from parent/.style={draw=none}},
  line/.style={draw, thick, -latex', shorten >=0pt}]

    \matrix [column sep=5mm,row sep=4mm] {     
     \node [comments](initial_sources){};
     & \node [block_center](root){740 objects in the \Chandra\ \\ 4Ms source catalog};\\
     \node[comments](coverage){};
     & \node [block_center](visual_preselection){Visual preselection: Is source covered by\\ $B$, $V$, $i$, $z$, $J$, $H$?}; 
     & \node[decision, fill=purple!20](no1){No (366)};
     & \node[discarded](discard1){Discard object};\\
     \node[comments]{};
     & \node[decision, fill=green!20](yes1){Yes (374)};\\
     \node[comments](aftercoverage){};
     & \node[block_center](Hbandimage){Images intact?};
     & \node[decision, fill=purple!20](no2){No (3)};
     & \node[discarded](discard2){Discard object};\\
     \node[comments]{};
     & \node[decision,fill=green!20](yes2){Yes (371)};\\
     \node[comments]{};
     & \node[block_center](allbands){Object clearly visible in all bands ($z<4$)?};
     & \node[decision, fill=green!20](yes3){Yes (305)};
     & \node[discarded](discard3){Discard object};\\
     \node[comments]{};
     & \node[decision, fill=purple!20](no3){No (66)};\\
     \node[comments]{};
     & \node [block_center, fill=cyan!10](mystery){Classification according to Lyman\\ Break Technique possible?};
     & \node [decision, fill=purple!20](no5){No (8)};
     & \node [block_center, text width=7em, minimum height=.6em, fill=orange!20](mystery2){'low significance objects'};\\

     \node[comments]{};
     & \node [decision, fill=green!20](yes5){Yes (58)};\\

     \node[comments]{};
     & \node [block_center, fill=orange!20](mainsample){58 objects in main sample};\\

     \node[block_tree, draw = none, fill = white, minimum height = .5em]{};\\

     \node[block_tree, minimum height = 3em](photoz){Photometric redshift code};
     & \node[block_tree, minimum height = 3em](stacking){Stacking Method};
     & \node[block_tree, minimum height = 3em](CC){Colour-Colour-\\Diagram};
     & \node[block_tree, minimum height = 3em](HR){Hardness Ratio};\\

     \node[block_tree_noborder](photoz_r){6 $z$ > 5 sources, 52 $z\leq5$ sources};
     & \node[block_tree_noborder](stacking_r){35 $z\lesssim4$,\\ 5 $z\sim5$, 3 $z\sim6$, 13 $z\gtrsim7$ sources};
     & \node[block_tree_noborder](CC_r){49 sources cannot be classified,\\ 2 z $\sim$ 4, \\2 z $\sim$ 5 and \\5 z $\sim$ 6 object};
     & \node[block_tree_noborder](HR_r){46 HR > 0\\ (z < 4.3) sources, 10 HR $\leq$ 0\\ (z $\geq$ 4.3) sources, 2 sources cannot be classified};\\

     \node[block_tree, fill=cyan!10, minimum height = 3em](photoz_v){$z_\mathrm{phot}$ (6/58)};
     & \node[block_tree, fill=cyan!10, minimum height = 3em](stacking_v){$z_\mathrm{stacking}$ (21/56)};
     & \node[block_tree, fill=cyan!10, minimum height = 3em](CC_v){$z_\mathrm{colour-colour}$ (7/9)};
     & \node[block_tree, fill=cyan!10, minimum height = 3em](HR_v){HR (10/56)};\\

     \node[block_tree, draw = none, fill = white, minimum height = .5em]{};\\

    \node[block_tree, draw = none, fill = white, anchor = north]{}; 

    & \node[block_center, fill=orange!20, anchor = center](final_crit){$z_\mathrm{colour-colour}$ > 5 and $z_\mathrm{phot}$ > 5 and HR $\leq$ 0 and $z_\mathrm{stacking}$ > 5?}
    child{node[decision, anchor = north,  fill=green!20, yshift = 2em](finalyes){Yes (3)}
    child{node[block_center, anchor = north, draw = none, fill = white, yshift = 3em](HUGS){additional data: VLT/Hawk-I HUGS $K_S$-band + HST/WFC3 HUDF $Y$-band}
    child{node[block_center, fill=orange!20, anchor = north, yshift = 2em](low_z){all 3 candidates: most likely low-redshift}}}};

    & \node[decision, anchor = center, fill=purple!20](no8){No (55)};
    
    & \node[discarded, anchor = center](discarded4){Discard object};\\
    };
          
    \begin{scope}[every path/.style=line]
       \path (root) -- (visual_preselection);
       \path (visual_preselection) -- (yes1);
       \path (yes1) -- (Hbandimage);
       \path (Hbandimage) -- (yes2);
       \path (yes2) -- (allbands);
       \path (allbands) -- (no3);
       \path (no3) -- (mystery);
       \path (mystery) --(yes5);
       \path (visual_preselection) -- (no1);
       \path (no1) -- (discard1);
       \path (Hbandimage) -- (no2);
       \path (no2) -- (discard2);
       \path (allbands) -- (yes3);
       \path (yes3) -- (discard3);
       \path (mystery) -- (no5);
       \path (no5) -- (mystery2);
       \path (yes5) -- (mainsample);

       \path (mainsample.south) -- ++ (0, 0) -- ++ (0, -.5) -| (photoz);
       \path (mainsample.south) -- (stacking);
       \path (mainsample.south) -- ++ (0, 0) -- ++ (0, -.5) -| (CC);
       \path (mainsample.south) -- ++ (0, 0) -- ++ (0, -.5) -| (HR);

       \path (photoz) -- (photoz_r);
       \path (stacking) -- (stacking_r);
       \path (CC) -- (CC_r);
       \path (HR) -- (HR_r);

       \path (photoz_r) -- (photoz_v);
       \path (stacking_r) -- (stacking_v);
       \path (CC_r) -- (CC_v);
       \path (HR_r) -- (HR_v);

       \path (photoz_v.south) -- ++ (0, 0) -- ++ (0, -.5) -| (final_crit);
       \path (stacking_v.south) -- (final_crit);
       \path (CC_v.south) -- ++ (0, 0) -- ++ (0, -.5) -| (final_crit);
       \path (HR_v.south) -- ++ (0, 0) -- ++ (0, -.5) -| (final_crit);

       \path (final_crit) -- (finalyes);
       \path (finalyes) -- (HUGS);
       \path (HUGS) -- (low_z);

       \path (final_crit) -- (no8);
       \path (no8) -- (discarded4);

    \end{scope} 
  \end{tikzpicture}
\end{center}
\caption{\label{fig:flowchart} Flowchart illustrating the analysis steps that we take to determine possible $z\gtrsim5$ candidates. The number of objects that pass each step is given in brackets. After executing this analysis for all 740 Chandra sources, three $z\gtrsim5$ candidates remain in our sample. Additional $K_S$-band and deep $Y$-band data does however show that they are most likely low-redshift sources.}
\end{centering}
\end{figure*}
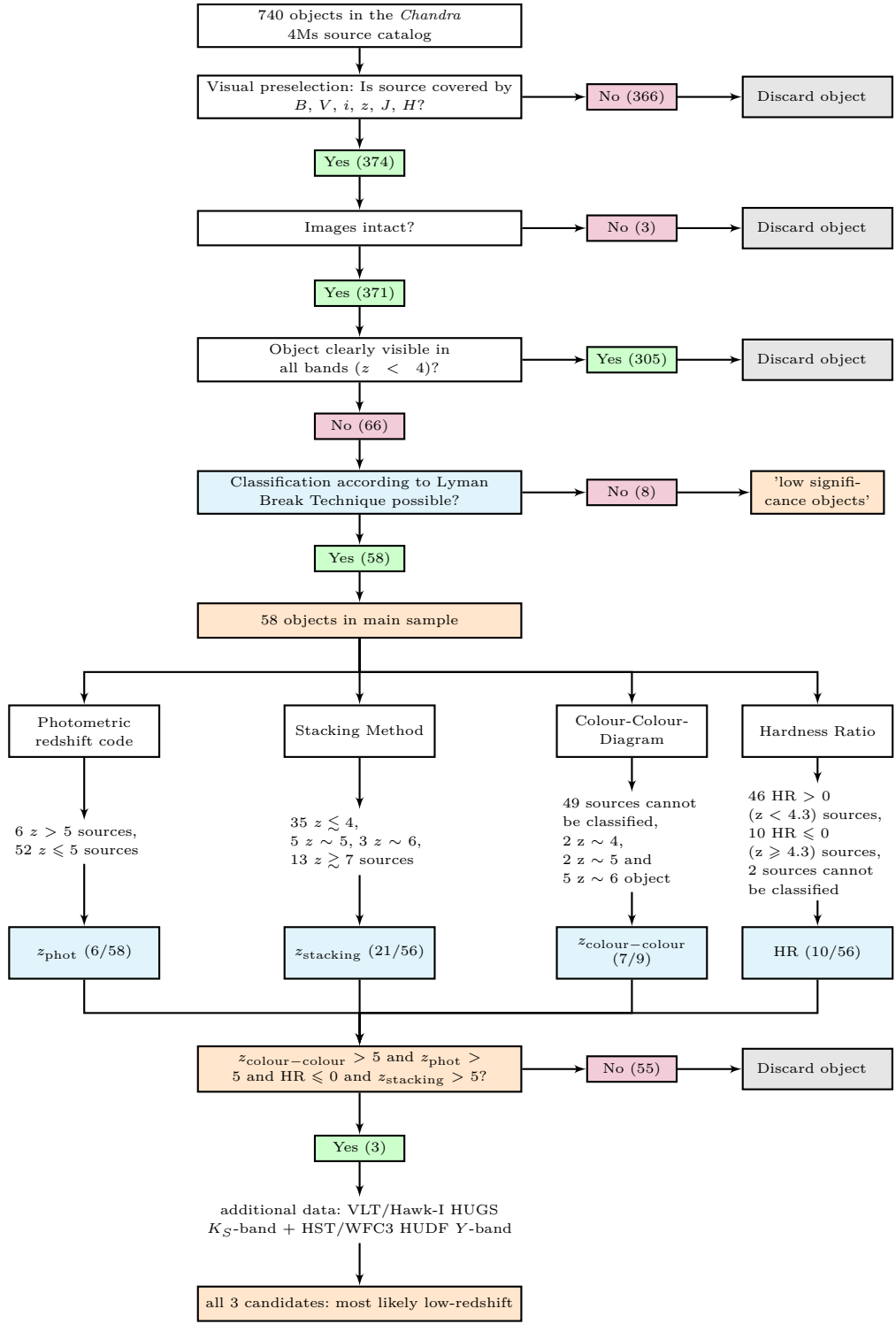

In this work we carefully examine the \Chandra\ 4-Ms catalog for possible $z\gtrsim5$ AGN. We combine the deep \Chandra\ observations with optical and infrared data from GOODS, CANDELS and \Spitzer. We use the Lyman Break Technique and a photometric redshift code to estimate the redshift of our targets. We also use colour criteria, stacking and the X-ray Hardness Ratio. In contrast to \citealt{T13}, we base this analysis on detected X-ray sources instead of trying to determine if a high-redshift object possesses a X-ray counterpart. We therefore also analyse X-ray sources that would not be classified as Lyman Break Galaxies because they are heavily obscured in the optical and the infrared. 

We know that the CDF-S contains hundreds of well constrained $z\gtrsim5$ Lyman Break Galaxies (see e.g. \citealt{Stark:2009aa, Vanzella:2009aa, Wilkins:2010aa, Bouwens:2014aa, Duncan:2014aa}). These should all pass our manual inspection, stacking, colour criteria and photometric redshift measurement. To be considered as a high-redshift AGN candidate, they must however also be detected in the X-rays and pass our X-ray Hardness Ratio test.

\cite{Volonteri:2010ab} showed that the expected number density of high-redshift AGN depends on the assumed seed formation model. Both, a detection and a non-detection, of high-redshift AGN gives us a lower limit on this number density. Our search thus constrains possible seed formation scenarios and sheds light on BH growth modes.  

\cite{Vito:2013aa} searched for $z > 3$ AGN in the 4-Ms CDF-S. They mainly analysed the evolution of obscuration and AGN space density with redshift. In contrast to this work, \cite{Vito:2013aa} based their analysis on already existing photometric and spectroscopic information on the \Chandra\ sources. We compare our results to \cite{Vito:2013aa} in Section \ref{sec:discussion}. 

This paper is organized as follows. Section \ref{sec:data} describes the data that is used in this work. Sections \ref{sec:analysis} and \ref{sec:combination} introduce our redshift tests and illustrate the results of their combination. We conclude with a discussion in Section \ref{sec:discussion} and a summary in section \ref{sec:summary}. Throughout this paper we assume a $\Lambda$CDM cosmology with $h_0$ = 0.7, $\Omega_\mathrm{m}$ = 0.3 and $\Omega_\Lambda$ = 0.7. All magnitudes are given in the AB system \citep{Oke:1983aa}.

\begin{figure}
\begin{centering}
\includegraphics[width=0.49\textwidth]{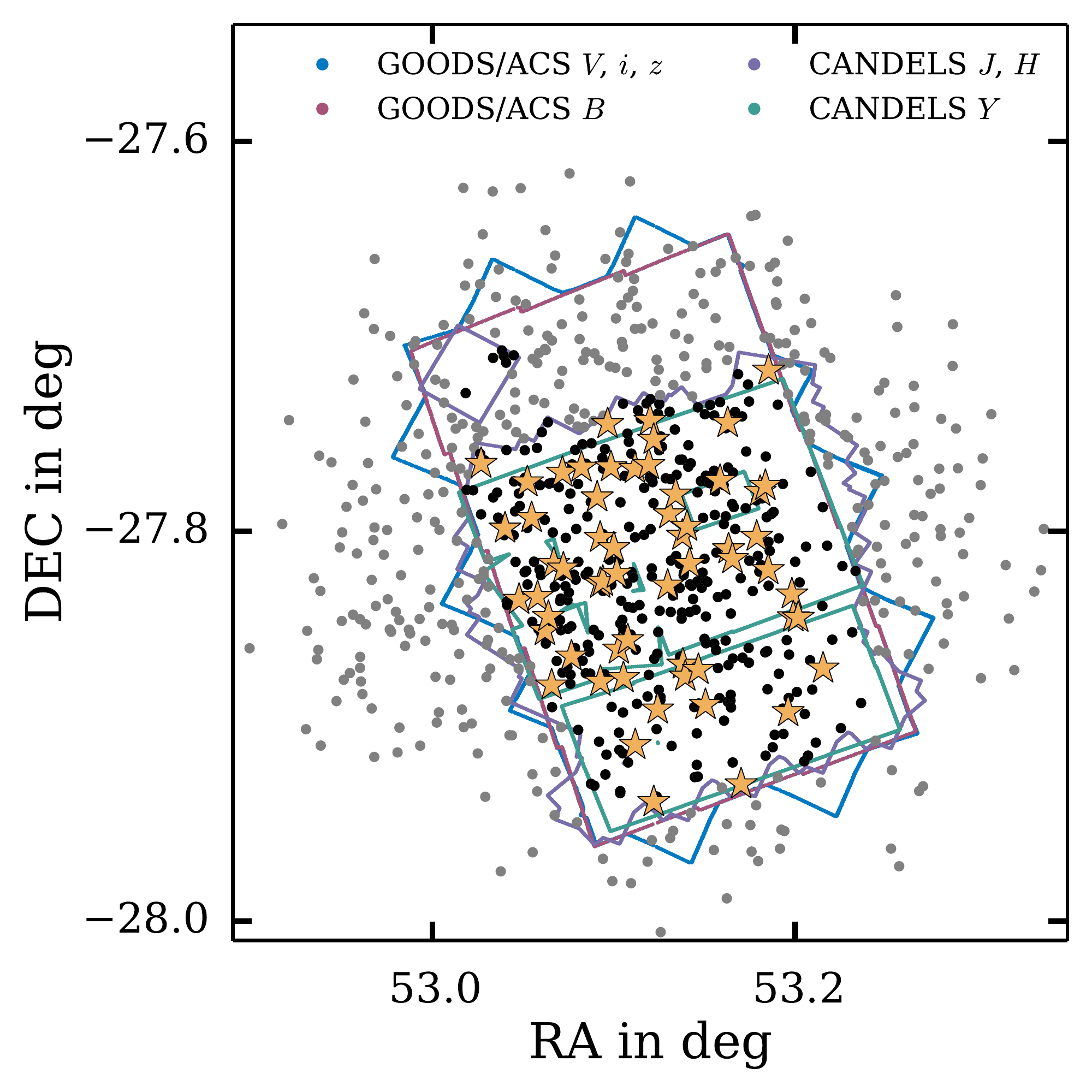}
\end{centering}
\caption{\label{fig:fields}Overview of the area covered by the HST/ACS filters $B$ (red), $V$, $i$, $z$ (blue), the HST/WFC3 bands $Y$ (green), $J$, $H$ (purple) and our 740 \Chandra\ 4-Ms sources. Grey points mark sources that are not of interest for this work because they are not covered by enough bands ($B$, $V$, $i$, $z$, $J$, $H$) to use the Lyman Break Technique. Black points indicate objects with enough filter coverage that were eliminated because the objects are clearly visible in all bands. According to the Lyman Break Technique this indicates $z < 4$. Yellow stars mark the positions of the 58 potential high-redshift AGN that we analyse more closely.}
\end{figure}

\begin{figure}
\begin{centering}
\includegraphics[width=\columnwidth]{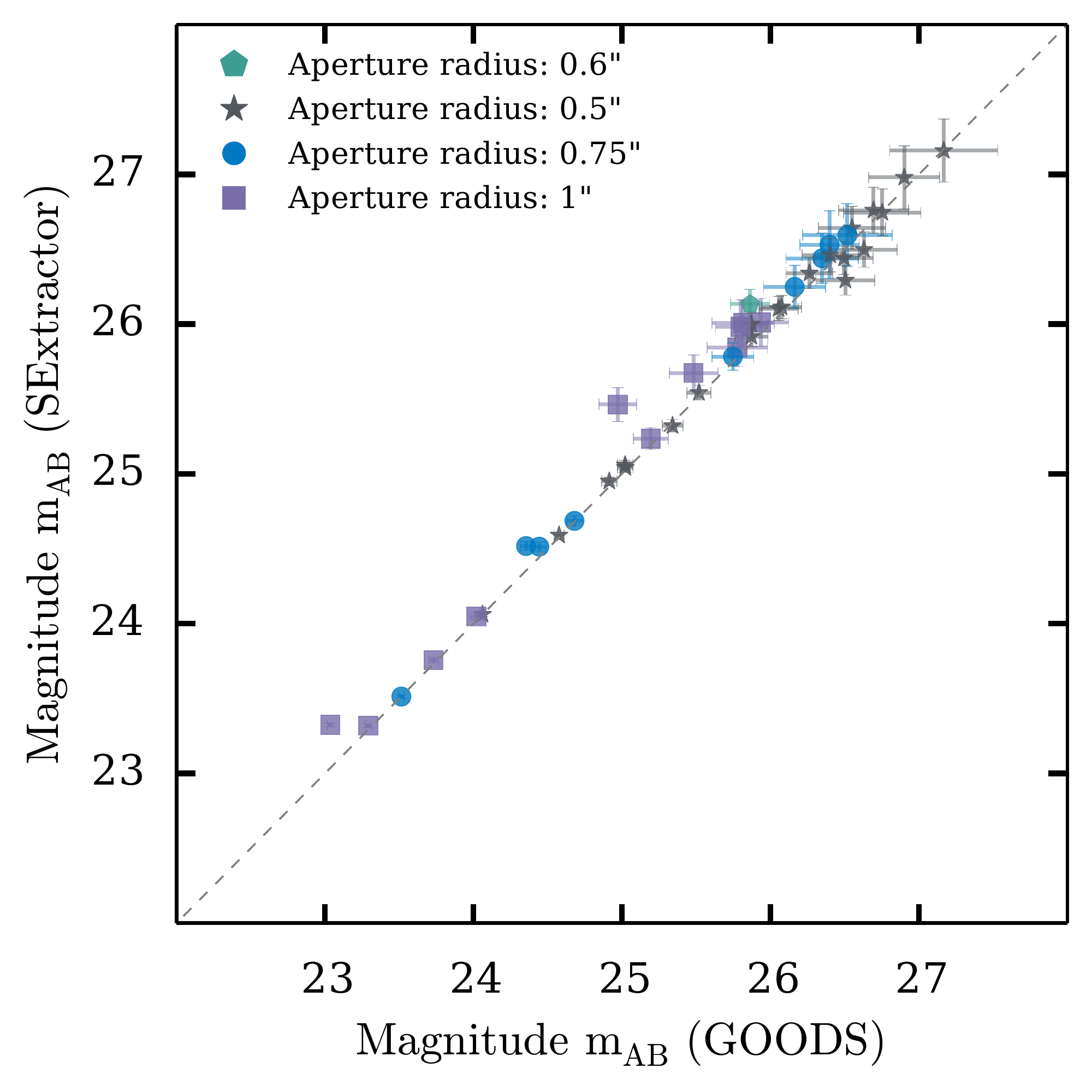}
\caption{\label{fig:comp_phot} Comparison between our own photometry and the flux values reported in the GOODS catalog  \citep{Giavalisco:2004aa}. We show the AB magnitude values for the $z$-band. Note that the GOODS/ACS catalog only contains flux values for an aperture with a 0.707$^{\prime\prime}$ and not a 0.75$^{\prime\prime}$ radius. Nonetheless, we compare these values to the brightness we measured within a 0.75$^{\prime\prime}$ radius aperture. We use a small 0.3$^{\prime\prime}$ radius aperture for two objects only. Since these two sources are not detected in the GOODS/ACS catalog, they are not part of this comparison.}
\end{centering}
\end{figure}

\section{Data}
\label{sec:data}

The \Chandra\ 4-Ms source catalog by \cite{Xue:2011aa} is the starting point of this work. It contains 740 sources and provides counts and observed frame fluxes in the soft (0.5 keV - 2 keV), hard (2 keV - 8 keV) and full (0.5 keV - 8 keV) band. All object IDs used in this work refer to the source numbers listed in the \cite{Xue:2011aa} \Chandra\ 4-Ms catalog.
We make use of HST/ACS data from the GOODS-south survey in the optical wavelength range. We use catalogs and images for filters F435W ($B$), F606W ($V$), F775W ($i$) and 850LP ($z$) from the second GOODS/ACS data release (v2.0) \citep{Giavalisco:2004aa}. 
We use CANDELS WFC3/IR data from the first data release (v1.0) for passbands F105W ($Y$), F125W ($J$) and F160W ($H$) \citep{Grogin:2011aa,Koekemoer:2011aa}. 
To determine which objects are red, dusty, low-redshift interlopers, we also include the 3.6 micron and 4.5 micron \Spitzer\ IRAC channels. We use SIMPLE  image data from the first data release (DR1, \citealt{van-Dokkum:2005aa}) and the first version of the extended SIMPLE catalog by \cite{Damen:2011aa}.  

When comparing \Chandra, GOODS/ACS and CANDELS object positions, a clear offset in the \Chandra\ coordinates is apparent. We illustrate this inconsistency in Figure \ref{fig:offset}. To correct for this discrepancy, we calculate  the mean displacement between the \Chandra\ and the GOODS/ACS catalog. We determine a mean offset of $\mathrm{RA}_{Chandra}-\mathrm{RA}_\mathrm{ACS}=0.128^{\prime\prime}$ and $\mathrm{DEC}_{Chandra}-\mathrm{DEC}_\mathrm{ACS}=-0.237^{\prime\prime}$.  We adjust the GOODS/ACS and CANDELS positions of each object by subtracting the mean displacement from the originally given catalog position. 
 
\section{Analysis}
\label{sec:analysis}
In the following section we describe in detail the set of criteria we employed to the data in order to identify $z\gtrsim5$ candidates. We first exclude objects with insufficient filter coverage, perform our own aperture photometry and determine the dropout band of each source by manual inspection. We  run a photometric redshift code, stack the GOODS/ACS data and apply colour criteria. In addition, we use the X-ray data as a photometric redshift indicator. We combine all redshift tests in section \ref{sec:combination}. Figure \ref{fig:flowchart} illustrates and summarizes the complete analysis that is detailed in the following subsections. 

\cite{Dahlen:2010aa}, \cite{McLure:2011aa} and \cite{Duncan:2014aa} showed that when selecting high-redshift objects, a selection based only on colour criteria is not as reliable as calculating photometric redshifts. Especially for faint, low signal-to-noise objects, errors and upper limits can lead to scattering out of the colour selection region. Similar to colour criteria, photometric redshifts strongly depend on the position of the Lyman Break. A photometric redshift code does however consider all filter information, including upper limits and errors. Furthermore, low-redshift interlopers can be identified by including the filters redward of the Lyman Break. \cite{Dahlen:2010aa} illustrated the discrepancy between colour criteria and photometric redshifts for the GOODS-S field. Only 50$\%$ of their photometrically selected $z\sim4$ sources were also classified as $B$-dropouts according to colour criteria. We therefore primarily use a photometric redshift code. The colour criteria, our visual classification, the stacking of the GOODS/ACS data and the X-ray Hardness Ratio provide additional redshift indications. 

\subsection{Initial Sample Selection}
Our initial sample consists  of the 740 objects given in the \Chandra\ 4-Ms source catalog. Figure \ref{fig:fields} shows that not all \Chandra\ targets are covered by the GOODS/ACS and CANDELS images. However, adjacent filter coverage is necessary for the application of the Lyman Break Technique. We therefore narrow the number of possible candidates down to 374 by removing sources that are not covered by $B$, $V$, $i$, $z$, $J$ and $H$. The $Y$-band area is small compared to the other filters. We therefore also include sources that are not covered by the $Y$-band provided that they are covered by all other GOODS/ACS and CANDELS filters. The optical and infrared counterpart detection is primarily based upon the $H$-band image since this is the deepest band. The $H$-band images for objects 105 and 521 show significant artifacts, we therefore discard them. Source 366 is eliminated due to the object's position being at the edge of the GOODS/ACS images. Hence, 371 possible candidates remain after this first visual preselection.  

\begin{figure*}
\begin{minipage}{\textwidth}
\includegraphics[width=\textwidth]{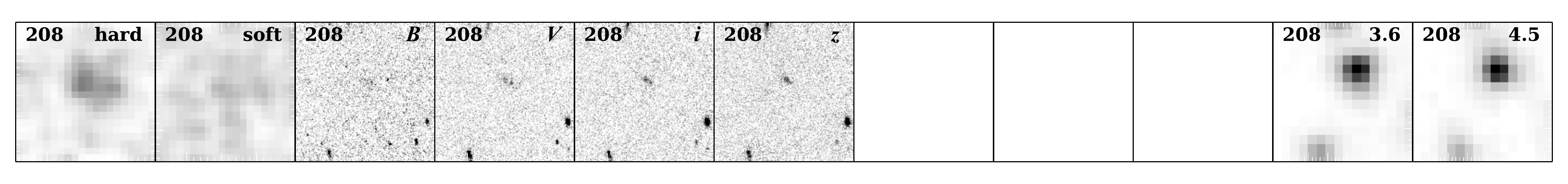}
\end{minipage}
\begin{minipage}{\textwidth}
$\rightarrow \textit{eliminated: lack of CANDELS coverage}$
\end{minipage}

\begin{minipage}{\textwidth}
\includegraphics[width=\textwidth]{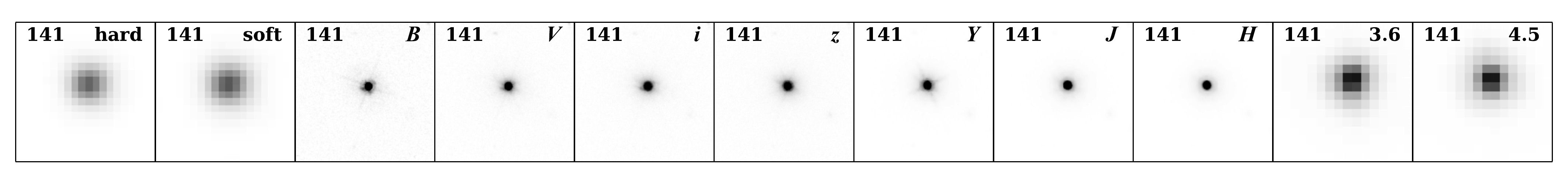}
\end{minipage}
\begin{minipage}{\textwidth}
$\rightarrow \textit{eliminated: source is clearly visible by eye in the optical and the infrared, this indicates $z < 4$ according to}$\\
$\textit{the Lyman Break Technique}$
\end{minipage}

\begin{minipage}{\textwidth}
\includegraphics[width=\textwidth]{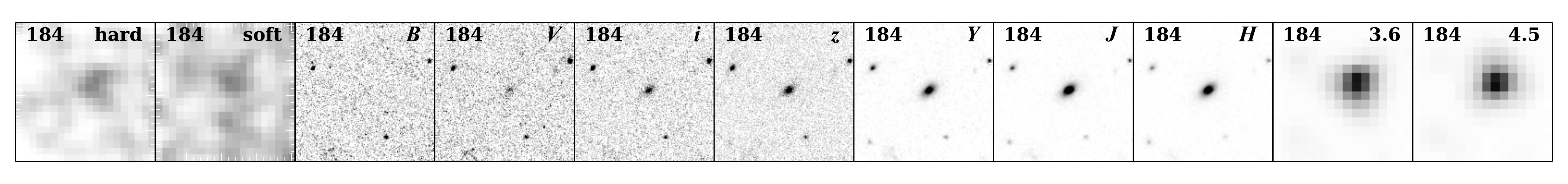}
\end{minipage}
\begin{minipage}{\textwidth}
$\rightarrow  \textit{kept in sample: by eye classified as a $B$-dropout, this indicates $z\sim4$ according to the Lyman Break Technique}$
\end{minipage}

\caption{\label{fig:visclass}Classification examples. We only kept source 184 in our sample. The images are 10$^{\prime\prime}$ x 10$^{\prime\prime}$ in size and were colour inverted. }
\end{figure*}

\begin{figure*}

\begin{minipage}{\textwidth}
\includegraphics[width=\textwidth]{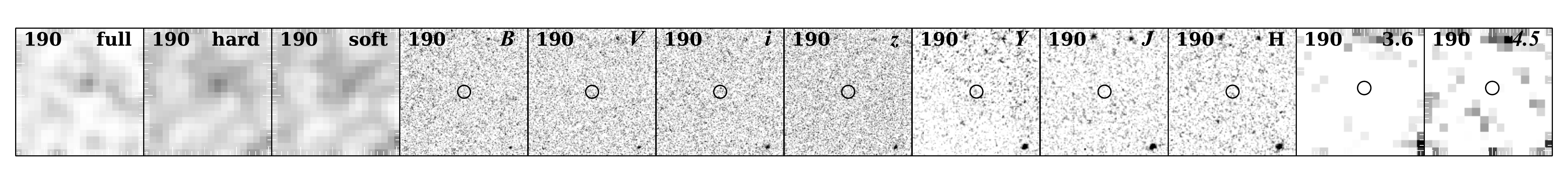}
\end{minipage}

\begin{minipage}{\textwidth}
\includegraphics[width=\textwidth]{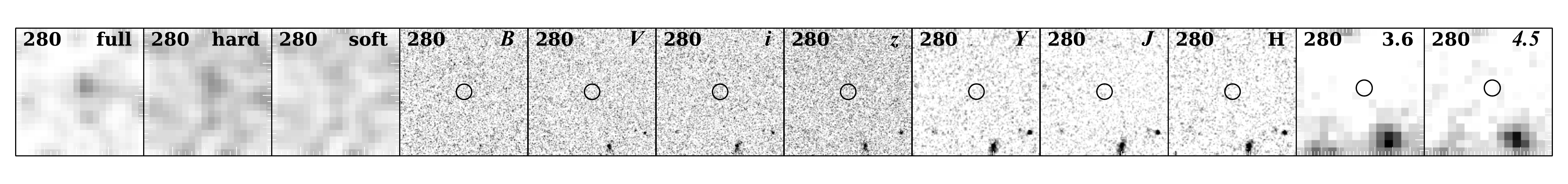}
\end{minipage}

\begin{minipage}{\textwidth}
\includegraphics[width=\textwidth]{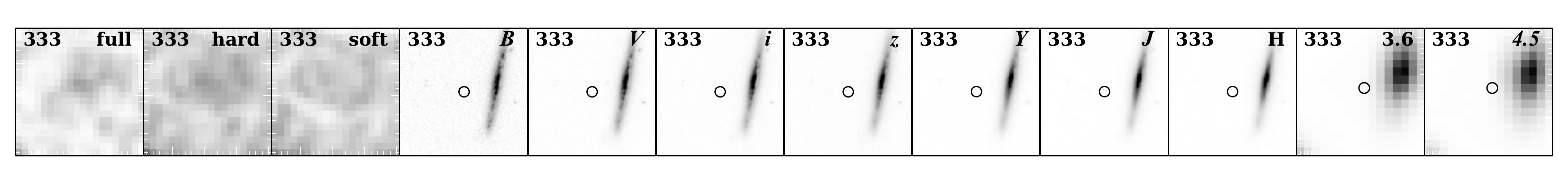}
\end{minipage}

\begin{minipage}{\textwidth}
\includegraphics[width=\textwidth]{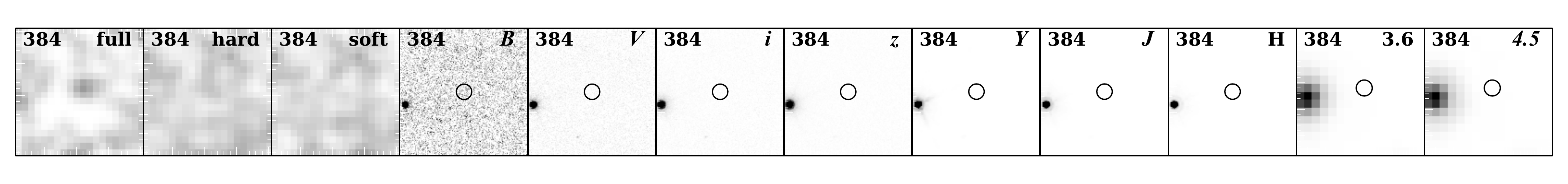}
\end{minipage}

\begin{minipage}{\textwidth}
\includegraphics[width=\textwidth]{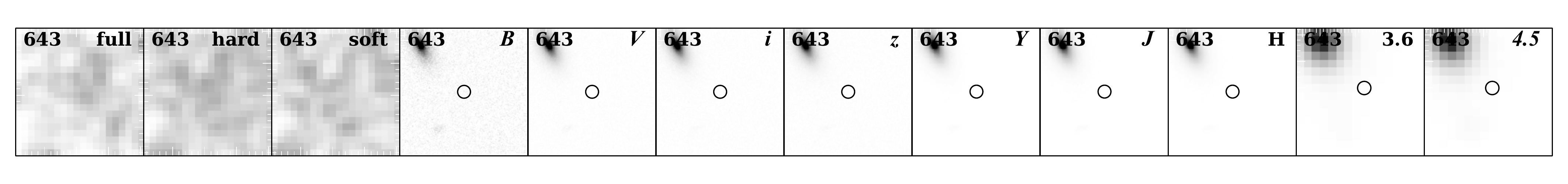}
\end{minipage}

\begin{center}
\line(1,0){450}
\end{center}

\begin{minipage}{\textwidth}
\includegraphics[width=\textwidth]{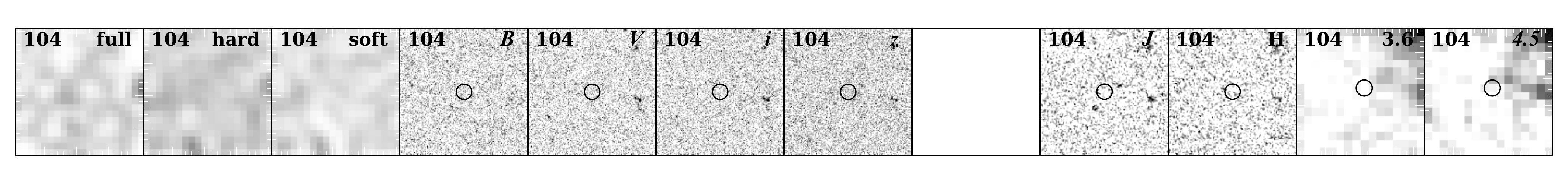}
\end{minipage}

\begin{minipage}{\textwidth}
\includegraphics[width=\textwidth]{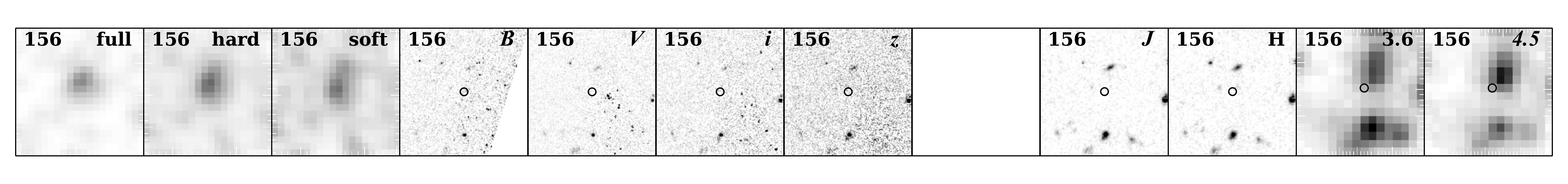}
\end{minipage}

\begin{minipage}{\textwidth}
\includegraphics[width=\textwidth]{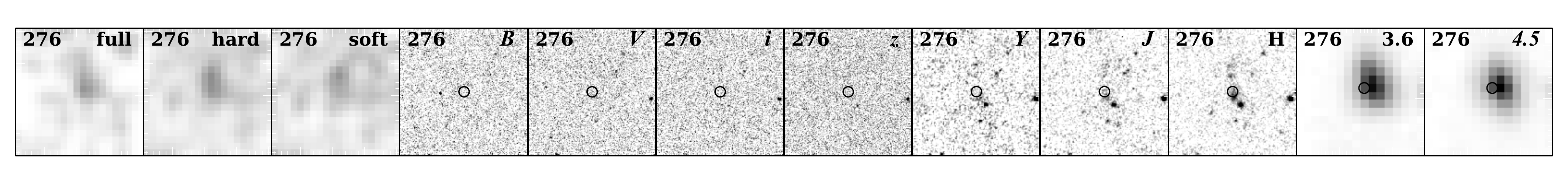}
\end{minipage}

\caption{\label{fig:mysteryobjects}Sources that can not be classified according to their dropout band. $Top$: Sources that are in the \Chandra\ 4-Ms catalog but do not show a clear counterpart in the optical and infrared, \textit{'low significance objects'}. 333, 384 and 643 are close to a bright galaxy which is why we might not be able to detect the counterpart in the optical and infrared. Deeper observations would be needed to detect possible counterparts. $Bottom$: Sources with multiple possible counterparts. Out of these eight objects only 156, 276 and 333 are simultaneously detected in the hard and in the soft band. We are unable to gain redshift estimates for these eight objects and are thus unable to determine if they are low or high-redshift sources or if they might be spurious detections. The black circles are centered on the original \Chandra\ position and illustrate the positional uncertainty given in the \protect\cite{Xue:2011aa} catalog. All images were colour inverted and are 10$^{\prime\prime}$ x 10$^{\prime\prime}$ in size.} 
\end{figure*}

\subsection{Aperture Photometry}
\label{sec:aperturephot}
We perform our own aperture photometry on the GOODS/ACS and CANDELS images to gain flux values and to estimate parameters such as detection threshold and aperture size. We compare our results to the GOODS/ACS catalog (Figure \ref{fig:comp_phot}). 

To perform aperture photometry we use Source-Extractor (SExtractor, \citealt{Bertin:1996aa, 2002Terapix, Holwerda:2005aa}). We determine the counterpart position in the $H$ band in a first SExtractor run. We then run SExtractor on the remaining optical and infrared images to establish if the counterpart is present at the same location. The flux measurements are carried out within circular apertures with radii between 0.3$^{\prime\prime}$ and 1$^{\prime\prime}$. We alter the aperture size for faint sources and to prevent contamination through nearby objects. For sources with a signal-to-noise ratio $<1$ we use the $1\sigma$ sensitivity limit of the corresponding filter as an upper limit. The SExtractor parameter values and aperture sizes are summarized in Table \ref{tab:SE_para} and Table \ref{tab:SE_ID}. For the 3.6 micron and 4.5 micron bands we rely on the flux values reported in the SIMPLE catalog \citep{Damen:2011aa}. We make use of the flux values reported for a 1.5$^{\prime\prime}$ radius aperture. We show flux and error values for our main sample in Tables \ref{tab:apphot1} and \ref{tab:apphot2}.

\subsection{Lyman Break Technique and visual classification}
\label{sec:LBT}
The Lyman Break Technique \citep{Steidel:1999fj, Giavalisco:2002aa, Dunlop:2013zr} employs the pronounced feature of the Lyman continuum discontinuity in the spectral energy distribution (SED) of young, star forming galaxies. We use the common terminology of referring to Lyman Break Galaxies as 'dropouts'. If a source is not detected in the $B$-band or any bluer passbands, but is visible in all redder filters, this indicates $z\sim4$ and we refer to it as a '$B$-dropout'. $V$-dropouts, $i$-dropouts and $z$-dropouts correspond to redshifts of $\sim5$, $\sim6$ and $\sim7$ respectively. 

We classify the 371 possible candidates by eye according to their dropout band. If an object is clearly visible in all bands, this indicates $z < 4$. We exclude such sources from our sample. Figure \ref{fig:visclass} illustrates the conditions sources have to fulfill to be included in the further analysis. \\

\begin{figure}
\begin{centering}
\includegraphics[width=\columnwidth]{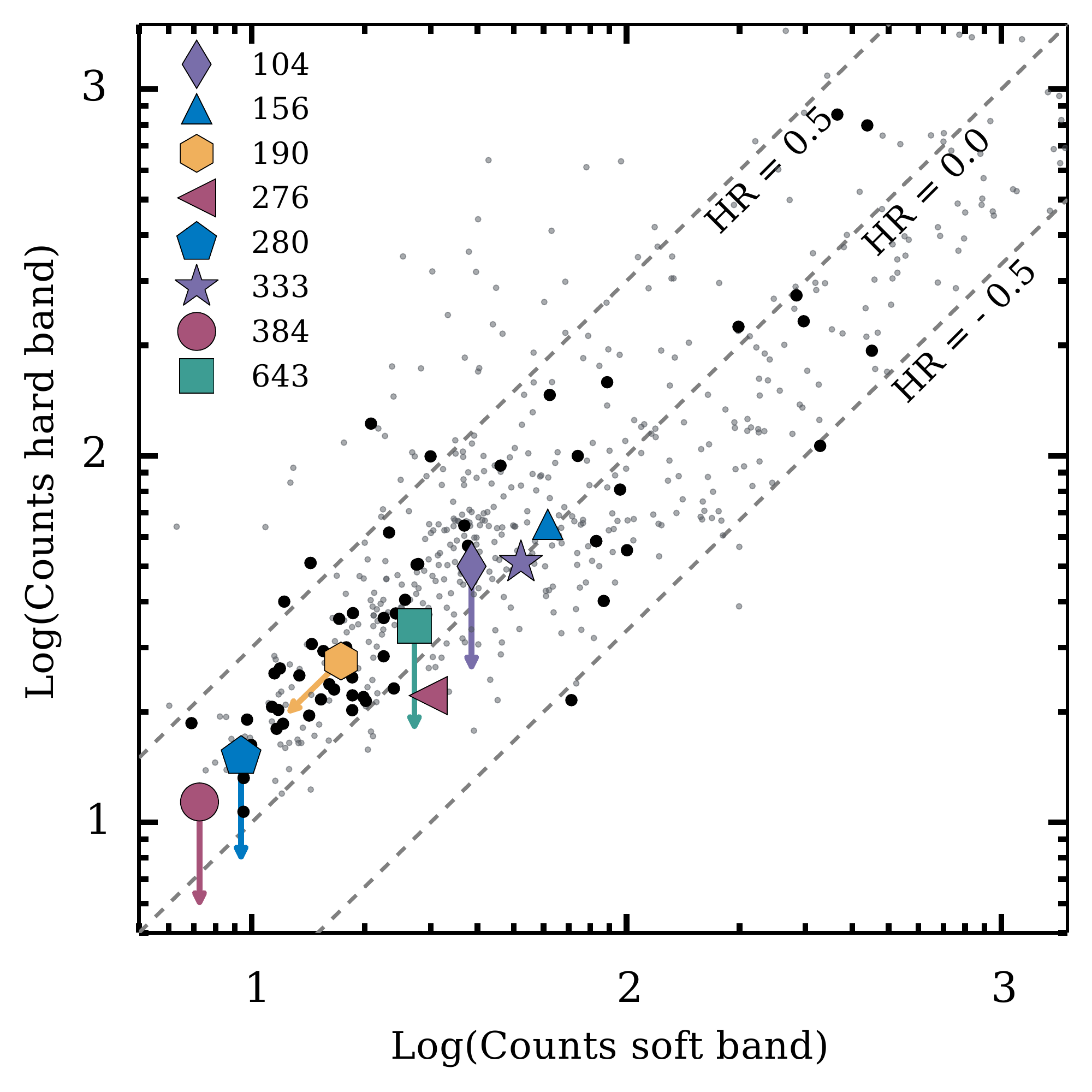}
\caption{\label{fig:counts} Counts in the hard and in the soft band for the \Chandra\ 4-Ms sources. We highlight the eight objects that could not be classified according to the Lyman Break Technique. These eight sources might be real sources or spurious detections. The black points illustrate the 58 objects that are left in our sample. The grey points show the positions of the additional 502 \Chandra\ 4-Ms objects that are also classified as AGN. They were excluded because of insufficient filter coverage or quality or because they are clear low-redshift dropouts. \protect\cite{Xue:2011aa} categorize all of these eight objects and all 58 sample sources as AGN.}
\end{centering}
\end{figure}

Eight sources are not classified according to their dropout band. These objects are shown in Figures \ref{fig:mysteryobjects}. Figure \ref{fig:counts} shows the hard and soft band counts for all eight sources in comparison to the entire sample. The hard, soft and full band counts are given in Table \ref{tab:mysobscounts}.\\ 

\begin{table*}
	\begin{center}
		\begin{tabular}{llllllllll}
		\toprule
			{ID} & {Hard counts} & {$\sigma_\mathrm{Hard}$} & {$\mathrm{SNR}_\mathrm{Hard}$} & {Soft counts} & {$\sigma_\mathrm{Soft}$} & {$\mathrm{SNR}_\mathrm{Soft}$} & {Full counts} & {$\sigma_\mathrm{Full}$} & {$\mathrm{SNR}_\mathrm{Full}$}\\

			\midrule
			$190$ & $27.55$ & $-1.00$ & $\mathrm{-}$ & $17.29$ & $-1.00$ & $\mathrm{-}$ & $21.99$ & $10.12$ & $2.17$\\
			$280$ & $15.13$ & $-1.00$ & $\mathrm{-}$ & $9.36$ & $4.94$ & $1.89$ & $18.36$ & $-1.00$ & $\mathrm{-}$\\
			$333$ & $51.15$ & $16.91$ & $3.02$ & $52.23$ & $11.59$ & $4.51$ & $103.21$ & $19.66$ & $5.25$\\
			$384$ & $11.37$ & $-1.00$ & $\mathrm{-}$ & $7.25$ & $4.44$ & $1.63$ & $14.98$ & $-1.00$ & $\mathrm{-}$\\
			$643$ & $34.34$ & $-1.00$ & $\mathrm{-}$ & $27.16$ & $8.72$ & $3.11$ & $33.54$ & $13.21$ & $2.54$\\
			\midrule
			$104$ & $49.95$ & $-1.00$ & $\mathrm{-}$ & $38.57$ & $12.16$ & $3.17$ & $57.78$ & $-1.00$ & $\mathrm{-}$\\
			$156$ & $65.01$ & $13.43$ & $4.84$ & $61.59$ & $10.59$ & $5.82$ & $126.33$ & $16.40$ & $7.70$\\
			$276$ & $22.17$ & $9.54$ & $2.32$ & $29.55$ & $7.95$ & $3.72$ & $51.61$ & $11.70$ & $4.41$\\

		\bottomrule
		\end{tabular}
	\end{center}
	\caption{\label{tab:mysobscounts}Hard, soft and full band counts for sources that can not be classified according to their dropout band. We refer to the top 5 sources as \textit{'low significance objects'}. For objects that are not detected we give an upper limit on the counts. We set $\sigma$ to -1.00 and mark the signal-to-noise ratio with a dash. All values were directly extracted from the \Chandra\ 4-Ms catalog \protect\citep{Xue:2011aa}.}
\end{table*}

Five (190, 280, 333, 384, 643) of these eight objects are especially interesting since they do not have a counterpart in the optical or the infrared. We refer to these sources as \textit{'low significance objects'}. We show these five objects in the upper panel of Figure \ref{fig:mysteryobjects}. For three (104, 156, 276) of these eight objects it is unclear which source represents the counterpart in the optical and infrared since multiple objects are visible in the GOODS/ACS and CANDELS bands. These objects are shown in the lower panel of Figure \ref{fig:mysteryobjects}. 

Only three of the eight objects (333, 156, 276) are detected in the full, the hard and the soft band. 104 and 280 are detected in the soft band only. 384 and 643 are found in the soft and in the full band. 190 is detected in the full band only. Only one of the low significance objects (333) and one of the three sources with multiple counterparts in the optical and infrared (156) show a signal-to-noise ratio $\geq5$ in at least one of the bands. Out of the 371 objects with enough filter coverage and intact images 140 (37.7\%) have a signal-to-noise ratio $\geq5$ in the soft, hard or full band. See Table \ref{tab:SN} for the number of sources with signal-to-noise ratios $\geq1$ and $\geq5$ in the individual bands.

The objects for which we do not detect a counterpart in the optical and the infrared could be spurious detections. On the one hand, \cite{Xue:2011aa} report that, for the entire catalog, the probability of a source not being real is < 0.004. The entire catalog does therefore contain up to three spurious sources per band. We do however only consider sources that are also covered by GOODS and CANDELS. The CANDELS wide and deep survey fields are $~0.03$ $\mathrm{deg}^2$ in size and hence only make up $\sim27\%$ of the CDF-S ($0.11$ $\mathrm{deg}^2$) area. According to this, we would expect to find $\sim0.8$ spurious detections per band. So, we can not rule out the fact that we may have found one or more false detections. On the other hand, we find many sources of comparable X-ray brightness that do possess an optical and/or infrared counterpart and that are detected in both bands (Figure \ref{fig:counts}). 333, 384 and 643 are also close to a bright galaxy which might be why the counterpart remains undetected. This could indicate that at least some of the sources are real.  

The Lyman Break Technique is not applicable to the low significance objects and we are unable to measure a photometric redshift without a detection in the optical and the infrared. Since 333 is detected in the hard and the soft band, it is the only object for which we can apply our Hardness Ratio test (see Section \ref{sec:HR}). With $HR=-0.01$ 333 could be a potential high-redshift AGN candidate. A negative Hardness Ratio alone does however not convince us of 333 indeed being at high-redshift. At this point we can thus not determine if our sources are real high-redshift AGN candidates, false detections or low-redshift objects that are optically faint. Hopefully, the forthcoming 7-Ms observations of the CDF-S (PI: William Brandt, Proposal ID: 15900132) will shed more light on our five low significance objects. We eliminate all eight sources from our sample. We stress that these targets could still be high-redshift AGN.   

After discarding targets that are clearly visible in all bands and eliminating the eight objects that could not be classified according to the Lyman Break Technique, 58 $B$, $V$, $i$, $z$ and $Y$-dropouts remain in our main sample. For sources that we visually classify as $B$-dropouts the signal-to-noise in the $V$-band might be too low for a detection by SExtractor. By eliminating visually classified $B$-dropouts we could hence be missing objects that might be classified as $z\sim5$ sources by other redshift tests. We therefore keep $B$-dropouts in our sample.  
  
\subsection{Photometric Redshift measurements}
\label{sec:photoredcode}
Even though the Lyman Break Technique provides a fast and easy way of identifying possible candidates, it is not without caveats. Dust in red, low-redshift galaxies can produce a sharp break in the SED that might be mistaken for the Lyman Break \citep{Dunlop:2007aa, McLure:2010aa, Finkelstein:2012aa}. 
Applying the Lyman Break Technique therefore only produces a sample that may still contain low-redshift interlopers. 
Yet, even the results of a photometric redshift code have to be treated carefully. While sources with a low photometric redshift are most likely indeed nearby objects, high photometric redshift results are less reliable. This is mainly due to template incompleteness and large photometric errors for faint high-redshift sources.   

The photometric redshifts we determine for our objects prove to be highly dependent on the filters used as input. We use the photometric redshift code EAZY \citep{Brammer:2008aa}. We apply the default template set by \cite{Brammer:2008aa} which is part of the EAZY distribution. Five of these six EAZY templates were created by using the \cite{Blanton:2007aa} algorithm to reduce the template set by \cite{Grazian:2006aa}. The sixth template describes a young starburst with a dust screen following the Calzetti law with $A_\mathrm{v}$ = 2.75 \citep{Calzetti:2000aa}.
Due to large uncertainties in the luminosity function of high-redshift AGN, we do not include a luminosity prior when running EAZY.
To gain reliable redshift values we take the \Spitzer/IRAC images into account. We carefully compare the $H$ band CANDELS images to the \Spitzer\ 3.6 micron images to determine which of our candidates possesses a \Spitzer\ counterpart and for which sources the \Spitzer\ flux values can not be used due to source confusion. We include the \Spitzer\ images for 30 of our 58 objects. We show the $H$ band and \Spitzer\ stamps for these sources in the appendix.  We do not perform aperture photometry on the \Spitzer\ images, but simply extract the $1.5^{\prime\prime}$ aperture radius flux values from the \Spitzer\ catalog by \cite{Damen:2011aa}.

If a source is not detected in a passband, we use the $1\sigma$ sensitivity limit of this filter as an upper limit in the EAZY run. For the GOODS/ACS bands $B$, $V$, $i$ and $z$ we determine the sensitivity limits by measuring the mean background flux. We measure the number of background counts for six of our objects. For each source, we determine the flux within apertures of varying size at five different positions. For the CANDELS bands $Y$, $J$ and $H$ we rely on the sensitivity limits reported in \cite{Grogin:2011aa}. Table \ref{tab:fluxlimits} shows the flux limits that we used in our analysis.

\begin{table}
	\begin{center}
		\begin{tabular}{lc}
			\toprule
			{} & {$1\sigma$ flux limit } \\
			{} & {in $\mu$Jy / $\mathrm{arcsec}^2$} \\ 
			\midrule
			$B$ & $4.636\e{-2}$\\
			$V$ & $4.160\e{-2}$\\
			$i$ & $8.255\e{-2}$\\
			$z$ & $1.487\e{-2}$\\ 
			\bottomrule
		\end{tabular}
	\end{center}
	\caption{\label{tab:fluxlimits}GOODS/ACS flux limits. We determine these mean sensitivity limits by measuring the mean background flux for six different objects. For each object we determine the background counts within five apertures of varying size.}
\end{table}  

\begin{figure*}
\includegraphics[width=\textwidth]{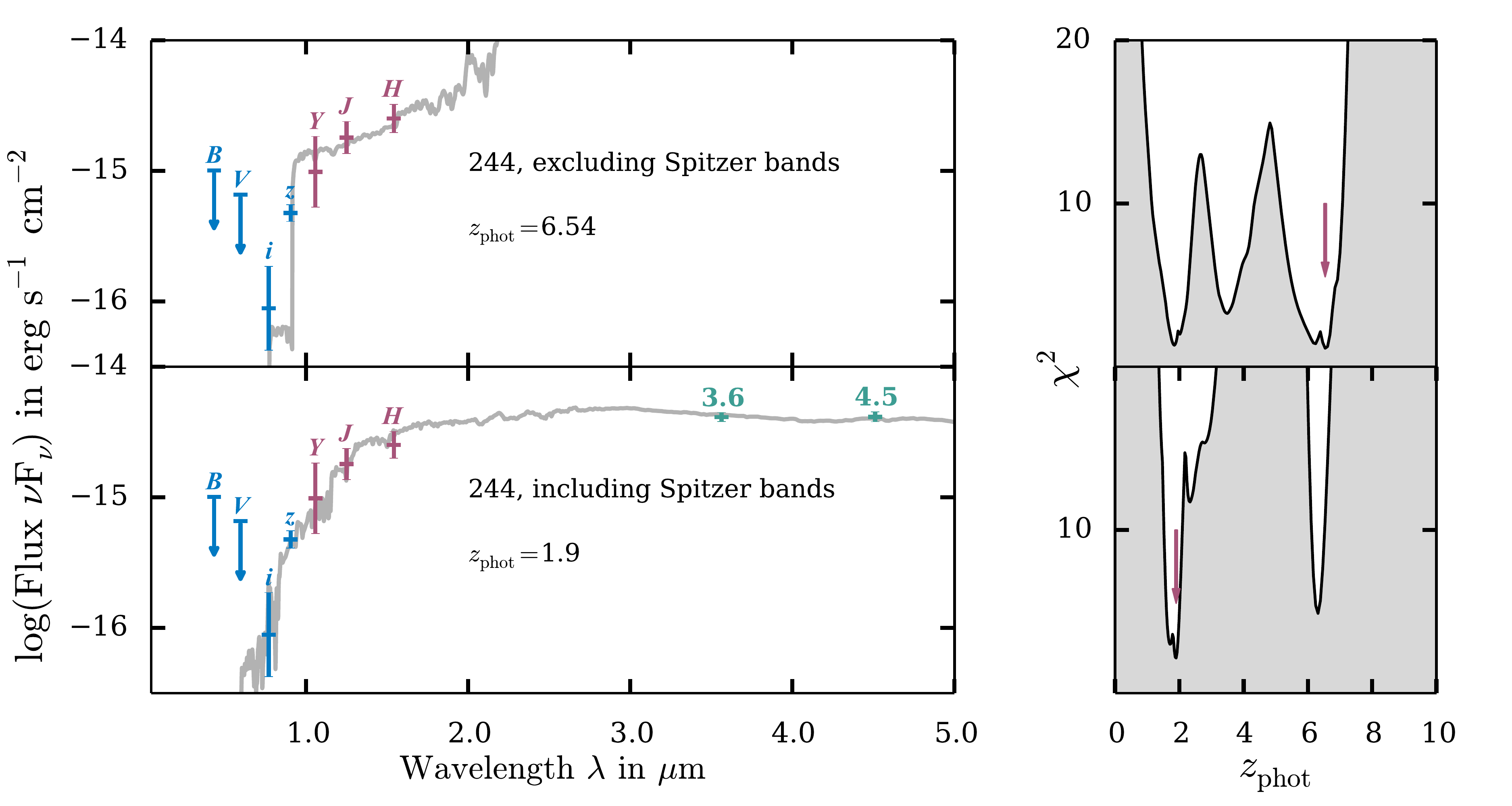}
\caption{\label{fig:244} Best fit SED for 244 determined by running EAZY without the \Spitzer\ data ($top$) and with the \Spitzer\ data ($bottom$). If we only use the GOODS/ACS and CANDELS filters as input for EAZY, the photometric redshift code will classify 402 as a $z\sim6.54$ source. Even though the flux values are fit well by this SED, the shape of the SED seems unphysical for a $z\sim6.54$ source. We would expect the continuum flux to be lower and bluer. A comparison to the $z\sim7$ UV luminosity function by \protect\cite{Bouwens:2014aa} shows that 244 would indeed be bright if it was at  $z\sim6.54$ ($M_{H}=-24.6$). The $\chi^2$ distribution on the right shows multiple secondary low redshift solutions. SED shape and $\chi^2$ distribution thus suggest that 244 might be a low-redshift object. In the bottom panel we illustrate the best fit SED determined by including the \Spitzer\ 3.6 and 4.5 micron IRAC channels. 244 is now exposed as a low-redshift source. Based upon $z\sim1.9$ we reject 244 as a possible high-redshift candidate. The shown limits correspond to the 1$\sigma$ sensitivity limits. }
\end{figure*}

Including the \Spitzer\ data turns out to be crucial for our purposes. As an example we show the photometric redshift code results we get for source 244 in Figure \ref{fig:244}. By running EAZY without the \Spitzer\ flux values we gain $z_\mathrm{phot}$ = 6.54. Assuming that this photometric redshift value is correct, we determine an absolute magnitude of $M_{H}=-24.63$ for the $H$-band. \cite{Bouwens:2014aa} however give $M_\mathrm{UV}^{*}=-21.2$ for $z\sim6$ Lyman Break Galaxies. 244 would therefore be very bright if it indeed was at $z\sim6$. Including the 3.6 and 4.5 micron \Spitzer\ flux values in the photometric redshift code analysis results in  $z_\mathrm{phot}$ = 1.9. Including the \Spitzer\ infrared data hence proves to be crucial in revealing low-redshift interlopers.  

\subsection{Stacking GOODS/ACS data }
We combine the GOODS/ACS images of each object into stacks. We generated three stacks per source: (1) combines $B$ and $V$, (2) $B$, $V$ and $i$ and (3) $B$, $V$, $i$, $z$. We examine these deeper stacks for detections by running SExtractor on them. We use a detection threshold of 1.5 $\sigma$ (DETECT$\_$TRESH = 1.5) and a minimum detection area of 15 pixels above the threshold (DETECT$\_$MINAREA = 15). The remaining SExtractor parameter values are left at their default values. A detection in the first stack indicates a source that drops out before or in the $B$-band and therefore implies $z\lesssim4$. Sources that are detected in (1), (2) and (3) are hence of no interest to us. We assume $z\sim5$ if an object is detected in (2) and (3), but not in (1). Only being detected in (3) indicates $z\sim6$. Finally, no detection in any stack signals $z\gtrsim7$. Using this approach we find \stackingfour\ $z\lesssim4$, \stackingfive\ $z\sim5$, \stackingsix\ $z\sim6$ and \stackingseven\ $z\gtrsim7$ sources. The stacking analysis proves to be inconclusive for two objects (303, 651). Source 303 is detected in (1) and (3), but not in (2). 651 is only found in stack (2).

\subsection{Colour criteria}
\label{sec:cc}
We obtain an additional redshift indication by applying colour criteria based upon population synthesis models \citep{Guhathakurta:1990aa,Steidel:1992gf}. These colour criteria do not only depend on the position of the Lyman Break, but also use the overall SED shape. The criteria we apply here are based on \cite{Vanzella:2009aa} who used 114 Lyman Break Galaxies from the GOODS field. The sample consisted of 51 $z\sim4$, 31 $z\sim5$ and 32 $z\sim6$ Lyman Break Galaxies. These objects were first chosen by applying the Lyman Break Technique and then followed up spectroscopically.  

Objects that fulfill the following condition are classified as $z\sim4$ sources:\\
\begin{equation}
(B-V)\geq(V-z)\,\wedge\,(B-V)\geq1.1\,\wedge\,(V-z)\leq1.6
\end{equation}

\noindent Here $\wedge$ and $\vee$ represent the logical 'and' and 'or' operators respectively. $z\sim5$ objects need to satisfy the constraints given here:

\begin{align}
\begin{split}
[(V-i) >1.5+0.9\times(i-z)]\vee[(V-i)>2.0]\wedge\\
(V-i)\geq1.2\wedge (i-z)\leq1.3\wedge\mathrm{(S/N)}_{B}<2
\end{split}
\end{align}

\noindent For $z\sim6$ galaxies these conditions apply:

\begin{equation}
(i-z)>1.3\wedge[(\mathrm{S/N)_{B}<2\vee(\mathrm{S/N)}_{V}<2}]
\end{equation}
To classify the 58 possible candidates according to colour criteria, we use the magnitude values from our aperture photometry (see \ref{sec:aperturephot}). If a source is not detected, we use the 1$\sigma$ sensitivity limit as an upper limit. \ccnotclass\ of the 58 sources do not fulfill any colour criteria. Note that a source that simultaneously possesses upper limits in the $B$ and $V$ or $V$ and $z$ band is not included in the $z\sim4$ diagram since its position cannot be determined. The same applies for the $z\sim5$ and $z\sim6$ diagrams. Furthermore, we only use colour criteria to determine $z\sim4$, $z\sim5$ and $z\sim6$ dropouts. The \ccnotclass\ objects that can not be classified are hence not all necessarily at $z < 4$. Figure \ref{fig:cc} shows the colour-colour diagrams that illustrate these criteria. Based on colour criteria we find \ccfour\ $z\sim4$ (373, 444), \ccfive\ $z\sim5$ (303, 321) and \ccsix\ $z\sim6$ (226, 244, 296, 522, 589) objects.\\

\subsection{X-rays as a photometric redshift indicator}
\label{sec:HR}

\begin{figure*}
\begin{centering}
\begin{minipage}[t]{0.49\textwidth}
\begin{centering}
\includegraphics[scale=0.31]{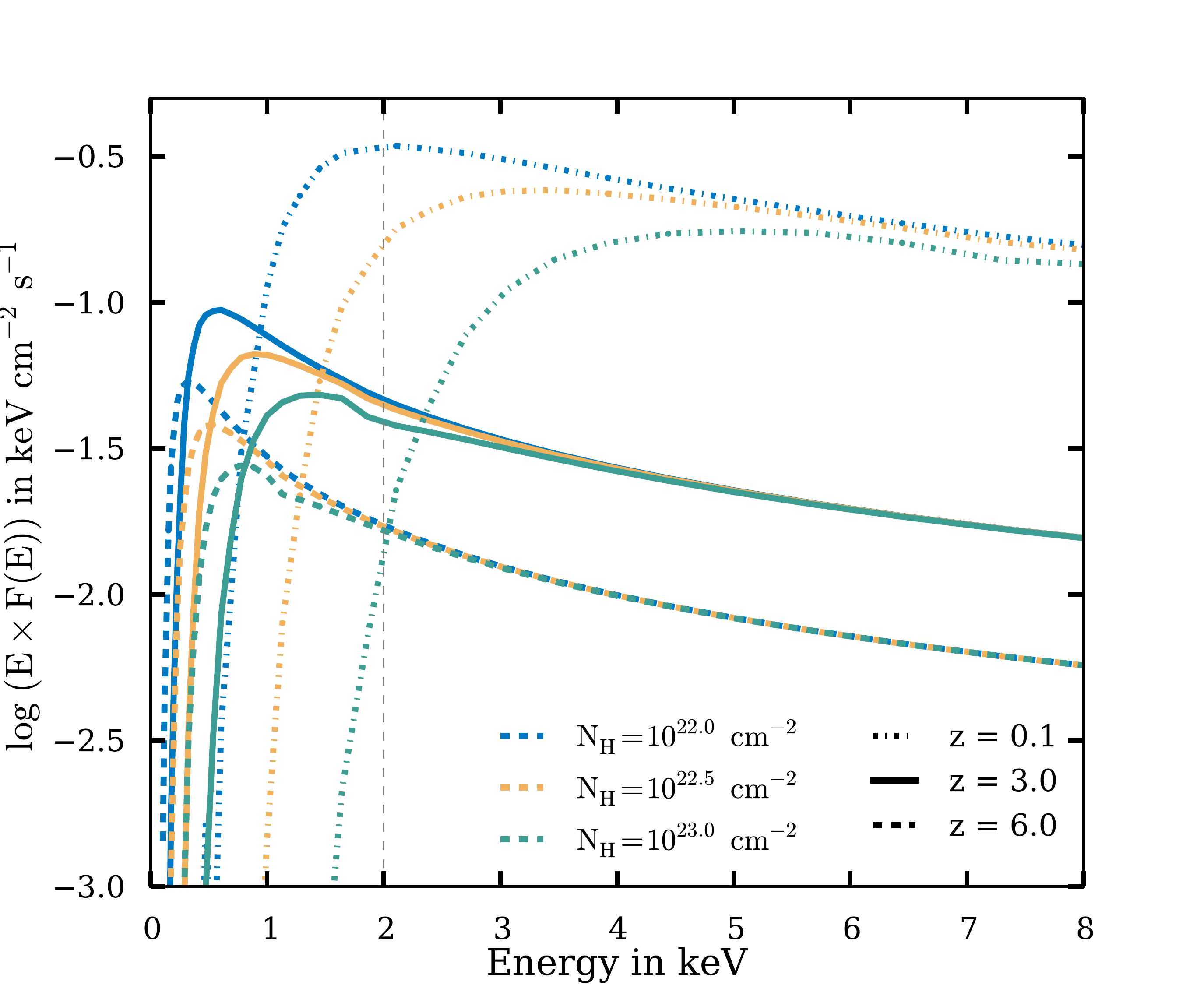}
\end{centering}
\end{minipage}
\begin{minipage}[t]{0.49\textwidth}
\begin{centering}
\includegraphics[scale=0.31]{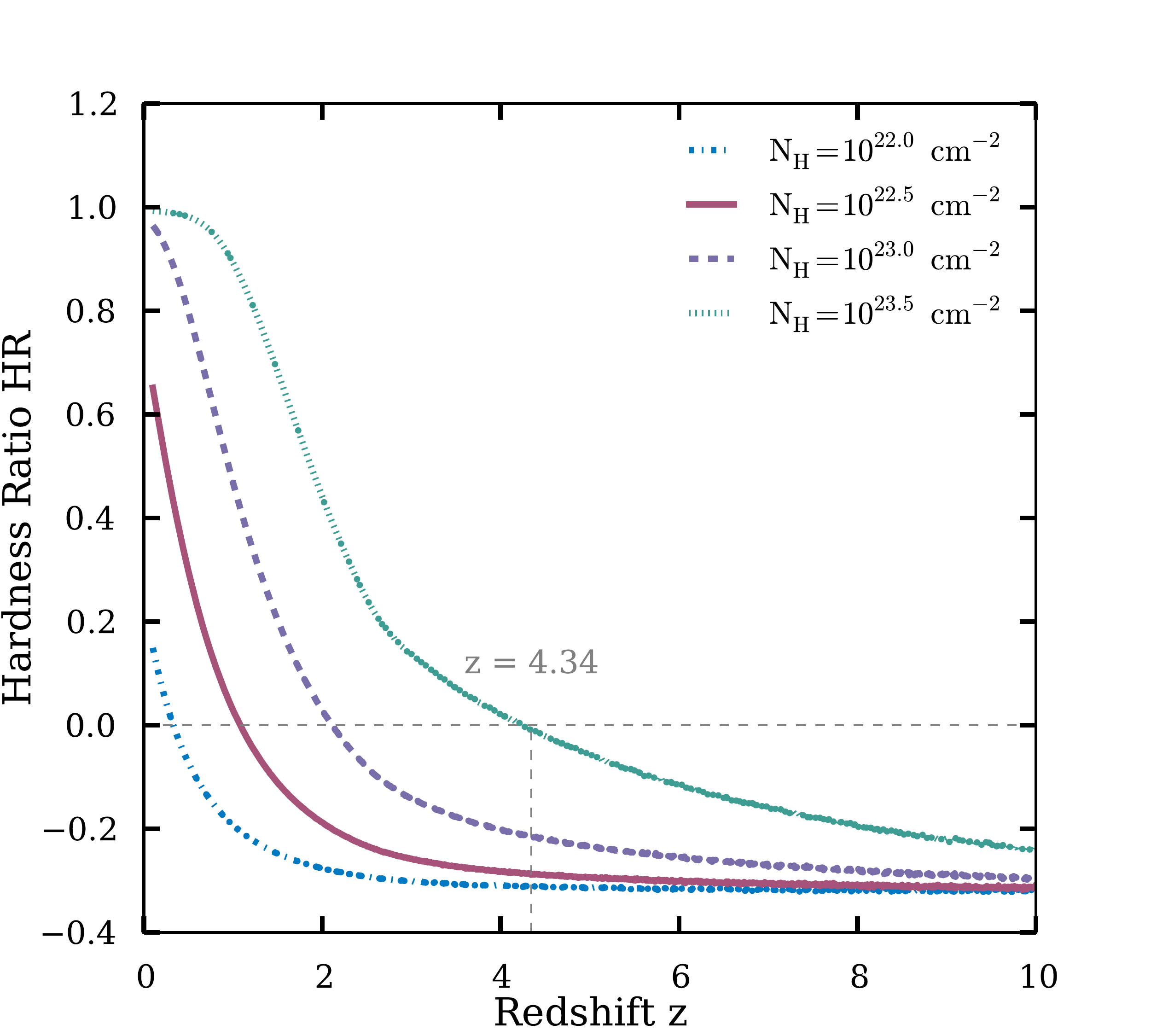}
\end{centering}
\end{minipage}
\caption{\label{fig:HR}AGN X-ray spectrum simulation. Shown on the $left$ is our model that we generated by running Xspec \citep{Arnaud:1996aa}. We assume a power law with a slope of 1.8 and a turnover due to photoelectric absorption (Xspec \texttt{zphabs*zpow} model). This left plot illustrates the model SED for moderately high column densities (line colours) at several redshifts (line types). With increasing redshift the spectrum gets shifted to lower energies and the number of counts in the hard and in the soft band changes. $Right$: measured Hardness Ratio HR as a function of redshift ($HR=\frac{H-S}{H+S}$, $H$ and $S$ representing the counts in the hard and in the soft band respectively). Based on our model, a HR > 0 indicates $z < 4.34$ for column densities up to \NH\ = $10^{23.5} \mathrm{cm}^{-2}$. In our analysis we thus dismiss objects with an HR > 0 as obscured AGN at low redshift.}
\end{centering}
\end{figure*}

The Hardness Ratio (HR), sometimes denoted as X-ray colour, represents an additional indicator for high-redshift AGN. The Hardness Ratio is defined as: 
\begin{equation}\label{eq:HR}
HR=\frac{H-S}{H+S}
\end{equation} 
Here H and S represent the observed frame hard (2 - 10 keV) and soft (0.5 - 2 keV) band counts respectively.

To zeroth-order, an AGN X-ray spectrum follows a powerlaw with an obscuration dependent turnover at lower energies. In the Compton-thin regime a higher column density \NH\  means that, relative to the hard band, we will detect fewer counts in the soft band. In a galaxy's restframe the X-ray spectrum hence appears harder for higher obscuration. With increasing redshift the spectrum gets shifted to lower energies and the number of counts that we observe in the hard and in the soft band changes accordingly. We can therefore use the Hardness Ratio as an additional redshift indicator.

A soft X-ray spectrum is then expected for Compton-thin (\NH\ $< 10^{24}\mathrm{cm}^{-2}$) objects. For sources in which $10^{24}\lesssim$ \NH\ $\mathrm{H}\lesssim 10^{25}\mathrm{cm}^{-2}$, i.e. transmission-dominated Compton-thick AGN, the direct emission is still visible, although the E $<$ 10~keV radiation is completely obscured by photoelectric absorption, while the detected emission at higher energies is reduced by Compton scattering \citep{Comastri:2010aa, Murphy:2009aa}. Hence, in these cases we expect to observe a hard X-ray spectrum even for sources at $z > 5$. When \NH\ $ > 10^{25}\mathrm{cm}^{-2}$, i.e. for reflection-dominated Compton-thick AGN, we only observe the small fraction of the initial emission which is reflected off the accretion disk or the obscuring material \citep{Ajello:2009aa,2012MNRAS.422.1166B}. We can hence not make the same assumptions as for Compton-thin sources. Nevertheless, it is safe to assume that all $z\gtrsim5$ sources that can be detected individually in the X-rays are Compton-thin objects. Compton-thick $z\gtrsim5$ AGN would simply be too faint to be detected individually and are therefore not part of our sample.

\begin{figure}
\begin{centering}
\includegraphics[width=0.49\textwidth]{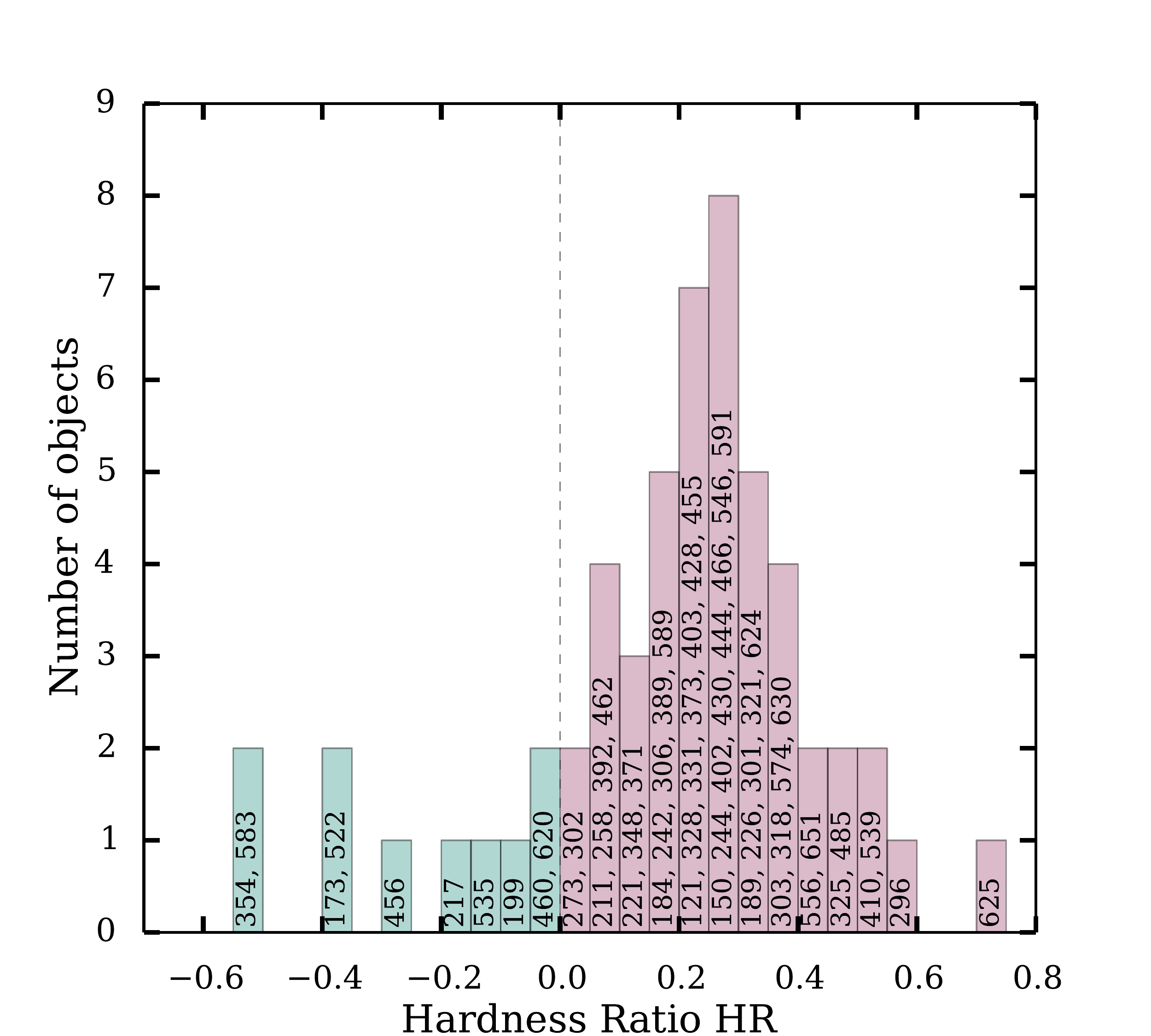}
\end{centering}
\caption{\label{fig:HRhist}Number of objects per HR bin. This figure illustrates the distribution of our 54 objects in terms of HR. HR < 0  indicates that this source might be at $z > 4.34$. A positive HR suggests that this source might be at $z\leq$ 4.34. Thus ten of our 54 objects might be at $z > 4.34$. Not shown are objects 496 and 578. These two sources are only detected in the full band and can therefore not be constrained by the Hardness Ratio. The numbers in each bar coincide with the object IDs in the corresponding HR bin.}
\end{figure}

To quantify the HR($z$) relation we use the X-ray spectral fitting tool Xspec to simulate X-ray spectra at different redshifts \citep{Arnaud:1996aa}. We use a \texttt{zphabs*zpow} model and assume a power law slope of 1.8 \citep{Turner:1997aa, Tozzi:2006aa}. Figure \ref{fig:HR} summarizes our results. The left panel illustrates our model for \NH\ $ = 10^{22} \mathrm{cm}^{-2}$, $10^{22.5} \mathrm{cm}^{-2}$ and $10^{23.5} \mathrm{cm}^{-2}$  (blue, yellow and green, respectively) at $z$ = 0.1, 3 and 6 (dot dashed, solid and dashed line, respectively). It is evident that the number of counts in the soft and hard band changes with redshift. After including the \Chandra\ Redistribution Matrix (RMF) and Auxiliary Response Files (ARF) for on-axis sources, we measure the spectral counts in the hard and in the soft band and determine the HR. The right panel of Figure \ref{fig:HR} shows our results. According to our simulations HR > 0 signifies $z < 4.34$ for sources with \NH\ up to $10^{23.5} \mathrm{cm}^{-2}$. Allowing for a small amount of transmission, e.g. 1$\%$ by using a \texttt{zpcfabs*zpow} model does not change our results significantly (HR = 0 for $z$ = 4.3). In our analysis we hence discard objects with HR > 0. Figure \ref{fig:HRhist}, which illustrates our results, shows that based solely on the HR, ten of our 58 candidates might be at $z > 4.34$. For 496 and 583 the HR can not be used to constrain $z$ since they are only detected in the full band. Table \ref{tab:counts} summarizes the X-ray counts, signal-to-noise ratios and Hardness Ratios for each of our main sample sources.

Our results are in good agreement with a similar analysis by \cite{Wang:2004aa}. They also used Xspec \citep{Arnaud:1996aa} to simulate the HR at different redshifts. Yet, \cite{Wang:2004aa} chose a power law slope of 1.9 and used a \texttt{wabs} model. Our analysis shows that $z > 5$ sources with \NH\ up to $10^{23} \mathrm{cm}^{-2}$ should have a HR of $\sim-0.3$, \cite{Wang:2004aa} find HR$\sim-0.5$. For \NH\ = $10^{23} \mathrm{cm}^{-2}$ we determine HR = 0 at $z\sim2$, \cite{Wang:2004aa} find HR = 0 at $z\sim1.5$.  

\newcommand\low{\textcolor{red}{$\times$}}
\newcommand\high{\textcolor{green}{\checkmark}}
\definecolor{green}{HTML}{3D9D93}
\definecolor{red}{HTML}{A75379}

\begin{table*}
	\begin{center}
		\begin{tabular}{llccccc}
		\toprule
			{} & {ID} & {visual classification} & {colour criteria} & {Stacking} & {Hardness Ratio} & {photo-z}\\
			\midrule
			{} & $121$ & \low & 0 & \low & \low & \low \\
			{} & $150$ & \low & 0 & \low & \low & \low \\
			{} & $173$ & \low & 0 & \low & \high & \low \\
			{} & $184$ & \low & 0 & \low & \low & \low \\
			{} & $189$ & \low & 0 & \low & \low & \low \\
			{} & $199$ & \high & 0 & \high & \high & \low \\
			{} & $211$ & \low & 0 & \low & \low & \low \\
			{} & $217$ & \high & 0 & \high & \high & \low \\
			{} & $221$ & \low & 0 & \low & \low & \low\\
			{} & $226$ & \low & \high & \low & \low & \high \\
			{} & $242$ & \low & 0 & \low & \low & \low \\
			{} & $244$ & \high & \high & \high & \low & \low \\
			{} & $258$ & \high & 0 & \high & \low & \low \\
			{} & $273$ & \high & 0 & \high & \low & \low \\
			{} & $296$ & \high & \high & \high & \low & \low \\
			{} & $301$ & \low & 0 & \low & \low & \low \\
			{} & $302$ & \low & 0 & \low & \low & \low \\
			{} & $303$ & \high & \high & 0 & \low & \low \\
			{} & $306$ & \high & 0 & \high & \low & \low \\
			{} & $318$ & \low & 0 & \low & \low & \low \\
			{} & $321$ & \low & \high & \low & \low & \low \\
			{} & $325$ & \low & 0 & \low & \low & \low \\
			{} & $328$ & \low & 0 & \low & \low & \low \\
			{} & $331$ & \low & 0 & \low & \low & \low\\
			{} & $348$ & \high & 0 & \high & \low & \low \\
			{} & $354$ & \high & 0  & \high & \high & \low \\
			{} & $371$ & \high & 0  & \high & \low & \high \\
			{} & $373$ & \low & \low  & \low & \low & \low \\
			{} & $389$ & \low & 0  & \low & \low & \low \\
			{} & $392$ & \high & 0  & \high & \low & \low \\
			{} & $402$ & \high & 0  & \high & \low & \low \\
			{} & $403$ & \low & 0 & \low & \low & \low \\
			{} & $410$ & \low & 0  & \low & \low & \low \\
			{} & $428$ & \low & 0  & \low & \low & \low \\
			{} & $430$ & \high & 0 & \high & \low & \high \\
 			{} & $444$ & \low & \low  & \low & \low & \low \\
			{} & $455$ & \low & 0  & \low & \low & \low\\
			$\rightarrow$ & $456$ & \high & 0  & \high & \high & \high \\
			{} & $460$ & \high & 0  & \high & \high & \low \\
			{} & $462$ & \low & 0  & \low & \low & \low \\
			{} & $466$ & \low & 0  & \low & \low & \low \\
			{} & $485$ & \high & 0  & \high & \low & \low \\
			{} & $496$ & \low & 0  & \low & 0 & \low \\
			{} & $522$ & \low & \high  & \high & \high& \low \\
			{} & $535$ & \low & 0  & \low & \high & \low \\
			{} & $539$ & \low & 0  & \low & \low & \low \\
			{} & $546$ & \low & 0  & \low & \low & \low\\
			{} & $556$ & \low & 0  & \low & \low & \low \\
			{} & $574$ & \high & 0  & \high & \low & \low \\
			$\rightarrow$ & $578$ & \high & 0  & \high & 0 & \high \\
			$\rightarrow$ & $583$ & \high & 0  & \high & \high & \high \\
			{} & $589$ & \low & \high & \low & \low & \low \\
			{} & $591$ & \low & 0 & \low & \low & \low \\
			{} & $620$ & \low & 0  & \low & \high & \low \\
			{} & $624$ & \high & 0  & \high & \low & \low \\
			{} & $625$ & \low & 0  & \low & \low & \low \\
			{} & $630$ &  \high & 0  & \low & \low & \low \\
			{} & $651$ & \high & 0 & 0 & \low & \low \\
		\bottomrule
		\end{tabular}
	\end{center}
\caption{\label{tab:allresults} Overview showing all five redshift test results. Sources for which the test indicates $z < 5$ are marked with \low. If the redshift test results in $z\geq$ 5, we show a \high. Objects that could not be classified via colour criteria, stacking or the Hardness Ratio are marked with 0. For the Hardness Ratio we can only distinguish between $z < 4.3$ (HR > 0) and $z\geq$ 4.3 (HR $\leq$ 0). Hence, for the Hardness Ratio \high means $z\geq$ 4.3. After combining all redshift tests we are left with three possible candidates (arrows). See Table \ref{tab:overview} and Table \ref{tab:counts} for a detailed overview of the results and the X-ray counts of each individual source.}
\end{table*}
\newpage

\section{Combining all redshift tests}
\label{sec:combination}

\begin{figure*}
\includegraphics[width=\textwidth]{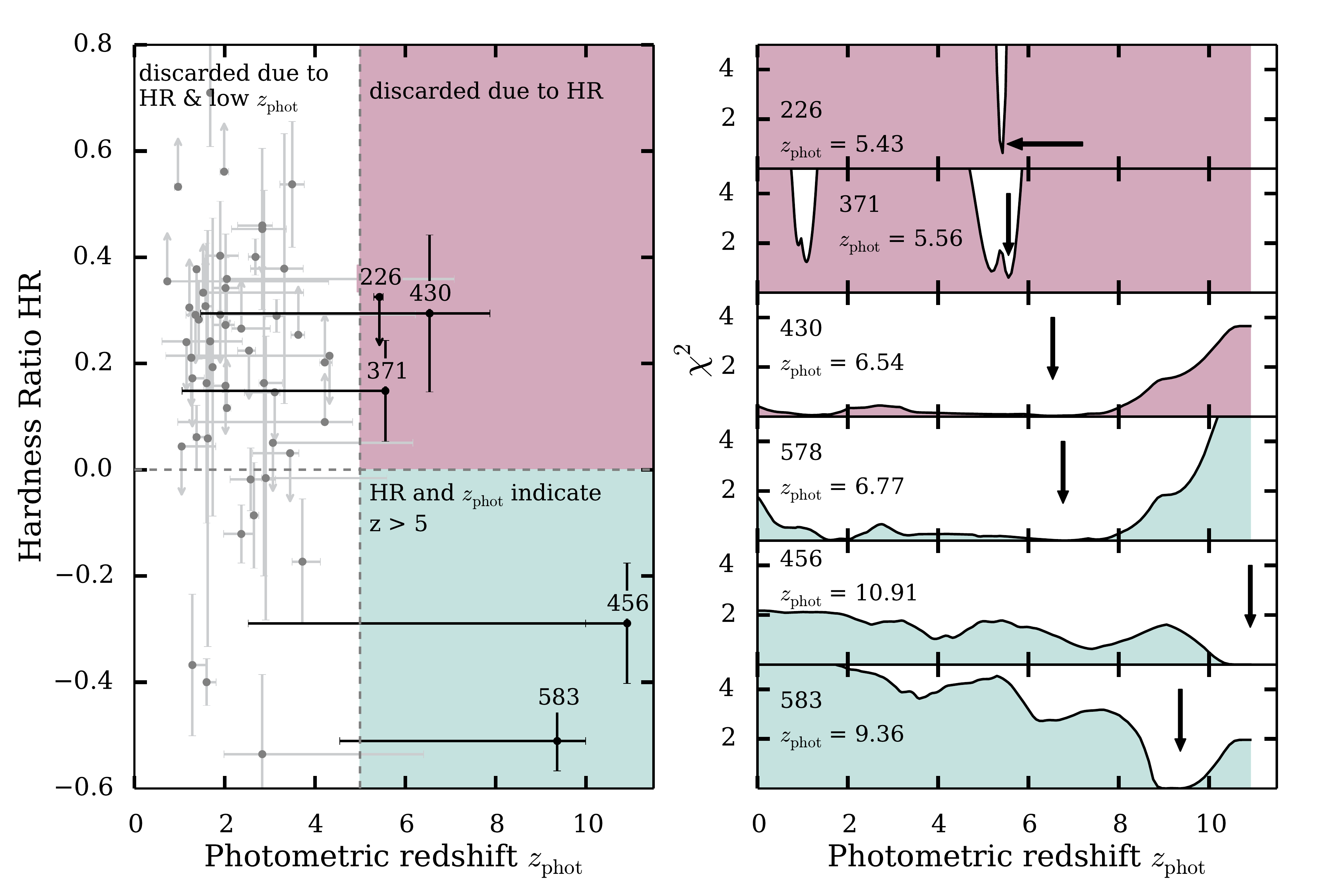}
\caption{\label{fig:HR_photz_chi2}After we combine our visual classifications with our $z_\mathrm{phot}$, stacking and colour criteria results, we are left with seven possible candidates. We show their Hardness Ratio and photometric redshift values in the $left$ panel. The $right$ figure illustrates their $z_\mathrm{phot}$ $\chi^2$ distributions. By also eliminating objects with HR > 0, we are left with three final candidates (456, 578, 583). The left panel does not show sources 496 and 578. These two objects are only detected in the full band and can therefore not be constrained by the Hardness Ratio. While we find a low $z_\mathrm{phot}$ for 496, 578 has $z_\mathrm{phot}=6.77$ and is therefore one of our final high $z$ candidates. We show $1\sigma$ error bars. For the X-ray counts and signal-to-noise ratios of each individual source see Table \ref{tab:counts}.}
\end{figure*}

We now combine our stacking, colour criteria and photometric redshift code results. We exclude objects with  $z_\mathrm{phot}<5$, $z_{\mathrm{stacking}}<5$ and $z_{\mathrm{colour}}<5$. Without the Hardness Ratio constraint six $z>5$ objects remain (Figure \ref{fig:HR_photz_chi2}). Neither stacking, colour criteria nor our visual classification contradict $z>5$ for these six sources (226, 371, 456, 578, 583, 430). For 371, 456, 578, 583 and 430 the $1\sigma$ error bars on $z_\mathrm{phot}$ are asymmetric and reach below $z\sim5$. The $\chi^{2}$ distributions (Figure \ref{fig:HR_photz_chi2}) also do not show clear global minima. For 430 and 578 the $\chi^2$ distributions are flat and allow for a wide range of lower redshift solutions. The $z_\mathrm{phot}$ solution for 456 is at the upper end of our allowed $z_\mathrm{phot}$ range ($z_\mathrm{phot}=$ 0 to 11). 371 shows additional minima at $z_\mathrm{phot}\sim1$. For 226 we determine an absolute magnitude of $M_H = -25.6$ assuming a photometric redshift of 5.43. 226 would therefore be extremely bright if it indeed was at $z \sim 5$ (see section \ref{sec:photoredcode}). There is thus only little evidence supporting the fact that these objects might be at high redshift. Only 226, 456 and 583 show global minima and are thus our most promising candidates.

\subsection{Our three final candidates}
\label{sec:final_three}

\begin{figure*}
\begin{minipage}{\textwidth}
\includegraphics[width=\textwidth]{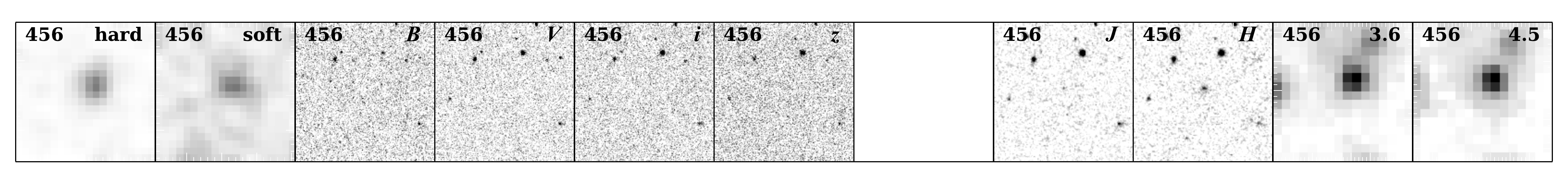}
\end{minipage}

\begin{minipage}{\textwidth}
\includegraphics[width=\textwidth]{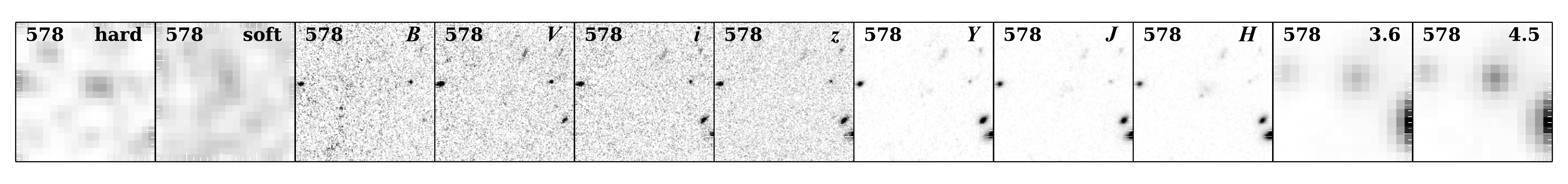}
\end{minipage}

\begin{minipage}{\textwidth}
\includegraphics[width=\textwidth]{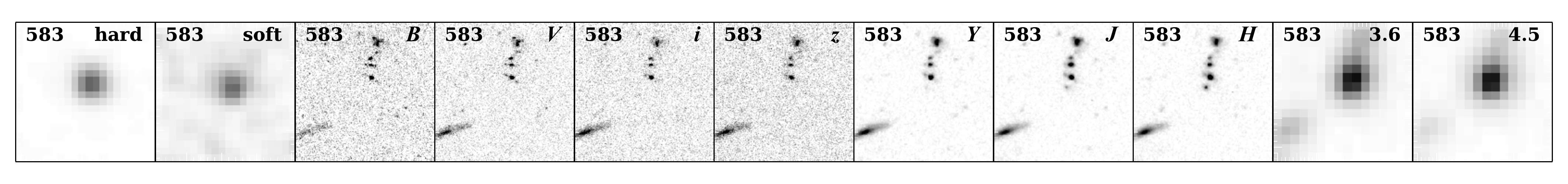}
\end{minipage}
\caption{\label{fig:candidates}Candidate $z > 5$ AGN that remain after combining all redshift tests. 456, 578 and 583 are the only sources that remain after we combine our stacking ($z_{\mathrm{stacking}}\sim7$ for all), colour criteria (all not classified), Hardness Ratio (456: $HR=-0.289$, 578: not classified, 583: $HR=-0.511$ ) and photometric redshift code results (456: $z_{\mathrm{phot}}=10.913$, 578: $z_{\mathrm{phot}}=6.766$, 583: $z_{\mathrm{phot}}=9.364$). Visually we classify 456, 578 and 583 as $z$ dropouts. Due to source confusion, we do not use the \Spitzer\ 3.6 and 4.5 micron images when running the photometric redshift code for these objects. All images are colour inverted and are $10^{\prime\prime}$ x $10^{\prime\prime}$ in size.}
\end{figure*}

We now also take the Hardness Ratio information into account and exclude three of the six remaining sources based on a positive HR value. After combining all of our redshift tests we are hence left with three final high-redshift candidates (456, 578, 583, Figure \ref{fig:candidates}). 456 and 583 have a negative Hardness Ratio whereas 578 cannot be constrained by HR since it is only detected in the full band.  

For 456 we find $z_\mathrm{phot} = 10.91^{-0.92}_{-8.39}$ ($\chi^2 \sim 0$) and HR = -0.29. For 578 we determine $z_\mathrm{phot} = 6.77^{+0.97}_{-5.10}$ ($\chi^2 \sim 0$). 583 has HR = -0.51 and $z_\mathrm{phot}= 9.36^{+0.63}_{-4.82}$ ($\chi^2 \sim 0$). 456, 578 and 583 could not be classified according to our colour criteria. Nonetheless, our visual classification ($z$ dropouts) and our stacking analysis ($z_{\mathrm{stacking}}\sim7$ for all) indicate that these sources might be $z > 5$ AGN.   

\subsection{Expected number densities}

We note that 456, 578 and 583 seem bright for $z\sim7$ and $z\sim10$ sources. We use UV rest-frame galaxy luminosity functions by \cite{Bouwens:2014aa} to quantify this statement. To justify our comparison to the galaxy, and not quasar luminosity function, we calculate $\alpha_\mathrm{ox}$ for our candidates. $\alpha_\mathrm{ox}$, the X-ray to optical-UV ratio, is defined as $\alpha_\mathrm{ox}=\log\big[{L_{2500\AA}/L_{2keV}}\big]/2.605$ (e.g. \citealt{Tananbaum:1979aa, Wilkes:1994aa, Vignali:2003aa, Steffen:2006aa}). \cite{Lusso:2010aa} analyse a sample of 545 X-ray selected Type 1 AGN from the XMM-COSMOS survey and find a mean $\alpha_\mathrm{ox}$ value of $\sim1.37$ and a weak redshift dependence out to $z\sim4$. We use the $H$-band and soft band flux values to calculate $\alpha_\mathrm{ox}$ for 456, 578 and 583. We find $\alpha_{\mathrm{ox}}=0.11$, $0.35$ and $-0.02$ for 456, 578 and 583, respectively. Assuming that $\alpha_\mathrm{ox}$ has no strong redshift dependence for $z>4$, this indicates that 456, 578 and 583 are not high-redshift quasars and justifies our comparison to the results by \cite{Bouwens:2014aa}.     

We estimate the number of sources similar to 456, 578 and 583 that we expect to find in our field. Since we only analyse objects for which GOODS/ACS and CANDELS data is available, we are only considering the area of the CANDELS deep and wide surveys (0.03 $\mathrm{deg}^2$) and not the entire CDF-S (0.11 $\mathrm{deg}^2$). Based on the surface density of $z\sim7$ and $z\sim10$ galaxies by \cite{Bouwens:2014aa}, we expect to find $0.14\pm0.14$ sources as bright as 456 at $z\sim10$, $3.72\pm0.80$ objects similar to 578 and less than $0.15$ $z\sim10$ objects as bright as 583. We hence expect high-redshift sources as bright as 456, 578 and 583 to be rare. 

\subsection{Deeper Y-band imaging for 583}
\label{sec:HUDF}
\begin{figure*}
\begin{minipage}{\textwidth}
\includegraphics[width = \textwidth]{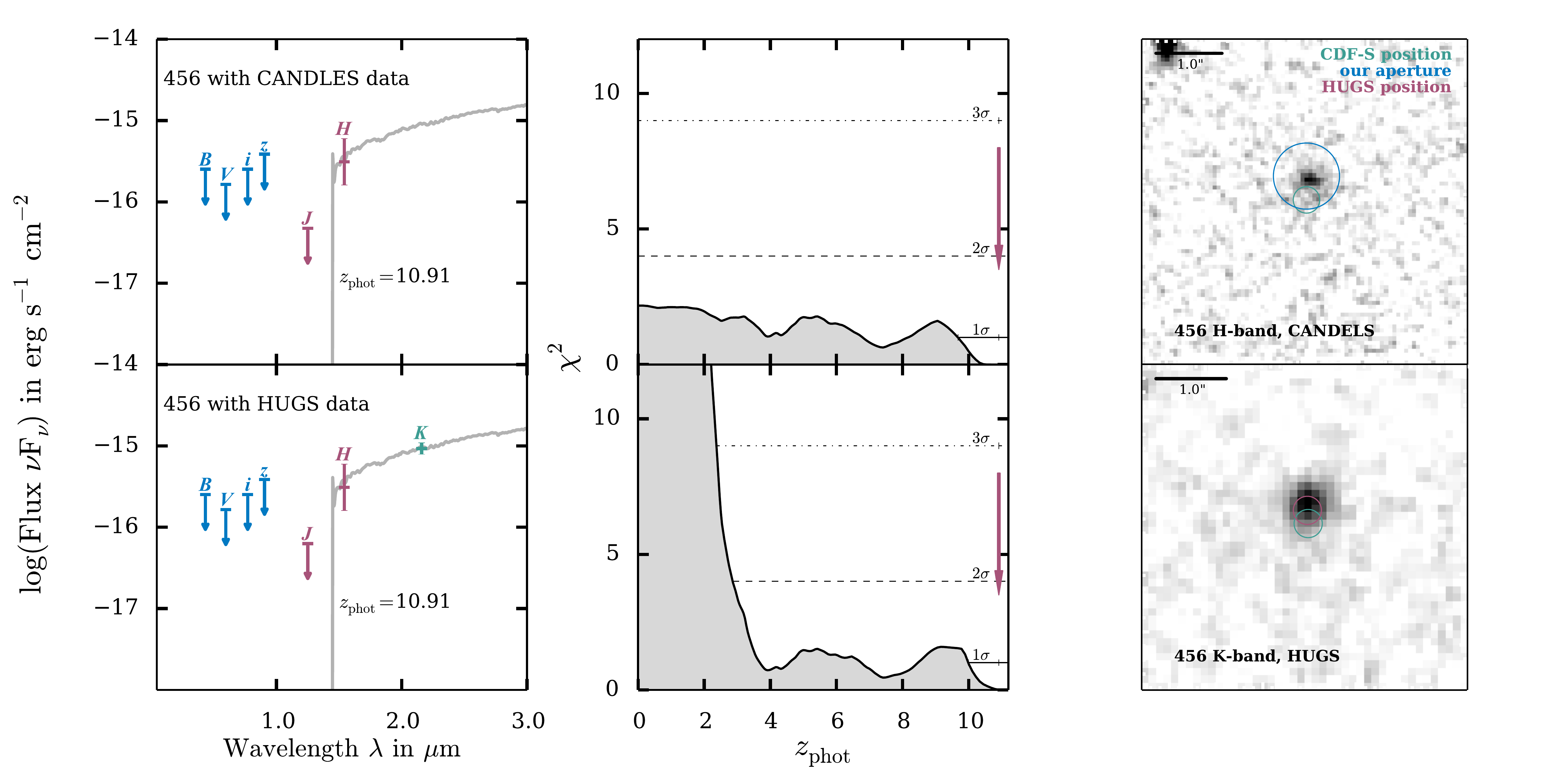}
\end{minipage}

\begin{minipage}{\textwidth}
\includegraphics[width = \textwidth]{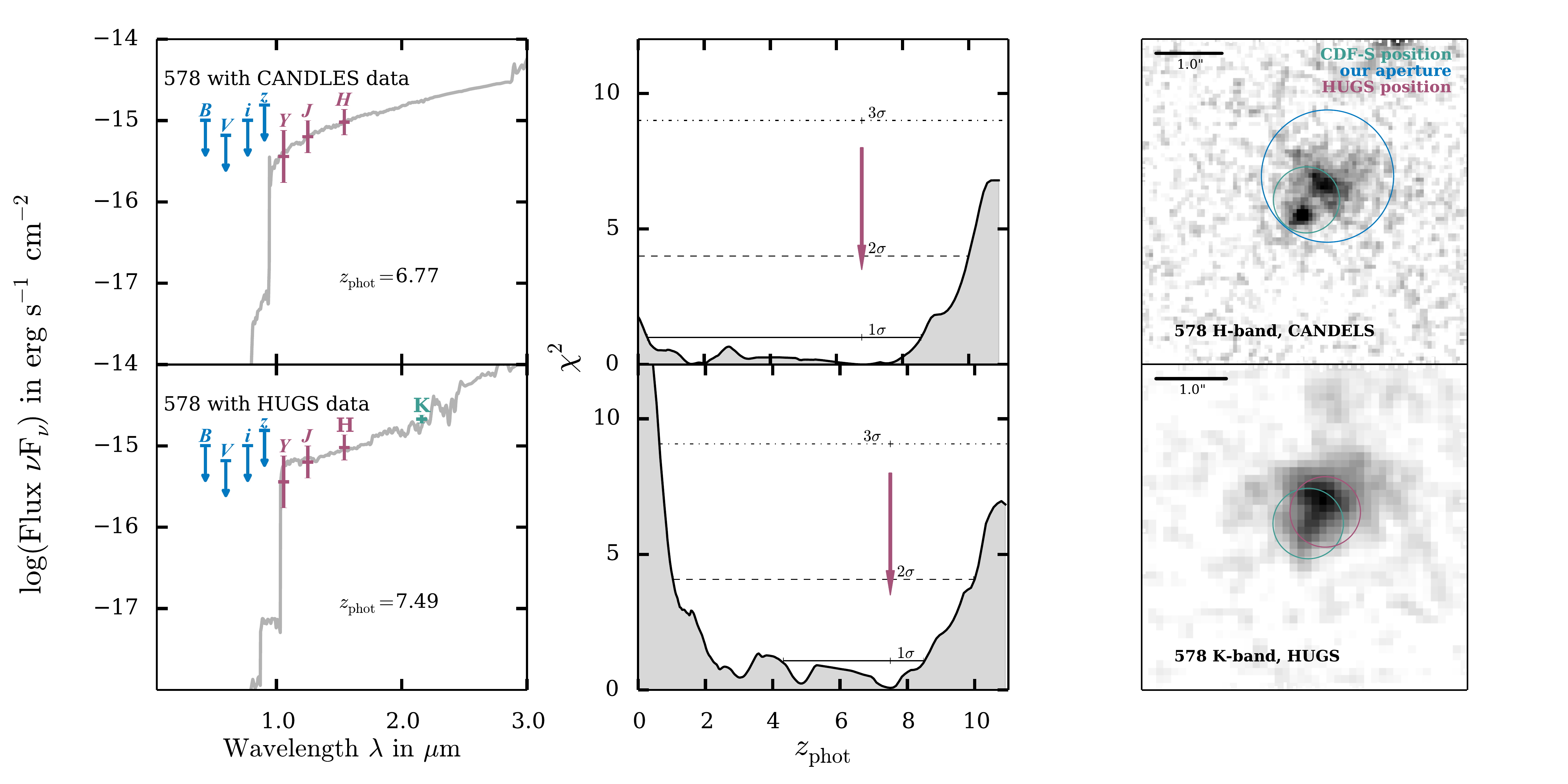}
\end{minipage}

\caption{\label{fig:456_578}Photometric redshift code results for objects 456 and 578 before ($top\ panels$) and after including the HUGS $K_S$-band ($bottom\ panels $). After combining all of our redshift tests, we are left with three final candidates (for 583 see Figure \ref{fig:583}). For these we include the Hawk-I UDS and GOODS Survey (HUGS) \protect\citep{Fontana:2014aa} $K_S$-band values when determining the photometric redshift. We show the best fit template SEDs ($left$), the $\chi^2$ distributions ($middle$) and the corresponding images ($right$), before and after including the $K_S$-band. For 456 we redetermine the $J$-band upper limit. Our result is close to the value reported by \protect\cite{Grogin:2011aa} and does not change the $z_\mathrm{phot}$ result significantly. For both, 456 and 578, a low redshift solution is allowed within the $2\sigma$ error. These two final candidates are hence likely to be at low redshift. We use the $\Delta\chi^2$ method under the assumption of one free parameter ($z_\mathrm{phot}$) to compute the $1, 2$ and $3\sigma$ error bars that are shown in the middle panels. All images are colour inverted and are $5^{\prime\prime} \times 5^{\prime\prime}$ in size.} 
\end{figure*}

\begin{figure*}
\begin{minipage}{\textwidth}
\includegraphics[width = \textwidth]{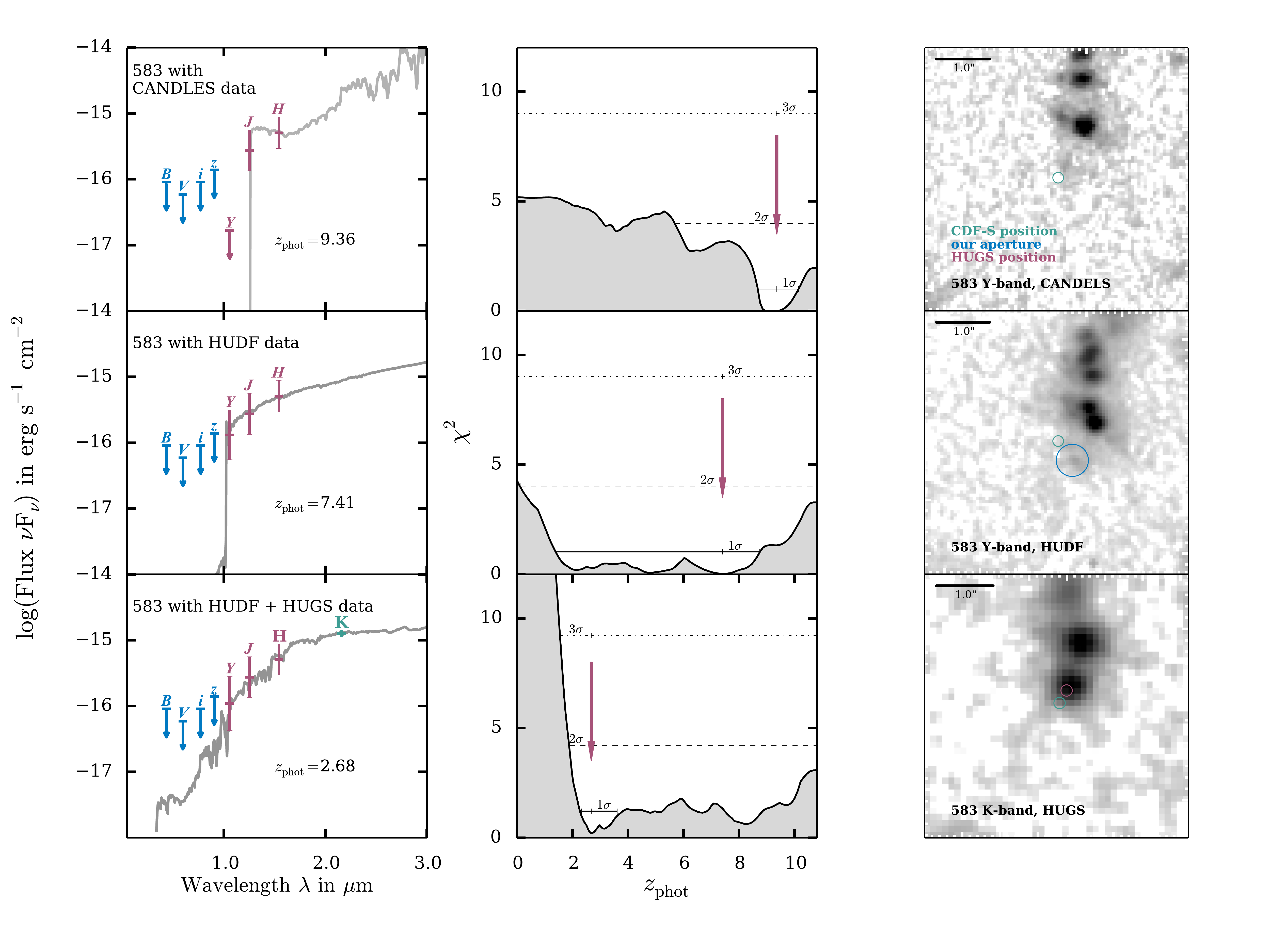}
\end{minipage}
\caption{\label{fig:583} Photometric redshift code results for object 583 before ($top\ panels$) and after including the HUDF $Y$-band ($middle\ panels$) and the HUGS $K_S$-band ($bottom\ panels$), analogue to Figure \ref{fig:456_578}. After combining all of our redshift tests, we are left with three final candidates (for 456 and 578 see Figure \ref{fig:456_578}). Fortunately, 583 is covered by the HUDF. We thus perform aperture photometry on the deep HST/WFC3 HUDF $Y$-band and replace the CANDELS $Y$-band value with this flux ($middle\ panels$). We also include the Hawk-I UDS and GOODS Survey (HUGS) \protect\citep{Fontana:2014aa} $K_S$-band value when determining the photometric redshift ($bottom\ row$). With the HUDF $Y$-band and the HUGS $K_S$-band 583's photometric redshift value drops to $2.68$. It is hence most likely a low-redshift source. We show the best fit template SEDs ($left$), the $\chi^2$ distributions ($middle$) and the corresponding images ($right$). We use the $\Delta\chi^2$ method under the assumption of one free parameter ($z_\mathrm{phot}$) to compute the shown $1, 2$ and $3\sigma$ error bars. All images are colour inverted and are $5^{\prime\prime} \times 5^{\prime\prime}$ in size.}
\end{figure*}
  
583, one of our three final candidates, has a high photometric redshift ($z_\mathrm{phot} = 9.36$) and shows a clear global minimum in the $\chi^2$ distribution. In terms of the $\chi^2$ distribution it is thus our most promising candidate. Fortunately, 583 is part of the $Hubble$ Ultra Deep Field (HUDF) \citep{Beckwith:2006aa}. We are especially interested in deeper $Y$-band imaging since 583's high photometric redshift hinges on the upper limit in this band. We show our results in Figure \ref{fig:583}. The top panels illustrate, that the upper limit in the $Y$-band, for which we used the official sensitivity limit given by \cite{Grogin:2011aa}, is very low and thus forces a strong break in 583's SED. We hence combine all available HST/WFC3 HUDF $Y$-band images (\citealt{Koekemoer:2013aa}, HST Program ID 12498, PI: R. Ellis; \citealt{Illingworth:2013aa}, HST Program ID 11563, PI: G. Illingworth, HST Program ID 12099, PI: A. Riess) using the HST DrizzlePac package. We measure the flux of 583 using simple aperture photometry and a $0.3^{\prime\prime}$ radius aperture (middle panels) and rerun EAZY. We use the $\Delta\chi^2$ method under the assumption of one free parameter ($z_\mathrm{phot}$) to compute the $1, 2$ and $3\sigma$ error bars. We still determine a high photometric redshift of $z_\mathrm{phot}=7.41$, a low redshift solution at $z_\mathrm{phot}\sim2$ is however allowed within $1\sigma$. 583 is hence unlikely to be at high redshift.  

\subsection{HUGS $K_S$-band data}
\label{sec:HUGS}

To gain more reliable photometric redshift values for all three of our final candidates, we now also take VLT/Hawk-I $K_S$-band data into account. We take the $K_S$-band flux values for 456, 578 and 583 from the Hawk-I UDS and GOODS Survey (HUGS) catalog (v1.1) \citep{Fontana:2014aa} and rerun EAZY. Our results are illustrated in Figures \ref{fig:456_578} and \ref{fig:583}. The photometric redshift value for 456 remains at $10.91^{-0.92}_{+8.39}$ ($\chi^2 \sim 0$). However, the $z_\mathrm{phot}$ value seems highly dependent on the $J$-band upper limit. We thus redetermine the $J$-band sensitivity limit by measuring the background flux within 35 apertures which are scattered across the CANDELS field. We determine a flux limit close to the value reported by \cite{Grogin:2011aa} (measured: $2.62 \times 10^{-2} \mu$Jy, reported: $1.97 \times 10^{-2} \mu$Jy for a $0.5^{\prime\prime}$ radius aperture). We rerun EAZY and find a value almost identical to the previous result ($z_\mathrm{phot} = 10.91^{-1.14}_{-7.11}$, $\chi^2 = 0.01$). 456 thus remains a source with a high photometric redshift. The $2\sigma$ error bar, computed through the $\Delta\chi^2$ method, does however allow for a low-redshift solution at $z\sim4$. For 578 the $K_S$-band causes the photometric redshift to increase from 6.77 to $7.49^{+0.54}_{-4.60}$ ($\chi^2 = 0.08$). Nonetheless, similar to 456, 578's $2\sigma$ error bar permits a $z\sim3$ solution. For 583 we determine $z_\mathrm{phot} = 2.68^{+5.61}_{+0.37}$ ($\chi^2=0.21$) by including the HUDF deep $Y$-band and the HUGS $K_S$-band flux values. This $z_\mathrm{phot}$ value matches what has previously been reported by \cite{Szokoly:2004aa} for the galaxy next to 583 ( 53.1833, -27.7764). We thus suspect, that 583 might be part of a large clumpy galaxy at low redshift \citep{Schawinski:2011aa}. With the X-ray emission being offset from what could be the main galaxy, this does remain a very interesting object. 

In summary, the HUGS $K_S$-band and HUDF $Y$-band data causes the photometric redshifts of our three final candidates to either drop to low-redshift or to allow for a low-redshift solution within a $2\sigma$ error bar. Considering these photometric redshift code results and how rare objects as bright as our candidates should be at high-redshift, we conclude that 456, 578 and 583 are unlikely to be at high redshift. They do however remain compelling candidates for follow-up observations. 456 remains a very interesting candidate since the $1\sigma$ $z_\mathrm{phot}$ error bar does not allow for lower redshift solutions. 578 could still be at high redshift and 583 is intriguing since the X-ray emission seems to be offset from the main galaxy. 

\section{Discussion}
\label{sec:discussion}

\begin{figure*}
\begin{centering}
\includegraphics[width=\textwidth]{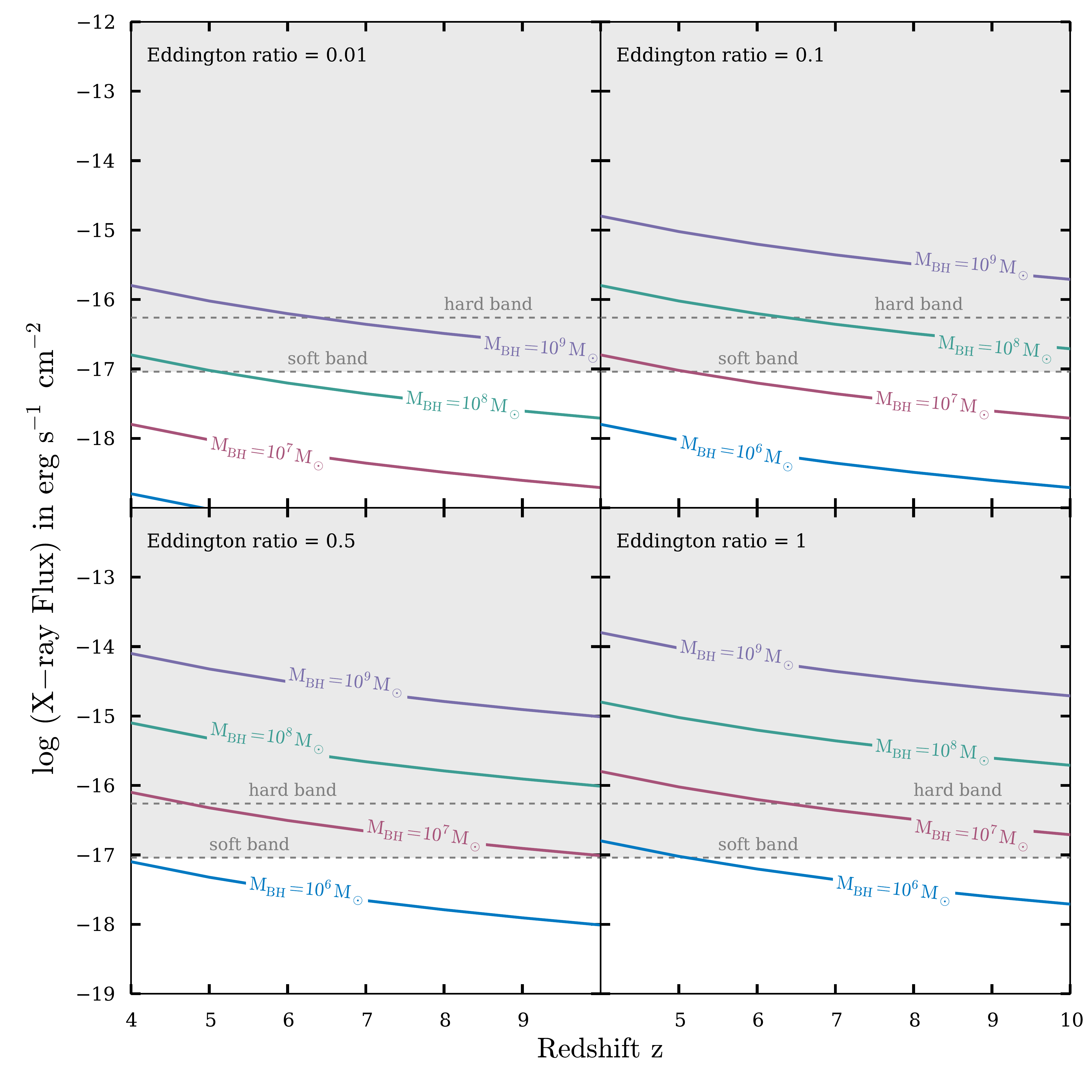}
\end{centering}
\caption{\label{fig:BHmasses}Expected X-ray flux as a function of BH mass, Eddington ratio and redshift. We assume spherical accretion, \NH\ $=10^{22}\mathrm{cm}^{-2}$ and a bolometric correction of $k_\mathrm{corr}=25$ to calculate the expected X-ray flux for BH masses ranging from $10^{6}\ M_{\odot}$ to $10^{9}\ M_{\odot}$, Eddington ratios between 0.01 and 1. and redshifts between $z=4$ to $z=10$. The dashed lines show the \Chandra\ 4-Ms flux limits in the soft, full and hard bands. The BH mass in a typical Lyman Break Galaxy at $z\sim5$ is $\sim10^{7}$\Msun. This figure illustrates that we are sensitive enough to detect such sources if we assume low obscuration and an Eddington ratio > 0.1. }
\end{figure*}

\subsection{Sensitivity and AGN number density}
In this section we show that our analysis should have been sensitive enough to detect an active BH in a typical high-redshift Lyman Break Galaxy and why we expected to find at least some high $z$ AGN. Furthermore, we determine an upper limit on the AGN number density. 

In a first step we determine the BH masses and accretion rates we are sensitive to by calculating the expected X-ray flux as a function of BH mass, Eddington ratio  and redshift. We calculate the X-ray luminosity and translate it to observed flux. We use Eddington ratios between 0.01 and 1. and a redshift range from $z=4$ to $z=10$. For simplicity we assume a constant bolometric correction of $k_\mathrm{corr}=25$ ($L_\mathrm{bol}=k_\mathrm{corr} L_\mathrm{X}$, \citealt{Vasudevan:2009aa}) and \NH\ $=10^{22}\mathrm{cm}^{-2}$. Figure \ref{fig:BHmasses} illustrates our results. We also show the \Chandra\ 4-Ms flux limits which lie at $9.1\times10^{-18}\ \mathrm{erg}\ \mathrm{s}^{-1}\ \mathrm{cm}^{-2}$ for the soft band and at $5.5\times10^{-17}\ \mathrm{erg}\ \mathrm{s}^{-1}\ \mathrm{cm}^{-2}$ for the hard band \citep{Xue:2011aa}. At $z\sim5$ we are sensitive to luminosities as low as $\sim10^{42}\mathrm{erg}\ \mathrm{s}^{-1}$ in the soft band and $\sim10^{43}\mathrm{erg}\ \mathrm{s}^{-1}$ in the hard band.

In the GOODS field, the typical stellar mass of a Lyman Break Galaxy at $z\sim5$ is $2.82\times10^{10}$\Msun\ \citep{Lee:2012aa}. If we use the local BH-stellar mass relation ($M_{\star, \mathrm{total}}{/}$ \MBH\ $=562$, \citealt{Jahnke:2009aa}) to determine the corresponding BH mass, we find \MBH\ $=5\times10^{7}$\Msun. Figure \ref{fig:BHmasses} shows that, assuming low obscuration and Eddington ratios $>0.1$, we should have been capable of detecting such AGN at $z>5$.

To estimate the number of high $z$ AGN in the CDF-S we use the results by \cite{Bouwens:2014aa}. They find 680 $z\sim5$, 252 $z\sim6$ and 113 $z\sim7$ Lyman Break Galaxies in the CANDELS/Deep and CANDELS/Wide surveys for the GOODS-S field. \cite{Nandra:2002aa} report an AGN fraction of $3\%$ for $z\sim3$ Lyman Break Galaxies in the Hubble Deep Field North. Assuming that this fraction does not evolve with redshift, we would expect the volume we looked at to contain $\sim20$ AGN at $z\sim5$, $\sim8$ at $z\sim6$ and $\sim3$ at $z\sim7$. Since we are sensitive enough to detect Compton-thin AGN in typical Lyman Break Galaxies, we would have expected to find at least some convincing high-redshift sources. Even if all of our three final candidates (456, 578, 583) would prove to be at high-redshift, the number of high-redshift AGN in our field would still be lower than expected. 

We use the estimated number of high redshift AGN to determine an upper limit on the AGN number density. The combined CANDELS/Wide and CANDELS/Deep survey covers an area of 0.03 $\mathrm{deg}^{2}$. At $z\sim5$ the faint X-ray selected AGN number density is hence fewer than 655 AGN per $\mathrm{deg}^{2}$. At $z\sim6$ and $z\sim7$ we would expect to find fewer than 262 and 98 AGN per $\mathrm{deg}^{2}$, respectively. Here, we assumed the $z\sim3$ Lyman Break Galaxy AGN fraction by \cite{Nandra:2002aa}. This AGN fraction could however be especially low for high redshift Lyman Break Galaxies or generally evolve with redshift. We discuss this in more detail in Section \ref{sec:explanations}.

\subsection{Comparison to existing work}

We compare our results to \cite{Vito:2013aa}. They found three $z > 5$ AGN (139, 197, 485) in the CDF-S. Source 139 has a photometric redshift of $z_\mathrm{phot}$ = 5.73 based on \cite{Luo:2010aa}. Since 139 is not covered by $B$, $Y$, $J$ and $H$ it was not part of our analysis. Object 197 has $z_\mathrm{phot}$ = 6.07 with a secondary solution at $z_\mathrm{phot}$ = 4.39 \citep{Luo:2010aa}. 197 is not in the GOODS field and was therefore immediately excluded by us. Source 485 has a photometric redshift of $z_\mathrm{phot}$ = 7.62 with a secondary solution at $z_\mathrm{phot}$ = 3.31 based on  \cite{Luo:2010aa} and has $z_\mathrm{phot}$ = 4.42 according to \cite{Santini:2009aa}. Both, \cite{Luo:2010aa} and \cite{Santini:2009aa}, did not take CANDELS WFC3/IR data into account when determining photometric redshifts. Instead they used the GOODS - MUSIC catalog \citep{Grazian:2006aa} which contains the VLT/ISAAC $J$, $H$, and $K$ bands. Taking CANDELS and \Spitzer\ data into account, we determined $z_\mathrm{phot} = 2.83^{+0.54}_{-0.68}$ for object 485.

\cite{Vito:2013aa} also report spectroscopic and photometric redshifts for 31 additional AGN at lower redshifts. Seven of these 31 sources are also part of our sample. For three objects our photometric redshifts lie within $1\sigma$ from the redshift reported by \cite{Vito:2013aa} (331: $z_\mathrm{phot,\/Vito} = 3.780$, $z_\mathrm{phot} = 4.32^{+0.02}_{-3.63}$; 371: $z_\mathrm{phot,\/Vito} = 3.10$, $z_\mathrm{phot} = 5.56^{+0.04}_{-4.50}$; 546: $z_\mathrm{spec,\/Vito} = 3.06$, $z_\mathrm{phot} = 3.15^{+0.18}_{-0.25}$). For four sources our photometric redshifts do not lie within $1\sigma$ (150: $z_\mathrm{phot,\/Vito} = 3.34$, $z_\mathrm{phot} = 3.63^{+0.14}_{-0.16}$; 403: $z_\mathrm{spec,\/Vito} = 4.76$, $z_\mathrm{phot} = 4.22^{+0.17}_{-0.12}$; 556: $z_\mathrm{phot,\/Vito} = 3.53$, $z_\mathrm{phot} = 2.68^{+0.17}_{-0.16}$; 651: $z_\mathrm{phot,\/Vito} = 4.66$, $z_\mathrm{phot} = 1.90^{+0.41}_{-0.34}$). The remaining 24 low-redshift AGN were not part of our sample because they were not covered by enough bands ($B$, $V$, $i$, $z$, $J$, $H$), because they were clearly visible in all bands and therefore discarded as $z<5$ sources or because their images were disturbed by artefacts.     
Except for source 485 and the four low-redshift AGN (150, 403, 556, 651) our findings do hence agree with the results by \cite{Vito:2013aa}.

\cite{T13} searched for X-ray emission of $z = 6 - 8$ Lyman Break dropout and photometrically selected sources.  None of the $\sim600$ $z\sim6$, $\sim150$ $z\sim7$ or $\sim80$ $z\sim8$ sources could be detected individually in the X-rays. Stacking the X-ray data in redshift bins did not generate a significant detection either. In the stacks the $3\sigma$ upper limits on the X-ray emission lay at $\sim10^{41}\mathrm{erg}\ \mathrm{s}^{-1}$ (soft) and $\sim10^{42}\mathrm{erg}\ \mathrm{s}^{-1}$ (hard) for the $z\sim6$ and $z\sim7$ bins. Assuming a bolometric correction of $k_\mathrm{corr}=25$ \citep{Vasudevan:2009aa} and an Eddington ratio of 0.1, these upper limits correspond to $\sim10^{5}$ \Msun\ (soft) and $\sim10^{6}$ \Msun\ (hard) in terms of BH mass. 
In comparison to our work, \cite{T13} based their search on a sample of optically selected sources, whereas we selected our objects in the X-rays. Nonetheless, the results are in agreement. 

\subsection{$z\gtrsim5$ AGN in the CDF-S}

In this work we searched for possible $z > 5$ AGN candidates in the CDF-S. In contrast to \cite{T13} we started out with a sample of confirmed X-ray sources. We used visual classification, colour criteria, stacking, a photometric redshift code and the Hardness Ratio to obtain multiple redshift indications. After dismissing sources for which our redshift tests indicated $z < 5$, three final candidates with $z_\mathrm{phot} \sim 7$ (578), $z_\mathrm{phot} \sim 9$ (583) and $z_\mathrm{phot} \sim 10$ (456) remained in our sample. Our comparison to the galaxy luminosity function showed that $z\sim7$, $z\sim9$ and $z\sim10$ objects as bright as 578, 583 and 456 are rare. By including the Hawk-I UDS and GOODS Survey (HUGS) $K_S$-band \citep{Fontana:2014aa} and the HUDF $Y$-band data, the photometric redshifts for our three final candidates either dropped to low-redshift (583) or allowed for a low-redshift solution within $2\sigma$ error bars (578, 583). Our three final candidates did hence not pass this extended redshift test. We conclude that considering our photometric redshift code results and the rarity of such high-redshift objects, 456, 578 and 583 are most likely low-redshift sources. We also found five low significance objects. These sources are detected in the X-rays, but they do not seem to possess a counterpart in the optical or infrared (including the \Spitzer/IRAC 3.6 and 4.5 micron channels and the VLT/Hawk-I $K_S$-band). The currently available data does not allow us to constrain their redshifts or determine if these objects are spurious detections.  

Based upon currently available GOODS/ACS, CANDELS and \Spitzer\ data, the analysis did therefore yield three final $z > 5$ candidates and five low significance objects for this deep, but narrow field. Including HUGS and HUDF data did however show, that 456, 578 and 583 are likely to be at low redshift. Follow-up observations are necessary to gain more reliable redshifts for our three final candidates and to constrain the nature of our five low significance objects. 

\subsection{Possible explanations}
\label{sec:explanations}

Both, the approach by \cite{T13} and our approach, assumed that X-ray emission is a valid tracer for BH growth. If at high redshift BHs primarily grow through BH mergers instead of accretion, electromagnetic radiation might not be emitted during the growth process. X-ray emission might hence not be a indicator for active BHs.

At high redshift the number of actively accreting BHs could also be generally low. This could be caused by a low BH occupation fraction, a low AGN fraction or BH growth through short, super-Eddington episodes. We stress the difference between the BH occupation fraction and the AGN fraction since they describe different scenarios. 

If the BH occupation fraction is low only very few haloes are seeded with BHs. Our sample could therefore be too small to not only contain a BH, but to contain a BH that is also actively accreting. 

Even if the BH occupation fraction is high, the AGN fraction could still be low. So, even if there are plenty of BHs in our field, only few of them could be active. For instance, BHs at high redshift could only grow in optically faint galaxies. In our analysis these faint galaxies could correspond to the low significance objects that do not seem to possess an optical or infrared counterpart (190, 280, 333, 384, 643, see Section \ref{sec:LBT}). The currently available data does not allow us to investigate these sources further. We are however hopeful that the forthcoming 7-Ms survey for the CDF-S (PI: William Brandt, Proposal ID: 15900132) will show if these objects are real.  

BHs could also grow through short, super-Eddington accretion phases \citep{Madau:2014aa, Volonteri:2014aa}. \cite{Madau:2014aa} illustrated that a duty cycle of $20\%$ is enough to grow a  $100$ \Msun\ non-rotating seed BH into a $\sim10^9$ \Msun\ BH by $z\sim7$. This could, for instance, be realized through five $20\ \mathrm{Myr}$ long growth episodes with $\dot{m}/\dot{m}_\mathrm{Edd}=4$, each followed by a $100\ \mathrm{Myr}$ phase of quiescence. The \cite{T13} and our sample, could thus not contain any BHs that are actively accreting at the time of observation. 

Simulations suggest that the BH occupation fraction should be high enough for our field to contain high-redshift BHs. \cite{Menou:2001aa} ran Monte Carlo simulations of the merger history of dark matter haloes. They showed that to reproduce the local BH distribution, $>3\%$ of the $M_\mathrm{halo}\gtrsim10^8$ \Msun\ haloes should be seeded with BHs at $z\sim5$.  

 \cite{Bellovary:2011aa} ran SPH+$N$-body simulations in which only the local gas properties, such as density, temperature and metallicity, influence the BH formation and evolution. They showed that the BH occupation fraction is halo mass dependent. At $z\sim5$ they found a BH occupation fraction of $\sim50\%$ for $10^8$ \Msun\ $<M_\mathrm{halo}<10^{9}$ \Msun\ haloes and a fraction of $100\%$ for $M_\mathrm{halo}>10^{9}$ \Msun\ haloes. 

\cite{Alexander:2014aa} presented a sophisticated model in which a BH seed is being fed by dense cold gas flows while it is embedded in a nuclear star cluster. This can lead to supra-exponential accretion and could explain how a light ($\sim10$ \Msun) Pop III remnant  BH seed could grow into a $\gtrsim10^4$ \Msun\ seed within $\sim10^7$ years. Nevertheless, these $\gtrsim10^4$ \Msun\ seeds still need to grow into the massive $10^9$ \Msun\ sources that we find at $z>6$. This most likely happens via Eddington limited accretion. The distribution of high-redshift quasars that we observe at $z>6$ can be reproduced if the supra-exponential accretion and the subsequent Eddington-limited growth work efficiently in at least $1-5\%$ of the dark matter haloes.     

\cite{Stark:2009aa}, \cite{Vanzella:2009aa}, \cite{Wilkins:2010aa}, \cite{Bouwens:2014aa}, \cite{Duncan:2014aa}  and many more have shown that the GOODS-south field contains hundreds of $z\gtrsim5$ Lyman Break Galaxies. These high-redshift sources should have passed our manual inspection, the colour criteria, the stacking and the photometric redshift measurement. So, according to the simulations and the number of high-redshift Lyman Break Galaxies in our field, we would expect our sample to include high-redshift BHs. The number of actively accreting BHs could however still be low. For instance, it would be possible that only the most massive haloes host AGN.  

Finding one or more high-redshift AGN would have opened up the window to the early BH growth era. \cite{Treister:2011aa} showed that the massive and luminous quasars we observe at $z>6$ are most likely not representative of the entire high-redshift BH population. Such quasars are rare and presumably only constitute the high mass end of the entire BH population \citep{Volonteri:2010aa}. At $z\sim6$ we expect to find only $\sim2$ in a 1000 $\mathrm{deg}^{2}$ field \citep{Fan:2001aa, Fan:2000aa}. Furthermore, these objects only allow us limited insight into seed formation scenarios. For typical seed masses $\lesssim10^{6}$\Msun, these objects have to undergo multiple Salpeter times \citep{Salpeter:1964aa} to reach $M_\mathrm{BH}\sim10^9$ \Msun. By the time we observe them as quasars, all initial seed information will be lost. We are hence especially interested in the population of more abundant, less massive, less luminous AGN that will end up in galaxies similar to the Milky Way. If our analysis had yielded a convincing high-redshift AGN candidate, we would have been able to put first constraints on this more representative and revealing BH population.         

To constrain the explanations for our results and to gain further insights into BH formation and growth at high redshift, this analysis needs to be repeated for a larger sample. Especially constraining the BH occupation fraction and the short, super-Eddington growth scenario requires a larger field. The new field does not need to be deeper, but wider than the $0.11\ \mathrm{deg}^{2}$ CDF-S \citep{Luo:2008aa}. The 2.8-Ms \Chandra\ COSMOS Legacy Survey, for which the data will be available soon, covers a $2.2\ \mathrm{deg}^{2}$\ area \citep{Civano:2014aa}. Not being as deep as the 4-Ms \Chandra\ data, the \Chandra\ COSMOS Legacy Survey will probe a slightly different parameter space (see \citealt{Treister:2011aa} for an illustration of the high-redshift number density that is necessary for an individual detection). Nonetheless, it will provide data for a much wider field and will thus enable us to repeat this analysis for a larger sample. ATHENA, which is meant to be launched in 2028, will allow us to constrain the BH occupation fraction to new accuracy. With its X-ray Wide Field Imager, ATHENA is meant to detect over 400 $z = 6 -8$ and over 30 $z > 8$ X-ray selected active BHs \citep{Nandra:2013aa,Aird:2013aa}. JWST data could help to investigate the nature of our five low significance objects.

\section{Summary}
\label{sec:summary}

We searched for $z\gtrsim5$ AGN in the \Chandra\ Deep Field South (CDF-S). We used the \Chandra\ 4-Ms catalog and combined it with GOODS/ACS, CANDELS/WFC3 and \Spitzer/IRAC data. Our main sample contained 58 sources. We ran a photometric redshift code, stacked the GOODS/ACS data, applied colour criteria and the Lyman Break Technique. Furthermore, we used the X-ray Hardness Ratio as a redshift indicator. After combining all redshift tests, three final $z\gtrsim5$ AGN candidates remained. Redetermining their photometric redshifts with additional VLT/Hawk-I HUGS $K_S$-band and HST/WFC3 HUDF $Y$-band data showed, that they are most likely low-redshift sources. We also found five sources that are detected in the X-rays, but that do not seem to possess a counterpart in the optical or infrared (low significance objects). The currently available data did not allow us to determine if these objects are possible high-redshift AGN candidates, spurious detections or optically faint low-redshift sources. Assuming that our three final candidates are indeed low-redshift sources and that our five low significance objects are either spurious detections or also at low redshift, we concluded that the CDF-S does not contain any high-redshift AGN. We also showed that we should have been able to detect active BHs in typical $z\sim5$ Lyman Break Galaxies and why we would have expected to find at least some high $z$ AGN. 
Our results could be explained by:    
\begin{itemize}
\item a low BH occupation fraction or a low AGN fraction. If at high redshift only very few haloes contain a BH or only very few BHs are actively accreting, our sample could be too small to contain an AGN.   
\item BH growth via short, super-Eddington growth modes. If BHs primarily grow through short accretion episodes, the number of actively accreting BHs in our sample might be zero.
\item BH growth in optically faint galaxies. Our five low significance objects could indicate that high-redshift AGN primarily grow in galaxies that we do not detect in the optical. We were however unable to constrain their redshifts and noted that these could be spurious detections.  
\item BH growth via BH-BH mergers. If at high $z$ BHs primarily grow through mergers instead of accretion, X-rays might not trace BH growth.  
\end{itemize}

\begin{table*}
\tiny
	\begin{center}
		\begin{tabular}{lllllllllllllll}
		\toprule
			{ } & {ID} & {$\mathrm{RA}$} & {$\mathrm{DEC}$} & {$\mathrm{RA}_{H}$} & {$\mathrm{DEC}_{H}$} & {visual} & {stacking} & {colour} & {Hardness} & {$\sigma_{\mathrm{HR}}$} & {$z_\mathrm{phot}$} & {$\sigma_{\mathrm{phot},{+}}$} & {$\sigma_{\mathrm{phot},{-}}$} &{$\chi^2$}\\
			{ } & { } & {(\Chandra)} & {(\Chandra)} & { } & { } & { } & { } & {criteria} & {Ratio (HR) } & { } & { } & { }\\ 
			\midrule
			$*$ & $121$ & $53.0268$ & $-27.7653$ & $53.0267$ & $-27.7653$ & $4$ & $4$ & $0$ & $0.21^{l}$ & $-$ & $1.26$ & $0.64$ & $0.04$ & $14.31$\\
			$*$ & $150$ & $53.0400$ & $-27.7985$ & $53.0398$ & $-27.7985$ & $4$ & $4$ & $0$ & $0.25^{l}$ & $-$ & $3.63$ & $0.14$ & $0.16$ & $1.66$\\
			$ $ & $173$ & $53.0477$ & $-27.8351$ & $53.0477$ & $-27.8351$ & $4$ & $4$ & $0$ & $-0.37$ & $0.13$ & $1.28$ & $0.30$ & $0.03$ & $4.11$\\
			$*$ & $184$ & $53.0523$ & $-27.7748$ & $53.0522$ & $-27.7747$ & $4$ & $4$ & $0$ & $0.19$ & $0.28$ & $1.73$ & $0.05$ & $0.05$ & $2.19$\\
			$*$ & $189$ & $53.0546$ & $-27.7931$ & $53.0544$ & $-27.7931$ & $4$ & $4$ & $0$ & $0.31^{l}$ & $-$ & $1.57$ & $0.17$ & $0.07$ & $9.34$\\
			$*$ & $199$ & $53.0579$ & $-27.8336$ & $53.0579$ & $-27.8336$ & $6$ & $7$ & $0$ & $-0.09$ & $0.10$ & $2.65$ & $0.11$ & $0.08$ & $16.61$\\
			$ $ & $211$ & $53.0620$ & $-27.8511$ & $53.0625$ & $-27.8508$ & $4$ & $4$ & $0$ & $0.06$ & $0.06$ & $1.38$ & $0.29$ & $0.07$ & $11.58$\\
			$*$ & $217$ & $53.0639$ & $-27.8438$ & $53.0638$ & $-27.8433$ & $0$ & $7$ & $0$ & $-0.17$ & $0.12$ & $3.72$ & $0.40$ & $0.23$ & $38.86$\\
			$ $ & $221$ & $53.0657$ & $-27.8790$ & $53.0660$ & $-27.8787$ & $4$ & $4$ & $0$ & $0.12^{l}$ & $-$ & $2.05$ & $0.09$ & $0.06$ & $8.84$\\
			$ $ & $226$ & $53.0668$ & $-27.8166$ & $53.0668$ & $-27.8165$ & $4$ & $4$ & $6$ & $0.33^{u}$ & $-$ & $5.43$ & $0.08$ & $0.12$ & $0.64$\\
			$*$ & $242$ & $53.0716$ & $-27.7699$ & $53.0713$ & $-27.7696$ & $4$ & $4$ & $0$ & $0.16$ & $0.26$ & $1.60$ & $0.06$ & $0.04$ & $16.75$\\
			$*$ & $244$ & $53.0721$ & $-27.8190$ & $53.0717$ & $-27.8187$ & $6$ & $7$ & $6$ & $0.29^{u}$ & $-$ & $1.90$ & $4.32$ & $0.22$ & $2.15$\\
			$ $ & $258$ & $53.0766$ & $-27.8641$ & $53.0766$ & $-27.8644$ & $6$ & $6$ & $0$ & $0.09^{l}$ & $-$ & $4.22$ & $0.62$ & $3.26$ & $12.49$\\
			$*$ & $273$ & $53.0821$ & $-27.7673$ & $53.0826$ & $-27.7681$ & $6$ & $7$ & $0$ & $0.03^{u}$ & $-$ & $3.45$ & $0.20$ & $0.83$ & $1.01$\\
			$ $ & $296$ & $53.0907$ & $-27.7825$ & $53.0913$ & $-27.7820$ & $6$ & $7$ & $6$ & $0.56^{l}$ & $-$ & $1.99$ & $0.10$ & $0.09$ & $12.71$\\
			$ $ & $301$ & $53.0924$ & $-27.8033$ & $53.0918$ & $-27.8028$ & $4$ & $4$ & $0$ & $0.34$ & $0.10$ & $2.02$ & $0.30$ & $0.29$ & $6.78$\\
			$*$ & $302$ & $53.0924$ & $-27.8268$ & $53.0923$ & $-27.8260$ & $4$ & $4$ & $0$ & $0.04^{u}$ & $-$ & $1.05$ & $0.75$ & $0.05$ & $26.22$\\
			$ $ & $303$ & $53.0925$ & $-27.8771$ & $53.0921$ & $-27.8767$ & $5$ & $0$ & $5$ & $0.35^{l}$ & $-$ & $0.73$ & $3.57$ & $-0.07$ & $0.37$\\
			$ $ & $306$ & $53.0939$ & $-27.8258$ & $53.0944$ & $-27.8259$ & $7$ & $7$ & $0$ & $0.16$ & $0.36$ & $2.87$ & $0.41$ & $0.11$ & $12.30$\\
			$*$ & $318$ & $53.0965$ & $-27.7449$ & $53.0966$ & $-27.7447$ & $4$ & $4$ & $0$ & $0.38^{u}$ & $-$ & $1.38$ & $0.06$ & $0.05$ & $34.03$\\
			$*$ & $321$ & $53.0984$ & $-27.7671$ & $53.0983$ & $-27.7667$ & $4$ & $4$ & $5$ & $0.31^{l}$ & $-$ & $1.22$ & $0.05$ & $0.03$ & $18.91$\\
			$ $ & $325$ & $53.1000$ & $-27.8086$ & $53.0995$ & $-27.8085$ & $4$ & $4$ & $0$ & $0.46^{u}$ & $-$ & $2.83$ & $0.23$ & $0.55$ & $7.27$\\
			$ $ & $328$ & $53.1016$ & $-27.8217$ & $53.1012$ & $-27.8224$ & $4$ & $4$ & $0$ & $0.24^{u}$ & $-$ & $1.68$ & $0.71$ & $1.06$ & $6.38$\\
			$ $ & $331$ & $53.1027$ & $-27.8606$ & $53.1028$ & $-27.8610$ & $4$ & $4$ & $0$ & $0.21^{u}$ & $-$ & $4.32$ & $0.02$ & $3.63$ & $3.33$\\
			$*$ & $348$ & $53.1052$ & $-27.8752$ & $53.1058$ & $-27.8753$ & $6$ & $6$ & $0$ & $0.15^{u}$ & $-$ & $3.11$ & $0.12$ & $0.67$ & $3.85$\\
			$*$ & $354$ & $53.1076$ & $-27.8558$ & $53.1079$ & $-27.8558$ & $7$ & $7$ & $0$ & $-0.54$ & $0.15$ & $2.83$ & $3.58$ & $0.85$ & $0.38$\\
			$ $ & $371$ & $53.1116$ & $-27.7679$ & $53.1118$ & $-27.7680$ & $5$ & $5$ & $0$ & $0.15$ & $0.10$ & $5.56$ & $0.04$ & $4.50$ & $0.61$\\
			$ $ & $373$ & $53.1118$ & $-27.9096$ & $53.1113$ & $-27.9094$ & $4$ & $4$ & $4$ & $0.22^{u}$ & $-$ & $2.54$ & $0.15$ & $0.25$ & $12.99$\\
			$*$ & $389$ & $53.1193$ & $-27.7659$ & $53.1186$ & $-27.7658$ & $4$ & $4$ & $0$ & $0.16^{u}$ & $-$ & $2.02$ & $0.05$ & $0.48$ & $7.31$\\
			$*$ & $392$ & $53.1199$ & $-27.7432$ & $53.1198$ & $-27.7436$ & $7$ & $5$ & $0$ & $0.05^{u}$ & $-$ & $3.07$ & $3.10$ & $0.00$ & $1.31$\\
			$*$ & $402$ & $53.1219$ & $-27.7529$ & $53.1214$ & $-27.7531$ & $5$ & $5$ & $0$ & $0.27^{l}$ & $-$ & $2.37$ & $0.64$ & $0.21$ & $2.91$\\
			$ $ & $403$ & $53.1220$ & $-27.9388$ & $53.1224$ & $-27.9381$ & $4$ & $4$ & $0$ & $0.20^{l}$ & $-$ & $4.22$ & $0.17$ & $0.12$ & $6.97$\\
			$ $ & $410$ & $53.1241$ & $-27.8913$ & $53.1242$ & $-27.8917$ & $4$ & $4$ & $0$ & $0.54$ & $0.12$ & $3.49$ & $0.27$ & $0.27$ & $2.63$\\
			$*$ & $428$ & $53.1296$ & $-27.8278$ & $53.1295$ & $-27.8276$ & $4$ & $4$ & $0$ & $0.24^{u}$ & $-$ & $1.15$ & $0.07$ & $0.04$ & $17.71$\\
			$ $ & $430$ & $53.1305$ & $-27.7912$ & $53.1310$ & $-27.7911$ & $0$ & $7$ & $0$ & $0.29$ & $0.15$ & $6.54$ & $1.34$ & $5.07$ & $0.03$\\			
			$*$ & $444$ & $53.1340$ & $-27.7811$ & $53.1340$ & $-27.7809$ & $4$ & $4$ & $4$ & $0.28$ & $0.07$ & $1.42$ & $0.10$ & $0.06$ & $41.98$\\
			$*$ & $455$ & $53.1378$ & $-27.8022$ & $53.1378$ & $-27.8021$ & $4$ & $4$ & $0$ & $0.21^{u}$ & $-$ & $1.26$ & $0.05$ & $0.03$ & $5.63$\\
			$ $ & $456$ & $53.1380$ & $-27.8683$ & $53.1381$ & $-27.8684$ & $7$ & $7$ & $0$ & $-0.29$ & $0.11$ & $10.91$ & $-0.92$ & $8.39$ & $0.00$\\
			$*$ & $460$ & $53.1393$ & $-27.8745$ & $53.1394$ & $-27.8746$ & $0$ & $5$ & $0$ & $-0.02$ & $0.27$ & $2.91$ & $2.68$ & $0.09$ & $0.73$\\
			$*$ & $462$ & $53.1403$ & $-27.7976$ & $53.1405$ & $-27.7973$ & $4$ & $4$ & $0$ & $0.06$ & $0.39$ & $1.63$ & $0.06$ & $0.04$ & $5.12$\\
			$ $ & $466$ & $53.1417$ & $-27.8167$ & $53.1416$ & $-27.8166$ & $4$ & $4$ & $0$ & $0.27$ & $0.13$ & $2.02$ & $0.20$ & $0.31$ & $5.73$\\
			$*$ & $485$ & $53.1466$ & $-27.8710$ & $53.1460$ & $-27.8711$ & $7$ & $7$ & $0$ & $0.45$ & $0.15$ & $2.83$ & $0.54$ & $0.68$ & $14.74$\\
			$ $ & $496$ & $53.1505$ & $-27.8890$ & $53.1507$ & $-27.8886$ & $4$ & $4$ & $0$ & $0.00$ & $-$ & $2.50$ & $0.31$ & $2.11$ & $1.44$\\
			$*$ & $522$ & $53.1585$ & $-27.7741$ & $53.1583$ & $-27.7738$ & $4$ & $5$ & $6$ & $-0.40$ & $0.04$ & $1.60$ & $0.21$ & $0.05$ & $4.06$\\
			$ $ & $535$ & $53.1627$ & $-27.7443$ & $53.1622$ & $-27.7442$ & $4$ & $4$ & $0$ & $-0.12$ & $0.06$ & $2.37$ & $0.30$ & $0.39$ & $3.44$\\
			$*$ & $539$ & $53.1632$ & $-27.8091$ & $53.1621$ & $-27.8097$ & $4$ & $4$ & $0$ & $0.53^{l}$ & $-$ & $0.97$ & $0.06$ & $0.08$ & $44.03$\\
			$ $ & $546$ & $53.1653$ & $-27.8142$ & $53.1648$ & $-27.8144$ & $4$ & $4$ & $0$ & $0.29$ & $0.03$ & $3.15$ & $0.18$ & $0.25$ & $8.09$\\
			$ $ & $556$ & $53.1701$ & $-27.9298$ & $53.1699$ & $-27.9304$ & $4$ & $4$ & $0$ & $0.40$ & $0.03$ & $2.68$ & $0.17$ & $0.16$ & $12.11$\\
			$*$ & $574$ & $53.1787$ & $-27.8027$ & $53.1782$ & $-27.8027$ & $6$ & $7$ & $0$ & $0.38$ & $0.25$ & $3.32$ & $0.42$ & $0.75$ & $2.72$\\
			$ $ & $578$ & $53.1806$ & $-27.7797$ & $53.1807$ & $-27.7796$ & $7$ & $7$ & $0$ & $0.00$ & $-$ & $6.77$ & $0.97$ & $5.10$ & $0.00$\\
			$ $ & $583$ & $53.1835$ & $-27.7766$ & $53.1834$ & $-27.7764$ & $7$ & $7$ & $0$ & $-0.51$ & $0.06$ & $9.36$ & $0.63$ & $4.82$ & $0.00$\\
			$ $ & $589$ & $53.1850$ & $-27.8198$ & $53.1851$ & $-27.8196$ & $4$ & $4$ & $6$ & $0.17^{u}$ & $-$ & $1.28$ & $0.26$ & $0.08$ & $5.77$\\
			$*$ & $591$ & $53.1852$ & $-27.7174$ & $53.1848$ & $-27.7173$ & $4$ & $4$ & $0$ & $0.29^{u}$ & $-$ & $1.35$ & $0.17$ & $0.21$ & $21.54$\\
			$ $ & $620$ & $53.1960$ & $-27.8927$ & $53.1957$ & $-27.8928$ & $4$ & $4$ & $0$ & $-0.02$ & $0.06$ & $2.57$ & $0.55$ & $0.45$ & $4.34$\\
			$ $ & $624$ & $53.1981$ & $-27.8323$ & $53.1979$ & $-27.8319$ & $5$ & $6$ & $0$ & $0.33^{l}$ & $-$ & $1.52$ & $2.23$ & $0.07$ & $0.56$\\
			$*$ & $625$ & $53.1989$ & $-27.8440$ & $53.1989$ & $-27.8439$ & $4$ & $4$ & $0$ & $0.71$ & $0.10$ & $1.68$ & $0.07$ & $0.04$ & $16.22$\\
			$*$ & $630$ & $53.2016$ & $-27.8443$ & $53.2027$ & $-27.8448$ & $7$ & $4$ & $0$ & $0.36^{u}$ & $-$ & $2.05$ & $5.04$ & $0.16$ & $0.37$\\
			$*$ & $651$ & $53.2153$ & $-27.8703$ & $53.2150$ & $-27.8695$ & $7$ & $0$ & $0$ & $0.40$ & $0.10$ & $1.90$ & $0.41$ & $0.34$ & $16.22$\\
		\bottomrule
		\end{tabular}
		\end{center}
	\caption{\label{tab:overview}Overview over the redshift estimates that we gained for each object in the course of this analysis. All photometric redshifts were gained by using GOODS/ACS, CANDELS and in some cases also \Spitzer\ data (marked with an asterisk). For our three final candidates (456, 578, 583) we also computed photometric redshifts using HUDF and HUGS data. Please see sections \ref{sec:HUDF} and \ref{sec:HUGS} for the corresponding values.  Note that the stacking procedure only gives a redshift indication. Hence, if '4' is given as the stacking result this corresponds to $z\lesssim4$,  '7' indicates $z\gtrsim7$. We mark sources for which we only gained upper or lower limits on the Hardness Ratio with '$u$' and '$l$'. '0' indicates that a source could not be classified by the corresponding redshift test.}	
\end{table*}
\newpage

\section*{Acknowledgements}
We thank  Andreas Faisst for helpful discussions and Richard Ellis for suggesting the stacking analysis. We also thank the anonymous referee for helpful comments. AKW, KS and MK gratefully acknowledge support from Swiss National Science Foundation Grant PP00P2\_138979/1. Support for the work of ET was provided by the Center of Excellence in Astrophysics and Associated Technologies (PFB 06), by the FONDECYT regular grant 1120061 and by the CONICYT Anillo project ACT1101. This research has made use of NASA's ADS Service. This research made use of Astropy, a community-developed core Python package for Astronomy (Astropy Collaboration, 2013).
\newpage

\bibliographystyle{mn2e}
\bibliography{Bib_Lib.bib}

\begin{thebibliography}{94}
\expandafter\ifx\csname natexlab\endcsname\relax\def\natexlab#1{#1}\fi

\bibitem[{{Aird} {et~al}\mbox{.}(2013){Aird}, {Comastri}, {Brusa},
  {Cappelluti}, {Moretti}, {Vanzella}, {Volonteri}, {Alexander}, {Afonso},
  {Fiore}, {Georgantopoulos}, {Iwasawa}, {Merloni}, {Nandra}, {Salvaterra},
  {Salvato}, {Severgnini}, {Schawinski}, {Shankar}, {Vignali}, \&
  {Vito}}]{Aird:2013aa}
{Aird} J. {et~al.}, 2013, ArXiv e-prints, 1306.2325

\bibitem[{{Ajello}(2009)}]{Ajello:2009aa}
{Ajello} M., 2009, ArXiv e-prints, 0902.3033

\bibitem[{{Alexander} \& {Natarajan}(2014)}]{Alexander:2014aa}
{Alexander} T., {Natarajan} P., 2014

\bibitem[{{Alvarez} {et~al}\mbox{.}(2009){Alvarez}, {Wise}, \&
  {Abel}}]{Alvarez:2009aa}
{Alvarez} M.~A., {Wise} J.~H., {Abel} T., 2009, apjl, 701, L133

\bibitem[{{Arnaud}(1996)}]{Arnaud:1996aa}
{Arnaud} K.~A., 1996, in Astronomical Society of the Pacific Conference Series,
  Vol. 101, Astronomical Data Analysis Software and Systems V, {Jacoby} G.~H.,
  {Barnes} J., eds., p.~17

\bibitem[{{Barth} {et~al}\mbox{.}(2003){Barth}, {Martini}, {Nelson}, \&
  {Ho}}]{Barth:2003ab}
{Barth} A.~J., {Martini} P., {Nelson} C.~H., {Ho} L.~C., 2003, apjl, 594, L95

\bibitem[{{Beckwith} {et~al}\mbox{.}(2006){Beckwith}, {Stiavelli}, {Koekemoer},
  {Caldwell}, {Ferguson}, {Hook}, {Lucas}, {Bergeron}, {Corbin}, {Jogee},
  {Panagia}, {Robberto}, {Royle}, {Somerville}, \& {Sosey}}]{Beckwith:2006aa}
{Beckwith} S.~V.~W. {et~al.}, 2006, aj, 132, 1729

\bibitem[{{Bellovary} {et~al}\mbox{.}(2011){Bellovary}, {Volonteri},
  {Governato}, {Shen}, {Quinn}, \& {Wadsley}}]{Bellovary:2011aa}
{Bellovary} J., {Volonteri} M., {Governato} F., {Shen} S., {Quinn} T.,
  {Wadsley} J., 2011, ArXiv e-prints, 1104.3858

\bibitem[{{Bertin} \& {Arnouts}(1996)}]{Bertin:1996aa}
{Bertin} E., {Arnouts} S., 1996, aaps, 117, 393

\bibitem[{{Bertin} {et~al}\mbox{.}(2002){Bertin}, {Mellier}, {Radovich},
  {Missonnier}, {Didelon}, \& {Morin}}]{2002Terapix}
{Bertin} E., {Mellier} Y., {Radovich} M., {Missonnier} G., {Didelon} P.,
  {Morin} B., 2002, in Astronomical Society of the Pacific Conference Series,
  Vol. 281, Astronomical Data Analysis Software and Systems XI, {Bohlender}
  D.~A., {Durand} D., {Handley} T.~H., eds., p. 228

\bibitem[{{Blanton} \& {Roweis}(2007)}]{Blanton:2007aa}
{Blanton} M.~R., {Roweis} S., 2007, aj, 133, 734

\bibitem[{{Bouwens} {et~al}\mbox{.}(2014){Bouwens}, {Illingworth}, {Oesch},
  {Trenti}, {Labbe'}, {Bradley}, {Carollo}, {van Dokkum}, {Gonzalez},
  {Holwerda}, {Franx}, {Spitler}, {Smit}, \& {Magee}}]{Bouwens:2014aa}
{Bouwens} R.~J. {et~al.}, 2014, ArXiv e-prints, 1403.4295

\bibitem[{{Brammer} {et~al}\mbox{.}(2008){Brammer}, {van Dokkum}, \&
  {Coppi}}]{Brammer:2008aa}
{Brammer} G.~B., {van Dokkum} P.~G., {Coppi} P., 2008, apj, 686, 1503

\bibitem[{{Brightman} \& {Nandra}(2012)}]{2012MNRAS.422.1166B}
{Brightman} M., {Nandra} K., 2012, mnras, 422, 1166

\bibitem[{{Bromm} {et~al}\mbox{.}(2002){Bromm}, {Coppi}, \&
  {Larson}}]{Bromm:2002aa}
{Bromm} V., {Coppi} P.~S., {Larson} R.~B., 2002, apj, 564, 23

\bibitem[{{Bromm} \& {Yoshida}(2011)}]{Bromm:2011aa}
{Bromm} V., {Yoshida} N., 2011, araa, 49, 373

\bibitem[{{Bromm} {et~al}\mbox{.}(2009){Bromm}, {Yoshida}, {Hernquist}, \&
  {McKee}}]{Bromm:2009aa}
{Bromm} V., {Yoshida} N., {Hernquist} L., {McKee} C.~F., 2009, nat, 459, 49

\bibitem[{{Calzetti} {et~al}\mbox{.}(2000){Calzetti}, {Armus}, {Bohlin},
  {Kinney}, {Koornneef}, \& {Storchi-Bergmann}}]{Calzetti:2000aa}
{Calzetti} D., {Armus} L., {Bohlin} R.~C., {Kinney} A.~L., {Koornneef} J.,
  {Storchi-Bergmann} T., 2000, apj, 533, 682

\bibitem[{{Civano} \& {the Chandra COSMOS Legacy Team}(2014)}]{Civano:2014aa}
{Civano} F.~M., {the Chandra COSMOS Legacy Team}, 2014, in American
  Astronomical Society Meeting Abstracts, Vol. 223, American Astronomical
  Society Meeting Abstracts, p. 254.46

\bibitem[{{Comastri} {et~al}\mbox{.}(2010){Comastri}, {Iwasawa}, {Gilli},
  {Vignali}, {Ranalli}, {Matt}, \& {Fiore}}]{Comastri:2010aa}
{Comastri} A., {Iwasawa} K., {Gilli} R., {Vignali} C., {Ranalli} P., {Matt} G.,
  {Fiore} F., 2010, apj, 717, 787

\bibitem[{{Dahlen} {et~al}\mbox{.}(2010){Dahlen}, {Mobasher}, {Dickinson},
  {Ferguson}, {Giavalisco}, {Grogin}, {Guo}, {Koekemoer}, {Lee}, {Lee},
  {Nonino}, {Riess}, \& {Salimbeni}}]{Dahlen:2010aa}
{Dahlen} T. {et~al.}, 2010, apj, 724, 425

\bibitem[{{Damen} {et~al}\mbox{.}(2011){Damen}, {Labb{\'e}}, {van Dokkum},
  {Franx}, {Taylor}, {Brandt}, {Dickinson}, {Gawiser}, {Illingworth}, {Kriek},
  {Marchesini}, {Muzzin}, {Papovich}, \& {Rix}}]{Damen:2011aa}
{Damen} M. {et~al.}, 2011, apj, 727, 1

\bibitem[{{Duncan} {et~al}\mbox{.}(2014){Duncan}, {Conselice}, {Mortlock},
  {Hartley}, {Guo}, {Ferguson}, {Dav{\'e}}, {Lu}, {Ownsworth}, {Ashby},
  {Dekel}, {Dickinson}, {Faber}, {Giavalisco}, {Grogin}, {Kocevski},
  {Koekemoer}, {Somerville}, \& {White}}]{Duncan:2014aa}
{Duncan} K. {et~al.}, 2014

\bibitem[{{Dunlop}(2013)}]{Dunlop:2013zr}
{Dunlop} J.~S., 2013, in Astrophysics and Space Science Library, Vol. 396,
  Astrophysics and Space Science Library, {Wiklind} T., {Mobasher} B., {Bromm}
  V., eds., p. 223

\bibitem[{{Dunlop} {et~al}\mbox{.}(2007){Dunlop}, {Cirasuolo}, \&
  {McLure}}]{Dunlop:2007aa}
{Dunlop} J.~S., {Cirasuolo} M., {McLure} R.~J., 2007, mnras, 376, 1054

\bibitem[{{Fan} {et~al}\mbox{.}(2001){Fan}, {Narayanan}, {Lupton}, {Strauss},
  {Knapp}, {Becker}, {White}, {Pentericci}, {Leggett}, {Haiman}, {Gunn},
  {Ivezi{\'c}}, {Schneider}, {Anderson}, {Brinkmann}, {Bahcall}, {Connolly},
  {Csabai}, {Doi}, {Fukugita}, {Geballe}, {Grebel}, {Harbeck}, {Hennessy},
  {Lamb}, {Miknaitis}, {Munn}, {Nichol}, {Okamura}, {Pier}, {Prada},
  {Richards}, {Szalay}, \& {York}}]{Fan:2001aa}
{Fan} X. {et~al.}, 2001, aj, 122, 2833

\bibitem[{{Fan} {et~al}\mbox{.}(2000){Fan}, {White}, {Davis}, {Becker},
  {Strauss}, {Haiman}, {Schneider}, {Gregg}, {Gunn}, {Knapp}, {Lupton},
  {Anderson}, {Anderson}, {Annis}, {Bahcall}, {Boroski}, {Brunner}, {Chen},
  {Connolly}, {Csabai}, {Doi}, {Fukugita}, {Hennessy}, {Hindsley}, {Ichikawa},
  {Ivezi{\'c}}, {Loveday}, {Meiksin}, {McKay}, {Munn}, {Newberg}, {Nichol},
  {Okamura}, {Pier}, {Sekiguchi}, {Shimasaku}, {Stoughton}, {Szalay},
  {Szokoly}, {Thakar}, {Vogeley}, \& {York}}]{Fan:2000aa}
{Fan} X. {et~al.}, 2000, aj, 120, 1167

\bibitem[{{Finkelstein} {et~al}\mbox{.}(2012){Finkelstein}, {Papovich},
  {Salmon}, {Finlator}, {Dickinson}, {Ferguson}, {Giavalisco}, {Koekemoer},
  {Reddy}, {Bassett}, {Conselice}, {Dunlop}, {Faber}, {Grogin}, {Hathi},
  {Kocevski}, {Lai}, {Lee}, {McLure}, {Mobasher}, \&
  {Newman}}]{Finkelstein:2012aa}
{Finkelstein} S.~L. {et~al.}, 2012, apj, 756, 164

\bibitem[{{Fiore} {et~al}\mbox{.}(2009){Fiore}, {Puccetti}, {Brusa}, {Salvato},
  {Zamorani}, {Aldcroft}, {Aussel}, {Brunner}, {Capak}, {Cappelluti}, {Civano},
  {Comastri}, {Elvis}, {Feruglio}, {Finoguenov}, {Fruscione}, {Gilli},
  {Hasinger}, {Koekemoer}, {Kartaltepe}, {Ilbert}, {Impey}, {Le Floc'h},
  {Lilly}, {Mainieri}, {Martinez-Sansigre}, {McCracken}, {Menci}, {Merloni},
  {Miyaji}, {Sanders}, {Sargent}, {Schinnerer}, {Scoville}, {Silverman},
  {Smolcic}, {Steffen}, {Santini}, {Taniguchi}, {Thompson}, {Trump}, {Vignali},
  {Urry}, \& {Yan}}]{Fiore:2009aa}
{Fiore} F. {et~al.}, 2009, apj, 693, 447

\bibitem[{{Fontana} {et~al}\mbox{.}(2014){Fontana}, {Dunlop}, {Paris},
  {Targett}, {Boutsia}, {Castellano}, {Galametz}, {Grazian}, {McLure},
  {Merlin}, {Pentericci}, {Wuyts}, {Almaini}, {Caputi}, {Chary}, {Cirasuolo},
  {Conselice}, {Cooray}, {Daddi}, {Dickinson}, {Faber}, {Fazio}, {Ferguson},
  {Giallongo}, {Giavalisco}, {Grogin}, {Hathi}, {Koekemoer}, {Koo}, {Lucas},
  {Nonino}, {Rix}, {Renzini}, {Rosario}, {Santini}, {Scarlata}, {Sommariva},
  {Stark}, {van der Wel}, {Vanzella}, {Wild}, {Yan}, \&
  {Zibetti}}]{Fontana:2014aa}
{Fontana} A. {et~al.}, 2014, ArXiv e-prints, 1409.7082

\bibitem[{{Genzel} {et~al}\mbox{.}(1996){Genzel}, {Thatte}, {Krabbe}, {Kroker},
  \& {Tacconi-Garman}}]{Genzel:1996aa}
{Genzel} R., {Thatte} N., {Krabbe} A., {Kroker} H., {Tacconi-Garman} L.~E.,
  1996, apj, 472, 153

\bibitem[{{Ghez} {et~al}\mbox{.}(1998){Ghez}, {Klein}, {Morris}, \&
  {Becklin}}]{Ghez:1998aa}
{Ghez} A.~M., {Klein} B.~L., {Morris} M., {Becklin} E.~E., 1998, apj, 509, 678

\bibitem[{{Ghez} {et~al}\mbox{.}(2000){Ghez}, {Morris}, {Becklin}, {Tanner}, \&
  {Kremenek}}]{Ghez:2000aa}
{Ghez} A.~M., {Morris} M., {Becklin} E.~E., {Tanner} A., {Kremenek} T., 2000,
  nat, 407, 349

\bibitem[{{Ghez} {et~al}\mbox{.}(2008){Ghez}, {Salim}, {Weinberg}, {Lu}, {Do},
  {Dunn}, {Matthews}, {Morris}, {Yelda}, {Becklin}, {Kremenek},
  {Milosavljevic}, \& {Naiman}}]{Ghez:2008aa}
{Ghez} A.~M. {et~al.}, 2008, apj, 689, 1044

\bibitem[{{Giavalisco}(2002)}]{Giavalisco:2002aa}
{Giavalisco} M., 2002, araa, 40, 579

\bibitem[{{Giavalisco} {et~al}\mbox{.}(2004){Giavalisco}, {Ferguson},
  {Koekemoer}, {Dickinson}, {Alexander}, {Bauer}, {Bergeron}, {Biagetti},
  {Brandt}, {Casertano}, {Cesarsky}, {Chatzichristou}, {Conselice},
  {Cristiani}, {Da Costa}, {Dahlen}, {de Mello}, {Eisenhardt}, {Erben}, {Fall},
  {Fassnacht}, {Fosbury}, {Fruchter}, {Gardner}, {Grogin}, {Hook},
  {Hornschemeier}, {Idzi}, {Jogee}, {Kretchmer}, {Laidler}, {Lee}, {Livio},
  {Lucas}, {Madau}, {Mobasher}, {Moustakas}, {Nonino}, {Padovani}, {Papovich},
  {Park}, {Ravindranath}, {Renzini}, {Richardson}, {Riess}, {Rosati},
  {Schirmer}, {Schreier}, {Somerville}, {Spinrad}, {Stern}, {Stiavelli},
  {Strolger}, {Urry}, {Vandame}, {Williams}, \& {Wolf}}]{Giavalisco:2004aa}
{Giavalisco} M. {et~al.}, 2004, apjl, 600, L93

\bibitem[{{Grazian} {et~al}\mbox{.}(2006){Grazian}, {Fontana}, {de Santis},
  {Nonino}, {Salimbeni}, {Giallongo}, {Cristiani}, {Gallozzi}, \&
  {Vanzella}}]{Grazian:2006aa}
{Grazian} A. {et~al.}, 2006, aap, 449, 951

\bibitem[{{Grogin} {et~al}\mbox{.}(2011){Grogin}, {Kocevski}, {Faber},
  {Ferguson}, {Koekemoer}, {Riess}, {Acquaviva}, {Alexander}, {Almaini},
  {Ashby}, {Barden}, {Bell}, {Bournaud}, {Brown}, {Caputi}, {Casertano},
  {Cassata}, {Castellano}, {Challis}, {Chary}, {Cheung}, {Cirasuolo},
  {Conselice}, {Roshan Cooray}, {Croton}, {Daddi}, {Dahlen}, {Dav{\'e}}, {de
  Mello}, {Dekel}, {Dickinson}, {Dolch}, {Donley}, {Dunlop}, {Dutton}, {Elbaz},
  {Fazio}, {Filippenko}, {Finkelstein}, {Fontana}, {Gardner}, {Garnavich},
  {Gawiser}, {Giavalisco}, {Grazian}, {Guo}, {Hathi}, {H{\"a}ussler},
  {Hopkins}, {Huang}, {Huang}, {Jha}, {Kartaltepe}, {Kirshner}, {Koo}, {Lai},
  {Lee}, {Li}, {Lotz}, {Lucas}, {Madau}, {McCarthy}, {McGrath}, {McIntosh},
  {McLure}, {Mobasher}, {Moustakas}, {Mozena}, {Nandra}, {Newman}, {Niemi},
  {Noeske}, {Papovich}, {Pentericci}, {Pope}, {Primack}, {Rajan},
  {Ravindranath}, {Reddy}, {Renzini}, {Rix}, {Robaina}, {Rodney}, {Rosario},
  {Rosati}, {Salimbeni}, {Scarlata}, {Siana}, {Simard}, {Smidt}, {Somerville},
  {Spinrad}, {Straughn}, {Strolger}, {Telford}, {Teplitz}, {Trump}, {van der
  Wel}, {Villforth}, {Wechsler}, {Weiner}, {Wiklind}, {Wild}, {Wilson},
  {Wuyts}, {Yan}, \& {Yun}}]{Grogin:2011aa}
{Grogin} N.~A. {et~al.}, 2011, apjs, 197, 35

\bibitem[{{Guhathakurta} {et~al}\mbox{.}(1990){Guhathakurta}, {Tyson}, \&
  {Majewski}}]{Guhathakurta:1990aa}
{Guhathakurta} P., {Tyson} J.~A., {Majewski} S.~R., 1990, apjl, 357, L9

\bibitem[{{Haiman} \& {Loeb}(2001)}]{Haiman:2001aa}
{Haiman} Z., {Loeb} A., 2001, apj, 552, 459

\bibitem[{{Holwerda}(2005)}]{Holwerda:2005aa}
{Holwerda} B.~W., 2005, ArXiv Astrophysics e-prints

\bibitem[{{Illingworth} {et~al}\mbox{.}(2013){Illingworth}, {Magee}, {Oesch},
  {Bouwens}, {Labb{\'e}}, {Stiavelli}, {van Dokkum}, {Franx}, {Trenti},
  {Carollo}, \& {Gonzalez}}]{Illingworth:2013aa}
{Illingworth} G.~D. {et~al.}, 2013, apjs, 209, 6

\bibitem[{{Jahnke} {et~al}\mbox{.}(2009){Jahnke}, {Bongiorno}, {Brusa},
  {Capak}, {Cappelluti}, {Cisternas}, {Civano}, {Colbert}, {Comastri}, {Elvis},
  {Hasinger}, {Ilbert}, {Impey}, {Inskip}, {Koekemoer}, {Lilly}, {Maier},
  {Merloni}, {Riechers}, {Salvato}, {Schinnerer}, {Scoville}, {Silverman},
  {Taniguchi}, {Trump}, \& {Yan}}]{Jahnke:2009aa}
{Jahnke} K. {et~al.}, 2009, apjl, 706, L215

\bibitem[{{Johnson} {et~al}\mbox{.}(2012){Johnson}, {Whalen}, {Fryer}, \&
  {Li}}]{Johnson:2012aa}
{Johnson} J.~L., {Whalen} D.~J., {Fryer} C.~L., {Li} H., 2012, apj, 750, 66

\bibitem[{{Koekemoer} {et~al}\mbox{.}(2013){Koekemoer}, {Ellis}, {McLure},
  {Dunlop}, {Robertson}, {Ono}, {Schenker}, {Ouchi}, {Bowler}, {Rogers},
  {Curtis-Lake}, {Schneider}, {Charlot}, {Stark}, {Furlanetto}, {Cirasuolo},
  {Wild}, \& {Targett}}]{Koekemoer:2013aa}
{Koekemoer} A.~M. {et~al.}, 2013, apjs, 209, 3

\bibitem[{{Koekemoer} {et~al}\mbox{.}(2011){Koekemoer}, {Faber}, {Ferguson},
  {Grogin}, {Kocevski}, {Koo}, {Lai}, {Lotz}, {Lucas}, {McGrath}, {Ogaz},
  {Rajan}, {Riess}, {Rodney}, {Strolger}, {Casertano}, {Castellano}, {Dahlen},
  {Dickinson}, {Dolch}, {Fontana}, {Giavalisco}, {Grazian}, {Guo}, {Hathi},
  {Huang}, {van der Wel}, {Yan}, {Acquaviva}, {Alexander}, {Almaini}, {Ashby},
  {Barden}, {Bell}, {Bournaud}, {Brown}, {Caputi}, {Cassata}, {Challis},
  {Chary}, {Cheung}, {Cirasuolo}, {Conselice}, {Roshan Cooray}, {Croton},
  {Daddi}, {Dav{\'e}}, {de Mello}, {de Ravel}, {Dekel}, {Donley}, {Dunlop},
  {Dutton}, {Elbaz}, {Fazio}, {Filippenko}, {Finkelstein}, {Frazer}, {Gardner},
  {Garnavich}, {Gawiser}, {Gruetzbauch}, {Hartley}, {H{\"a}ussler},
  {Herrington}, {Hopkins}, {Huang}, {Jha}, {Johnson}, {Kartaltepe},
  {Khostovan}, {Kirshner}, {Lani}, {Lee}, {Li}, {Madau}, {McCarthy},
  {McIntosh}, {McLure}, {McPartland}, {Mobasher}, {Moreira}, {Mortlock},
  {Moustakas}, {Mozena}, {Nandra}, {Newman}, {Nielsen}, {Niemi}, {Noeske},
  {Papovich}, {Pentericci}, {Pope}, {Primack}, {Ravindranath}, {Reddy},
  {Renzini}, {Rix}, {Robaina}, {Rosario}, {Rosati}, {Salimbeni}, {Scarlata},
  {Siana}, {Simard}, {Smidt}, {Snyder}, {Somerville}, {Spinrad}, {Straughn},
  {Telford}, {Teplitz}, {Trump}, {Vargas}, {Villforth}, {Wagner}, {Wandro},
  {Wechsler}, {Weiner}, {Wiklind}, {Wild}, {Wilson}, {Wuyts}, \&
  {Yun}}]{Koekemoer:2011aa}
{Koekemoer} A.~M. {et~al.}, 2011, apjs, 197, 36

\bibitem[{{Latif} {et~al}\mbox{.}(2013){Latif}, {Schleicher}, {Schmidt}, \&
  {Niemeyer}}]{Latif:2013ab}
{Latif} M.~A., {Schleicher} D.~R.~G., {Schmidt} W., {Niemeyer} J.~C., 2013,
  mnras, 436, 2989

\bibitem[{{Lee} {et~al}\mbox{.}(2012){Lee}, {Ferguson}, {Wiklind}, {Dahlen},
  {Dickinson}, {Giavalisco}, {Grogin}, {Papovich}, {Messias}, {Guo}, \&
  {Lin}}]{Lee:2012aa}
{Lee} K.-S. {et~al.}, 2012, apj, 752, 66

\bibitem[{{Loeb} \& {Rasio}(1994)}]{Loeb:1994aa}
{Loeb} A., {Rasio} F.~A., 1994, apj, 432, 52

\bibitem[{{Luo} {et~al}\mbox{.}(2008){Luo}, {Bauer}, {Brandt}, {Alexander},
  {Lehmer}, {Schneider}, {Brusa}, {Comastri}, {Fabian}, {Finoguenov}, {Gilli},
  {Hasinger}, {Hornschemeier}, {Koekemoer}, {Mainieri}, {Paolillo}, {Rosati},
  {Shemmer}, {Silverman}, {Smail}, {Steffen}, \& {Vignali}}]{Luo:2008aa}
{Luo} B. {et~al.}, 2008, apjs, 179, 19

\bibitem[{{Luo} {et~al}\mbox{.}(2010){Luo}, {Brandt}, {Xue}, {Brusa},
  {Alexander}, {Bauer}, {Comastri}, {Koekemoer}, {Lehmer}, {Mainieri},
  {Rafferty}, {Schneider}, {Silverman}, \& {Vignali}}]{Luo:2010aa}
{Luo} B. {et~al.}, 2010, apjs, 187, 560

\bibitem[{{Lusso} {et~al}\mbox{.}(2010){Lusso}, {Comastri}, {Vignali},
  {Zamorani}, {Brusa}, {Gilli}, {Iwasawa}, {Salvato}, {Civano}, {Elvis},
  {Merloni}, {Bongiorno}, {Trump}, {Koekemoer}, {Schinnerer}, {Le Floc'h},
  {Cappelluti}, {Jahnke}, {Sargent}, {Silverman}, {Mainieri}, {Fiore},
  {Bolzonella}, {Le F{\`e}vre}, {Garilli}, {Iovino}, {Kneib}, {Lamareille},
  {Lilly}, {Mignoli}, {Scodeggio}, \& {Vergani}}]{Lusso:2010aa}
{Lusso} E. {et~al.}, 2010, aap, 512, A34

\bibitem[{{Madau} {et~al}\mbox{.}(2014){Madau}, {Haardt}, \&
  {Dotti}}]{Madau:2014aa}
{Madau} P., {Haardt} F., {Dotti} M., 2014, ArXiv e-prints, 1402.6995

\bibitem[{{Madau} \& {Rees}(2001)}]{Madau:2001aa}
{Madau} P., {Rees} M.~J., 2001, apjl, 551, L27

\bibitem[{{Magorrian} {et~al}\mbox{.}(1998){Magorrian}, {Tremaine},
  {Richstone}, {Bender}, {Bower}, {Dressler}, {Faber}, {Gebhardt}, {Green},
  {Grillmair}, {Kormendy}, \& {Lauer}}]{Magorrian:1998aa}
{Magorrian} J. {et~al.}, 1998, aj, 115, 2285

\bibitem[{{McLure} {et~al}\mbox{.}(2010){McLure}, {Dunlop}, {Cirasuolo},
  {Koekemoer}, {Sabbi}, {Stark}, {Targett}, \& {Ellis}}]{McLure:2010aa}
{McLure} R.~J., {Dunlop} J.~S., {Cirasuolo} M., {Koekemoer} A.~M., {Sabbi} E.,
  {Stark} D.~P., {Targett} T.~A., {Ellis} R.~S., 2010, mnras, 403, 960

\bibitem[{{McLure} {et~al}\mbox{.}(2011){McLure}, {Dunlop}, {de Ravel},
  {Cirasuolo}, {Ellis}, {Schenker}, {Robertson}, {Koekemoer}, {Stark}, \&
  {Bowler}}]{McLure:2011aa}
{McLure} R.~J. {et~al.}, 2011, mnras, 418, 2074

\bibitem[{{Menou} {et~al}\mbox{.}(2001){Menou}, {Haiman}, \&
  {Narayanan}}]{Menou:2001aa}
{Menou} K., {Haiman} Z., {Narayanan} V.~K., 2001, apj, 558, 535

\bibitem[{{Mortlock} {et~al}\mbox{.}(2011){Mortlock}, {Warren}, {Venemans},
  {Patel}, {Hewett}, {McMahon}, {Simpson}, {Theuns}, {Gonz{\'a}les-Solares},
  {Adamson}, {Dye}, {Hambly}, {Hirst}, {Irwin}, {Kuiper}, {Lawrence}, \&
  {R{\"o}ttgering}}]{Mortlock:2011aa}
{Mortlock} D.~J. {et~al.}, 2011, nat, 474, 616

\bibitem[{{Murphy} \& {Yaqoob}(2009)}]{Murphy:2009aa}
{Murphy} K.~D., {Yaqoob} T., 2009, mnras, 397, 1549

\bibitem[{{Nandra} {et~al}\mbox{.}(2013){Nandra}, {Barret}, {Barcons},
  {Fabian}, {den Herder}, {Piro}, {Watson}, {Adami}, {Aird}, {Afonso}, \&
  et~al.}]{Nandra:2013aa}
{Nandra} K. {et~al.}, 2013, ArXiv e-prints, 1306.2307

\bibitem[{{Nandra} {et~al}\mbox{.}(2002){Nandra}, {Mushotzky}, {Arnaud},
  {Steidel}, {Adelberger}, {Gardner}, {Teplitz}, \&
  {Windhorst}}]{Nandra:2002aa}
{Nandra} K., {Mushotzky} R.~F., {Arnaud} K., {Steidel} C.~C., {Adelberger}
  K.~L., {Gardner} J.~P., {Teplitz} H.~I., {Windhorst} R.~A., 2002, apj, 576,
  625

\bibitem[{{Oke} \& {Gunn}(1983)}]{Oke:1983aa}
{Oke} J.~B., {Gunn} J.~E., 1983, apj, 266, 713

\bibitem[{{Rees} \& {Volonteri}(2007)}]{Rees:2007aa}
{Rees} M.~J., {Volonteri} M., 2007, in IAU Symposium, Vol. 238, IAU Symposium,
  {Karas} V., {Matt} G., eds., pp. 51--58

\bibitem[{{Salpeter}(1964)}]{Salpeter:1964aa}
{Salpeter} E.~E., 1964, apj, 140, 796

\bibitem[{{Santini} {et~al}\mbox{.}(2009){Santini}, {Fontana}, {Grazian},
  {Salimbeni}, {Fiore}, {Fontanot}, {Boutsia}, {Castellano}, {Cristiani}, {de
  Santis}, {Gallozzi}, {Giallongo}, {Menci}, {Nonino}, {Paris}, {Pentericci},
  \& {Vanzella}}]{Santini:2009aa}
{Santini} P. {et~al.}, 2009, aap, 504, 751

\bibitem[{{Schawinski} {et~al}\mbox{.}(2011){Schawinski}, {Urry}, {Treister},
  {Simmons}, {Natarajan}, \& {Glikman}}]{Schawinski:2011aa}
{Schawinski} K., {Urry} M., {Treister} E., {Simmons} B., {Natarajan} P.,
  {Glikman} E., 2011, apjl, 743, L37

\bibitem[{{Sch{\"o}del} {et~al}\mbox{.}(2003){Sch{\"o}del}, {Ott}, {Genzel},
  {Eckart}, {Mouawad}, \& {Alexander}}]{Schodel:2003aa}
{Sch{\"o}del} R., {Ott} T., {Genzel} R., {Eckart} A., {Mouawad} N., {Alexander}
  T., 2003, apj, 596, 1015

\bibitem[{{Stark} {et~al}\mbox{.}(2009){Stark}, {Ellis}, {Bunker}, {Bundy},
  {Targett}, {Benson}, \& {Lacy}}]{Stark:2009aa}
{Stark} D.~P., {Ellis} R.~S., {Bunker} A., {Bundy} K., {Targett} T., {Benson}
  A., {Lacy} M., 2009, apj, 697, 1493

\bibitem[{{Steffen} {et~al}\mbox{.}(2006){Steffen}, {Strateva}, {Brandt},
  {Alexander}, {Koekemoer}, {Lehmer}, {Schneider}, \&
  {Vignali}}]{Steffen:2006aa}
{Steffen} A.~T., {Strateva} I., {Brandt} W.~N., {Alexander} D.~M., {Koekemoer}
  A.~M., {Lehmer} B.~D., {Schneider} D.~P., {Vignali} C., 2006, aj, 131, 2826

\bibitem[{{Steidel} {et~al}\mbox{.}(1999){Steidel}, {Adelberger}, {Giavalisco},
  {Dickinson}, \& {Pettini}}]{Steidel:1999fj}
{Steidel} C.~C., {Adelberger} K.~L., {Giavalisco} M., {Dickinson} M., {Pettini}
  M., 1999, apj, 519, 1

\bibitem[{{Steidel} \& {Hamilton}(1992)}]{Steidel:1992gf}
{Steidel} C.~C., {Hamilton} D., 1992, aj, 104, 941

\bibitem[{{Szokoly} {et~al}\mbox{.}(2004){Szokoly}, {Bergeron}, {Hasinger},
  {Lehmann}, {Kewley}, {Mainieri}, {Nonino}, {Rosati}, {Giacconi}, {Gilli},
  {Gilmozzi}, {Norman}, {Romaniello}, {Schreier}, {Tozzi}, {Wang}, {Zheng}, \&
  {Zirm}}]{Szokoly:2004aa}
{Szokoly} G.~P. {et~al.}, 2004, apjs, 155, 271

\bibitem[{{Tananbaum} {et~al}\mbox{.}(1979){Tananbaum}, {Avni}, {Branduardi},
  {Elvis}, {Fabbiano}, {Feigelson}, {Giacconi}, {Henry}, {Pye}, {Soltan}, \&
  {Zamorani}}]{Tananbaum:1979aa}
{Tananbaum} H. {et~al.}, 1979, apjl, 234, L9

\bibitem[{{Tozzi} {et~al}\mbox{.}(2006){Tozzi}, {Gilli}, {Mainieri}, {Norman},
  {Risaliti}, {Rosati}, {Bergeron}, {Borgani}, {Giacconi}, {Hasinger},
  {Nonino}, {Streblyanska}, {Szokoly}, {Wang}, \& {Zheng}}]{Tozzi:2006aa}
{Tozzi} P. {et~al.}, 2006, aap, 451, 457

\bibitem[{{Trakhtenbrot} {et~al}\mbox{.}(2011){Trakhtenbrot}, {Netzer}, {Lira},
  \& {Shemmer}}]{Trakhtenbrot:2011aa}
{Trakhtenbrot} B., {Netzer} H., {Lira} P., {Shemmer} O., 2011, apj, 730, 7

\bibitem[{{Treister} {et~al}\mbox{.}(2013){Treister}, {Schawinski},
  {Volonteri}, \& {Natarajan}}]{T13}
{Treister} E., {Schawinski} K., {Volonteri} M., {Natarajan} P., 2013, apj, 778,
  130

\bibitem[{{Treister} {et~al}\mbox{.}(2011){Treister}, {Schawinski},
  {Volonteri}, {Natarajan}, \& {Gawiser}}]{Treister:2011aa}
{Treister} E., {Schawinski} K., {Volonteri} M., {Natarajan} P., {Gawiser} E.,
  2011, nat, 474, 356

\bibitem[{{Treister} {et~al}\mbox{.}(2009){Treister}, {Urry}, \&
  {Virani}}]{Treister:2009aa}
{Treister} E., {Urry} C.~M., {Virani} S., 2009, apj, 696, 110

\bibitem[{{Turner} {et~al}\mbox{.}(1997){Turner}, {George}, {Nandra}, \&
  {Mushotzky}}]{Turner:1997aa}
{Turner} T.~J., {George} I.~M., {Nandra} K., {Mushotzky} R.~F., 1997, apjs,
  113, 23

\bibitem[{{van Dokkum} {et~al}\mbox{.}(2005){van Dokkum}, {Bell}, {Bouwens},
  {Brandt}, {Dickinson}, {Franx}, {Gawiser}, {Huang}, {Illingworth}, {Labbe},
  {Lira}, {Marchesini}, {McCarthy}, {Papovich}, {Rix}, {Taylor}, {Urry}, \&
  {Yi}}]{van-Dokkum:2005aa}
{van Dokkum} P. {et~al.}, 2005, Spitzer Proposal, 20708

\bibitem[{{Vanzella} {et~al}\mbox{.}(2009){Vanzella}, {Giavalisco},
  {Dickinson}, {Cristiani}, {Nonino}, {Kuntschner}, {Popesso}, {Rosati},
  {Renzini}, {Stern}, {Cesarsky}, {Ferguson}, \& {Fosbury}}]{Vanzella:2009aa}
{Vanzella} E. {et~al.}, 2009, apj, 695, 1163

\bibitem[{{Vasudevan} \& {Fabian}(2009)}]{Vasudevan:2009aa}
{Vasudevan} R.~V., {Fabian} A.~C., 2009, mnras, 392, 1124

\bibitem[{{Vignali} {et~al}\mbox{.}(2003){Vignali}, {Brandt}, \&
  {Schneider}}]{Vignali:2003aa}
{Vignali} C., {Brandt} W.~N., {Schneider} D.~P., 2003, aj, 125, 433

\bibitem[{{Vito} {et~al}\mbox{.}(2013){Vito}, {Vignali}, {Gilli}, {Comastri},
  {Iwasawa}, {Brandt}, {Alexander}, {Brusa}, {Lehmer}, {Bauer}, {Schneider},
  {Xue}, \& {Luo}}]{Vito:2013aa}
{Vito} F. {et~al.}, 2013, mnras, 428, 354

\bibitem[{{Volonteri}(2010)}]{Volonteri:2010aa}
{Volonteri} M., 2010, aapr, 18, 279

\bibitem[{{Volonteri} \& {Begelman}(2010)}]{Volonteri:2010ab}
{Volonteri} M., {Begelman} M.~C., 2010, mnras, 409, 1022

\bibitem[{{Volonteri} \& {Rees}(2005)}]{Volonteri:2005aa}
{Volonteri} M., {Rees} M.~J., 2005, apj, 633, 624

\bibitem[{{Volonteri} \& {Silk}(2014)}]{Volonteri:2014aa}
{Volonteri} M., {Silk} J., 2014, ArXiv e-prints, 1401.3513

\bibitem[{{Wang} {et~al}\mbox{.}(2004){Wang}, {Malhotra}, {Rhoads}, \&
  {Norman}}]{Wang:2004aa}
{Wang} J.~X., {Malhotra} S., {Rhoads} J.~E., {Norman} C.~A., 2004, apjl, 612,
  L109

\bibitem[{{Wilkes} {et~al}\mbox{.}(1994){Wilkes}, {Tananbaum}, {Worrall},
  {Avni}, {Oey}, \& {Flanagan}}]{Wilkes:1994aa}
{Wilkes} B.~J., {Tananbaum} H., {Worrall} D.~M., {Avni} Y., {Oey} M.~S.,
  {Flanagan} J., 1994, apjs, 92, 53

\bibitem[{{Wilkins} {et~al}\mbox{.}(2010){Wilkins}, {Bunker}, {Ellis}, {Stark},
  {Stanway}, {Chiu}, {Lorenzoni}, \& {Jarvis}}]{Wilkins:2010aa}
{Wilkins} S.~M., {Bunker} A.~J., {Ellis} R.~S., {Stark} D., {Stanway} E.~R.,
  {Chiu} K., {Lorenzoni} S., {Jarvis} M.~J., 2010, mnras, 403, 938

\bibitem[{{Willott} {et~al}\mbox{.}(2003){Willott}, {McLure}, \&
  {Jarvis}}]{Willott:2003aa}
{Willott} C.~J., {McLure} R.~J., {Jarvis} M.~J., 2003, apjl, 587, L15

\bibitem[{{Xue} {et~al}\mbox{.}(2011){Xue}, {Luo}, {Brandt}, {Bauer}, {Lehmer},
  {Broos}, {Schneider}, {Alexander}, {Brusa}, {Comastri}, {Fabian}, {Gilli},
  {Hasinger}, {Hornschemeier}, {Koekemoer}, {Liu}, {Mainieri}, {Paolillo},
  {Rafferty}, {Rosati}, {Shemmer}, {Silverman}, {Smail}, {Tozzi}, \&
  {Vignali}}]{Xue:2011aa}
{Xue} Y.~Q. {et~al.}, 2011, apjs, 195, 10

\end{thebibliography}
\bsp

\clearpage
\newpage
\onecolumn
\begin{appendix}
\section*{Appendix: Additional figures and tables}

\begin{table*}
\begin{center}
	\begin{tabular}{ll}
	\toprule
	{Keyword} & {Value}\\
	\midrule
		{CATALOG$\_$TYPE} & {ASCII$\_$HEAD}\\
		{DETECT$\_$MINAREA} & {20 - 90}\\
		{DETECT$\_$THRESH} & {0.5}\\
		{DEBLEND$\_$NTHRESH} & {32}\\
		{DEBLEND$\_$MINCONT} & {0.001}\\
		{WEIGHT$\_$TYPE} & {MAP$\_$VAR}\\
		{PHOT$\_$APERTURES} & {0.6$^{\prime\prime}$ - 2.0$^{\prime\prime}$}\\
		{PHOT$\_$AUTOPARAMS} & {2.5, 3.5}\\
		{PHOT$\_$PETROPARAMS} & {2.0, 3.5}\\
	\bottomrule
	\end{tabular}
\end{center}
\caption{\label{tab:SE_para}SExtractor parameter values. We adjusted the aperture size (PHOT$\_$APERTURES) and the minimum number of pixels above the threshold for a detection (DETECT$\_$MINAREA) for each source individually. Table \ref{tab:SE_ID} summarizes these values for each of our main sample sources. Note that the values given for PHOT$\_$APERTURES corresponds to the aperture diameter, not radius.}
\end{table*}

\begin{table*}
\begin{center}
	\begin{tabular}{llllll}
	\toprule
	{ID} & {DETECT$\_$MINAREA} & {PHOT$\_$APERTURES} & {ID} & {DETECT$\_$MINAREA} & {PHOT$\_$APERTURES}\\
	\midrule
	$121$ & $20$ & $2.0^{\prime\prime}$ & $392$ & $20$ & $2.0^{\prime\prime}$\\
	$150$ & $20$ & $2.0^{\prime\prime}$ & $402$ & $20$ & $1.0^{\prime\prime}$\\
	$173$ & $20$ & $1.0^{\prime\prime}$ & $403$ & $20$ & $1.0^{\prime\prime}$\\
	$184$ & $40$ & $1.5^{\prime\prime}$ & $410$ & $20$ & $1.2^{\prime\prime}$\\
	$189$ & $20$ & $1.0^{\prime\prime}$ & $428$ & $20$ & $1.0^{\prime\prime}$\\
	$199$ & $60$ & $2.0^{\prime\prime}$ & $430$ & $80$ & $1.5^{\prime\prime}$\\
	$211$ & $50$ & $1.0^{\prime\prime}$ & $444$ & $30$ & $1.0^{\prime\prime}$\\
	$217$ & $40$ & $1.5^{\prime\prime}$ & $455$ & $20$ & $2.0^{\prime\prime}$\\
	$221$ & $30$ & $1.5^{\prime\prime}$ & $456$ & $20$ & $1.0^{\prime\prime}$\\
	$226$ & $20$ & $2.0^{\prime\prime}$ & $460$ & $20$ & $2.0^{\prime\prime}$\\
	$242$ & $50$ & $1.5^{\prime\prime}$ & $462$ & $20$ & $1.0^{\prime\prime}$\\
	$244$ & $20$ & $2.0^{\prime\prime}$ & $466$ & $20$ & $1.5^{\prime\prime}$\\
	$258$ & $20$ & $0.6^{\prime\prime}$ & $485$ & $70$ & $2.0^{\prime\prime}$\\
	$273$ & $70$ & $2.0^{\prime\prime}$ & $496$ & $20$ & $1.0^{\prime\prime}$\\
	$296$ & $20$ & $1.0^{\prime\prime}$ & $522$ & $20$ & $1.0^{\prime\prime}$\\
	$301$ & $30$ & $2.0^{\prime\prime}$ & $535$ & $40$ & $1.0^{\prime\prime}$\\
	$302$ & $20$ & $1.0^{\prime\prime}$ & $539$ & $30$ & $1.0^{\prime\prime}$\\
	$303$ & $80$ & $2.0^{\prime\prime}$ & $546$ & $30$ & $1.0^{\prime\prime}$\\
	$306$ & $90$ & $1.5^{\prime\prime}$ & $556$ & $20$ & $2.0^{\prime\prime}$\\
	$318$ & $20$ & $2.0^{\prime\prime}$ & $574$ & $20$ & $1.5^{\prime\prime}$\\
	$321$ & $30$ & $2.0^{\prime\prime}$ & $578$ & $90$ & $2.0^{\prime\prime}$\\
	$325$ & $20$ & $1.0^{\prime\prime}$ & $583$ & $50$ & $0.6^{\prime\prime}$\\
	$328$ & $20$ & $1.5^{\prime\prime}$ & $589$ & $50$ & $1.5^{\prime\prime}$\\
	$331$ & $20$ & $1.0^{\prime\prime}$ & $591$ & $60$ & $1.0^{\prime\prime}$\\
	$348$ & $20$ & $1.5^{\prime\prime}$ & $620$ & $20$ & $2.0^{\prime\prime}$\\
	$354$ & $20$ & $1.5^{\prime\prime}$ & $624$ & $20$ & $1.0^{\prime\prime}$\\
	$371$ & $20$ & $1.0^{\prime\prime}$ & $625$ & $20$ & $1.5^{\prime\prime}$\\
	$373$ & $50$ & $2.0^{\prime\prime}$ & $630$ & $90$ & $2.0^{\prime\prime}$\\
	$389$ & $20$ & $1.0^{\prime\prime}$ & $651$ & $90$ & $2.0^{\prime\prime}$\\					
	\bottomrule
	\end{tabular}
\end{center}
\caption{\label{tab:SE_ID}SExtratcor DETECT$\_$MINAREA and PHOT$\_$APERTURES parameter values for each of our main sample sources. Note that the values given for PHOT$\_$APERTURES corresponds to the aperture diameter, not radius.}
\end{table*}
\clearpage
\newpage

\newpage
\label{sec:ccdiagrams}
\begin{figure*}
\begin{centering}
\begin{minipage}[t]{0.49\textwidth}
\begin{center}
\includegraphics[scale=0.33]{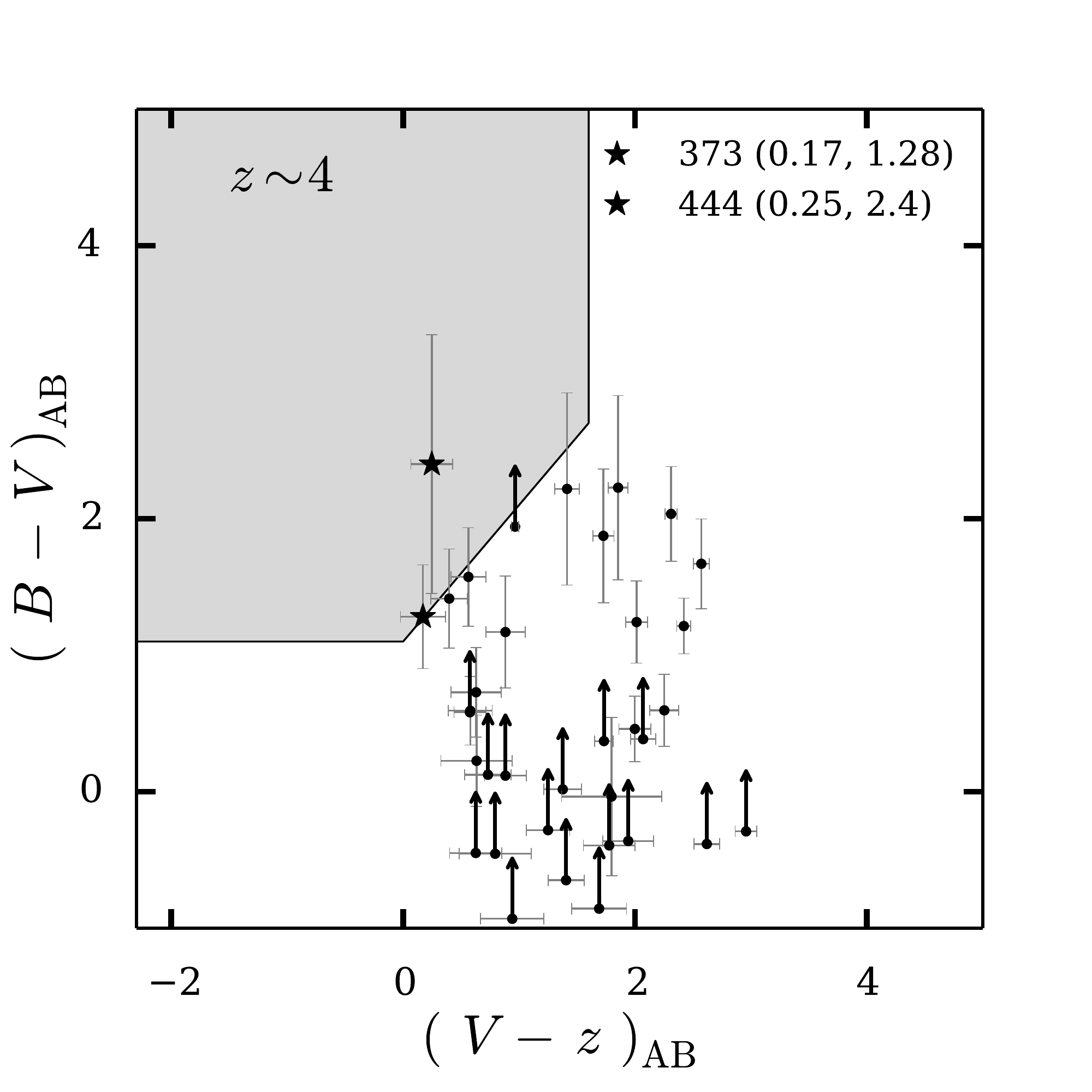}
\end{center}%
\end{minipage}%
\begin{minipage}[t]{0.49\textwidth}%
\begin{center}
\includegraphics[scale=0.33]{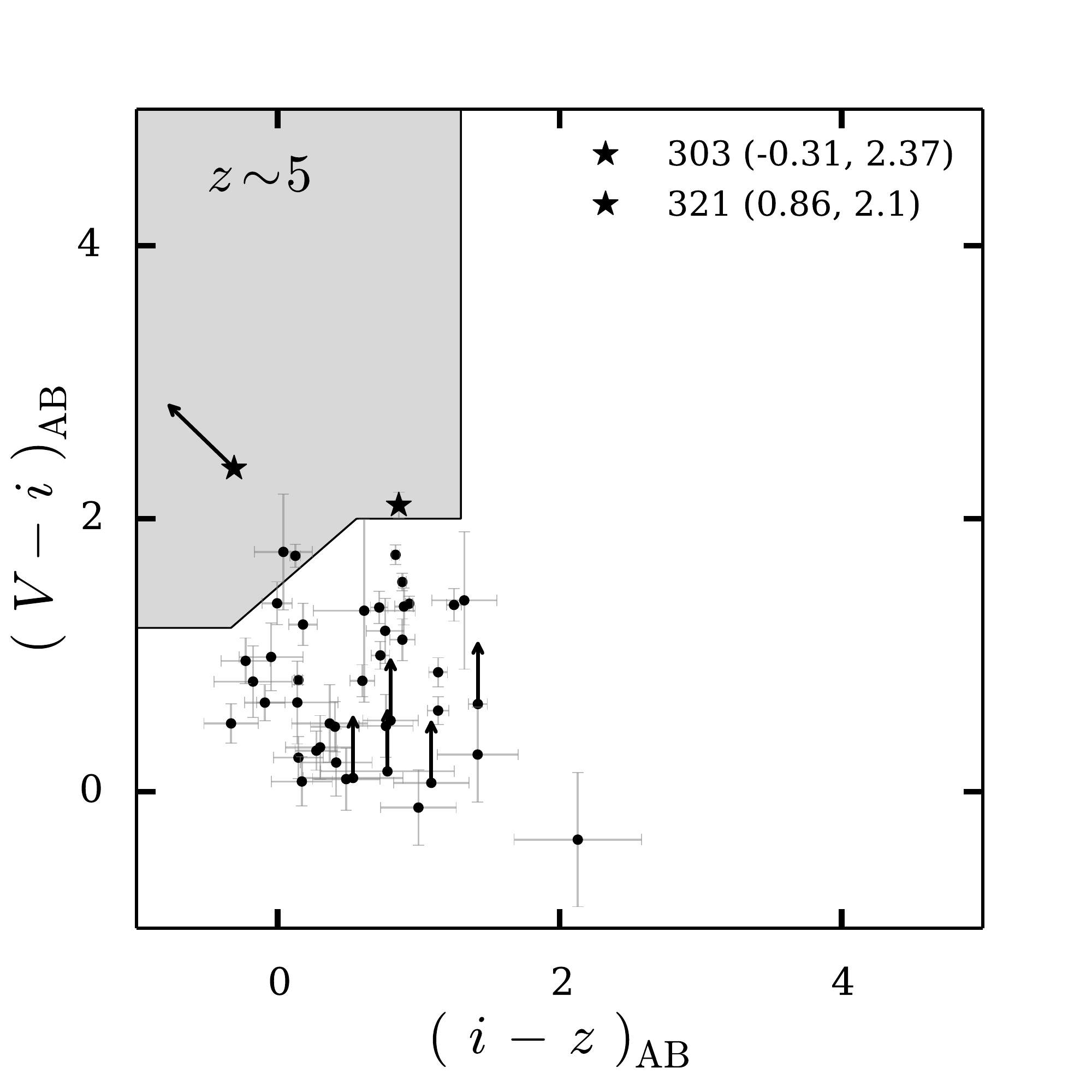}
\end{center}%
\end{minipage} 
\end{centering}
\begin{centering}
\begin{minipage}[t]{\textwidth}%
\begin{center}%
\includegraphics[scale=0.33]{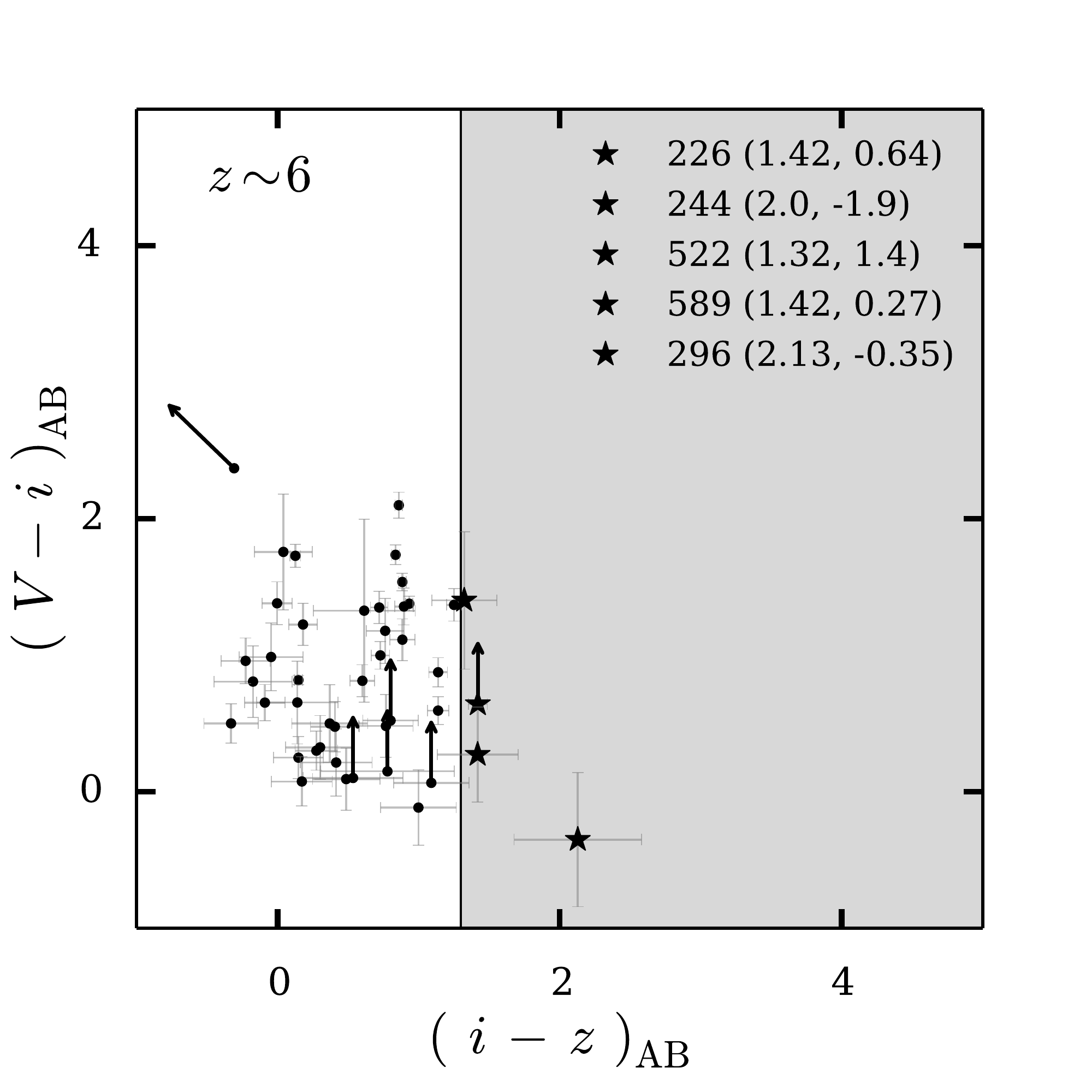}
\end{center}%
\end{minipage} \caption{\label{fig:cc} Colour-Colour Diagrams. The sources displayed here were categorized according to the conditions given in Section \ref{sec:cc}. Sources that meet the colour criteria lie in the grey shaded areas and are marked with an asterisk. The upper left plot illustrates $z\sim4$ sources. The arrows indicate upper limits in the $B$ and in the $z$ band. Note that sources with upper limits in the $B$ and the V or in the $V$ and the $z$ filter are not included, since their position can not be determined. The upper right panel and the figure at the bottom highlight the position of $z\sim5$ and $z\sim6$ sources respectively.}
\end{centering}
\end{figure*}

\newpage
\label{sec:Spitzercounterparts}
\begin{figure*}
\begin{minipage}[t]{0.49\textwidth}
\includegraphics[scale=0.5]{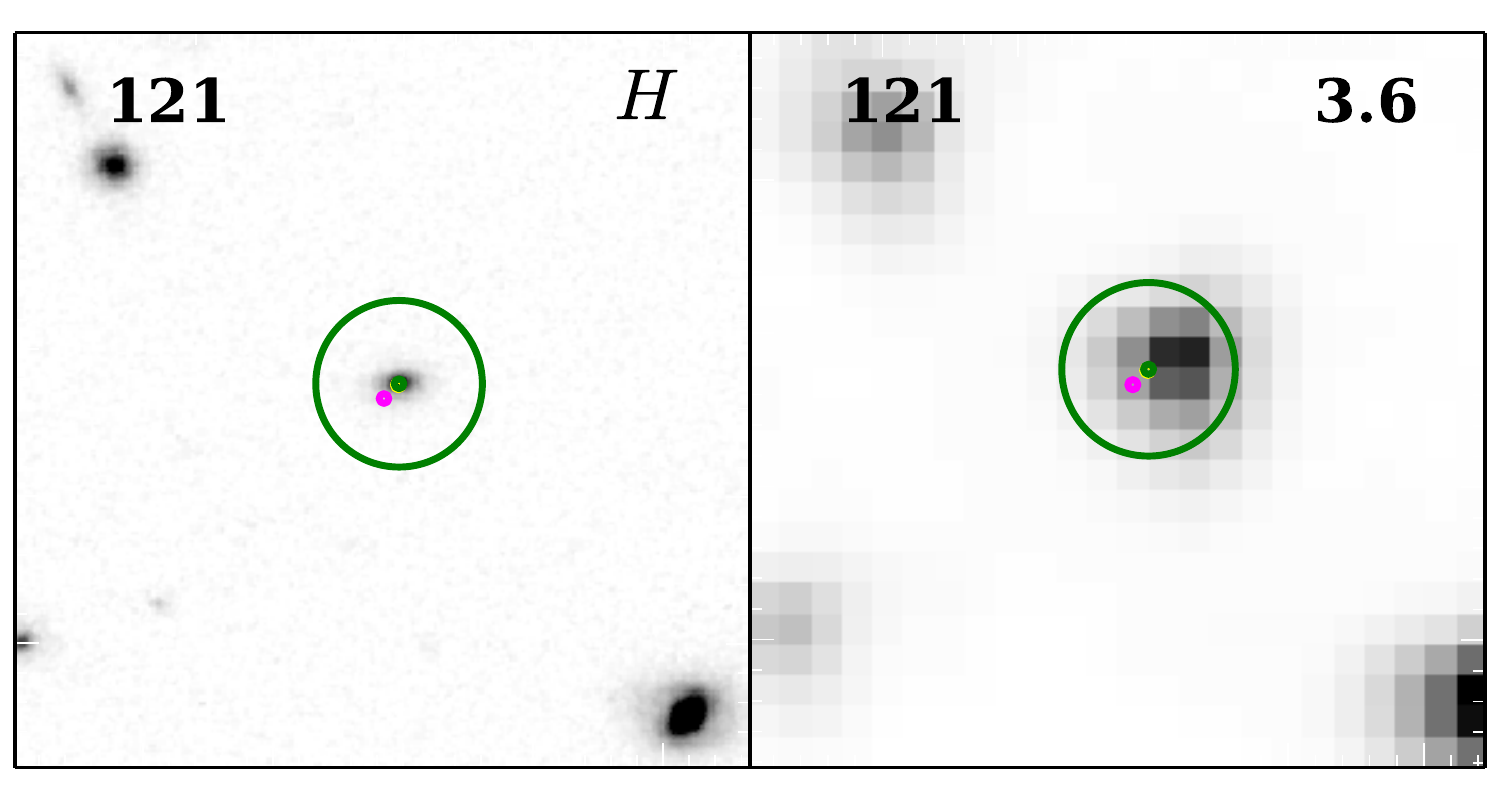}
\end{minipage}
\begin{minipage}[t]{0.49\textwidth}
\includegraphics[scale=0.5]{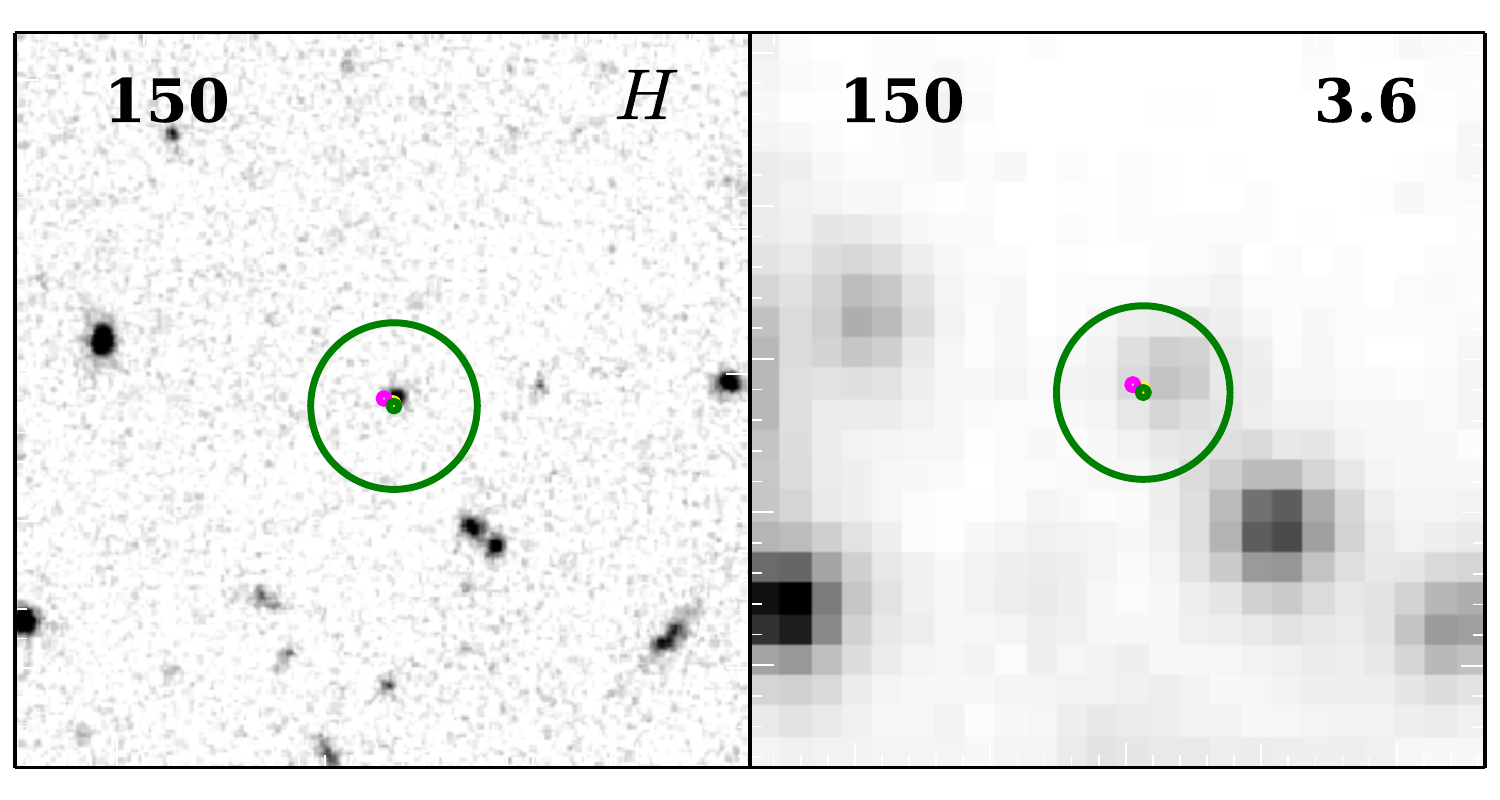}
\end{minipage}

\begin{minipage}[t]{0.49\textwidth}
\includegraphics[scale=0.5]{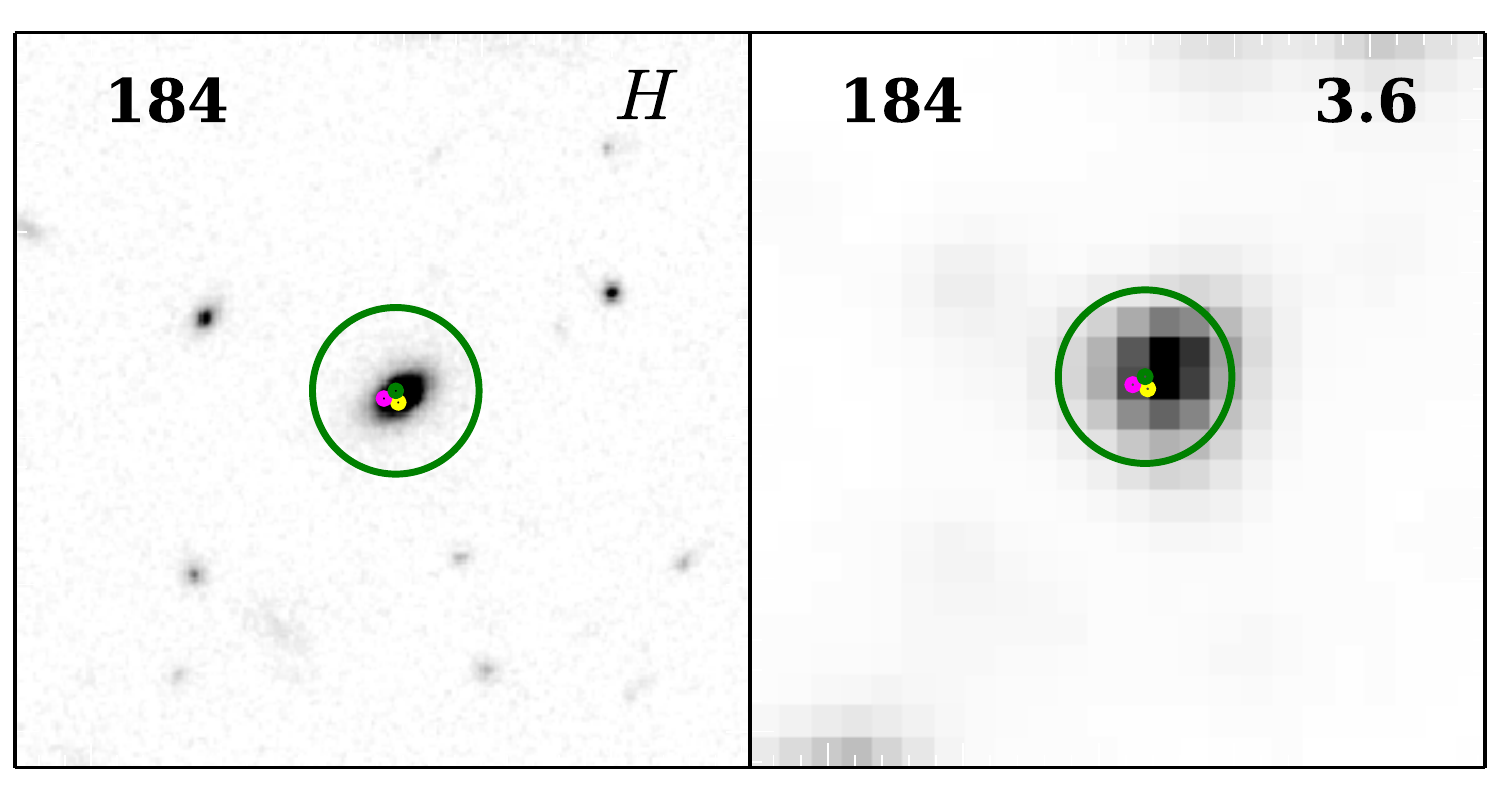}
\end{minipage}
\begin{minipage}[t]{0.49\textwidth}
\includegraphics[scale=0.5]{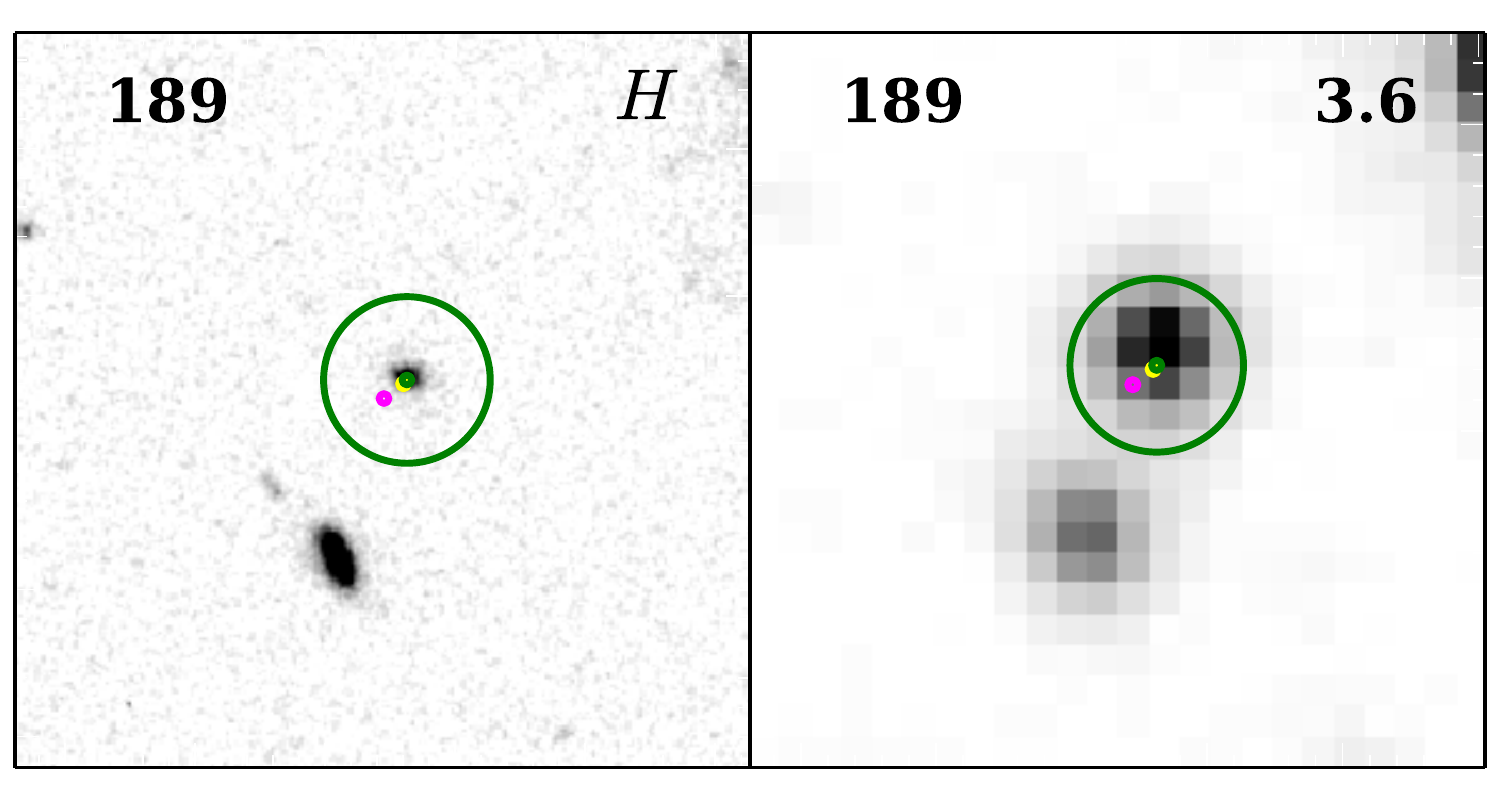}
\end{minipage}

\begin{minipage}[t]{0.49\textwidth}
\includegraphics[scale=0.5]{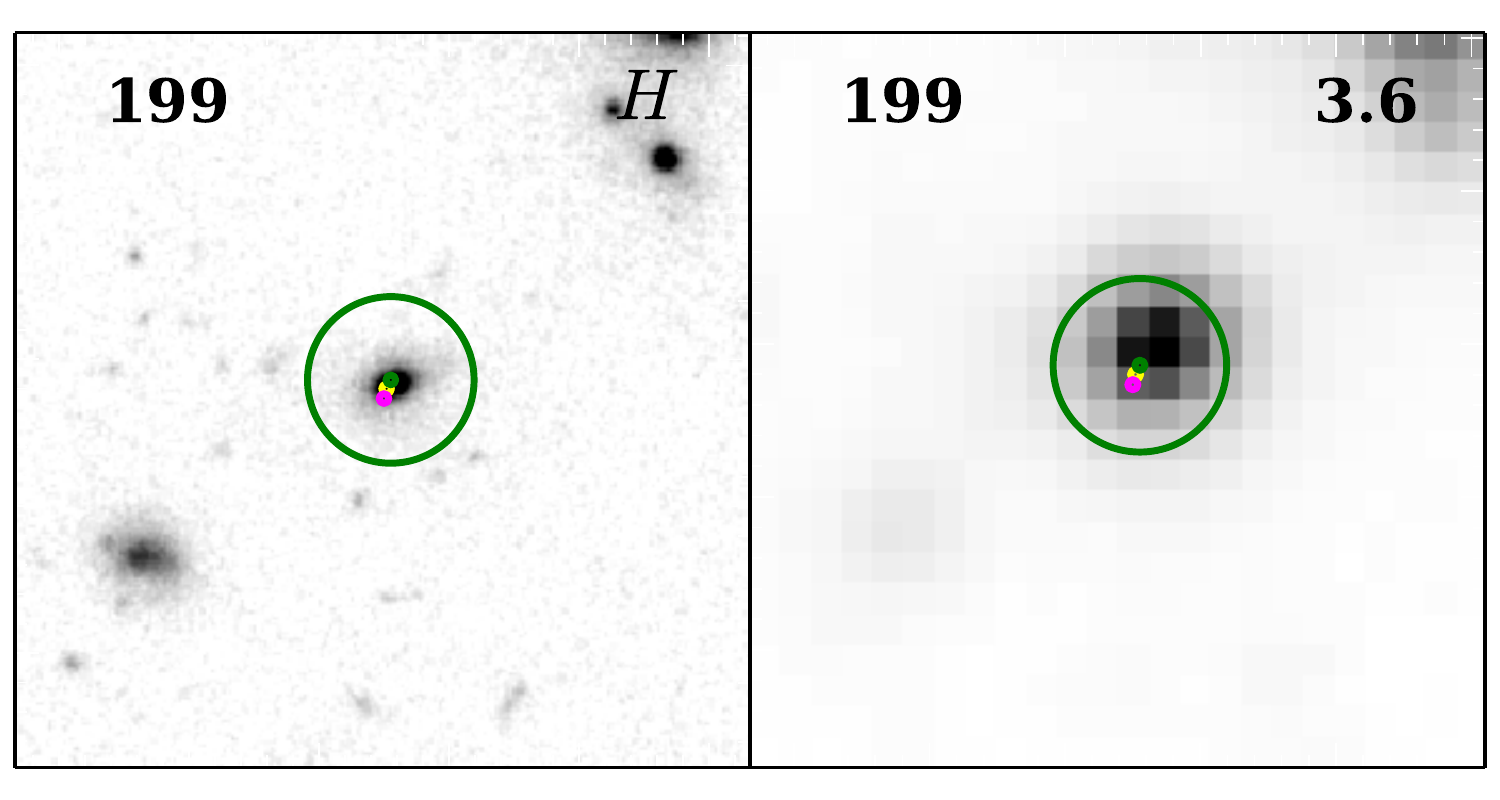}
\end{minipage}
\begin{minipage}[t]{0.49\textwidth}
\includegraphics[scale=0.5]{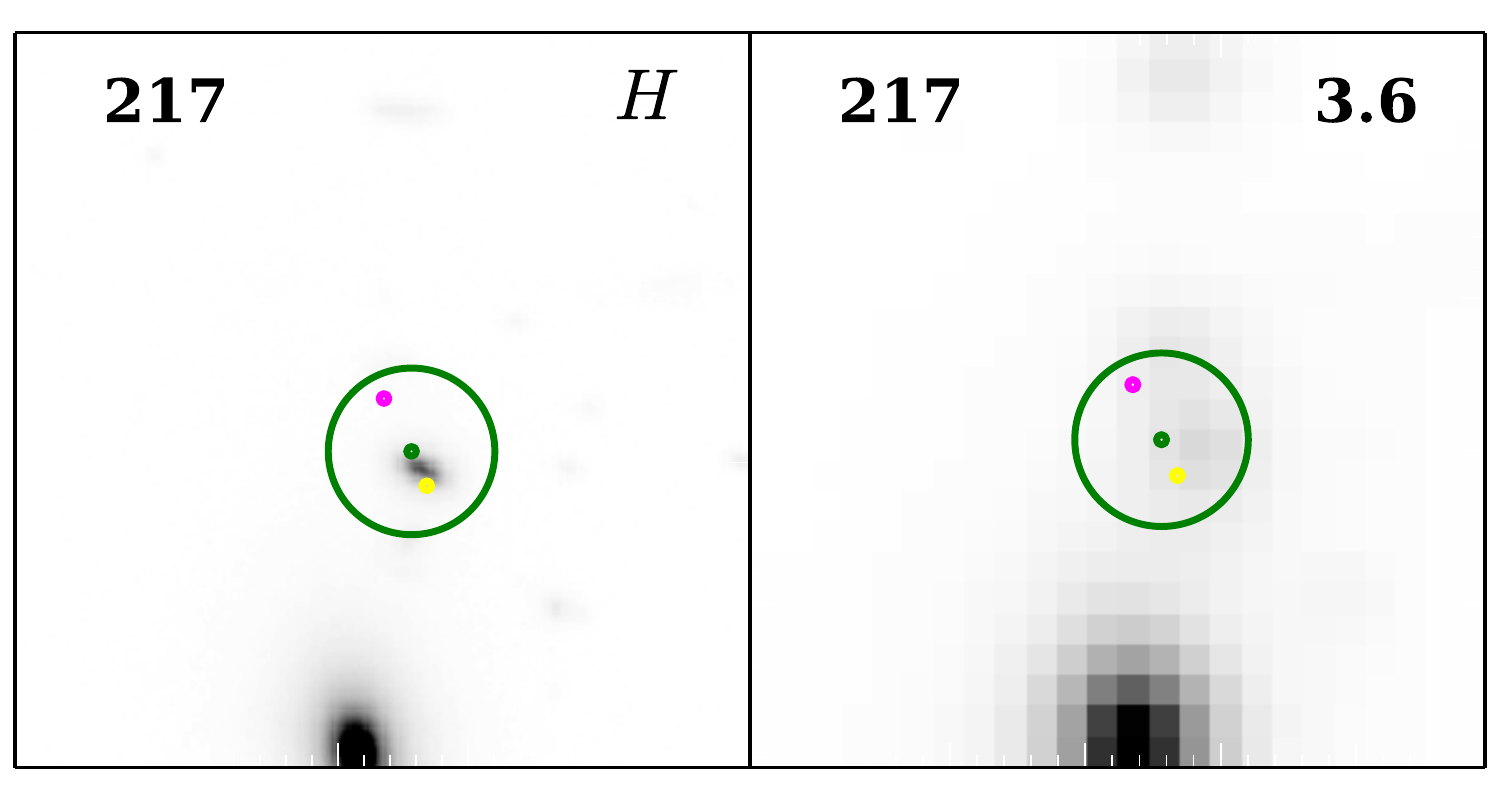}
\end{minipage}

\begin{minipage}[t]{0.49\textwidth}
\includegraphics[scale=0.5]{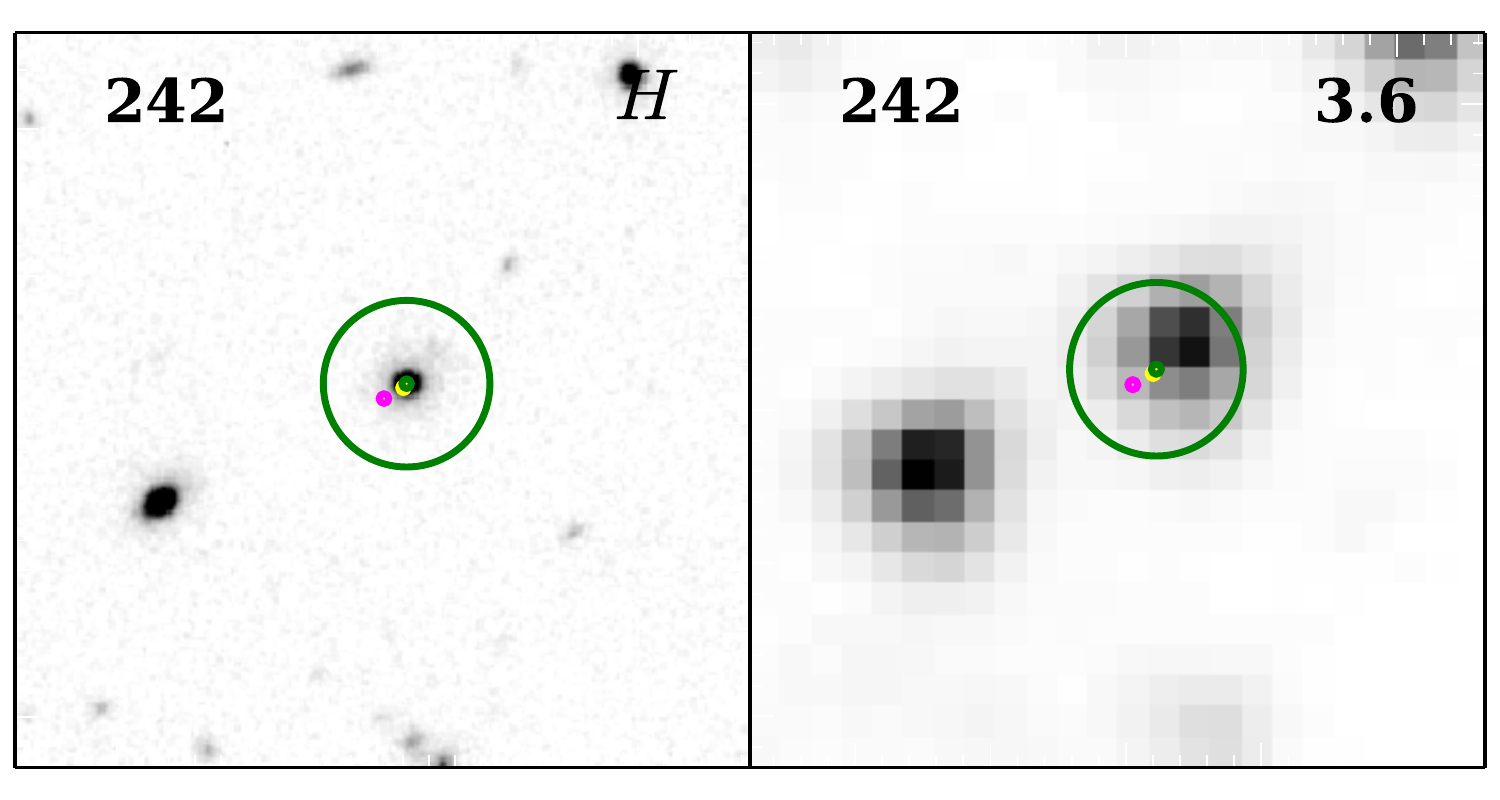}
\end{minipage}
\begin{minipage}[t]{0.49\textwidth}
\includegraphics[scale=0.5]{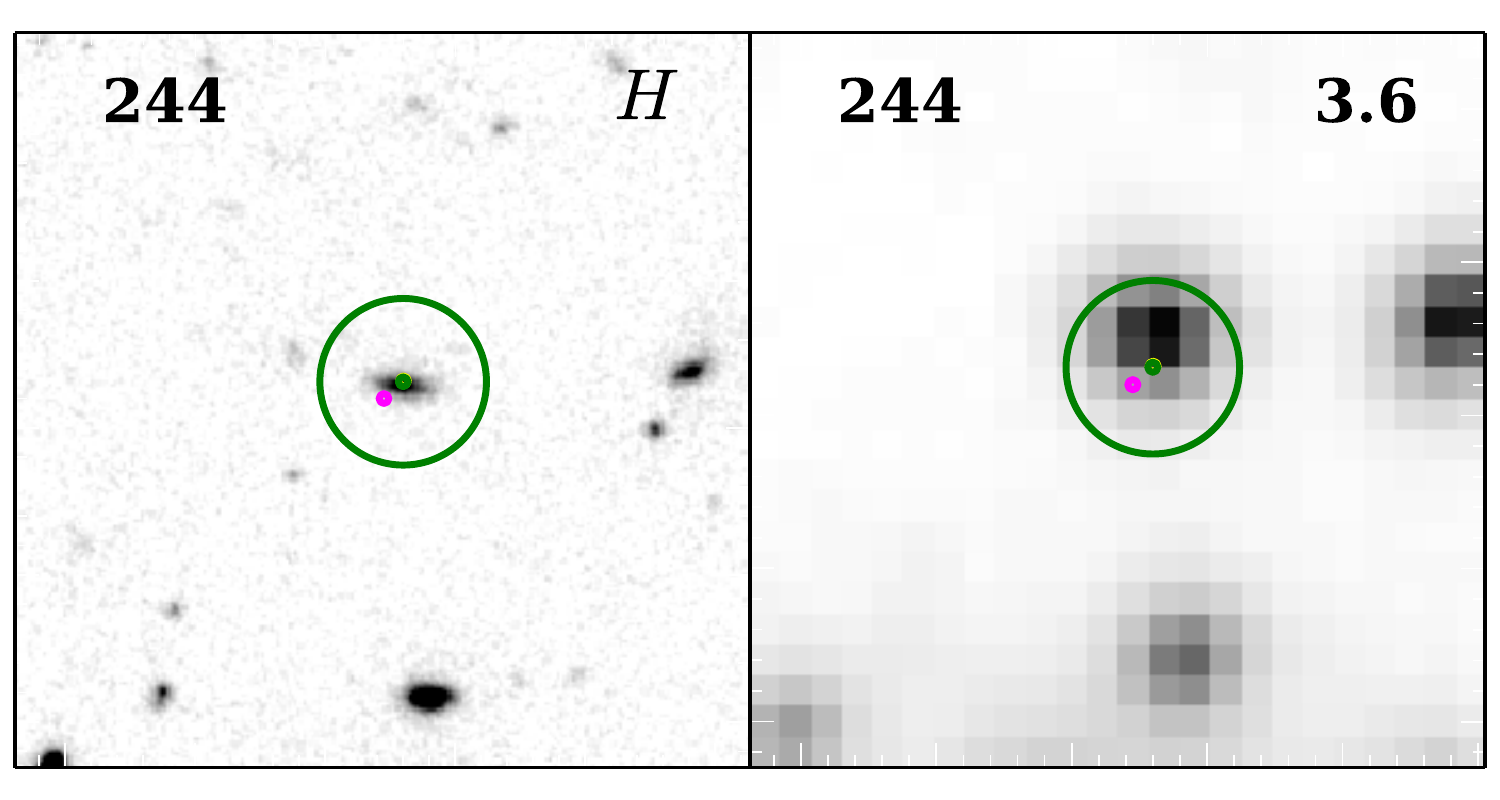}
\end{minipage}

\begin{minipage}[t]{0.49\textwidth}
\includegraphics[scale=0.5]{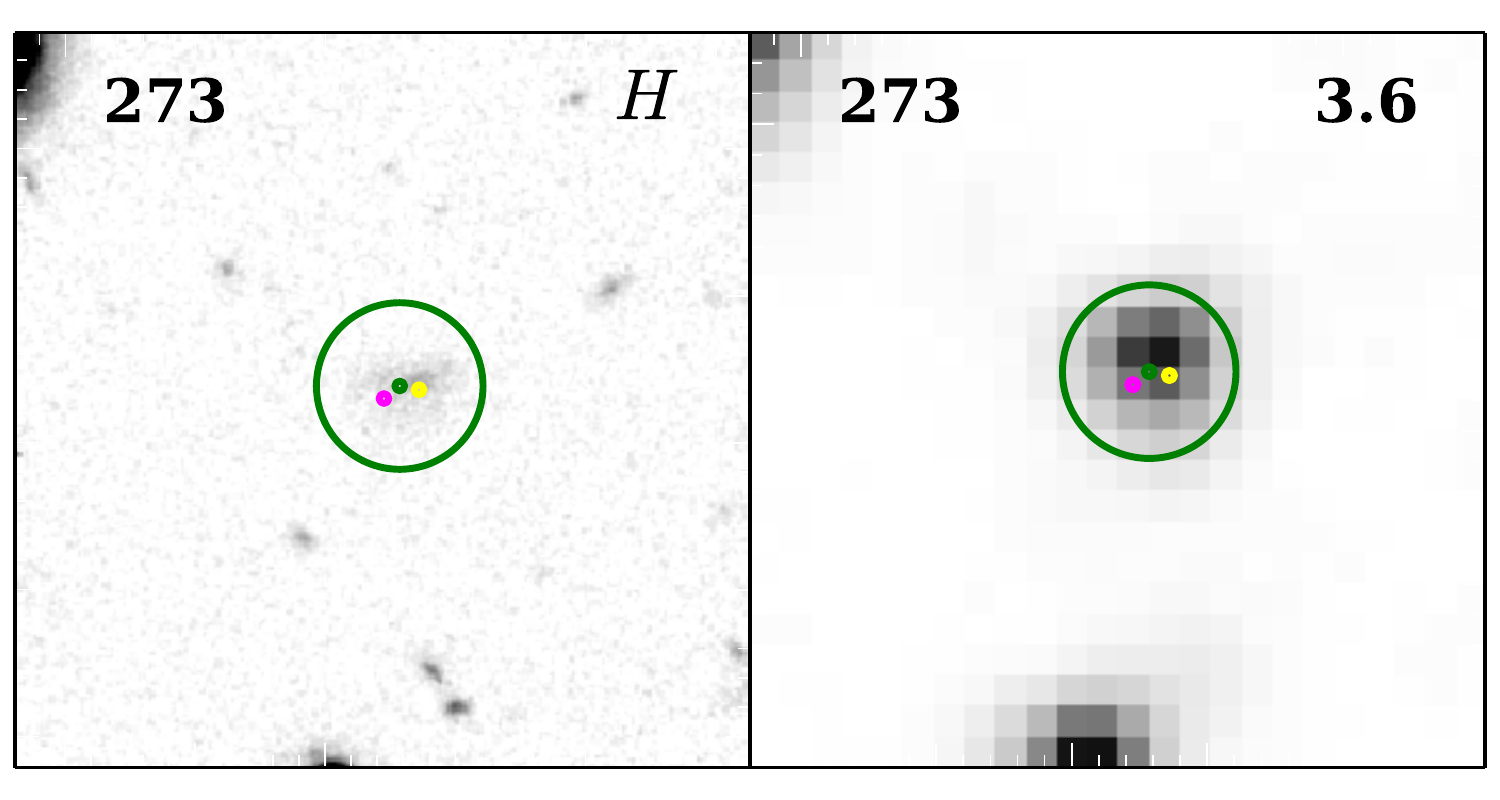}
\end{minipage}
\begin{minipage}[t]{0.49\textwidth}
\includegraphics[scale=0.5]{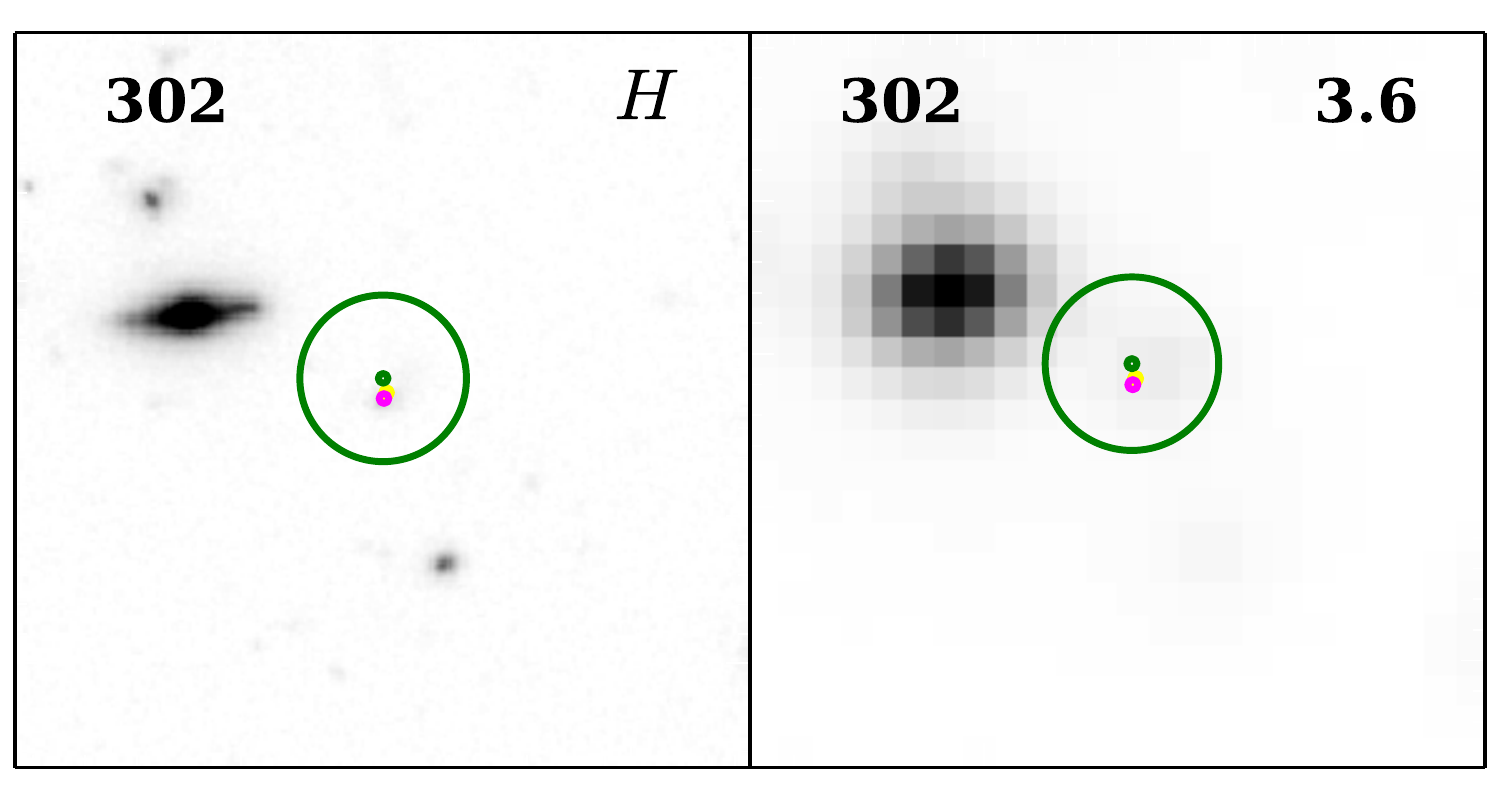}
\end{minipage}

\caption{$H$ band and 3.6 micron stamps for all objects for which the \Spitzer\ flux values are included in the photometric redshift determination, part 1/3. Shown in magenta is the original \Chandra\ 4-Ms catalog position. The yellow circle marks the object's $H$ band position that we determined by running SExtractor. The green point shows the position of the \Spitzer\ object closest to the $H$ band position. The green circle around this green point has a radius of 1.7$^{\prime\prime}$. It indicates the PSF size for the 3.6 micron IRAC channel. Due to source confusion the \Spitzer\ flux values could not be used for the whole sample. All stamps were colour inverted and are 15$^{\prime\prime}$ x 15$^{\prime\prime}$ in size.}
\end{figure*}

\begin{figure*}

\begin{minipage}[t]{0.49\textwidth}
\includegraphics[scale=0.5]{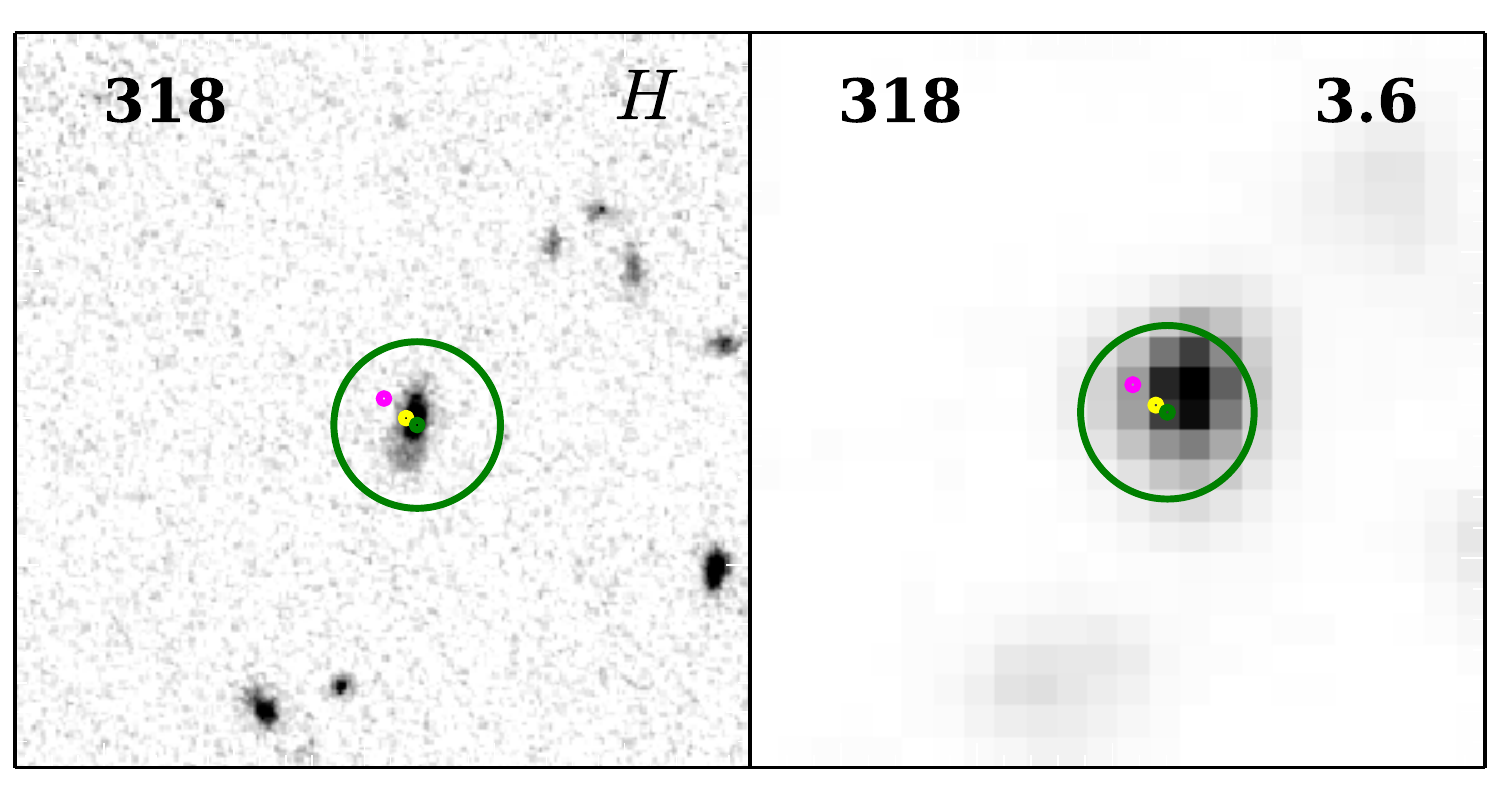}
\end{minipage}
\begin{minipage}[t]{0.49\textwidth}
\includegraphics[scale=0.5]{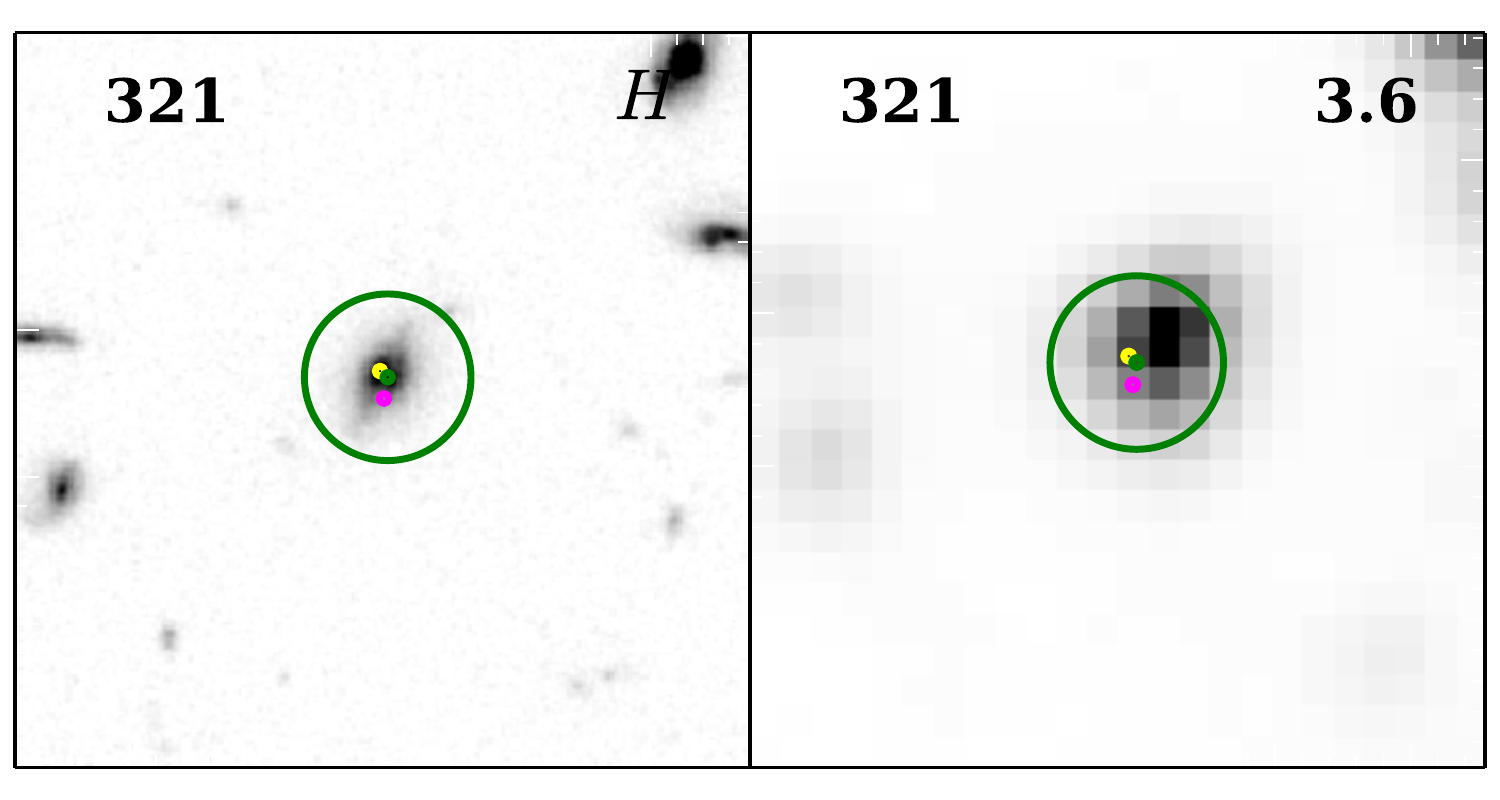}
\end{minipage}

\begin{minipage}[t]{0.49\textwidth}
\includegraphics[scale=0.5]{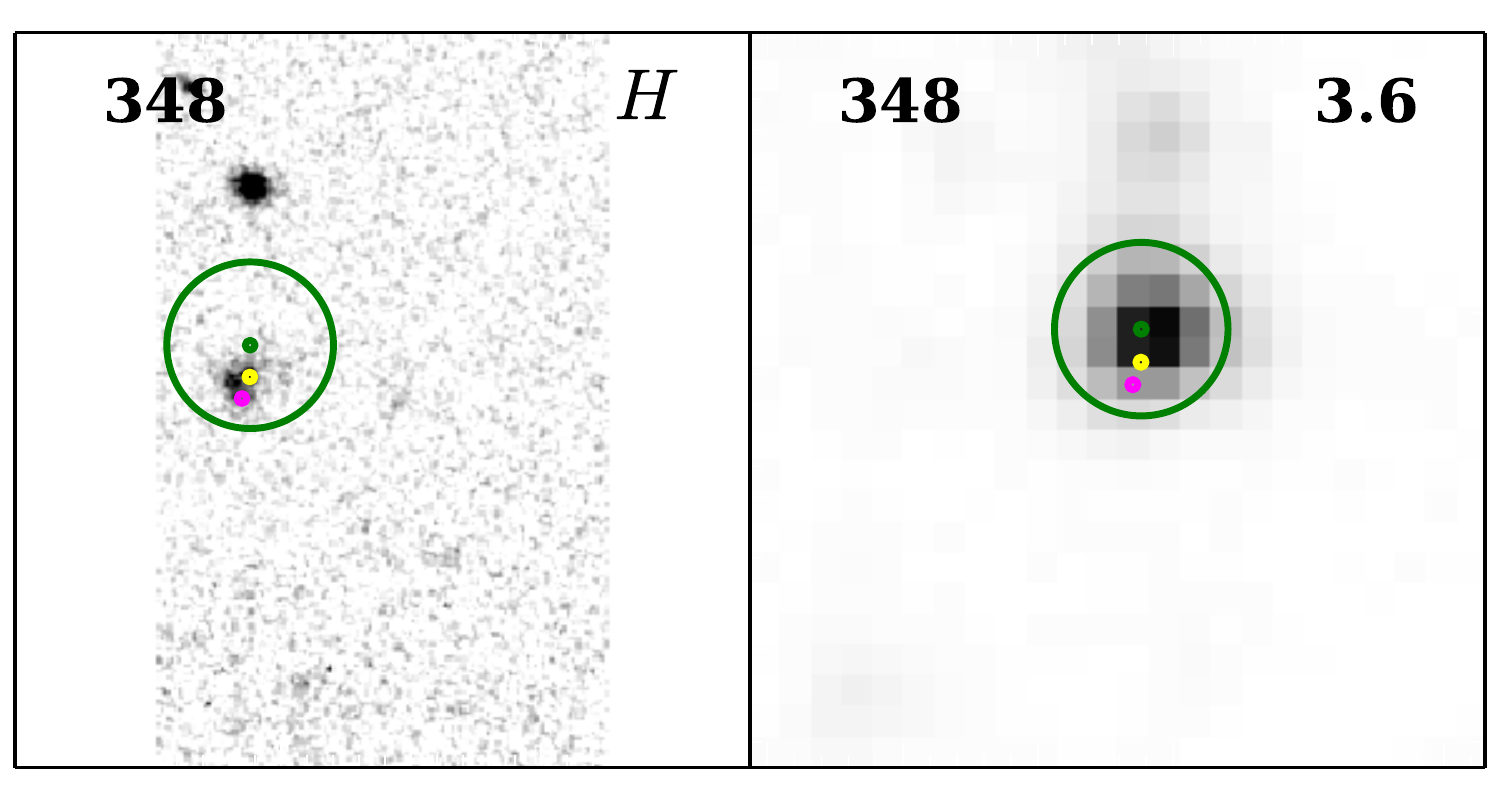}
\end{minipage}
\begin{minipage}[t]{0.49\textwidth}
\includegraphics[scale=0.5]{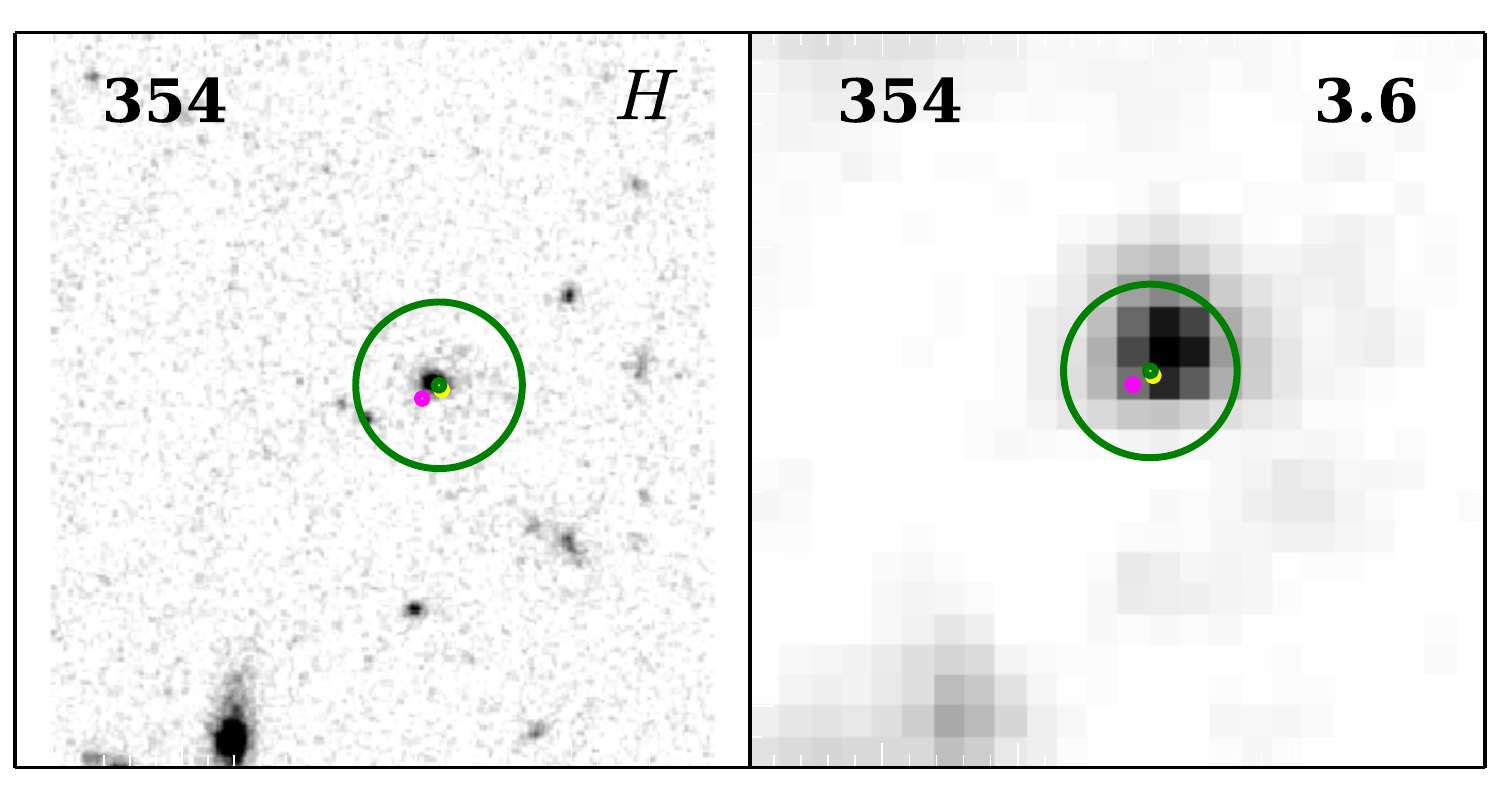}
\end{minipage}

\begin{minipage}[t]{0.49\textwidth}
\includegraphics[scale=0.5]{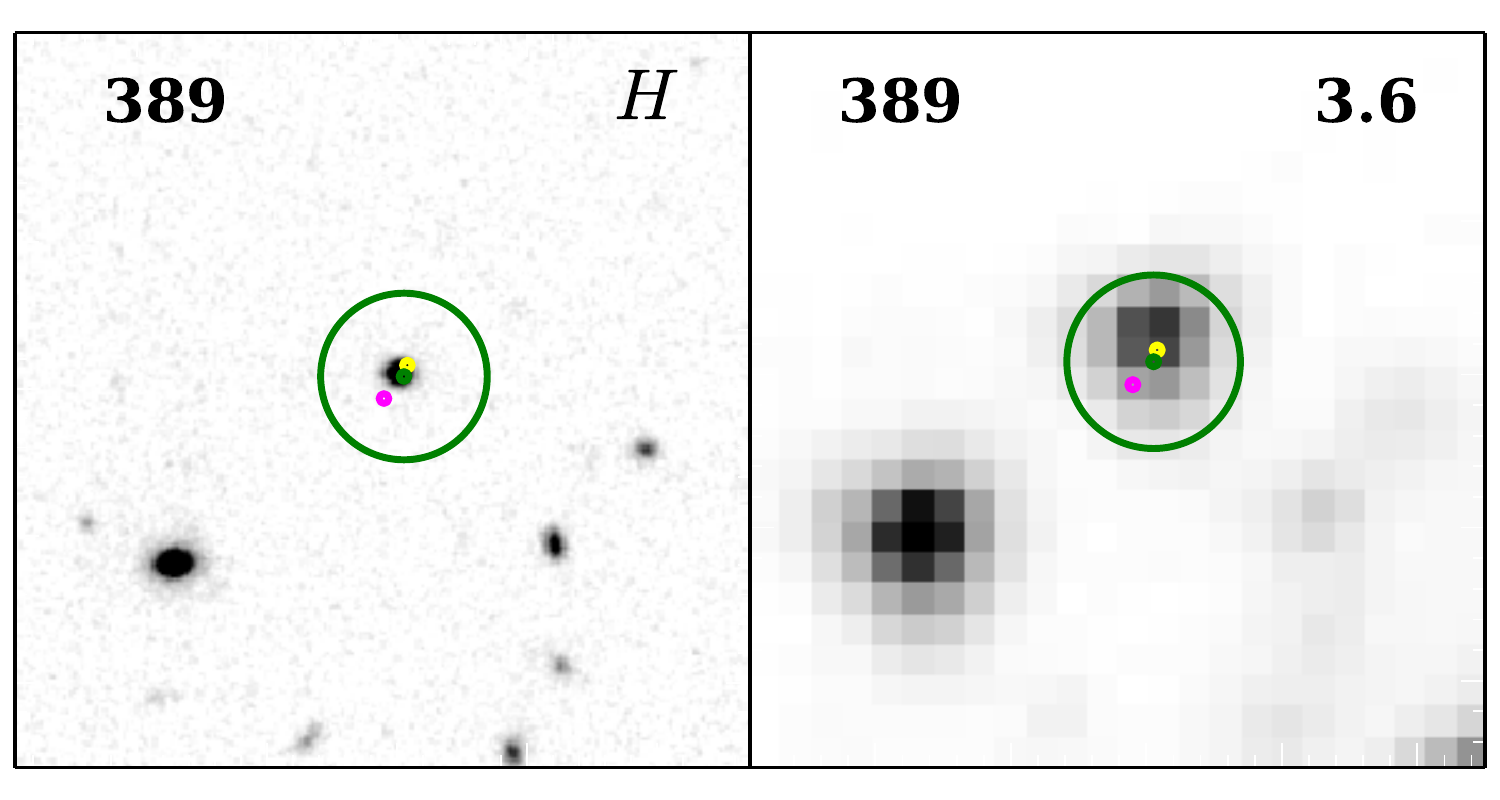}
\end{minipage}
\begin{minipage}[t]{0.49\textwidth}
\includegraphics[scale=0.5]{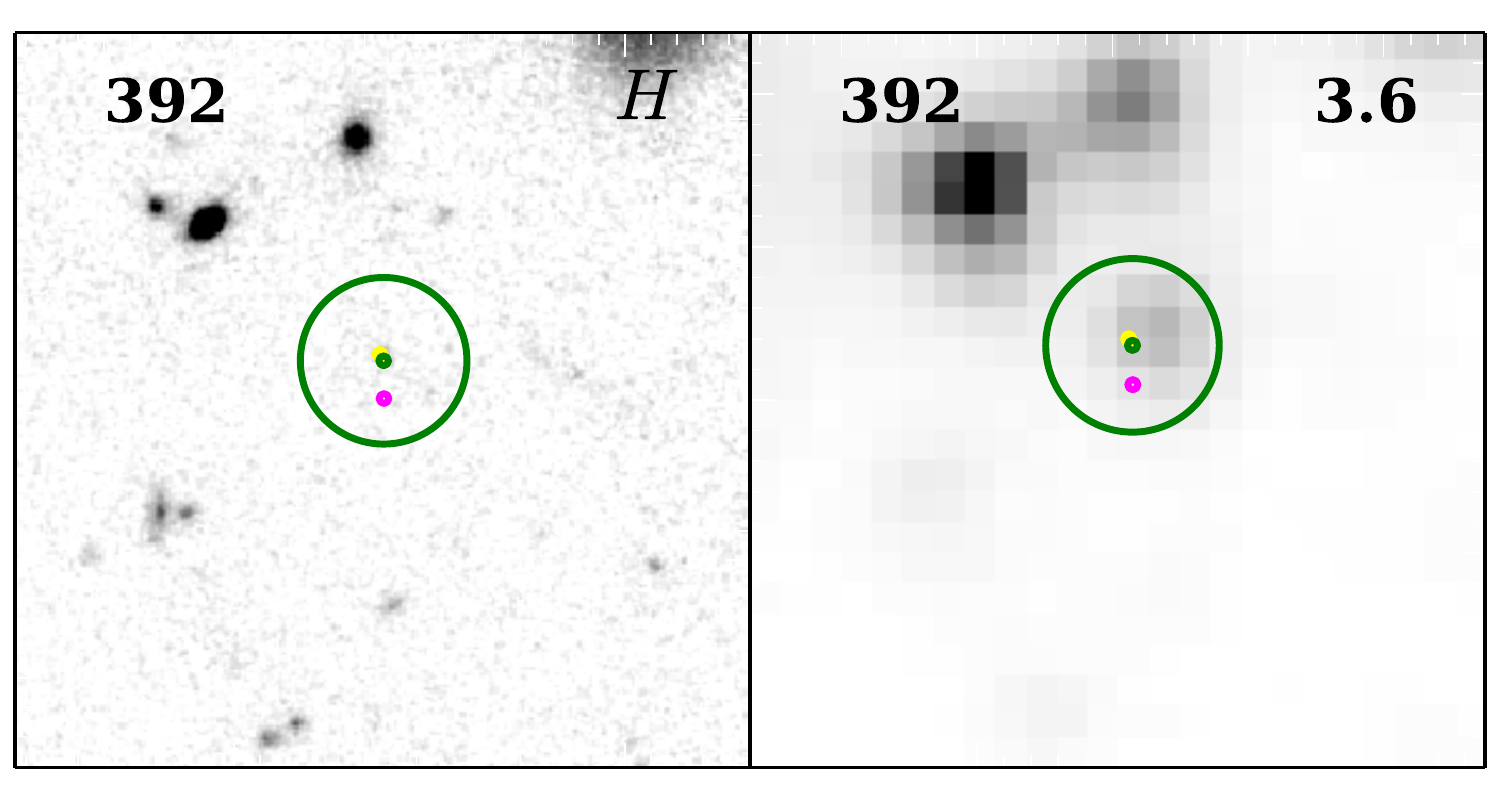}
\end{minipage}

\begin{minipage}[t]{0.49\textwidth}
\includegraphics[scale=0.5]{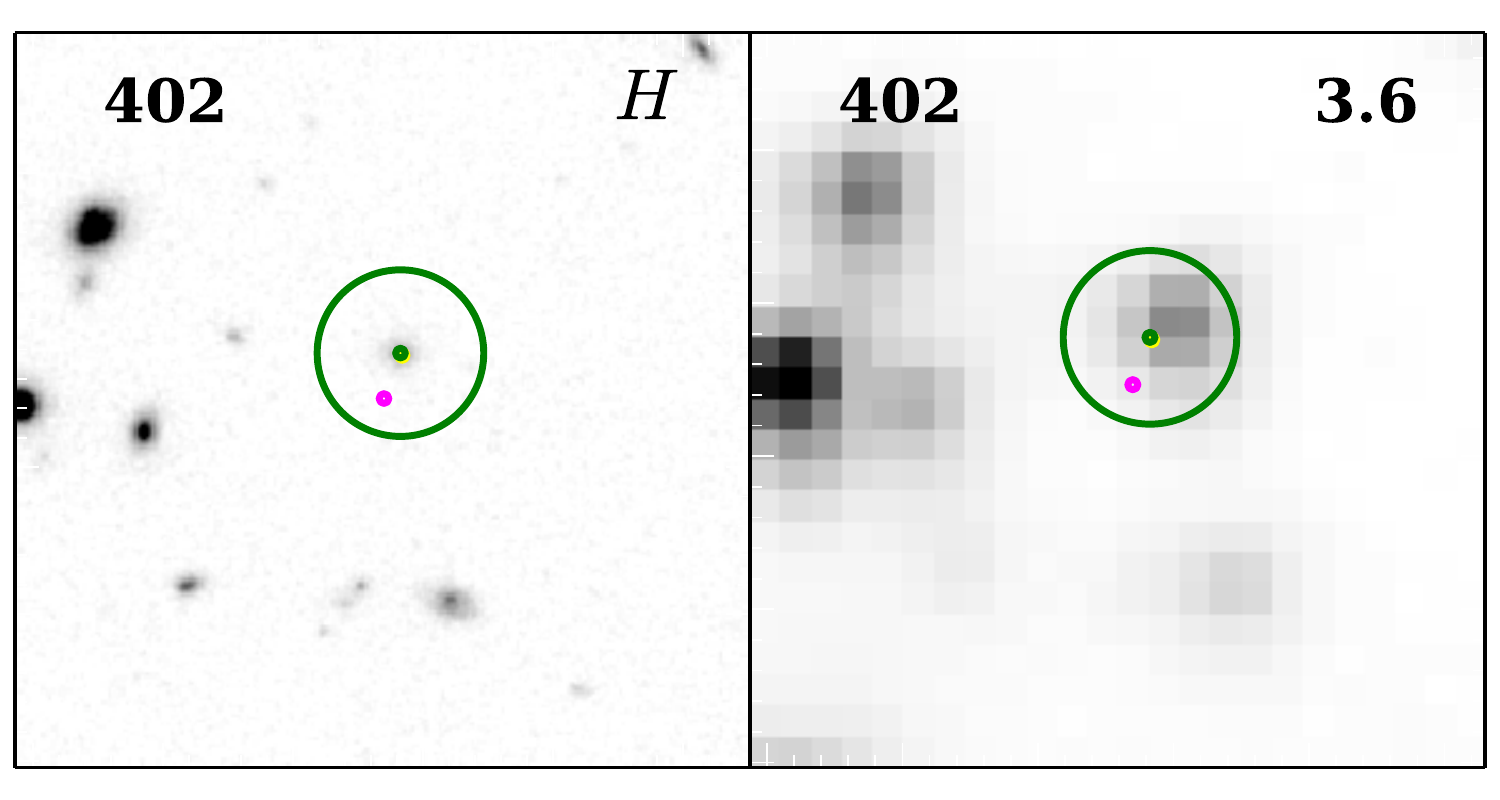}
\end{minipage}
\begin{minipage}[t]{0.49\textwidth}
\includegraphics[scale=0.5]{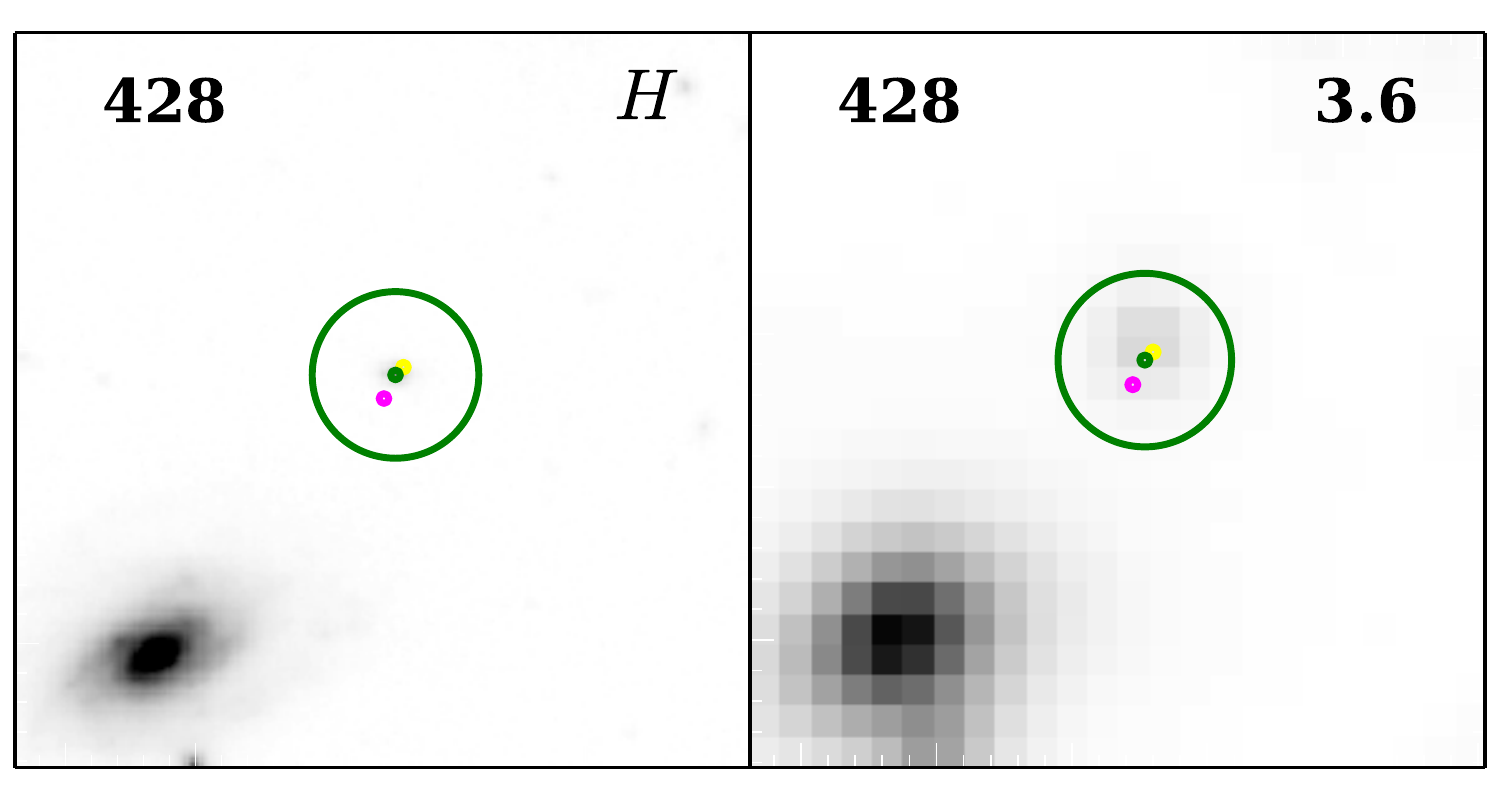}
\end{minipage}

\begin{minipage}[t]{0.49\textwidth}
\includegraphics[scale=0.5]{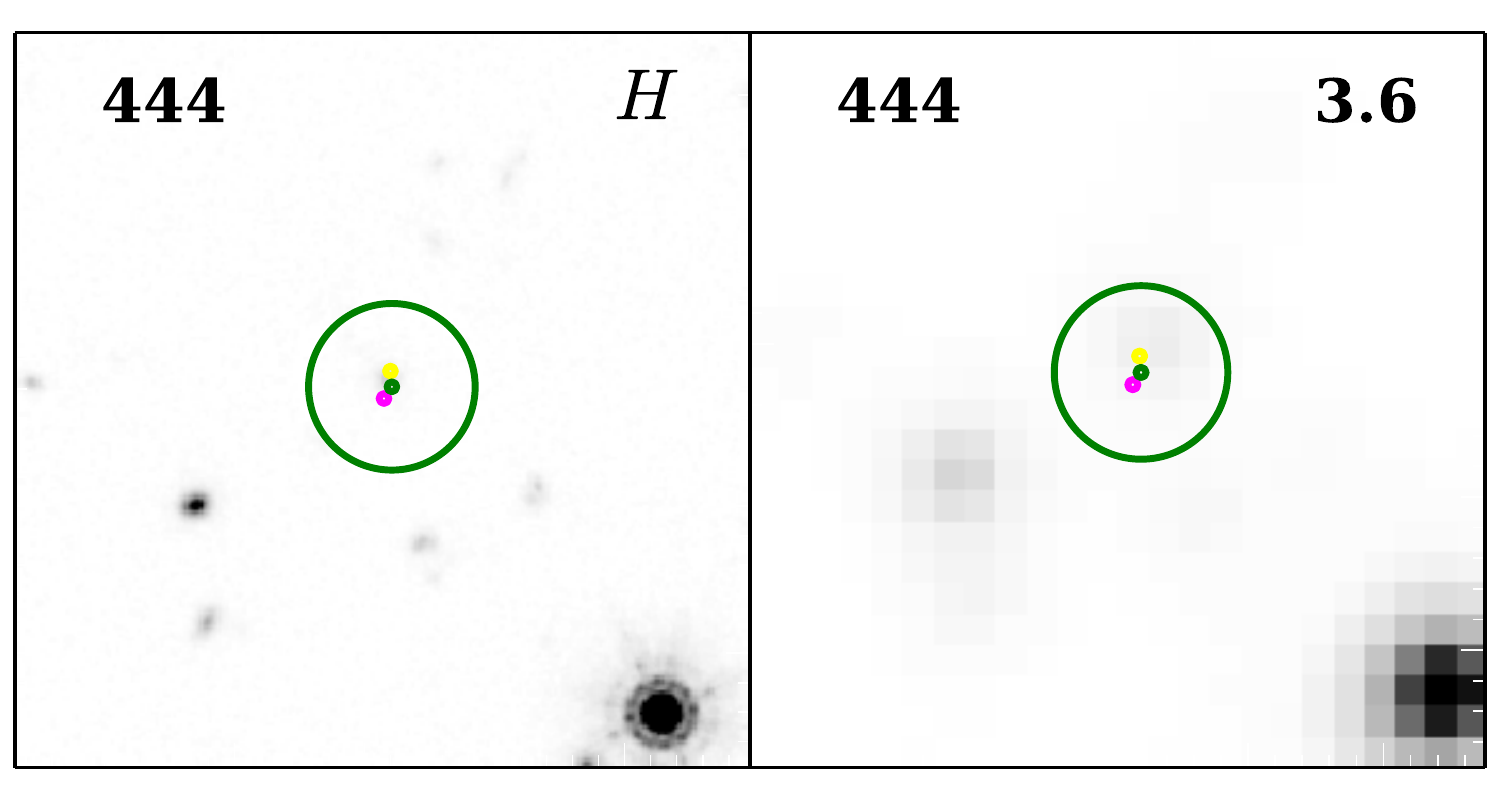}
\end{minipage}
\begin{minipage}[t]{0.49\textwidth}
\includegraphics[scale=0.5]{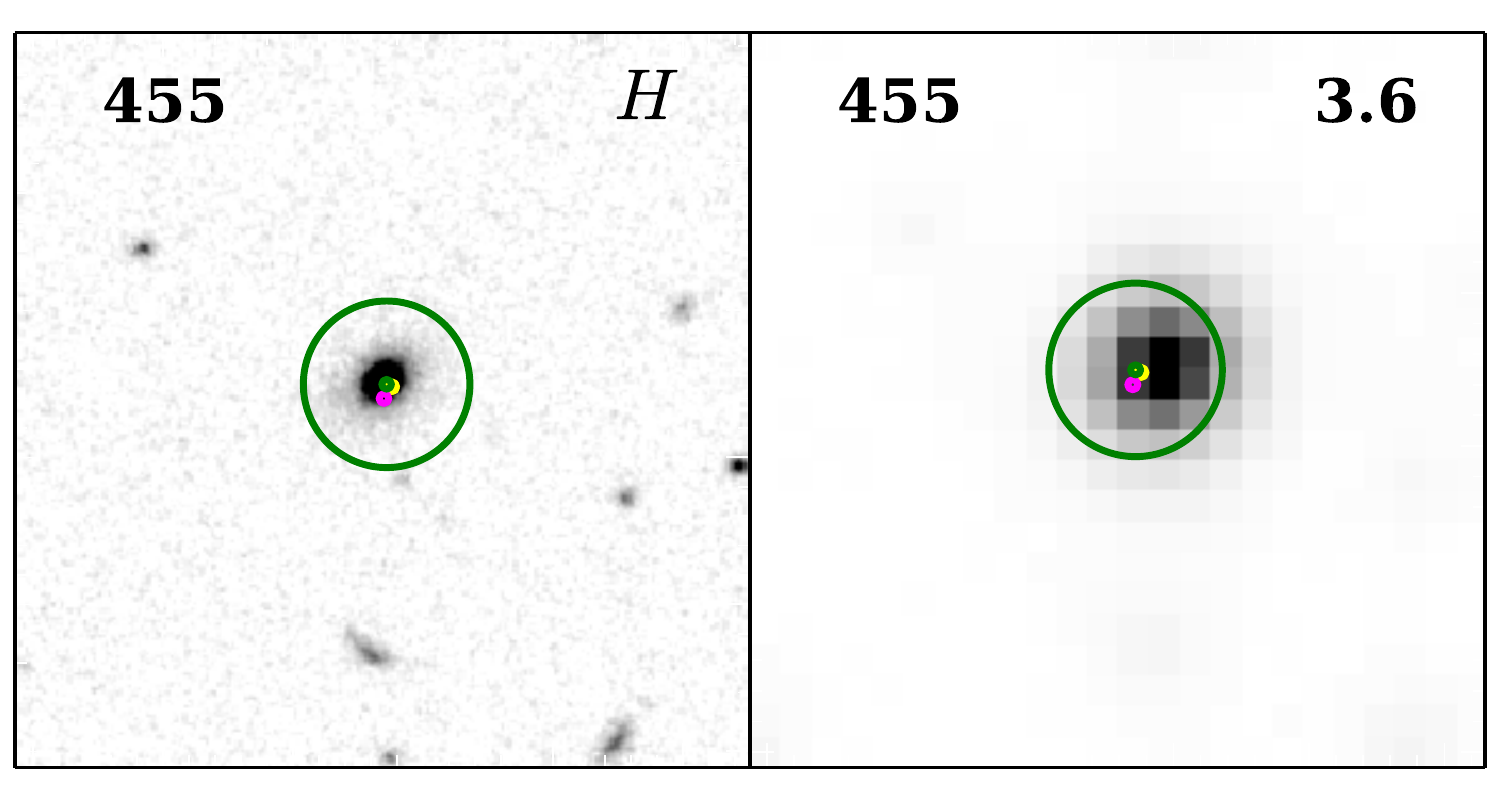}
\end{minipage}

\caption{$H$ band and 3.6 micron stamps for all objects for which the \Spitzer\ flux values are included in the photometric redshift determination, part 2/3. Shown in magenta is the original \Chandra\ 4-Ms catalog position. The yellow circle marks the object's $H$ band position that we determined by running SExtractor. The green point shows the position of the \Spitzer\ object closest to the $H$ band position. The green circle around this green point has a radius of 1.7$^{\prime\prime}$. It indicates the PSF size for the 3.6 micron IRAC channel. Due to source confusion the \Spitzer\ flux values could not be used for the whole sample. All stamps were colour inverted and are 15$^{\prime\prime}$ x 15$^{\prime\prime}$ in size.}
\end{figure*}

\begin{figure*}

\begin{minipage}[t]{0.49\textwidth}
\includegraphics[scale=0.5]{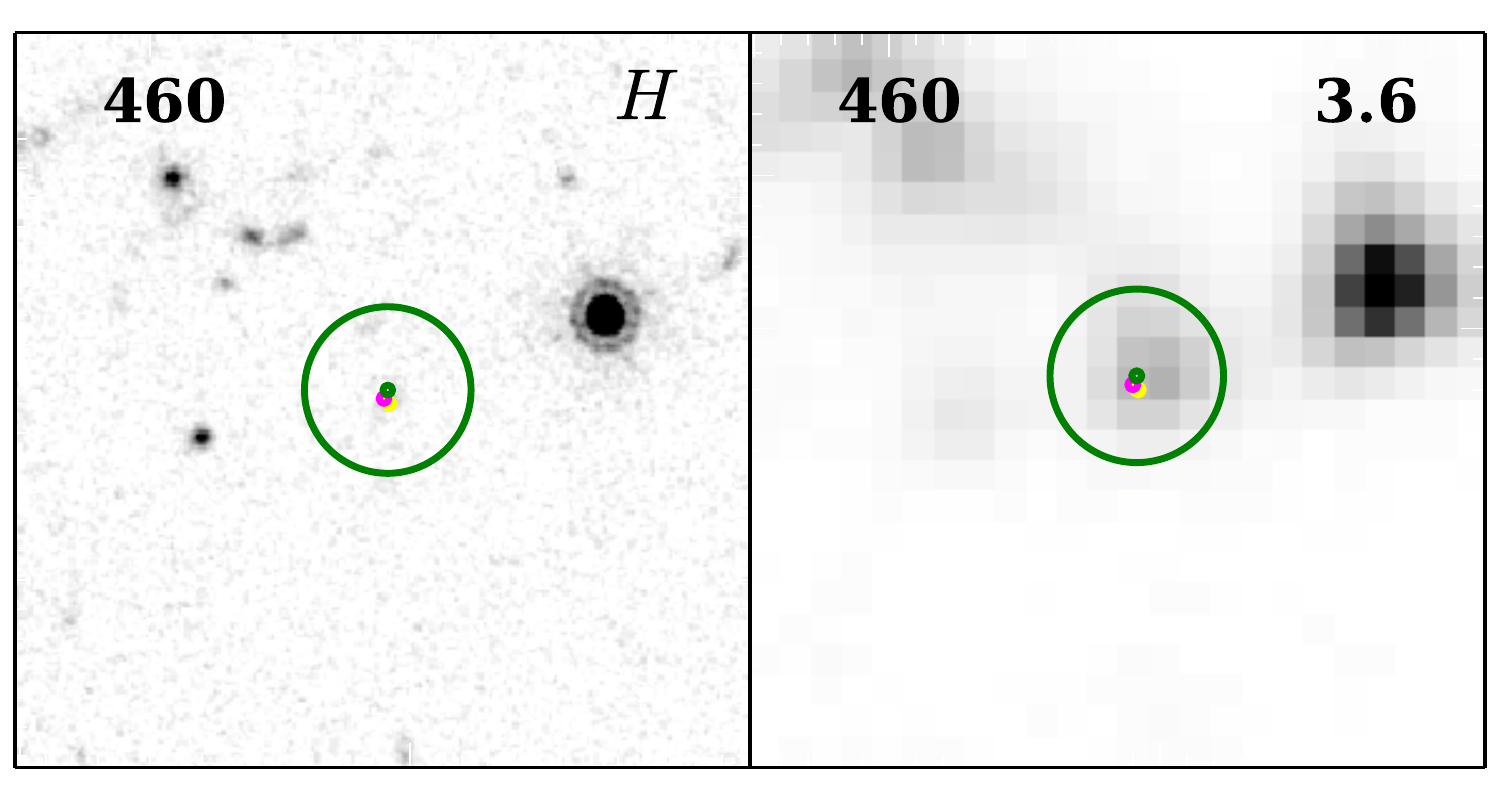}
\end{minipage}
\begin{minipage}[t]{0.49\textwidth}
\includegraphics[scale=0.5]{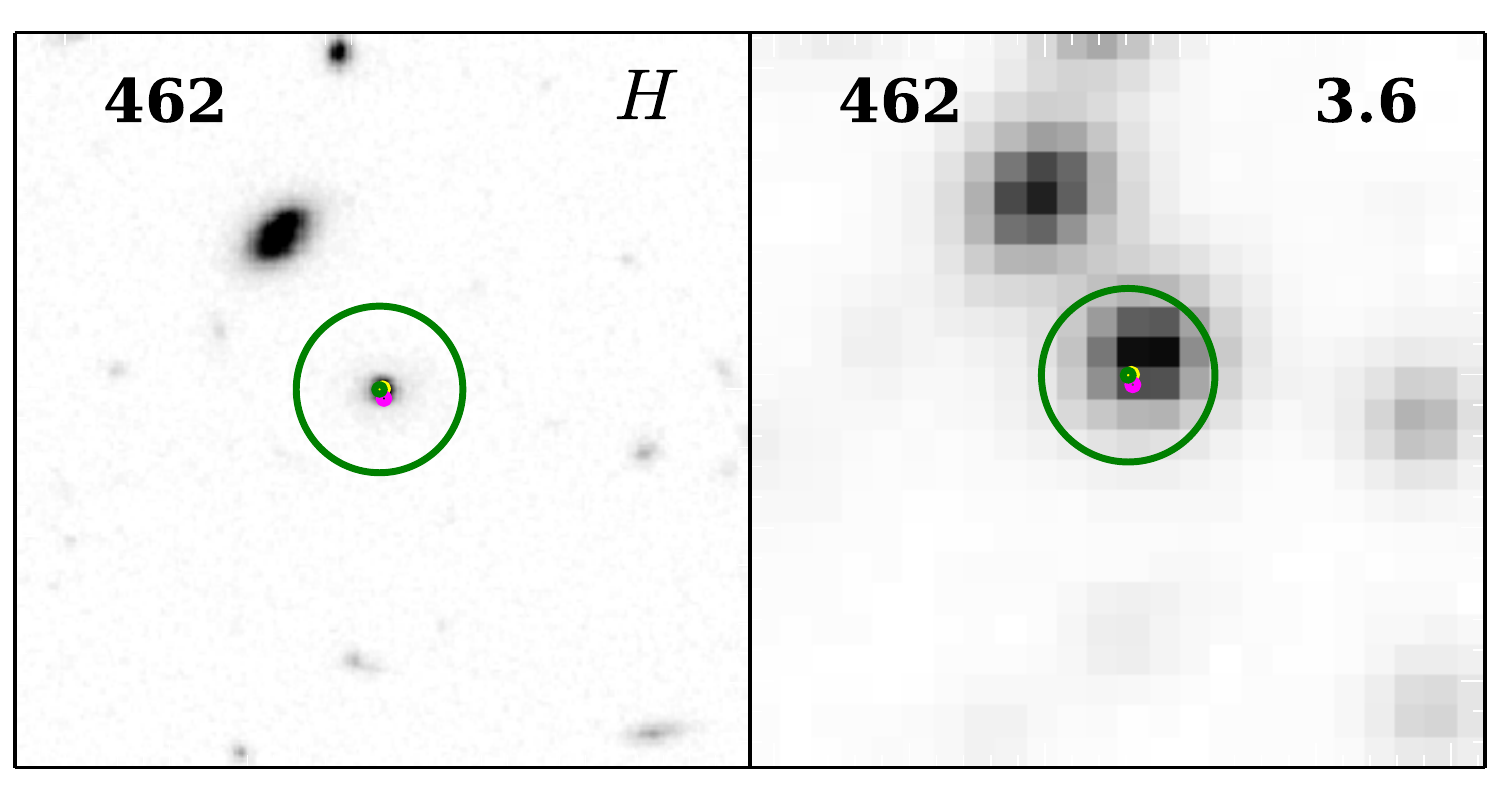}
\end{minipage}

\begin{minipage}[t]{0.49\textwidth}
\includegraphics[scale=0.5]{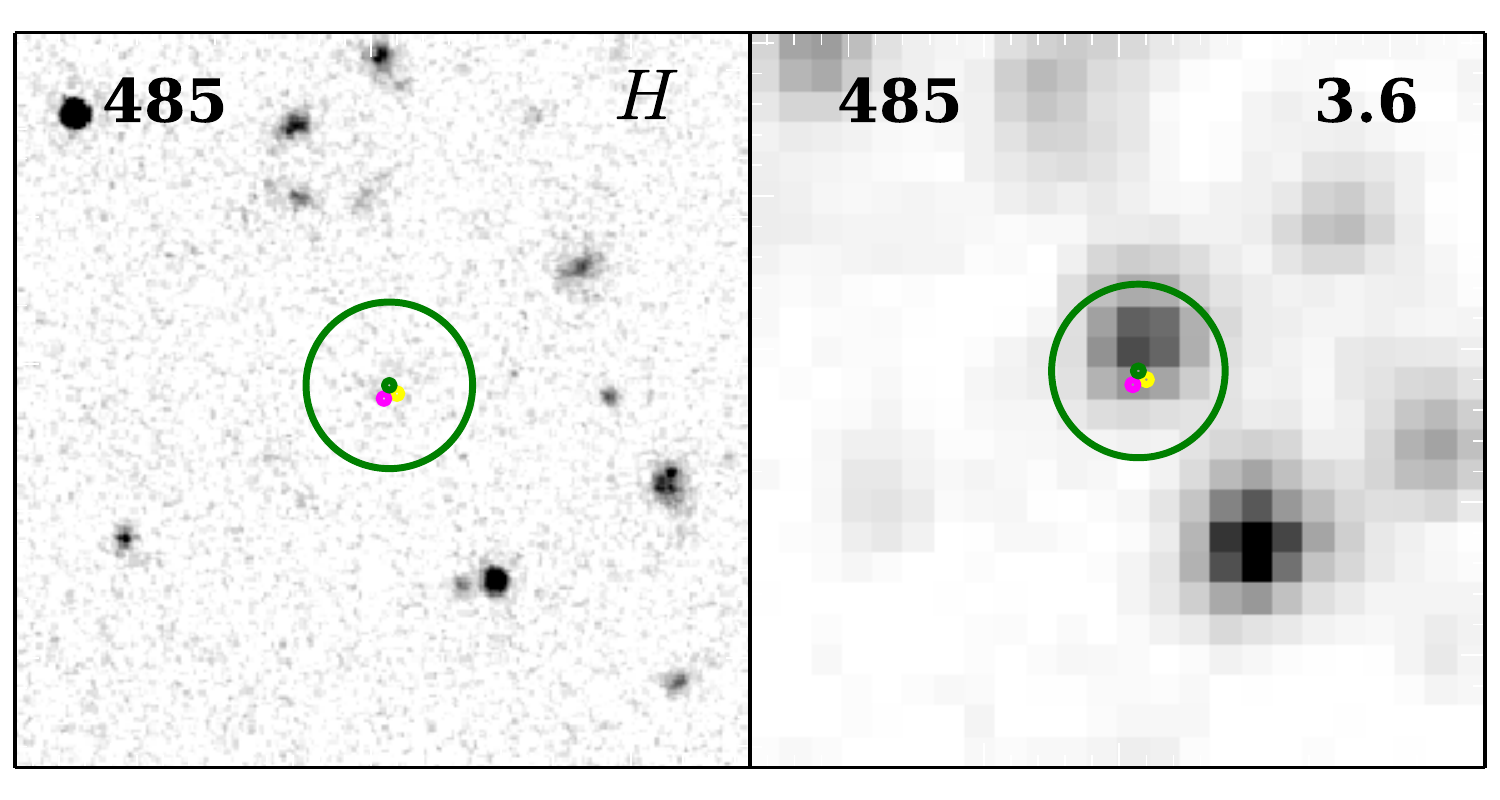}
\end{minipage}
\begin{minipage}[t]{0.49\textwidth}
\includegraphics[scale=0.5]{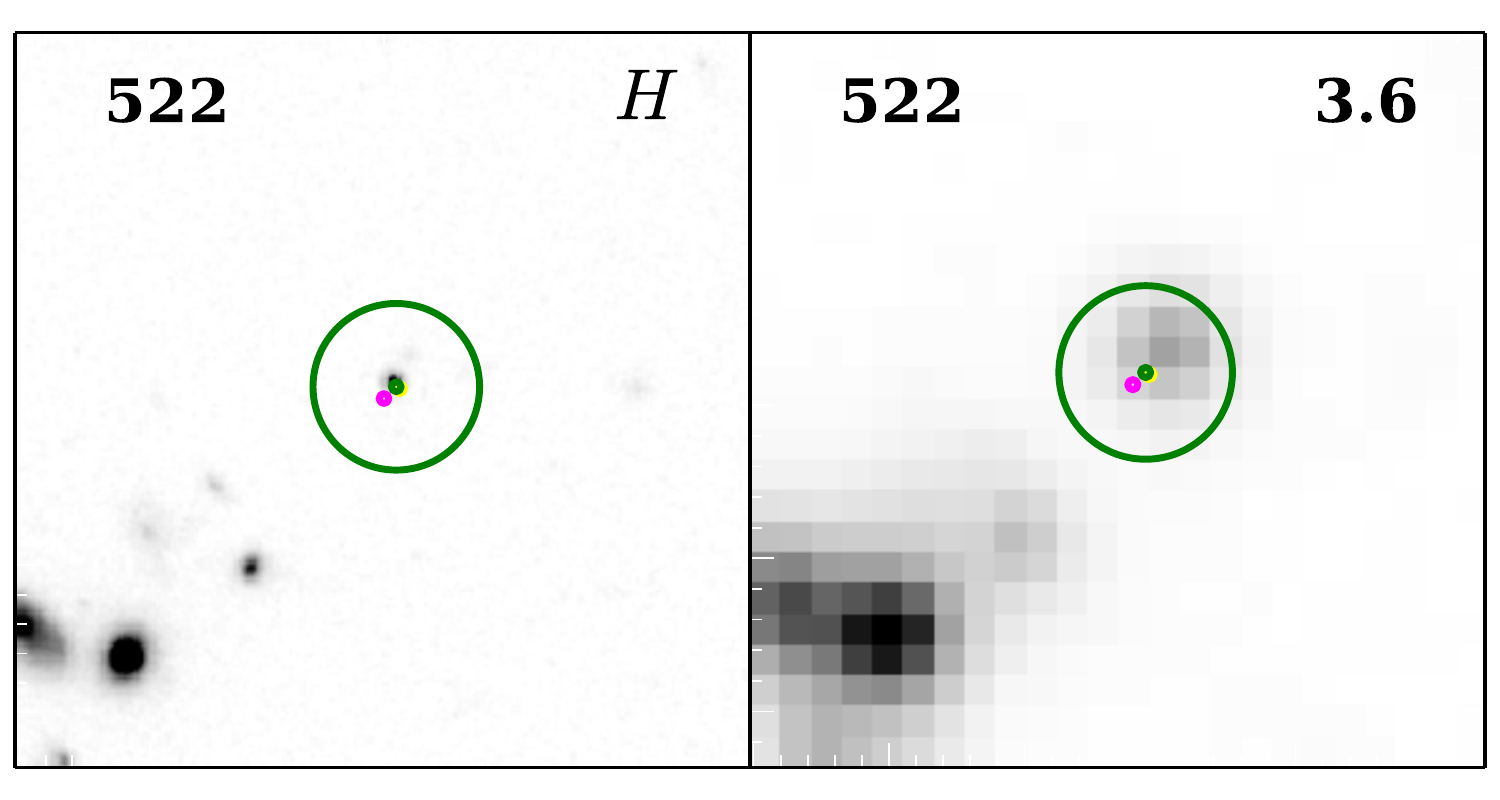}
\end{minipage}

\begin{minipage}[t]{0.49\textwidth}
\includegraphics[scale=0.5]{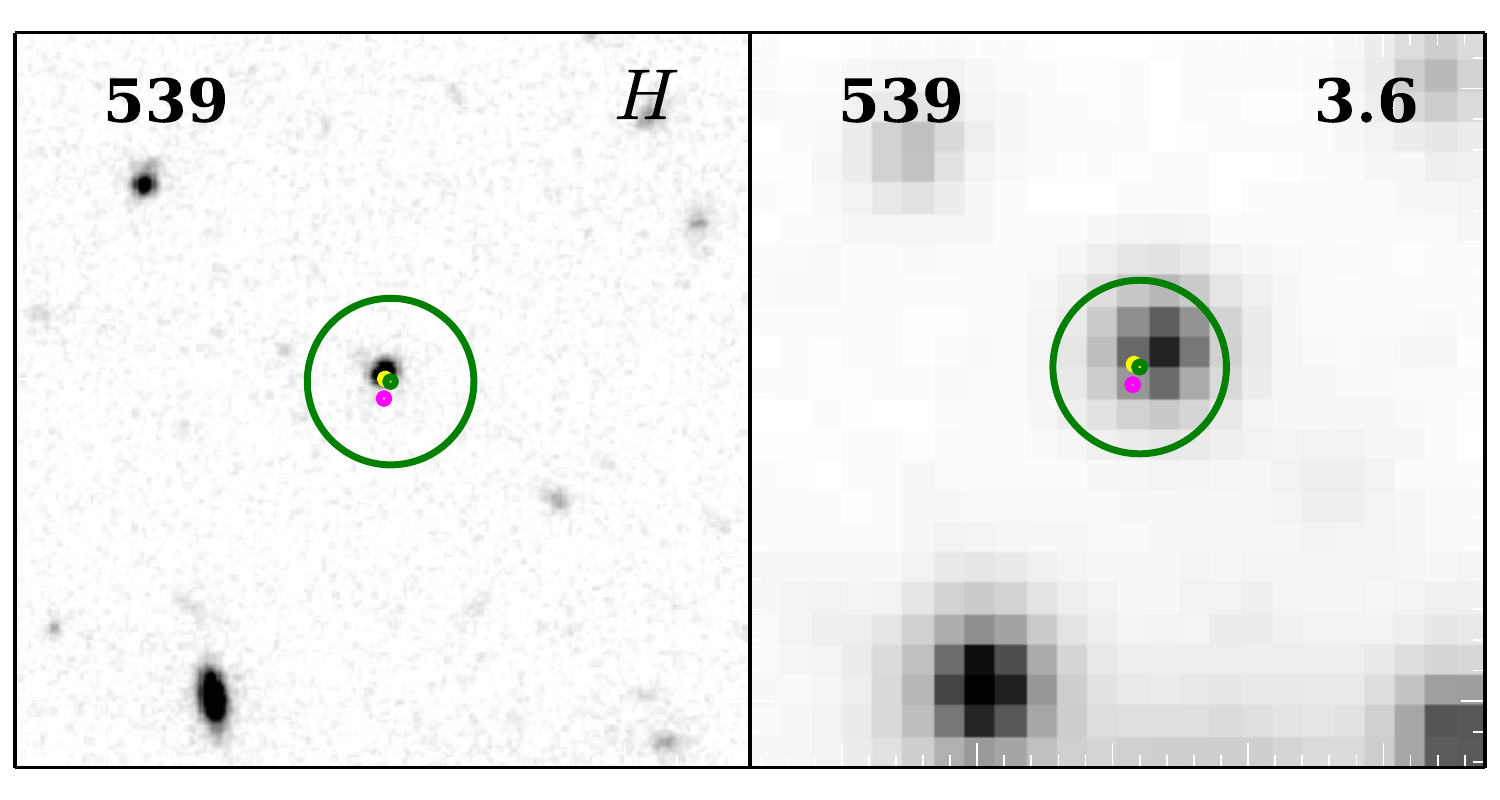}
\end{minipage}
\begin{minipage}[t]{0.49\textwidth}
\includegraphics[scale=0.5]{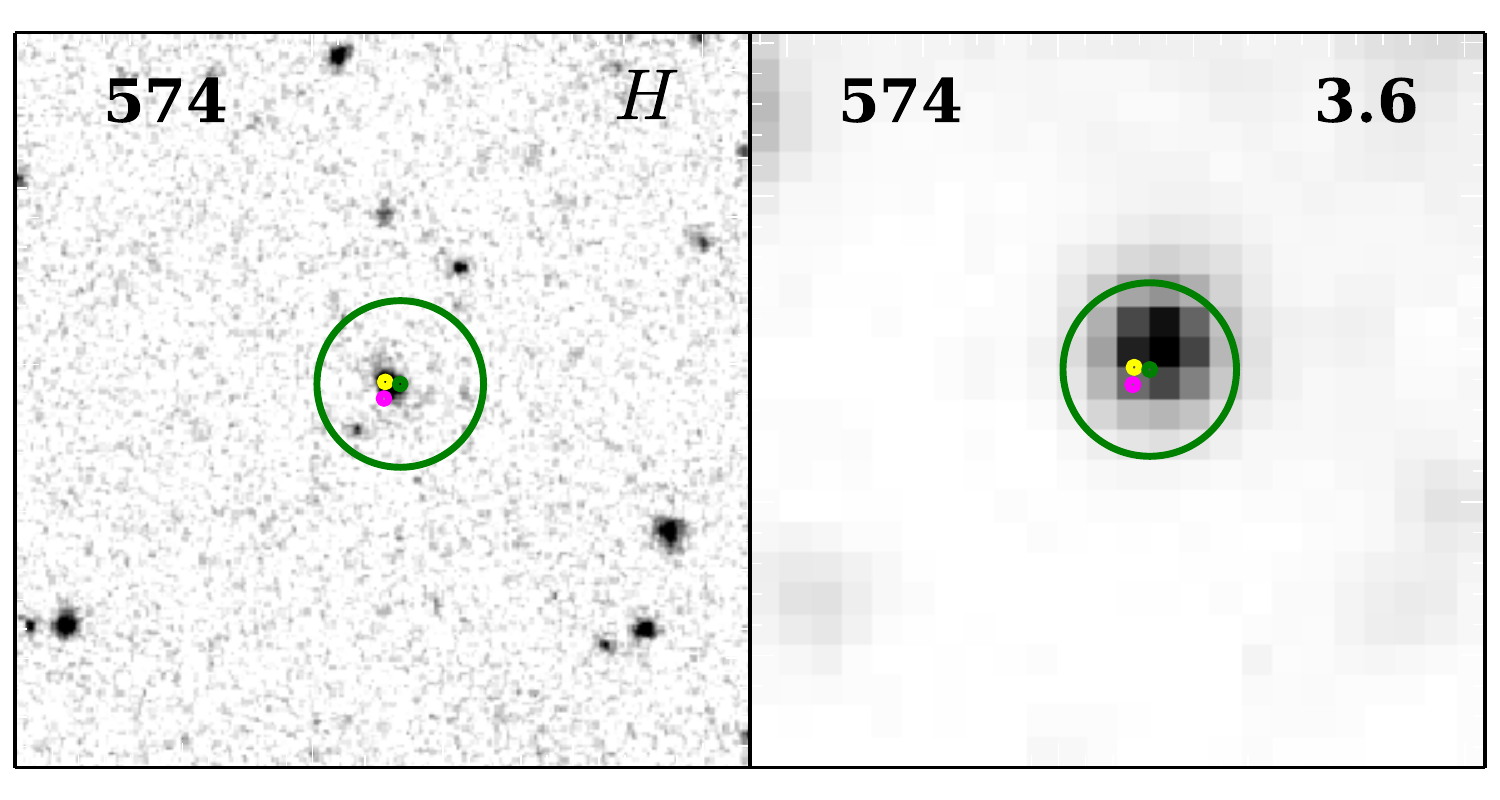}
\end{minipage}

\begin{minipage}[t]{0.49\textwidth}
\includegraphics[scale=0.5]{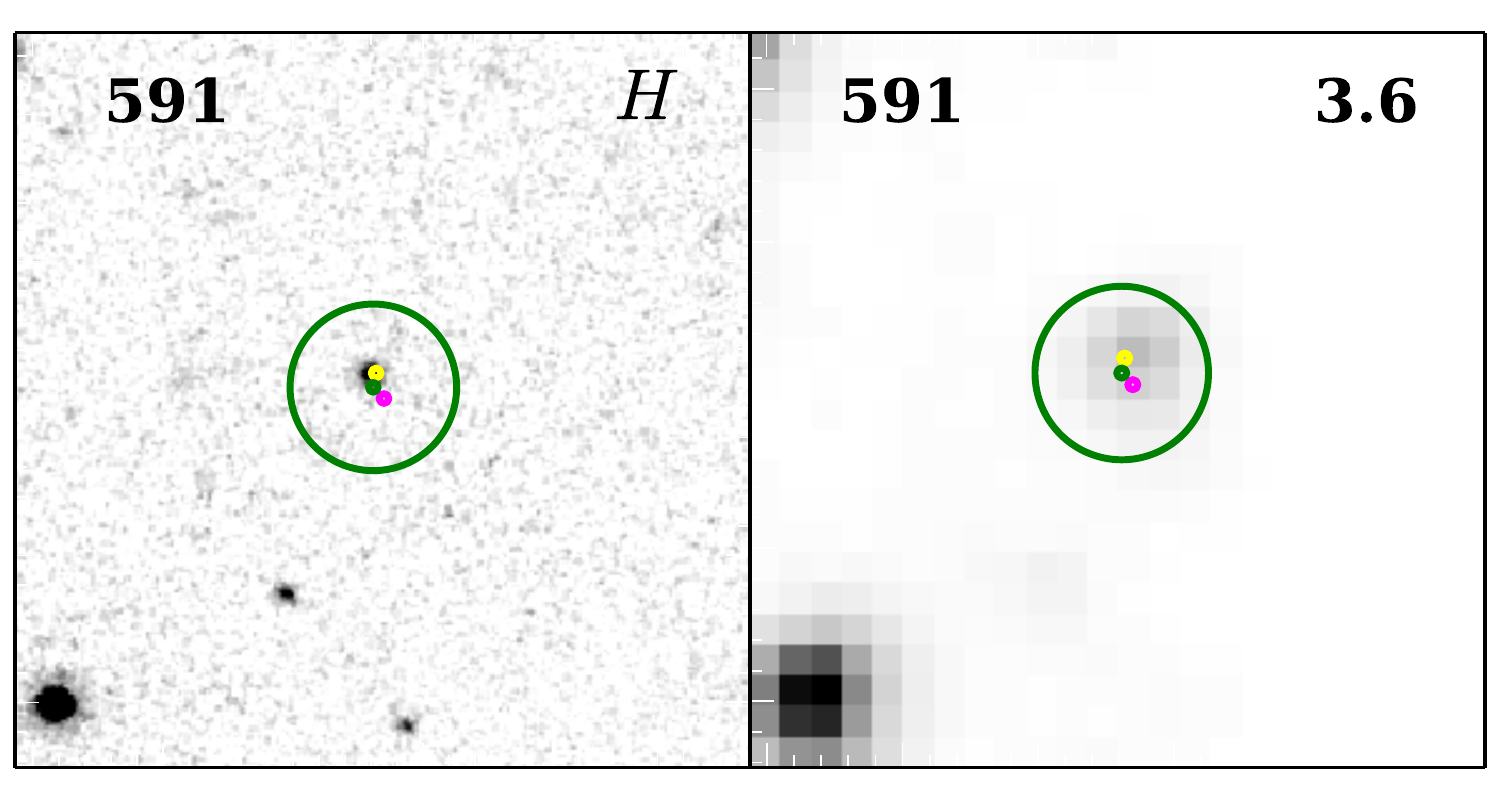}
\end{minipage}
\begin{minipage}[t]{0.49\textwidth}
\includegraphics[scale=0.5]{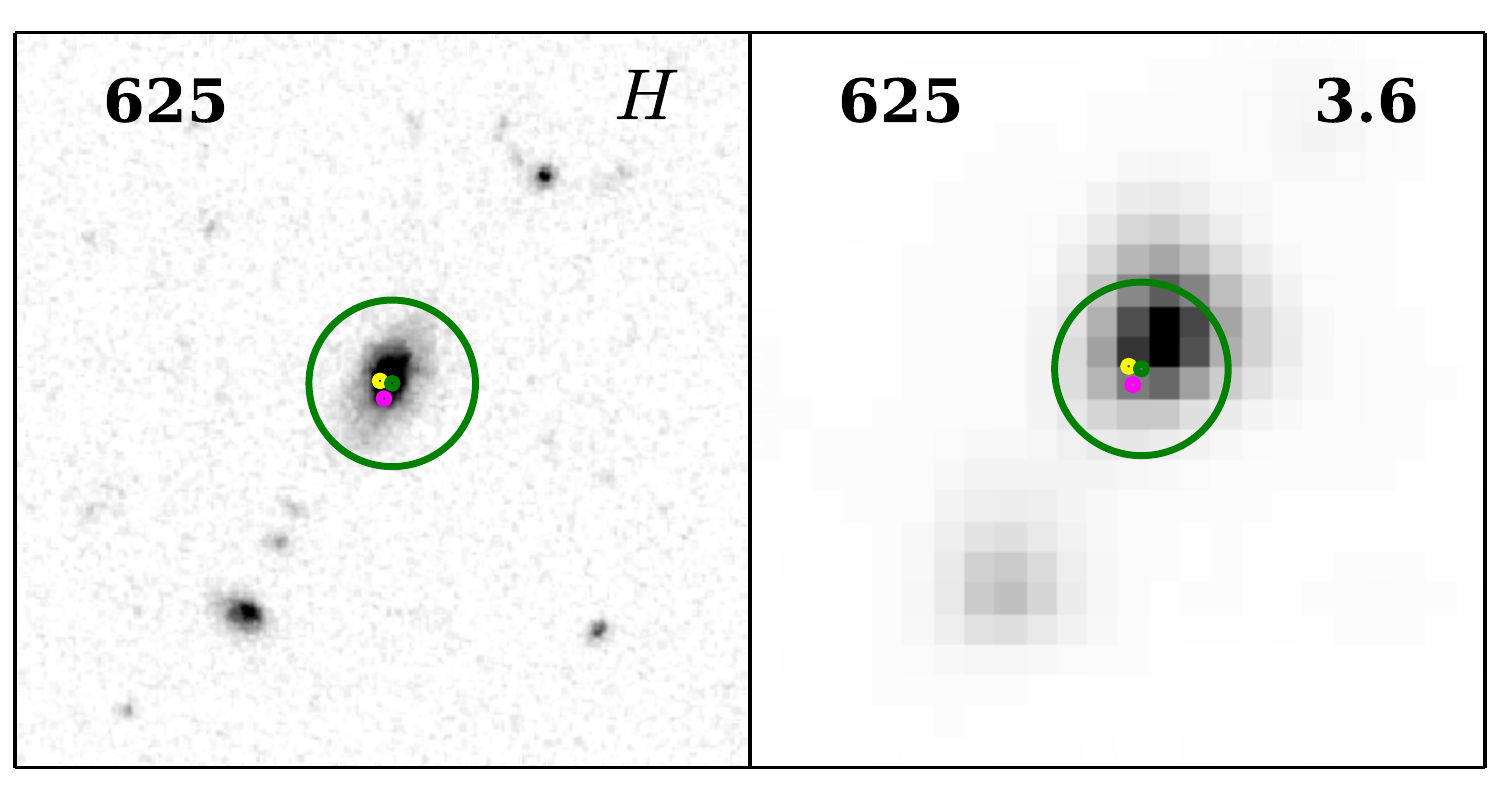}
\end{minipage}

\begin{minipage}[t]{0.49\textwidth}
\includegraphics[scale=0.5]{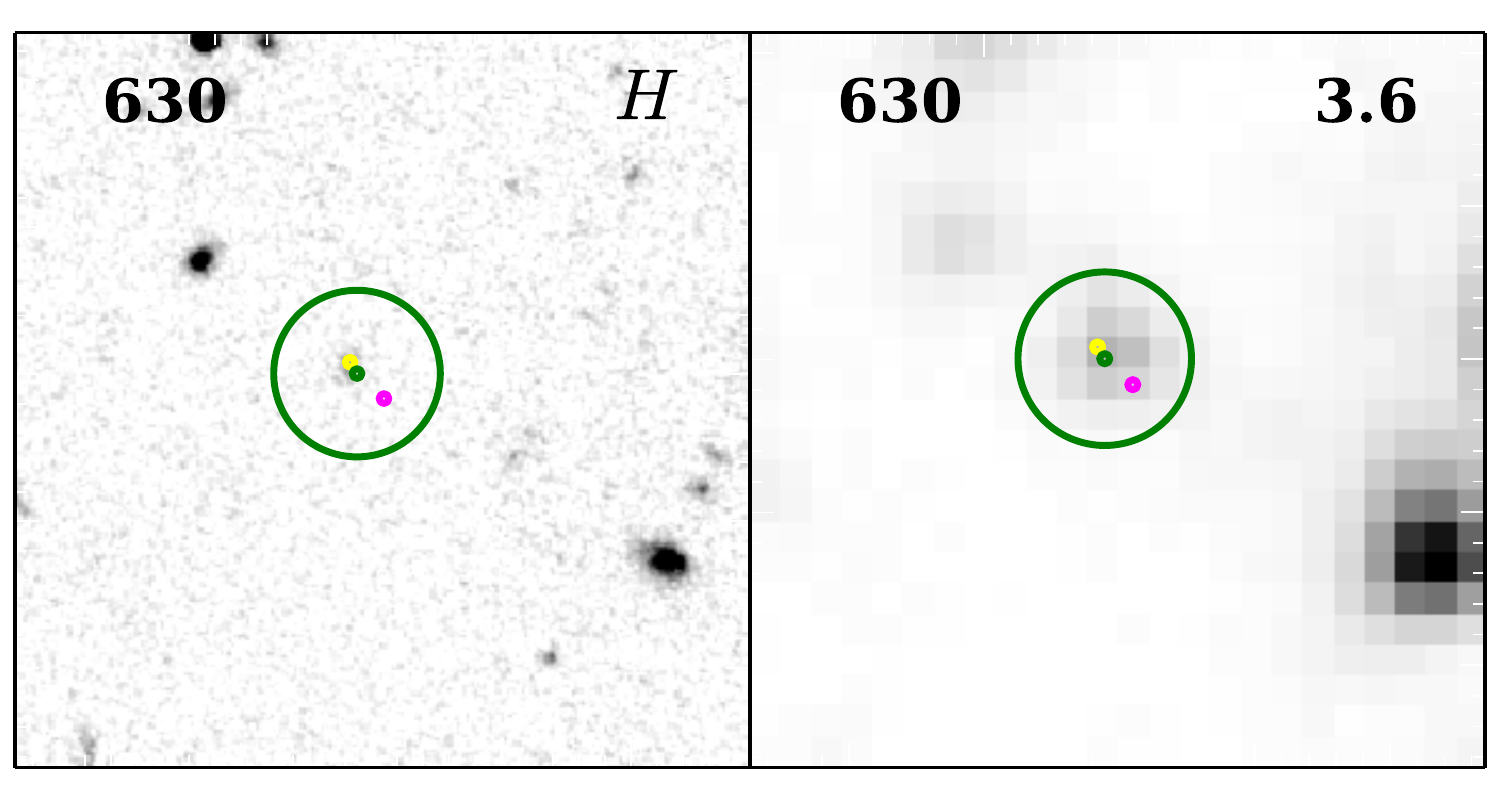}
\end{minipage}
\begin{minipage}[t]{0.49\textwidth}
\includegraphics[scale=0.5]{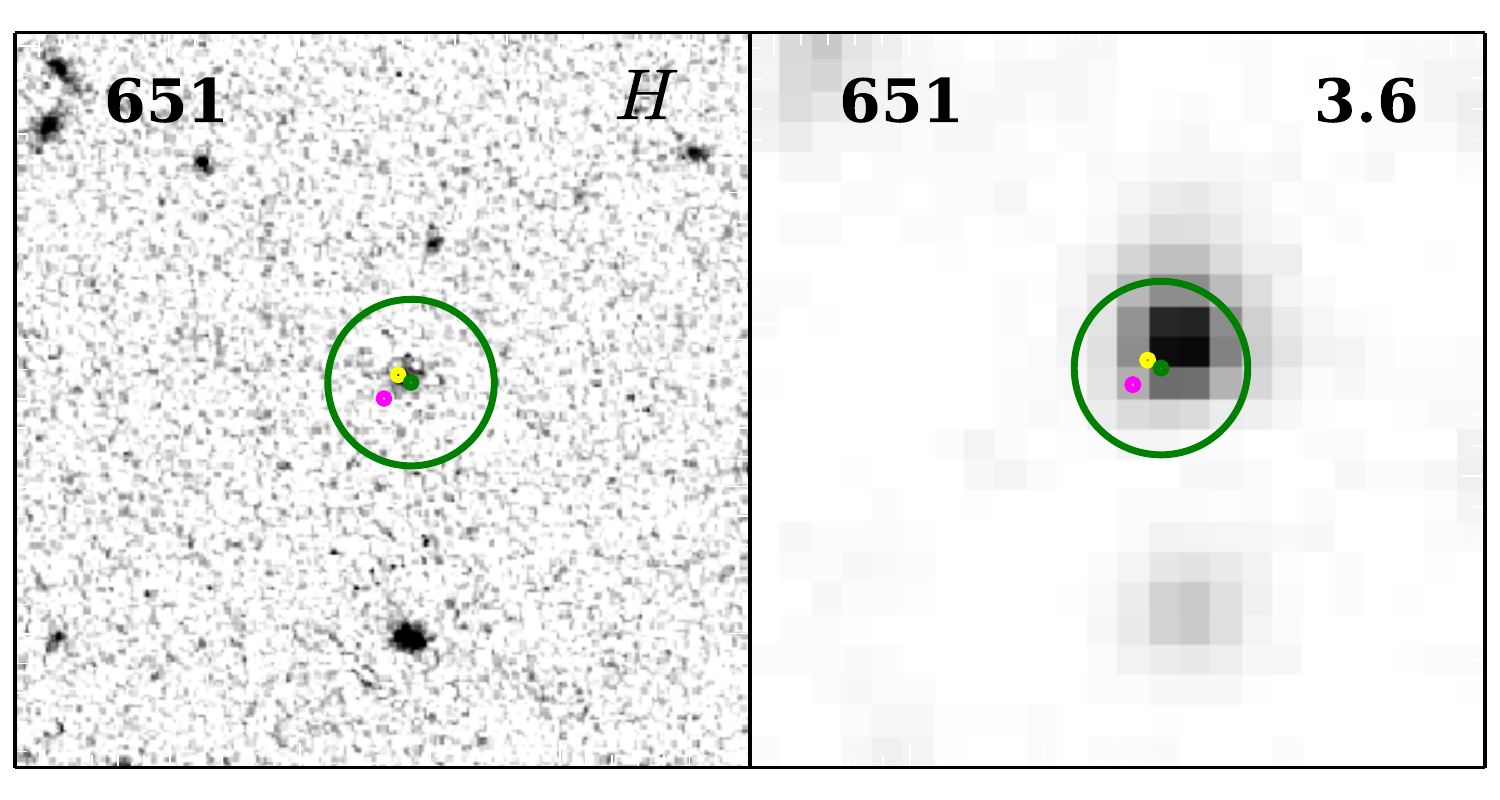}
\end{minipage}

\caption{$H$ band and 3.6 micron stamps for all objects for which the \Spitzer\ flux values are included in the photometric redshift determination, part 3/3. Shown in magenta is the original \Chandra\ 4-Ms catalog position. The yellow circle marks the object's $H$ band position that we determined by running SExtractor. The green point shows the position of the \Spitzer\ object closest to the $H$ band position. The green circle around this green point has a radius of 1.7$^{\prime\prime}$. It indicates the PSF size for the 3.6 micron IRAC channel. Due to source confusion the \Spitzer\ flux values could not be used for the whole sample. All stamps were colour inverted and are 15$^{\prime\prime}$ x 15$^{\prime\prime}$ in size.}
\end{figure*}

\clearpage
\newpage
\begin{table*}
\begin{center}
	\begin{tabular}{ll}
	\toprule
	{Number of objects with sufficient filter coverage and intact images: 371}\\
	{Out of these 371 sources, number of objects with a signal-to-noise ratio $\geq1$:} \\
	\midrule
		{in the soft band:} & {324 (87.3\%)}\\
		{in the hard band:} & {172 (46.4\%)}\\
		{in the full band:} & {303 (81.7\%)}\\
		{in the hard and the soft band:} & {139 (37.5\%)}\\
		{in the soft, hard or full band:} & {371 (100\%)}\\
	\midrule
	{Out of these 371 sources, number of objects with a signal-to-noise ratio $\geq5$:}\\
	\midrule
		{in the soft band:} & {110 (29.6\%)}\\
		{in the hard band:} & {96 (25.9\%)}\\
		{in the full band:} & {134 (36.1\%)}\\
		{in the hard and the soft band:} & {74 (19.9\%)}\\
		{in the soft, hard or full band:} & {140 (37.7\%)}\\
	\bottomrule
	\end{tabular}
\end{center}
\caption{\label{tab:SN}Signal-to-noise ratio statistics. Out of the 740 objects in the \Chandra\ 4-Ms catalog 371 have intact images and are covered by enough filters ($B$, $V$, $i$, $z$, $J$, $H$). We show the number of sources with signal-to-noise ratios $\geq1$ and $\geq5$}
\end{table*}

\begin{table*}
\tiny
	\begin{center}
		\begin{tabular}{llllllllllll}
		\toprule
			{ID} & {Hardness Ratio (HR)} & {$\sigma_\mathrm{HR}$} & {Hard counts} & {$\sigma_\mathrm{Hard}$} & {$\mathrm{SNR}_\mathrm{Hard}$} & {Soft counts} & {$\sigma_\mathrm{Soft}$} & {$\mathrm{SNR}_\mathrm{Soft}$} & {Full} & {$\sigma_\mathrm{Full}$} & {$\mathrm{SNR}_\mathrm{Full}$}\\
			\midrule
			$121$ & $0.21^{l}$ & $\mathrm{-}$ & $37.15$ & $13.41$ & $2.77$ & $24.21$ & $-1.00$ & $\mathrm{-}$ & $53.15$ & $15.51$ & $3.43$\\
			$150$ & $0.25^{l}$ & $\mathrm{-}$ & $30.02$ & $10.54$ & $2.85$ & $17.86$ & $-1.00$ & $\mathrm{-}$ & $38.59$ & $11.89$ & $3.25$\\
			$173$ & $-0.37$ & $0.13$ & $40.17$ & $11.14$ & $3.61$ & $86.87$ & $11.70$ & $7.42$ & $126.81$ & $15.39$ & $8.24$\\
			$184$ & $0.19$ & $0.28$ & $23.81$ & $9.82$ & $2.42$ & $16.10$ & $6.63$ & $2.43$ & $39.82$ & $11.21$ & $3.55$\\
			$189$ & $0.31^{l}$ & $\mathrm{-}$ & $29.33$ & $9.36$ & $3.13$ & $15.52$ & $-1.00$ & $\mathrm{-}$ & $37.01$ & $10.46$ & $3.54$\\
			$199$ & $-0.09$ & $0.10$ & $80.88$ & $12.71$ & $6.36$ & $96.07$ & $11.95$ & $8.04$ & $176.50$ & $16.78$ & $10.52$\\
			$211$ & $0.06$ & $0.06$ & $224.79$ & $19.43$ & $11.57$ & $198.79$ & $16.60$ & $11.98$ & $422.57$ & $24.89$ & $16.98$\\
			$217$ & $-0.17$ & $0.12$ & $58.49$ & $11.84$ & $4.94$ & $82.96$ & $11.29$ & $7.35$ & $141.11$ & $15.65$ & $9.02$\\
			$221$ & $0.12^{l}$ & $\mathrm{-}$ & $28.38$ & $12.27$ & $2.31$ & $22.47$ & $-1.00$ & $\mathrm{-}$ & $43.08$ & $14.21$ & $3.03$\\
			$226$ & $0.33^{u}$ & $\mathrm{-}$ & $19.07$ & $-1.00$ & $\mathrm{-}$ & $9.71$ & $5.14$ & $1.89$ & $21.36$ & $-1.00$ & $\mathrm{-}$\\
			$242$ & $0.16$ & $0.26$ & $23.04$ & $8.77$ & $2.63$ & $16.58$ & $6.38$ & $2.60$ & $39.52$ & $10.21$ & $3.87$\\
			$244$ & $0.29^{u}$ & $\mathrm{-}$ & $20.66$ & $-1.00$ & $\mathrm{-}$ & $11.32$ & $5.27$ & $2.15$ & $22.15$ & $7.56$ & $2.93$\\
			$258$ & $0.09^{l}$ & $\mathrm{-}$ & $22.21$ & $9.43$ & $2.36$ & $18.55$ & $-1.00$ & $\mathrm{-}$ & $33.35$ & $11.09$ & $3.01$\\
			$273$ & $0.03^{u}$ & $\mathrm{-}$ & $21.44$ & $-1.00$ & $\mathrm{-}$ & $20.14$ & $6.46$ & $3.12$ & $28.12$ & $8.58$ & $3.28$\\
			$296$ & $0.56^{l}$ & $\mathrm{-}$ & $51.01$ & $10.12$ & $5.04$ & $14.34$ & $-1.00$ & $\mathrm{-}$ & $59.21$ & $10.98$ & $5.39$\\
			$301$ & $0.34$ & $0.10$ & $94.01$ & $12.70$ & $7.40$ & $46.08$ & $8.57$ & $5.38$ & $139.58$ & $14.74$ & $9.47$\\
			$302$ & $0.04^{u}$ & $\mathrm{-}$ & $20.23$ & $-1.00$ & $\mathrm{-}$ & $18.53$ & $6.17$ & $3.00$ & $26.87$ & $8.06$ & $3.33$\\
			$303$ & $0.35^{l}$ & $\mathrm{-}$ & $35.90$ & $11.17$ & $3.21$ & $17.10$ & $-1.00$ & $\mathrm{-}$ & $41.77$ & $12.31$ & $3.39$\\
			$306$ & $0.16$ & $0.36$ & $13.22$ & $6.87$ & $1.92$ & $9.51$ & $5.09$ & $1.87$ & $22.65$ & $7.91$ & $2.86$\\
			$318$ & $0.38^{u}$ & $\mathrm{-}$ & $26.29$ & $-1.00$ & $\mathrm{-}$ & $11.88$ & $6.01$ & $1.98$ & $22.50$ & $9.49$ & $2.37$\\
			$321$ & $0.31^{l}$ & $\mathrm{-}$ & $25.16$ & $8.52$ & $2.95$ & $13.39$ & $-1.00$ & $\mathrm{-}$ & $30.41$ & $9.36$ & $3.25$\\
			$325$ & $0.46^{u}$ & $\mathrm{-}$ & $18.65$ & $-1.00$ & $\mathrm{-}$ & $6.90$ & $4.43$ & $1.56$ & $15.28$ & $6.48$ & $2.36$\\
			$328$ & $0.24^{u}$ & $\mathrm{-}$ & $14.56$ & $-1.00$ & $\mathrm{-}$ & $8.89$ & $4.78$ & $1.86$ & $17.55$ & $-1.00$ & $\mathrm{-}$\\
			$331$ & $0.21^{u}$ & $\mathrm{-}$ & $18.00$ & $-1.00$ & $\mathrm{-}$ & $11.64$ & $5.59$ & $2.08$ & $21.71$ & $-1.00$ & $\mathrm{-}$\\
			$348$ & $0.15^{u}$ & $\mathrm{-}$ & $24.86$ & $-1.00$ & $\mathrm{-}$ & $18.53$ & $6.82$ & $2.72$ & $21.72$ & $9.66$ & $2.25$\\
			$354$ & $-0.54$ & $0.15$ & $21.55$ & $8.50$ & $2.54$ & $71.22$ & $10.41$ & $6.84$ & $92.58$ & $12.69$ & $7.30$\\
			$371$ & $0.15$ & $0.10$ & $99.88$ & $13.25$ & $7.54$ & $74.06$ & $10.52$ & $7.04$ & $173.34$ & $16.29$ & $10.64$\\
			$373$ & $0.22^{u}$ & $\mathrm{-}$ & $40.47$ & $-1.00$ & $\mathrm{-}$ & $25.64$ & $9.80$ & $2.62$ & $46.27$ & $-1.00$ & $\mathrm{-}$\\
			$389$ & $0.16^{u}$ & $\mathrm{-}$ & $19.56$ & $-1.00$ & $\mathrm{-}$ & $14.22$ & $5.72$ & $2.49$ & $18.45$ & $7.61$ & $2.42$\\
			$392$ & $0.05^{u}$ & $\mathrm{-}$ & $21.97$ & $-1.00$ & $\mathrm{-}$ & $19.85$ & $6.95$ & $2.86$ & $27.31$ & $-1.00$ & $\mathrm{-}$\\
			$402$ & $0.27^{l}$ & $\mathrm{-}$ & $20.27$ & $8.72$ & $2.32$ & $11.76$ & $-1.00$ & $\mathrm{-}$ & $27.82$ & $-1.00$ & $\mathrm{-}$\\
			$403$ & $0.20^{l}$ & $\mathrm{-}$ & $56.82$ & $22.53$ & $2.52$ & $37.72$ & $-1.00$ & $\mathrm{-}$ & $82.03$ & $25.80$ & $3.18$\\
			$410$ & $0.54$ & $0.12$ & $99.62$ & $15.79$ & $6.31$ & $29.98$ & $8.78$ & $3.41$ & $129.43$ & $17.59$ & $7.36$\\
			$428$ & $0.24^{u}$ & $\mathrm{-}$ & $16.26$ & $-1.00$ & $\mathrm{-}$ & $9.96$ & $4.93$ & $2.02$ & $13.34$ & $6.32$ & $2.11$\\
			$430$ & $0.29$ & $0.15$ & $50.46$ & $10.11$ & $4.99$ & $27.51$ & $6.99$ & $3.94$ & $77.68$ & $11.68$ & $6.65$\\
			$444$ & $0.28$ & $0.07$ & $158.67$ & $15.81$ & $10.04$ & $88.73$ & $11.29$ & $7.86$ & $246.58$ & $18.84$ & $13.09$\\
			$455$ & $0.21^{u}$ & $\mathrm{-}$ & $18.58$ & $-1.00$ & $\mathrm{-}$ & $12.11$ & $5.26$ & $2.30$ & $20.22$ & $7.02$ & $2.88$\\
			$456$ & $-0.29$ & $0.11$ & $55.26$ & $11.88$ & $4.65$ & $100.21$ & $12.27$ & $8.17$ & $155.15$ & $16.35$ & $9.49$\\
			$460$ & $-0.02$ & $0.27$ & $23.19$ & $9.95$ & $2.33$ & $23.93$ & $7.62$ & $3.14$ & $47.01$ & $11.84$ & $3.97$\\
			$462$ & $0.06$ & $0.39$ & $10.69$ & $6.30$ & $1.70$ & $9.50$ & $4.94$ & $1.92$ & $20.10$ & $7.33$ & $2.74$\\
			$466$ & $0.27$ & $0.13$ & $64.51$ & $11.10$ & $5.81$ & $36.88$ & $7.86$ & $4.69$ & $101.03$ & $12.99$ & $7.78$\\
			$485$ & $0.45$ & $0.15$ & $61.74$ & $12.47$ & $4.95$ & $23.23$ & $7.53$ & $3.08$ & $84.81$ & $14.05$ & $6.04$\\
			$496$ & $0.00$ & $\mathrm{-}$ & $36.09$ & $-1.00$ & $\mathrm{-}$ & $22.48$ & $-1.00$ & $\mathrm{-}$ & $34.45$ & $13.44$ & $2.56$\\
			$522$ & $-0.40$ & $0.04$ & $193.40$ & $17.51$ & $11.05$ & $451.14$ & $23.72$ & $19.02$ & $642.92$ & $28.82$ & $22.31$\\
			$535$ & $-0.12$ & $0.06$ & $232.76$ & $20.51$ & $11.35$ & $296.76$ & $19.89$ & $14.92$ & $528.50$ & $27.85$ & $18.98$\\
			$539$ & $0.53^{l}$ & $\mathrm{-}$ & $40.01$ & $9.64$ & $4.15$ & $12.20$ & $-1.00$ & $\mathrm{-}$ & $42.86$ & $10.21$ & $4.20$\\
			$546$ & $0.29$ & $0.03$ & $796.39$ & $33.20$ & $23.99$ & $438.59$ & $23.37$ & $18.77$ & $1231.02$ & $39.92$ & $30.84$\\
			$556$ & $0.40$ & $0.03$ & $852.52$ & $39.38$ & $21.65$ & $364.81$ & $24.33$ & $14.99$ & $1216.20$ & $45.72$ & $26.60$\\
			$574$ & $0.38$ & $0.25$ & $25.49$ & $9.11$ & $2.80$ & $11.49$ & $5.44$ & $2.11$ & $36.89$ & $10.03$ & $3.68$\\
			$578$ & $0.00$ & $\mathrm{-}$ & $25.53$ & $-1.00$ & $\mathrm{-}$ & $17.39$ & $-1.00$ & $\mathrm{-}$ & $23.99$ & $9.57$ & $2.51$\\
			$583$ & $-0.51$ & $0.06$ & $106.38$ & $14.82$ & $7.18$ & $328.38$ & $20.50$ & $16.02$ & $434.10$ & $24.60$ & $17.65$\\
			$589$ & $0.17^{u}$ & $\mathrm{-}$ & $21.65$ & $-1.00$ & $\mathrm{-}$ & $15.29$ & $6.38$ & $2.40$ & $25.84$ & $-1.00$ & $\mathrm{-}$\\
			$591$ & $0.29^{u}$ & $\mathrm{-}$ & $50.67$ & $-1.00$ & $\mathrm{-}$ & $27.80$ & $10.44$ & $2.66$ & $62.46$ & $18.37$ & $3.40$\\
			$620$ & $-0.02$ & $0.06$ & $273.77$ & $25.37$ & $10.79$ & $283.93$ & $20.85$ & $13.62$ & $556.98$ & $32.07$ & $17.37$\\
			$624$ & $0.33^{l}$ & $\mathrm{-}$ & $37.22$ & $11.78$ & $3.16$ & $18.61$ & $-1.00$ & $\mathrm{-}$ & $45.69$ & $13.11$ & $3.49$\\
			$625$ & $0.71$ & $0.10$ & $122.38$ & $16.57$ & $7.39$ & $20.79$ & $7.99$ & $2.60$ & $143.04$ & $18.01$ & $7.94$\\
			$630$ & $0.36^{u}$ & $\mathrm{-}$ & $30.65$ & $-1.00$ & $\mathrm{-}$ & $14.45$ & $7.37$ & $1.96$ & $34.39$ & $-1.00$ & $\mathrm{-}$\\
			$651$ & $0.40$ & $0.10$ & $146.49$ & $20.77$ & $7.05$ & $62.35$ & $12.44$ & $5.01$ & $208.60$ & $23.66$ & $8.82$\\
		\bottomrule
		\end{tabular}
	\end{center}
\caption{\label{tab:counts}Hardness Ratio values, X-ray counts, errors on the X-ray counts and signal-to-noise ratio values for the main sample. The Hardness Ratio values were calculated using equation \ref{eq:HR}. Upper and lower limits on the Hardness Ratio are indicated with '$u$' and '$l$', respectively. Objects for which the Hardness Ratio could not be determined due to an upper limit in both the Hard and the Soft band, are marked with '0.00'. The X-ray count values were directly extracted from the \protect\cite{Xue:2011aa} catalog. For the errors on the X-ray counts we give the upper errors given in the 4-Ms catalog. For objects that are not detected we give an upper limit on the X-ray counts, set $\sigma$ to -1.00 and mark the signal-to-noise ratio with a dash. }
\end{table*}

\newpage
\begin{landscape}
\begin{table*}
\tiny
	\begin{center}
		\begin{tabular}{ll|llllllll}
		\toprule
		{} & {ID} & {$F_\nu$ ${(}B{)}$} & {$Ferr_\nu$ ${(}B{)}$ } & {$F_\nu$ ${(}V{)}$} & {$Ferr_\nu$ ${(}V{)}$} & {$F_\nu$ ${(}i{)}$} & {$Ferr_\nu$ ${(}i{)}$} & {$F_\nu$ ${(}z{)}$} & {$Ferr_\nu$ ${(}z{)}$}\\
		\midrule
			$*$ & $121$ & $0.000\e{+00}$ & $1.456\e{-01}$ & $1.023\e{-01}$ & $1.020\e{-02}$ & $3.608\e{-01}$ & $1.631\e{-02}$ & $1.142\e{+00}$ & $2.010\e{-02}$\\
			$*$ & $150$ & $3.147\e{-02}$ & $1.018\e{-02}$ & $1.158\e{-01}$ & $9.915\e{-03}$ & $1.457\e{-01}$ & $1.644\e{-02}$ & $1.670\e{-01}$ & $1.961\e{-02}$\\
			$ $ & $173$ & $1.732\e{-02}$ & $5.131\e{-03}$ & $8.063\e{-02}$ & $4.995\e{-03}$ & $3.987\e{-01}$ & $8.699\e{-03}$ & $8.616\e{-01}$ & $1.063\e{-02}$\\
			$*$ & $184$ & $2.607\e{-02}$ & $8.239\e{-03}$ & $1.699\e{-01}$ & $7.821\e{-03}$ & $6.041\e{-01}$ & $1.280\e{-02}$ & $1.428\e{+00}$ & $1.567\e{-02}$\\
			$*$ & $189$ & $0.000\e{+00}$ & $3.639\e{-02}$ & $2.394\e{-02}$ & $5.045\e{-03}$ & $4.371\e{-02}$ & $8.006\e{-03}$ & $4.974\e{-02}$ & $9.624\e{-03}$\\
			$*$ & $199$ & $0.000\e{+00}$ & $1.456\e{-01}$ & $5.441\e{-02}$ & $9.849\e{-03}$ & $8.621\e{-02}$ & $1.622\e{-02}$ & $1.212\e{-01}$ & $1.954\e{-02}$\\
			$ $ & $211$ & $2.494\e{-02}$ & $5.362\e{-03}$ & $4.317\e{-02}$ & $4.842\e{-03}$ & $1.506\e{-01}$ & $8.124\e{-03}$ & $3.438\e{-01}$ & $9.779\e{-03}$\\
			$*$ & $217$ & $0.000\e{+00}$ & $8.188\e{-02}$ & $0.000\e{+00}$ & $7.347\e{-02}$ & $0.000\e{+00}$ & $1.458\e{-01}$ & $0.000\e{+00}$ & $2.626\e{-01}$\\
			$ $ & $221$ & $0.000\e{+00}$ & $8.188\e{-02}$ & $1.152\e{-01}$ & $7.800\e{-03}$ & $1.991\e{-01}$ & $1.271\e{-02}$ & $5.681\e{-01}$ & $1.592\e{-02}$\\
			$ $ & $226$ & $0.000\e{+00}$ & $1.456\e{-01}$ & $0.000\e{+00}$ & $1.306\e{-01}$ & $2.360\e{-01}$ & $1.371\e{-02}$ & $8.728\e{-01}$ & $1.788\e{-02}$\\
			$*$ & $242$ & $1.757\e{-02}$ & $7.810\e{-03}$ & $9.874\e{-02}$ & $7.677\e{-03}$ & $2.477\e{-01}$ & $1.257\e{-02}$ & $4.847\e{-01}$ & $1.555\e{-02}$\\
			$*$ & $244$ & $0.000\e{+00}$ & $1.456\e{-01}$ & $0.000\e{+00}$ & $1.306\e{-01}$ & $2.279\e{-02}$ & $1.682\e{-02}$ & $1.435\e{-01}$ & $2.036\e{-02}$\\
			$ $ & $258$ & $0.000\e{+00}$ & $1.310\e{-02}$ & $0.000\e{+00}$ & $1.176\e{-02}$ & $1.349\e{-02}$ & $5.006\e{-03}$ & $2.765\e{-02}$ & $6.348\e{-03}$\\
			$*$ & $273$ & $0.000\e{+00}$ & $1.456\e{-01}$ & $0.000\e{+00}$ & $1.306\e{-01}$ & $0.000\e{+00}$ & $2.592\e{-01}$ & $6.494\e{-02}$ & $1.976\e{-02}$\\
			$ $ & $296$ & $0.000\e{+00}$ & $3.639\e{-02}$ & $2.533\e{-02}$ & $4.874\e{-03}$ & $1.833\e{-02}$ & $7.507\e{-03}$ & $1.302\e{-01}$ & $9.365\e{-03}$\\
			$ $ & $301$ & $4.095\e{-02}$ & $1.110\e{-02}$ & $8.010\e{-02}$ & $1.080\e{-02}$ & $9.752\e{-02}$ & $1.774\e{-02}$ & $1.430\e{-01}$ & $2.114\e{-02}$\\
			$*$ & $302$ & $0.000\e{+00}$ & $3.639\e{-02}$ & $4.055\e{-02}$ & $5.055\e{-03}$ & $3.643\e{-02}$ & $8.064\e{-03}$ & $9.140\e{-02}$ & $9.936\e{-03}$\\
			$ $ & $303$ & $0.000\e{+00}$ & $1.456\e{-01}$ & $2.923\e{-02}$ & $1.068\e{-02}$ & $0.000\e{+00}$ & $2.592\e{-01}$ & $1.952\e{-01}$ & $2.148\e{-02}$\\
			$ $ & $306$ & $0.000\e{+00}$ & $8.188\e{-02}$ & $5.415\e{-02}$ & $7.566\e{-03}$ & $7.303\e{-02}$ & $1.195\e{-02}$ & $9.642\e{-02}$ & $1.482\e{-02}$\\
			$*$ & $318$ & $0.000\e{+00}$ & $1.456\e{-01}$ & $8.013\e{-02}$ & $1.008\e{-02}$ & $2.476\e{-01}$ & $1.608\e{-02}$ & $2.924\e{-01}$ & $1.966\e{-02}$\\
			$*$ & $321$ & $0.000\e{+00}$ & $1.456\e{-01}$ & $1.114\e{-01}$ & $9.503\e{-03}$ & $7.698\e{-01}$ & $1.630\e{-02}$ & $1.699\e{+00}$ & $2.023\e{-02}$\\
			$ $ & $325$ & $7.765\e{-03}$ & $4.993\e{-03}$ & $5.990\e{-02}$ & $5.115\e{-03}$ & $1.266\e{-01}$ & $8.440\e{-03}$ & $2.203\e{-01}$ & $9.993\e{-03}$\\
			$ $ & $328$ & $4.020\e{-02}$ & $9.731\e{-03}$ & $4.951\e{-02}$ & $9.514\e{-03}$ & $1.041\e{-01}$ & $1.528\e{-02}$ & $8.870\e{-02}$ & $1.845\e{-02}$\\
			$ $ & $331$ & $0.000\e{+00}$ & $3.639\e{-02}$ & $4.080\e{-02}$ & $5.182\e{-03}$ & $9.863\e{-02}$ & $8.685\e{-03}$ & $8.005\e{-02}$ & $1.071\e{-02}$\\
			$*$ & $348$ & $0.000\e{+00}$ & $8.188\e{-02}$ & $1.400\e{-02}$ & $7.762\e{-03}$ & $4.749\e{-02}$ & $1.289\e{-02}$ & $8.363\e{-02}$ & $1.612\e{-02}$\\
			$*$ & $354$ & $0.000\e{+00}$ & $8.188\e{-02}$ & $0.000\e{+00}$ & $7.347\e{-02}$ & $0.000\e{+00}$ & $1.458\e{-01}$ & $0.000\e{+00}$ & $2.626\e{-01}$\\
			$ $ & $371$ & $0.000\e{+00}$ & $3.639\e{-02}$ & $0.000\e{+00}$ & $3.265\e{-02}$ & $3.466\e{-02}$ & $7.771\e{-03}$ & $9.459\e{-02}$ & $9.689\e{-03}$\\
			$ $ & $373$ & $3.412\e{-02}$ & $1.150\e{-02}$ & $1.111\e{-01}$ & $1.052\e{-02}$ & $1.761\e{-01}$ & $1.624\e{-02}$ & $1.300\e{-01}$ & $1.967\e{-02}$\\
			$*$ & $389$ & $0.000\e{+00}$ & $3.639\e{-02}$ & $2.609\e{-02}$ & $4.989\e{-03}$ & $7.725\e{-02}$ & $8.178\e{-03}$ & $1.560\e{-01}$ & $9.972\e{-03}$\\
			$*$ & $392$ & $0.000\e{+00}$ & $1.456\e{-01}$ & $0.000\e{+00}$ & $1.306\e{-01}$ & $0.000\e{+00}$ & $2.592\e{-01}$ & $0.000\e{+00}$ & $4.669\e{-01}$\\
			$*$ & $402$ & $0.000\e{+00}$ & $3.639\e{-02}$ & $0.000\e{+00}$ & $3.265\e{-02}$ & $3.582\e{-02}$ & $9.432\e{-03}$ & $5.864\e{-02}$ & $1.138\e{-02}$\\
			$ $ & $403$ & $8.854\e{-03}$ & $5.462\e{-03}$ & $6.888\e{-02}$ & $5.131\e{-03}$ & $3.383\e{-01}$ & $8.065\e{-03}$ & $3.800\e{-01}$ & $9.502\e{-03}$\\
			$ $ & $410$ & $0.000\e{+00}$ & $5.240\e{-02}$ & $4.042\e{-02}$ & $5.939\e{-03}$ & $6.298\e{-02}$ & $9.611\e{-03}$ & $1.278\e{-01}$ & $1.147\e{-02}$\\
			$*$ & $428$ & $2.806\e{-02}$ & $5.212\e{-03}$ & $4.285\e{-02}$ & $5.167\e{-03}$ & $1.195\e{-01}$ & $8.592\e{-03}$ & $2.700\e{-01}$ & $1.009\e{-02}$\\
			$ $ & $430$ & $0.000\e{+00}$ & $8.188\e{-02}$ & $0.000\e{+00}$ & $7.347\e{-02}$ & $0.000\e{+00}$ & $1.458\e{-01}$ & $0.000\e{+00}$ & $2.626\e{-01}$\\
			$*$ & $444$ & $6.266\e{-03}$ & $5.440\e{-03}$ & $5.713\e{-02}$ & $5.092\e{-03}$ & $6.121\e{-02}$ & $8.572\e{-03}$ & $7.174\e{-02}$ & $1.006\e{-02}$\\
			$*$ & $455$ & $6.007\e{-02}$ & $1.079\e{-02}$ & $1.837\e{-01}$ & $1.010\e{-02}$ & $7.563\e{-01}$ & $1.743\e{-02}$ & $1.708\e{+00}$ & $2.037\e{-02}$\\
			$ $ & $456$ & $0.000\e{+00}$ & $3.639\e{-02}$ & $0.000\e{+00}$ & $3.265\e{-02}$ & $0.000\e{+00}$ & $6.480\e{-02}$ & $0.000\e{+00}$ & $1.167\e{-01}$\\
			$*$ & $460$ & $0.000\e{+00}$ & $1.456\e{-01}$ & $0.000\e{+00}$ & $1.306\e{-01}$ & $0.000\e{+00}$ & $2.592\e{-01}$ & $0.000\e{+00}$ & $4.669\e{-01}$\\
			$*$ & $462$ & $0.000\e{+00}$ & $3.639\e{-02}$ & $5.188\e{-02}$ & $5.007\e{-03}$ & $1.798\e{-01}$ & $8.758\e{-03}$ & $3.492\e{-01}$ & $1.025\e{-02}$\\
			$ $ & $466$ & $3.907\e{-02}$ & $7.866\e{-03}$ & $6.747\e{-02}$ & $7.686\e{-03}$ & $7.346\e{-02}$ & $1.288\e{-02}$ & $1.150\e{-01}$ & $1.525\e{-02}$\\
			$*$ & $485$ & $0.000\e{+00}$ & $1.456\e{-01}$ & $4.446\e{-02}$ & $9.870\e{-03}$ & $0.000\e{+00}$ & $2.592\e{-01}$ & $0.000\e{+00}$ & $4.669\e{-01}$\\
			$ $ & $496$ & $1.456\e{-02}$ & $5.231\e{-03}$ & $4.275\e{-02}$ & $5.050\e{-03}$ & $6.628\e{-02}$ & $8.103\e{-03}$ & $9.636\e{-02}$ & $9.832\e{-03}$\\
			$ $ & $522$ & $0.000\e{+00}$ & $3.639\e{-02}$ & $1.176\e{-02}$ & $4.918\e{-03}$ & $4.276\e{-02}$ & $8.589\e{-03}$ & $1.447\e{-01}$ & $1.002\e{-02}$\\
			$ $ & $535$ & $0.000\e{+00}$ & $3.639\e{-02}$ & $6.218\e{-02}$ & $5.222\e{-03}$ & $8.202\e{-02}$ & $8.336\e{-03}$ & $1.057\e{-01}$ & $9.989\e{-03}$\\
			$*$ & $539$ & $0.000\e{+00}$ & $3.639\e{-02}$ & $3.697\e{-02}$ & $4.815\e{-03}$ & $1.318\e{-01}$ & $8.382\e{-03}$ & $1.314\e{-01}$ & $9.711\e{-03}$\\
			$ $ & $546$ & $0.000\e{+00}$ & $3.639\e{-02}$ & $2.176\e{-01}$ & $5.005\e{-03}$ & $4.621\e{-01}$ & $9.036\e{-03}$ & $5.296\e{-01}$ & $1.018\e{-02}$\\
			$ $ & $556$ & $3.312\e{-02}$ & $1.056\e{-02}$ & $1.411\e{-01}$ & $1.296\e{-02}$ & $2.574\e{-01}$ & $2.087\e{-02}$ & $2.370\e{-01}$ & $2.486\e{-02}$\\
			$*$ & $574$ & $0.000\e{+00}$ & $8.188\e{-02}$ & $0.000\e{+00}$ & $7.347\e{-02}$ & $0.000\e{+00}$ & $1.458\e{-01}$ & $4.035\e{-02}$ & $1.533\e{-02}$\\
			$ $ & $578$ & $0.000\e{+00}$ & $1.456\e{-01}$ & $0.000\e{+00}$ & $1.306\e{-01}$ & $0.000\e{+00}$ & $2.592\e{-01}$ & $0.000\e{+00}$ & $4.669\e{-01}$\\
			$ $ & $583$ & $0.000\e{+00}$ & $1.310\e{-02}$ & $0.000\e{+00}$ & $1.176\e{-02}$ & $0.000\e{+00}$ & $2.333\e{-02}$ & $0.000\e{+00}$ & $4.202\e{-02}$\\
			$ $ & $589$ & $0.000\e{+00}$ & $8.188\e{-02}$ & $3.719\e{-02}$ & $7.469\e{-03}$ & $4.779\e{-02}$ & $1.195\e{-02}$ & $1.765\e{-01}$ & $1.483\e{-02}$\\
			$*$ & $591$ & $1.435\e{-02}$ & $5.530\e{-03}$ & $1.389\e{-02}$ & $5.148\e{-03}$ & $7.005\e{-02}$ & $8.526\e{-03}$ & $7.277\e{-02}$ & $1.048\e{-02}$\\
			$ $ & $620$ & $0.000\e{+00}$ & $1.456\e{-01}$ & $6.177\e{-02}$ & $1.171\e{-02}$ & $1.533\e{-01}$ & $1.956\e{-02}$ & $1.470\e{-01}$ & $2.419\e{-02}$\\
			$ $ & $624$ & $0.000\e{+00}$ & $3.639\e{-02}$ & $0.000\e{+00}$ & $3.265\e{-02}$ & $5.278\e{-02}$ & $8.291\e{-03}$ & $1.104\e{-01}$ & $1.006\e{-02}$\\
			$*$ & $625$ & $2.820\e{-02}$ & $7.451\e{-03}$ & $8.857\e{-02}$ & $7.227\e{-03}$ & $1.983\e{-01}$ & $1.082\e{-02}$ & $5.659\e{-01}$ & $1.499\e{-02}$\\
			$*$ & $630$ & $0.000\e{+00}$ & $1.456\e{-01}$ & $0.000\e{+00}$ & $1.306\e{-01}$ & $0.000\e{+00}$ & $2.592\e{-01}$ & $0.000\e{+00}$ & $4.669\e{-01}$\\
			$*$ & $651$ & $0.000\e{+00}$ & $1.456\e{-01}$ & $5.578\e{-02}$ & $1.046\e{-02}$ & $0.000\e{+00}$ & $2.592\e{-01}$ & $0.000\e{+00}$ & $4.669\e{-01}$\\
		\end{tabular}
	\end{center}
\caption{\label{tab:apphot1} Flux densities for our 58 sample sources, part 1/2. All values are given in $\mu$Jy. For sources that are marked with an asterisk we included the \Spitzer\ 3.6 and 4.5 $\mu\mathrm{m}$ flux values in our photometric redshift analysis. The values given for the 3.6 and 4.5 micron channels correspond to the 1$^{\prime\prime}$ aperture radius flux values given in the  \protect\cite{Damen:2011aa} \Spitzer\ catalog. For the remaining sources the \Spitzer\ values could not be included due to source confusion. We performed our own aperture photometry for the GOODS/ACS ($B$, $V$, $i$, $z$) and CANDELS ($Y$, $J$, $H$) filters. If an object is not detected in an image (flux < detection threshold) we set $F_\nu$ = 0 and $Ferr_\nu$ to the sensitivity limit of the corresponding filter. If an object's position is not covered by the $Y$ band or we are not using the \Spitzer\ IRAC data we mark this with a dash. }
\end{table*}

\begin{table*}
\tiny
	\begin{center}
		\begin{tabular}{ll|llllllllll}
		\toprule
		{} & {ID} & {$\mathrm{F}_\nu$ ${(}Y{)}$} & {$\mathrm{Ferr}_\nu$ ${(}Y{)}$} & {$\mathrm{F}_\nu$ ${(}J{)}$} & {$\mathrm{Ferr}_\nu$ ${(}J{)}$} & {$\mathrm{F}_\nu$ ${(}H{)}$} & {$\mathrm{Ferr}_\nu$ ${(}H{)}$} & {$\mathrm{F}_\nu$ ${(}3.6\mu\mathrm{m}{)}$} & {$\mathrm{Ferr}_\nu$ ${(}3.6\mu\mathrm{m}{)}$} & {$\mathrm{F}_\nu$ ${(}4.5\mu\mathrm{m}{)}$} & {$\mathrm{Ferr}_\nu$ ${(}4.5\mu\mathrm{m}{)}$}\\
		\midrule
			$*$ & $121$ & $-$ & $-$ & $3.725\e{+00}$ & $4.184\e{-01}$ & $7.080\e{+00}$ & $6.655\e{-01}$ & $1.903\e{+01}$ & $2.303\e{-02}$ & $1.951\e{+01}$ & $3.215\e{-02}$\\
			$*$ & $150$ & $1.414\e{-01}$ & $1.011\e{-01}$ & $1.316\e{-01}$ & $1.027\e{-01}$ & $2.110\e{-01}$ & $1.419\e{-01}$ & $2.198\e{-01}$ & $1.864\e{-02}$ & $2.503\e{-01}$ & $2.675\e{-02}$\\
			$ $ & $173$ & $1.955\e{+00}$ & $2.968\e{-01}$ & $3.173\e{+00}$ & $3.816\e{-01}$ & $4.904\e{+00}$ & $5.432\e{-01}$ & $-$ & $-$ & $-$ & $-$\\
			$*$ & $184$ & $3.506\e{+00}$ & $3.671\e{-01}$ & $6.444\e{+00}$ & $5.487\e{-01}$ & $9.992\e{+00}$ & $7.844\e{-01}$ & $1.969\e{+01}$ & $2.173\e{-02}$ & $2.149\e{+01}$ & $3.043\e{-02}$\\
			$*$ & $189$ & $2.460\e{-01}$ & $1.037\e{-01}$ & $4.377\e{-01}$ & $1.505\e{-01}$ & $6.700\e{-01}$ & $2.119\e{-01}$ & $2.374\e{+00}$ & $1.810\e{-02}$ & $2.794\e{+00}$ & $2.589\e{-02}$\\
			$*$ & $199$ & $4.411\e{-01}$ & $1.415\e{-01}$ & $1.023\e{+00}$ & $2.168\e{-01}$ & $3.623\e{+00}$ & $4.670\e{-01}$ & $1.101\e{+01}$ & $1.636\e{-02}$ & $1.251\e{+01}$ & $2.382\e{-02}$\\
			$ $ & $211$ & $7.847\e{-01}$ & $1.956\e{-01}$ & $1.483\e{+00}$ & $2.775\e{-01}$ & $2.422\e{+00}$ & $4.072\e{-01}$ & $-$ & $-$ & $-$ & $-$\\
			$*$ & $217$ & $0.000\e{+00}$ & $3.690\e{-02}$ & $0.000\e{+00}$ & $4.437\e{-02}$ & $1.513\e{+00}$ & $3.104\e{-01}$ & $1.105\e{+01}$ & $1.756\e{-02}$ & $9.509\e{+00}$ & $2.579\e{-02}$\\
			$ $ & $221$ & $1.295\e{+00}$ & $2.458\e{-01}$ & $3.136\e{+00}$ & $3.735\e{-01}$ & $5.550\e{+00}$ & $5.733\e{-01}$ & $-$ & $-$ & $-$ & $-$\\
			$ $ & $226$ & $1.617\e{+00}$ & $2.685\e{-01}$ & $2.470\e{+00}$ & $3.400\e{-01}$ & $4.885\e{+00}$ & $5.431\e{-01}$ & $-$ & $-$ & $-$ & $-$\\
			$*$ & $242$ & $9.867\e{-01}$ & $2.119\e{-01}$ & $1.963\e{+00}$ & $3.022\e{-01}$ & $3.120\e{+00}$ & $4.334\e{-01}$ & $5.839\e{+00}$ & $2.194\e{-02}$ & $6.199\e{+00}$ & $3.032\e{-02}$\\
			$*$ & $244$ & $3.466\e{-01}$ & $2.151\e{-01}$ & $7.493\e{-01}$ & $2.027\e{-01}$ & $1.297\e{+00}$ & $3.026\e{-01}$ & $4.891\e{+00}$ & $1.617\e{-02}$ & $6.214\e{+00}$ & $2.290\e{-02}$\\
			$ $ & $258$ & $0.000\e{+00}$ & $5.904\e{-03}$ & $6.340\e{-02}$ & $6.057\e{-02}$ & $1.584\e{-01}$ & $1.061\e{-01}$ & $-$ & $-$ & $-$ & $-$\\
			$*$ & $273$ & $1.239\e{-01}$ & $9.358\e{-02}$ & $3.637\e{-01}$ & $1.348\e{-01}$ & $9.127\e{-01}$ & $2.382\e{-01}$ & $6.730\e{+00}$ & $2.202\e{-02}$ & $9.516\e{+00}$ & $3.018\e{-02}$\\
			$ $ & $296$ & $2.741\e{-01}$ & $1.138\e{-01}$ & $5.309\e{-01}$ & $1.589\e{-01}$ & $9.128\e{-01}$ & $2.402\e{-01}$ & $-$ & $-$ & $-$ & $-$\\
			$ $ & $301$ & $3.404\e{-01}$ & $1.313\e{-01}$ & $5.365\e{-01}$ & $1.617\e{-01}$ & $1.115\e{+00}$ & $2.615\e{-01}$ & $-$ & $-$ & $-$ & $-$\\
			$*$ & $302$ & $1.478\e{-01}$ & $8.496\e{-02}$ & $2.512\e{-01}$ & $1.111\e{-01}$ & $6.038\e{-01}$ & $1.943\e{-01}$ & $1.255\e{+00}$ & $1.634\e{-02}$ & $1.484\e{+00}$ & $2.333\e{-02}$\\
			$ $ & $303$ & $-$ & $-$ & $3.929\e{-01}$ & $1.307\e{-01}$ & $6.912\e{-01}$ & $1.977\e{-01}$ & $-$ & $-$ & $-$ & $-$\\
			$ $ & $306$ & $5.993\e{-01}$ & $1.672\e{-01}$ & $1.153\e{+00}$ & $2.333\e{-01}$ & $2.350\e{+00}$ & $3.808\e{-01}$ & $-$ & $-$ & $-$ & $-$\\
			$*$ & $318$ & $-$ & $-$ & $1.509\e{+00}$ & $3.468\e{-01}$ & $2.733\e{+00}$ & $6.181\e{-01}$ & $7.786\e{+00}$ & $2.103\e{-02}$ & $8.877\e{+00}$ & $2.899\e{-02}$\\
			$*$ & $321$ & $3.081\e{+00}$ & $3.662\e{-01}$ & $4.613\e{+00}$ & $4.736\e{-01}$ & $6.795\e{+00}$ & $6.491\e{-01}$ & $1.877\e{+01}$ & $2.169\e{-02}$ & $1.597\e{+01}$ & $2.999\e{-02}$\\
			$ $ & $325$ & $4.370\e{-01}$ & $1.395\e{-01}$ & $1.071\e{+00}$ & $2.191\e{-01}$ & $1.815\e{+00}$ & $3.293\e{-01}$ & $-$ & $-$ & $-$ & $-$\\
			$ $ & $328$ & $1.902\e{-01}$ & $9.893\e{-02}$ & $1.482\e{-01}$ & $9.912\e{-02}$ & $4.256\e{-01}$ & $1.799\e{-01}$ & $-$ & $-$ & $-$ & $-$\\
			$ $ & $331$ & $1.179\e{-01}$ & $7.724\e{-02}$ & $1.464\e{-01}$ & $8.201\e{-02}$ & $2.191\e{-01}$ & $1.132\e{-01}$ & $-$ & $-$ & $-$ & $-$\\
			$*$ & $348$ & $-$ & $-$ & $3.443\e{-01}$ & $1.377\e{-01}$ & $9.196\e{-01}$ & $2.578\e{-01}$ & $6.216\e{+00}$ & $2.168\e{-02}$ & $8.319\e{+00}$ & $3.125\e{-02}$\\
			$*$ & $354$ & $9.632\e{-02}$ & $7.838\e{-02}$ & $1.485\e{-01}$ & $9.213\e{-02}$ & $5.986\e{-01}$ & $1.910\e{-01}$ & $2.273\e{+00}$ & $2.086\e{-02}$ & $2.676\e{+00}$ & $3.151\e{-02}$\\
			$ $ & $371$ & $9.383\e{-02}$ & $6.683\e{-02}$ & $1.141\e{-01}$ & $8.705\e{-02}$ & $1.917\e{-01}$ & $1.234\e{-01}$ & $-$ & $-$ & $-$ & $-$\\
			$ $ & $373$ & $4.063\e{-01}$ & $1.654\e{-01}$ & $6.511\e{-01}$ & $1.886\e{-01}$ & $1.143\e{+00}$ & $2.762\e{-01}$ & $-$ & $-$ & $-$ & $-$\\
			$*$ & $389$ & $4.020\e{-01}$ & $1.380\e{-01}$ & $9.805\e{-01}$ & $2.127\e{-01}$ & $1.732\e{+00}$ & $3.308\e{-01}$ & $3.481\e{+00}$ & $2.054\e{-02}$ & $3.804\e{+00}$ & $2.885\e{-02}$\\
			$*$ & $392$ & $-$ & $-$ & $0.000\e{+00}$ & $7.887\e{-02}$ & $2.602\e{-01}$ & $1.267\e{-01}$ & $2.264\e{+00}$ & $2.079\e{-02}$ & $3.737\e{+00}$ & $2.948\e{-02}$\\
			$*$ & $402$ & $1.789\e{-01}$ & $9.189\e{-02}$ & $3.553\e{-01}$ & $1.352\e{-01}$ & $6.099\e{-01}$ & $2.008\e{-01}$ & $4.111\e{+00}$ & $2.116\e{-02}$ & $5.718\e{+00}$ & $2.951\e{-02}$\\
			$ $ & $403$ & $4.827\e{-01}$ & $1.603\e{-01}$ & $6.172\e{-01}$ & $1.771\e{-01}$ & $9.149\e{-01}$ & $2.434\e{-01}$ & $-$ & $-$ & $-$ & $-$\\
			$ $ & $410$ & $2.475\e{-01}$ & $1.170\e{-01}$ & $5.759\e{-01}$ & $1.256\e{-01}$ & $1.644\e{+00}$ & $2.453\e{-01}$ & $-$ & $-$ & $-$ & $-$\\
			$*$ & $428$ & $6.167\e{-01}$ & $1.653\e{-01}$ & $1.311\e{+00}$ & $2.508\e{-01}$ & $2.139\e{+00}$ & $3.710\e{-01}$ & $4.847\e{+00}$ & $1.642\e{-02}$ & $5.129\e{+00}$ & $2.409\e{-02}$\\
			$ $ & $430$ & $9.001\e{-02}$ & $7.286\e{-02}$ & $1.240\e{-01}$ & $8.494\e{-02}$ & $2.034\e{-01}$ & $1.143\e{-01}$ & $-$ & $-$ & $-$ & $-$\\
			$*$ & $444$ & $1.442\e{-01}$ & $9.279\e{-02}$ & $2.583\e{-01}$ & $1.093\e{-01}$ & $6.724\e{-01}$ & $2.041\e{-01}$ & $2.279\e{+00}$ & $1.648\e{-02}$ & $2.707\e{+00}$ & $2.408\e{-02}$\\
			$*$ & $455$ & $3.046\e{+00}$ & $3.809\e{-01}$ & $4.355\e{+00}$ & $4.483\e{-01}$ & $6.243\e{+00}$ & $6.130\e{-01}$ & $1.000\e{+01}$ & $1.539\e{-02}$ & $7.745\e{+00}$ & $2.226\e{-02}$\\
			$ $ & $456$ & $-$ & $-$ & $0.000\e{+00}$ & $1.972\e{-02}$ & $1.595\e{-01}$ & $1.042\e{-01}$ & $-$ & $-$ & $-$ & $-$\\
			$*$ & $460$ & $0.000\e{+00}$ & $6.560\e{-02}$ & $4.917\e{-02}$ & $4.868\e{-02}$ & $2.276\e{-01}$ & $1.178\e{-01}$ & $1.943\e{+00}$ & $2.092\e{-02}$ & $3.035\e{+00}$ & $3.044\e{-02}$\\
			$*$ & $462$ & $8.811\e{-01}$ & $1.941\e{-01}$ & $1.758\e{+00}$ & $2.863\e{-01}$ & $2.474\e{+00}$ & $3.865\e{-01}$ & $4.591\e{+00}$ & $1.567\e{-02}$ & $4.518\e{+00}$ & $2.253\e{-02}$\\
			$ $ & $466$ & $2.273\e{-01}$ & $1.109\e{-01}$ & $3.608\e{-01}$ & $1.244\e{-01}$ & $8.445\e{-01}$ & $2.213\e{-01}$ & $-$ & $-$ & $-$ & $-$\\
			$*$ & $485$ & $0.000\e{+00}$ & $6.560\e{-02}$ & $0.000\e{+00}$ & $7.887\e{-02}$ & $1.480\e{-01}$ & $9.532\e{-02}$ & $1.265\e{+00}$ & $2.031\e{-02}$ & $2.035\e{+00}$ & $3.015\e{-02}$\\
			$ $ & $496$ & $1.241\e{-01}$ & $8.545\e{-02}$ & $1.803\e{-01}$ & $9.386\e{-02}$ & $4.076\e{-01}$ & $1.585\e{-01}$ & $-$ & $-$ & $-$ & $-$\\
			$ $ & $522$ & $-$ & $-$ & $5.377\e{-01}$ & $1.595\e{-01}$ & $1.110\e{+00}$ & $2.593\e{-01}$ & $-$ & $-$ & $-$ & $-$\\
			$ $ & $535$ & $1.596\e{-01}$ & $8.098\e{-02}$ & $2.906\e{-01}$ & $1.086\e{-01}$ & $7.578\e{-01}$ & $1.998\e{-01}$ & $-$ & $-$ & $-$ & $-$\\
			$*$ & $539$ & $2.196\e{-01}$ & $1.058\e{-01}$ & $3.786\e{-01}$ & $1.309\e{-01}$ & $1.383\e{+00}$ & $2.879\e{-01}$ & $2.816\e{+00}$ & $1.619\e{-02}$ & $2.987\e{+00}$ & $2.334\e{-02}$\\
			$ $ & $546$ & $6.803\e{-01}$ & $1.789\e{-01}$ & $1.050\e{+00}$ & $2.335\e{-01}$ & $2.721\e{+00}$ & $4.399\e{-01}$ & $-$ & $-$ & $-$ & $-$\\
			$ $ & $556$ & $5.658\e{-01}$ & $1.947\e{-01}$ & $9.578\e{-01}$ & $2.224\e{-01}$ & $1.932\e{+00}$ & $3.544\e{-01}$ & $-$ & $-$ & $-$ & $-$\\
			$*$ & $574$ & $0.000\e{+00}$ & $3.690\e{-02}$ & $1.311\e{-01}$ & $8.630\e{-02}$ & $4.916\e{-01}$ & $1.764\e{-01}$ & $2.928\e{+00}$ & $1.630\e{-02}$ & $3.758\e{+00}$ & $2.329\e{-02}$\\
			$ $ & $578$ & $1.273\e{-01}$ & $9.351\e{-02}$ & $2.637\e{-01}$ & $1.184\e{-01}$ & $4.914\e{-01}$ & $1.703\e{-01}$ & $-$ & $-$ & $-$ & $-$\\
			$ $ & $583$ & $0.000\e{+00}$ & $5.904\e{-03}$ & $1.143\e{-01}$ & $8.153\e{-02}$ & $2.614\e{-01}$ & $1.420\e{-01}$ & $-$ & $-$ & $-$ & $-$\\
			$ $ & $589$ & $3.272\e{-01}$ & $1.224\e{-01}$ & $6.935\e{-01}$ & $1.810\e{-01}$ & $8.212\e{-01}$ & $2.184\e{-01}$ & $-$ & $-$ & $-$ & $-$\\
			$*$ & $591$ & $-$ & $-$ & $3.668\e{-01}$ & $1.347\e{-01}$ & $8.844\e{-01}$ & $2.340\e{-01}$ & $1.916\e{+00}$ & $2.339\e{-02}$ & $1.952\e{+00}$ & $3.256\e{-02}$\\
			$ $ & $620$ & $3.080\e{-01}$ & $1.459\e{-01}$ & $6.087\e{-01}$ & $1.814\e{-01}$ & $1.486\e{+00}$ & $3.091\e{-01}$ & $-$ & $-$ & $-$ & $-$\\
			$ $ & $624$ & $3.430\e{-01}$ & $1.603\e{-01}$ & $6.085\e{-01}$ & $1.904\e{-01}$ & $1.071\e{+00}$ & $2.685\e{-01}$ & $-$ & $-$ & $-$ & $-$\\
			$*$ & $625$ & $-$ & $-$ & $2.954\e{+00}$ & $3.335\e{-01}$ & $5.380\e{+00}$ & $5.413\e{-01}$ & $2.198\e{+01}$ & $2.106\e{-02}$ & $2.566\e{+01}$ & $3.072\e{-02}$\\
			$*$ & $630$ & $-$ & $-$ & $8.553\e{-02}$ & $6.361\e{-02}$ & $2.740\e{-01}$ & $1.292\e{-01}$ & $1.181\e{+00}$ & $2.124\e{-02}$ & $1.529\e{+00}$ & $3.052\e{-02}$\\
			$*$ & $651$ & $0.000\e{+00}$ & $6.560\e{-02}$ & $2.284\e{-01}$ & $1.311\e{-01}$ & $4.536\e{-01}$ & $2.005\e{-01}$ & $2.134\e{+00}$ & $2.122\e{-02}$ & $2.797\e{+00}$ & $3.177\e{-02}$\\
		\end{tabular}
	\end{center}
\caption{\label{tab:apphot2} Flux densities for our 58 sample sources, part 2/2. All values are given in $\mu$Jy. For sources that are marked with an asterisk we included the \Spitzer\ 3.6 and 4.5 $\mu\mathrm{m}$ flux values in our photometric redshift analysis. The values given for the 3.6 and 4.5 micron channels correspond to the 1.5$^{\prime\prime}$ aperture radius flux values given in the  \protect\cite{Damen:2011aa} \Spitzer\ catalog. For the remaining sources the \Spitzer\ values could not be included due to source confusion. We performed our own aperture photometry for the GOODS/ACS ($B$, $V$, $i$, $z$) and CANDELS ($Y$, $J$, $H$) filters. If an object is not detected in an image (flux < detection threshold) we set $\mathrm{F}_\nu$ = 0 and $\mathrm{Ferr}_\nu$ to the sensitivity limit of the corresponding filter. If an object's position is not covered by the $Y$ band or we are not using the \Spitzer\ IRAC data we mark this with a dash. }
\end{table*}

\begin{table*}
\tiny
	\begin{center}
		\begin{tabular}{ll|llllllllll}
		\toprule
		{} & {ID} & {$m_\mathrm{AB}$ ${(}B{)}$} & {$merr_\mathrm{AB}$ ${(}B{)}$ } & {$m_\mathrm{AB}$ ${(}V{)}$} & {$merr_\mathrm{AB}$ ${(}V{)}$} & {$m_\mathrm{AB}$ ${(}i{)}$} & {$merr_\mathrm{AB}$ ${(}i{)}$} & {$m_\mathrm{AB}$ ${(}z{)}$} & {$merr_\mathrm{AB}$ ${(}z{)}$}\\
		\midrule
			$*$ & $121$ & $0.000$ & $25.992$ & $26.375$ & $0.108$ & $25.007$ & $0.049$ & $23.756$ & $0.019$\\
			$*$ & $150$ & $27.655$ & $0.351$ & $26.241$ & $0.093$ & $25.991$ & $0.123$ & $25.843$ & $0.127$\\
			$ $ & $173$ & $28.304$ & $0.322$ & $26.634$ & $0.067$ & $24.898$ & $0.024$ & $24.062$ & $0.013$\\
			$*$ & $184$ & $27.860$ & $0.343$ & $25.825$ & $0.050$ & $24.447$ & $0.023$ & $23.513$ & $0.012$\\
			$*$ & $189$ & $0.000$ & $27.498$ & $27.952$ & $0.229$ & $27.299$ & $0.199$ & $27.158$ & $0.210$\\
			$*$ & $199$ & $0.000$ & $25.992$ & $27.061$ & $0.197$ & $26.561$ & $0.204$ & $26.191$ & $0.175$\\
			$ $ & $211$ & $27.908$ & $0.233$ & $27.312$ & $0.122$ & $25.955$ & $0.059$ & $25.059$ & $0.031$\\
			$*$ & $217$ & $0.000$ & $26.617$ & $0.000$ & $26.735$ & $0.000$ & $25.991$ & $0.000$ & $25.352$\\
			$ $ & $221$ & $0.000$ & $26.617$ & $26.246$ & $0.074$ & $25.652$ & $0.069$ & $24.514$ & $0.030$\\
			$ $ & $226$ & $0.000$ & $25.992$ & $0.000$ & $26.110$ & $25.468$ & $0.063$ & $24.048$ & $0.022$\\
			$*$ & $242$ & $28.288$ & $0.483$ & $26.414$ & $0.084$ & $25.415$ & $0.055$ & $24.686$ & $0.035$\\
			$*$ & $244$ & $0.000$ & $25.992$ & $0.000$ & $26.110$ & $28.006$ & $0.801$ & $26.008$ & $0.154$\\
			$ $ & $258$ & $0.000$ & $28.607$ & $0.000$ & $28.724$ & $28.575$ & $0.403$ & $27.796$ & $0.249$\\
			$*$ & $273$ & $0.000$ & $25.992$ & $0.000$ & $26.110$ & $0.000$ & $25.366$ & $26.869$ & $0.330$\\
			$ $ & $296$ & $0.000$ & $27.498$ & $27.891$ & $0.209$ & $28.242$ & $0.445$ & $26.113$ & $0.078$\\
			$ $ & $301$ & $27.369$ & $0.294$ & $26.641$ & $0.146$ & $26.427$ & $0.198$ & $26.012$ & $0.161$\\
			$*$ & $302$ & $0.000$ & $27.498$ & $27.380$ & $0.135$ & $27.496$ & $0.240$ & $26.498$ & $0.118$\\
			$ $ & $303$ & $0.000$ & $25.992$ & $27.735$ & $0.397$ & $0.000$ & $25.366$ & $25.674$ & $0.119$\\
			$ $ & $306$ & $0.000$ & $26.617$ & $27.066$ & $0.152$ & $26.741$ & $0.178$ & $26.440$ & $0.167$\\
			$*$ & $318$ & $0.000$ & $25.992$ & $26.641$ & $0.137$ & $25.416$ & $0.071$ & $25.235$ & $0.073$\\
			$*$ & $321$ & $0.000$ & $25.992$ & $26.283$ & $0.093$ & $24.184$ & $0.023$ & $23.325$ & $0.013$\\
			$ $ & $325$ & $29.175$ & $0.698$ & $26.956$ & $0.093$ & $26.144$ & $0.072$ & $25.542$ & $0.049$\\
			$ $ & $328$ & $27.389$ & $0.263$ & $27.163$ & $0.209$ & $26.356$ & $0.159$ & $26.530$ & $0.226$\\
			$ $ & $331$ & $0.000$ & $27.498$ & $27.373$ & $0.138$ & $26.415$ & $0.096$ & $26.642$ & $0.145$\\
			$*$ & $348$ & $0.000$ & $26.617$ & $28.535$ & $0.602$ & $27.208$ & $0.295$ & $26.594$ & $0.209$\\
			$*$ & $354$ & $0.000$ & $26.617$ & $0.000$ & $26.735$ & $0.000$ & $25.991$ & $0.000$ & $25.352$\\
			$ $ & $371$ & $0.000$ & $27.498$ & $0.000$ & $27.615$ & $27.550$ & $0.243$ & $26.460$ & $0.111$\\
			$ $ & $373$ & $27.567$ & $0.366$ & $26.286$ & $0.103$ & $25.786$ & $0.100$ & $26.115$ & $0.164$\\
			$*$ & $389$ & $0.000$ & $27.498$ & $27.859$ & $0.208$ & $26.680$ & $0.115$ & $25.917$ & $0.069$\\
			$*$ & $392$ & $0.000$ & $25.992$ & $0.000$ & $26.110$ & $0.000$ & $25.366$ & $0.000$ & $24.727$\\
			$*$ & $402$ & $0.000$ & $27.498$ & $0.000$ & $27.615$ & $27.515$ & $0.286$ & $26.980$ & $0.211$\\
			$ $ & $403$ & $29.032$ & $0.670$ & $26.805$ & $0.081$ & $25.077$ & $0.026$ & $24.951$ & $0.027$\\
			$ $ & $410$ & $0.000$ & $27.102$ & $27.384$ & $0.160$ & $26.902$ & $0.166$ & $26.134$ & $0.097$\\
			$*$ & $428$ & $27.780$ & $0.202$ & $27.320$ & $0.131$ & $26.207$ & $0.078$ & $25.322$ & $0.041$\\
			$ $ & $430$ & $0.000$ & $26.617$ & $0.000$ & $26.735$ & $0.000$ & $25.991$ & $0.000$ & $25.352$\\
			$*$ & $444$ & $29.408$ & $0.943$ & $27.008$ & $0.097$ & $26.933$ & $0.152$ & $26.761$ & $0.152$\\
			$*$ & $455$ & $26.953$ & $0.195$ & $25.740$ & $0.060$ & $24.203$ & $0.025$ & $23.319$ & $0.013$\\
			$ $ & $456$ & $0.000$ & $27.498$ & $0.000$ & $27.615$ & $0.000$ & $26.871$ & $0.000$ & $26.232$\\
			$*$ & $460$ & $0.000$ & $25.992$ & $0.000$ & $26.110$ & $0.000$ & $25.366$ & $0.000$ & $24.727$\\
			$*$ & $462$ & $0.000$ & $27.498$ & $27.113$ & $0.105$ & $25.763$ & $0.053$ & $25.042$ & $0.032$\\
			$ $ & $466$ & $27.420$ & $0.219$ & $26.827$ & $0.124$ & $26.735$ & $0.190$ & $26.248$ & $0.144$\\
			$*$ & $485$ & $0.000$ & $25.992$ & $27.280$ & $0.241$ & $0.000$ & $25.366$ & $0.000$ & $24.727$\\
			$ $ & $496$ & $28.492$ & $0.390$ & $27.323$ & $0.128$ & $26.847$ & $0.133$ & $26.440$ & $0.111$\\
			$ $ & $522$ & $0.000$ & $27.498$ & $28.724$ & $0.454$ & $27.322$ & $0.218$ & $25.999$ & $0.075$\\
			$ $ & $535$ & $0.000$ & $27.498$ & $26.916$ & $0.091$ & $26.615$ & $0.110$ & $26.340$ & $0.103$\\
			$*$ & $539$ & $0.000$ & $27.498$ & $27.480$ & $0.141$ & $26.100$ & $0.069$ & $26.104$ & $0.080$\\
			$ $ & $546$ & $0.000$ & $27.498$ & $25.556$ & $0.025$ & $24.738$ & $0.021$ & $24.590$ & $0.021$\\
			$ $ & $556$ & $27.600$ & $0.346$ & $26.026$ & $0.100$ & $25.373$ & $0.088$ & $25.463$ & $0.114$\\
			$*$ & $574$ & $0.000$ & $26.617$ & $0.000$ & $26.735$ & $0.000$ & $25.991$ & $27.385$ & $0.412$\\
			$ $ & $578$ & $0.000$ & $25.992$ & $0.000$ & $26.110$ & $0.000$ & $25.366$ & $0.000$ & $24.727$\\
			$ $ & $583$ & $0.000$ & $28.607$ & $0.000$ & $28.724$ & $0.000$ & $27.980$ & $0.000$ & $27.341$\\
			$ $ & $589$ & $0.000$ & $26.617$ & $27.474$ & $0.218$ & $27.202$ & $0.271$ & $25.783$ & $0.091$\\
			$*$ & $591$ & $28.508$ & $0.418$ & $28.543$ & $0.402$ & $26.786$ & $0.132$ & $26.745$ & $0.156$\\
			$ $ & $620$ & $0.000$ & $25.992$ & $26.923$ & $0.206$ & $25.936$ & $0.139$ & $25.982$ & $0.179$\\
			$ $ & $624$ & $0.000$ & $27.498$ & $0.000$ & $27.615$ & $27.094$ & $0.171$ & $26.293$ & $0.099$\\
			$*$ & $625$ & $27.774$ & $0.287$ & $26.532$ & $0.089$ & $25.657$ & $0.059$ & $24.518$ & $0.029$\\
			$*$ & $630$ & $0.000$ & $25.992$ & $0.000$ & $26.110$ & $0.000$ & $25.366$ & $0.000$ & $24.727$\\
			$*$ & $651$ & $0.000$ & $25.992$ & $27.034$ & $0.204$ & $0.000$ & $25.366$ & $0.000$ & $24.727$\\
		\end{tabular}
	\end{center}
\caption{\label{tab:apphotmag1} Magnitude values for our 58 sample sources, part 1/2. All values are given in AB magnitudes. For sources that are marked with an asterisk we included the \Spitzer\ 3.6 and 4.5 $\mu\mathrm{m}$ flux values in our photometric redshift analysis. The values given for the 3.6 and 4.5 micron channels correspond to the 1.5$^{\prime\prime}$ aperture radius flux values given in the  \protect\cite{Damen:2011aa} \Spitzer\ catalog. For the remaining sources the \Spitzer\ values could not be included due to source confusion. We performed our own aperture photometry for the GOODS/ACS ($B$, $V$, $i$, $z$) and CANDELS ($Y$, $J$, $H$) filters. If an object is not detected in an image (flux < detection threshold) we set $m_\mathrm{AB}$ = 0 and $merr_\mathrm{AB}$ to the sensitivity limit of the corresponding filter. If an object's position is not covered by the $Y$ band or we are not using the \Spitzer\ IRAC data we mark this with a dash. }
\end{table*}

\begin{table*}
\tiny
	\begin{center}
		\begin{tabular}{ll|llllllllll}
		\toprule
		{} & {ID} & {$m_\mathrm{AB}$ ${(}Y{)}$} & {$merr_\mathrm{AB}$ ${(}Y{)}$} & {$m_\mathrm{AB}$ ${(}J{)}$} & {$merr_\mathrm{AB}$ ${(}J{)}$} & {$m_\mathrm{AB}$ ${(}H{)}$} & {$merr_\mathrm{AB}$ ${(}H{)}$} & {$m_\mathrm{AB}$ ${(}3.6\mu\mathrm{m}{)}$} & {$merr_\mathrm{AB}$ ${(}3.6\mu\mathrm{m}{)}$} & {$m_\mathrm{AB}$ ${(}4.5\mu\mathrm{m}{)}$} & {$merr_\mathrm{AB}$ ${(}4.5\mu\mathrm{m}{)}$}\\
		\midrule
			$*$ & $121$ & $-$ & $-$ & $22.472$ & $0.122$ & $21.775$ & $0.102$ & $20.701$ & $0.001$ & $20.674$ & $0.002$\\
			$*$ & $150$ & $26.024$ & $0.776$ & $26.102$ & $0.847$ & $25.589$ & $0.730$ & $25.545$ & $0.092$ & $25.404$ & $0.116$\\
			$ $ & $173$ & $23.172$ & $0.165$ & $22.646$ & $0.131$ & $22.174$ & $0.120$ & $-$ & $-$ & $-$ & $-$\\
			$*$ & $184$ & $22.538$ & $0.114$ & $21.877$ & $0.092$ & $21.401$ & $0.085$ & $20.664$ & $0.001$ & $20.569$ & $0.002$\\
			$*$ & $189$ & $25.423$ & $0.458$ & $24.797$ & $0.373$ & $24.335$ & $0.343$ & $22.961$ & $0.008$ & $22.784$ & $0.010$\\
			$*$ & $199$ & $24.789$ & $0.348$ & $23.875$ & $0.230$ & $22.502$ & $0.140$ & $21.296$ & $0.002$ & $21.157$ & $0.002$\\
			$ $ & $211$ & $24.163$ & $0.271$ & $23.472$ & $0.203$ & $22.940$ & $0.183$ & $-$ & $-$ & $-$ & $-$\\
			$*$ & $217$ & $0.000$ & $27.482$ & $0.000$ & $27.282$ & $23.450$ & $0.223$ & $21.292$ & $0.002$ & $21.455$ & $0.003$\\
			$ $ & $221$ & $23.619$ & $0.206$ & $22.659$ & $0.129$ & $22.039$ & $0.112$ & $-$ & $-$ & $-$ & $-$\\
			$ $ & $226$ & $23.378$ & $0.180$ & $22.918$ & $0.149$ & $22.178$ & $0.121$ & $-$ & $-$ & $-$ & $-$\\
			$*$ & $242$ & $23.915$ & $0.233$ & $23.168$ & $0.167$ & $22.665$ & $0.151$ & $21.984$ & $0.004$ & $21.919$ & $0.005$\\
			$*$ & $244$ & $25.050$ & $0.674$ & $24.213$ & $0.294$ & $23.618$ & $0.253$ & $22.177$ & $0.004$ & $21.917$ & $0.004$\\
			$ $ & $258$ & $0.000$ & $29.472$ & $26.895$ & $1.037$ & $25.901$ & $0.727$ & $-$ & $-$ & $-$ & $-$\\
			$*$ & $273$ & $26.167$ & $0.820$ & $24.998$ & $0.402$ & $23.999$ & $0.283$ & $21.830$ & $0.004$ & $21.454$ & $0.003$\\
			$ $ & $296$ & $25.305$ & $0.451$ & $24.587$ & $0.325$ & $23.999$ & $0.286$ & $-$ & $-$ & $-$ & $-$\\
			$ $ & $301$ & $25.070$ & $0.419$ & $24.576$ & $0.327$ & $23.782$ & $0.255$ & $-$ & $-$ & $-$ & $-$\\
			$*$ & $302$ & $25.976$ & $0.624$ & $25.400$ & $0.480$ & $24.448$ & $0.349$ & $23.653$ & $0.014$ & $23.471$ & $0.017$\\
			$ $ & $303$ & $-$ & $-$ & $24.914$ & $0.361$ & $24.301$ & $0.311$ & $-$ & $-$ & $-$ & $-$\\
			$ $ & $306$ & $24.456$ & $0.303$ & $23.745$ & $0.220$ & $22.972$ & $0.176$ & $-$ & $-$ & $-$ & $-$\\
			$*$ & $318$ & $-$ & $-$ & $23.453$ & $0.250$ & $22.808$ & $0.246$ & $21.672$ & $0.003$ & $21.529$ & $0.004$\\
			$*$ & $321$ & $22.678$ & $0.129$ & $22.240$ & $0.111$ & $21.820$ & $0.104$ & $20.716$ & $0.001$ & $20.892$ & $0.002$\\
			$ $ & $325$ & $24.799$ & $0.347$ & $23.826$ & $0.222$ & $23.253$ & $0.197$ & $-$ & $-$ & $-$ & $-$\\
			$ $ & $328$ & $25.702$ & $0.565$ & $25.973$ & $0.726$ & $24.827$ & $0.459$ & $-$ & $-$ & $-$ & $-$\\
			$ $ & $331$ & $26.221$ & $0.711$ & $25.986$ & $0.608$ & $25.548$ & $0.561$ & $-$ & $-$ & $-$ & $-$\\
			$*$ & $348$ & $-$ & $-$ & $25.058$ & $0.434$ & $23.991$ & $0.304$ & $21.916$ & $0.004$ & $21.600$ & $0.004$\\
			$*$ & $354$ & $26.441$ & $0.884$ & $25.971$ & $0.674$ & $24.457$ & $0.346$ & $23.009$ & $0.010$ & $22.831$ & $0.013$\\
			$ $ & $371$ & $26.469$ & $0.773$ & $26.257$ & $0.828$ & $25.693$ & $0.699$ & $-$ & $-$ & $-$ & $-$\\
			$ $ & $373$ & $24.878$ & $0.442$ & $24.366$ & $0.314$ & $23.755$ & $0.262$ & $-$ & $-$ & $-$ & $-$\\
			$*$ & $389$ & $24.889$ & $0.373$ & $23.921$ & $0.236$ & $23.304$ & $0.207$ & $22.546$ & $0.006$ & $22.449$ & $0.008$\\
			$*$ & $392$ & $-$ & $-$ & $0.000$ & $26.658$ & $25.362$ & $0.529$ & $23.013$ & $0.010$ & $22.469$ & $0.009$\\
			$*$ & $402$ & $25.768$ & $0.558$ & $25.024$ & $0.413$ & $24.437$ & $0.357$ & $22.365$ & $0.006$ & $22.007$ & $0.006$\\
			$ $ & $403$ & $24.691$ & $0.361$ & $24.424$ & $0.312$ & $23.997$ & $0.289$ & $-$ & $-$ & $-$ & $-$\\
			$ $ & $410$ & $25.416$ & $0.513$ & $24.499$ & $0.237$ & $23.360$ & $0.162$ & $-$ & $-$ & $-$ & $-$\\
			$*$ & $428$ & $24.425$ & $0.291$ & $23.606$ & $0.208$ & $23.074$ & $0.188$ & $22.186$ & $0.004$ & $22.125$ & $0.005$\\
			$ $ & $430$ & $26.514$ & $0.879$ & $26.166$ & $0.744$ & $25.629$ & $0.610$ & $-$ & $-$ & $-$ & $-$\\
			$*$ & $444$ & $26.003$ & $0.699$ & $25.370$ & $0.459$ & $24.331$ & $0.330$ & $23.006$ & $0.008$ & $22.819$ & $0.010$\\
			$*$ & $455$ & $22.691$ & $0.136$ & $22.303$ & $0.112$ & $21.912$ & $0.107$ & $21.400$ & $0.002$ & $21.677$ & $0.003$\\
			$ $ & $456$ & $-$ & $-$ & $0.000$ & $28.163$ & $25.893$ & $0.709$ & $-$ & $-$ & $-$ & $-$\\
			$*$ & $460$ & $0.000$ & $26.858$ & $27.171$ & $1.075$ & $25.507$ & $0.562$ & $23.179$ & $0.012$ & $22.695$ & $0.011$\\
			$*$ & $462$ & $24.037$ & $0.239$ & $23.287$ & $0.177$ & $22.917$ & $0.170$ & $22.245$ & $0.004$ & $22.263$ & $0.005$\\
			$ $ & $466$ & $25.509$ & $0.530$ & $25.007$ & $0.374$ & $24.084$ & $0.285$ & $-$ & $-$ & $-$ & $-$\\
			$*$ & $485$ & $0.000$ & $26.858$ & $0.000$ & $26.658$ & $25.974$ & $0.699$ & $23.645$ & $0.017$ & $23.129$ & $0.016$\\
			$ $ & $496$ & $26.166$ & $0.748$ & $25.760$ & $0.565$ & $24.874$ & $0.422$ & $-$ & $-$ & $-$ & $-$\\
			$ $ & $522$ & $-$ & $-$ & $24.574$ & $0.322$ & $23.787$ & $0.254$ & $-$ & $-$ & $-$ & $-$\\
			$ $ & $535$ & $25.892$ & $0.551$ & $25.242$ & $0.406$ & $24.201$ & $0.286$ & $-$ & $-$ & $-$ & $-$\\
			$*$ & $539$ & $25.546$ & $0.523$ & $24.955$ & $0.375$ & $23.548$ & $0.226$ & $22.776$ & $0.006$ & $22.712$ & $0.008$\\
			$ $ & $546$ & $24.318$ & $0.286$ & $23.847$ & $0.241$ & $22.813$ & $0.176$ & $-$ & $-$ & $-$ & $-$\\
			$ $ & $556$ & $24.518$ & $0.374$ & $23.947$ & $0.252$ & $23.185$ & $0.199$ & $-$ & $-$ & $-$ & $-$\\
			$*$ & $574$ & $0.000$ & $27.482$ & $26.106$ & $0.715$ & $24.671$ & $0.390$ & $22.734$ & $0.006$ & $22.463$ & $0.007$\\
			$ $ & $578$ & $26.138$ & $0.798$ & $25.347$ & $0.487$ & $24.671$ & $0.376$ & $-$ & $-$ & $-$ & $-$\\
			$ $ & $583$ & $0.000$ & $29.472$ & $26.255$ & $0.774$ & $25.357$ & $0.590$ & $-$ & $-$ & $-$ & $-$\\
			$ $ & $589$ & $25.113$ & $0.406$ & $24.297$ & $0.283$ & $24.114$ & $0.289$ & $-$ & $-$ & $-$ & $-$\\
			$*$ & $591$ & $-$ & $-$ & $24.989$ & $0.399$ & $24.033$ & $0.287$ & $23.194$ & $0.013$ & $23.174$ & $0.018$\\
			$ $ & $620$ & $25.179$ & $0.514$ & $24.439$ & $0.324$ & $23.470$ & $0.226$ & $-$ & $-$ & $-$ & $-$\\
			$ $ & $624$ & $25.062$ & $0.507$ & $24.439$ & $0.340$ & $23.826$ & $0.272$ & $-$ & $-$ & $-$ & $-$\\
			$*$ & $625$ & $-$ & $-$ & $22.724$ & $0.123$ & $22.073$ & $0.109$ & $20.545$ & $0.001$ & $20.377$ & $0.001$\\
			$*$ & $630$ & $-$ & $-$ & $26.570$ & $0.807$ & $25.306$ & $0.512$ & $23.719$ & $0.020$ & $23.439$ & $0.022$\\
			$*$ & $651$ & $0.000$ & $26.858$ & $25.503$ & $0.623$ & $24.758$ & $0.480$ & $23.077$ & $0.011$ & $22.783$ & $0.012$\\		
			\end{tabular}
	\end{center}
\caption{\label{tab:apphotmag2} Magnitude values for our 58 sample sources, part 2/2. All values are given in AB magnitudes. For sources that are marked with an asterisk we included the \Spitzer\ 3.6 and 4.5 $\mu\mathrm{m}$ flux values in our photometric redshift analysis. The values given for the 3.6 and 4.5 micron channels correspond to the 1.5$^{\prime\prime}$ aperture radius flux values given in the  \protect\cite{Damen:2011aa} \Spitzer\ catalog. For the remaining sources the \Spitzer\ values could not be included due to source confusion. We performed our own aperture photometry for the GOODS/ACS ($B$, $V$, $i$, $z$) and CANDELS ($Y$, $J$, $H$) filters. If an object is not detected in an image (flux < detection threshold) we set $m_\mathrm{AB}$ = 0 and $merr_\mathrm{AB}$ to the sensitivity limit of the corresponding filter. If an object's position is not covered by the $Y$ band or we are not using the \Spitzer\ IRAC data we mark this with a dash. }
\end{table*}

\end{landscape}
\end{appendix}

\label{lastpage}
\end{document}